\documentclass[10pt]{article}
\usepackage{amsfonts}
\usepackage{amsthm}
\usepackage{amscd}
\usepackage{amsmath}

 \usepackage{verbatim}
\usepackage{enumerate}
\usepackage{enumitem}

\usepackage{natbib}

\usepackage{amssymb}

\usepackage{accents}

\usepackage{color}
\usepackage{graphicx}
\usepackage{graphics}
 \usepackage{url}

\usepackage{verbatim}
\usepackage{float}
\usepackage{algorithm}
\usepackage{algorithmic}
\usepackage{lscape}
\usepackage{array,arydshln}

\usepackage{geometry}

 \usepackage{xcolor}

\usepackage{hyperref} %PARA QUE AL PINCHAR VAYAS A LA ECUACION

\hypersetup{
    colorlinks,
    linkcolor= {blue!90!black}, %{red!80!black},
    citecolor={blue!60!black},
    urlcolor={blue!80!black}
}

\usepackage{color}

\usepackage{xr}

\newcommand\bb {\mathbf b}
\newcommand\bc {\mathbf c}
\newcommand\bd {\mathbf d}

\newcommand\bk {\mathbf k}

\newcommand\bx {\mathbf x}

\newcommand\bz {\mathbf z}

\newcommand\bV {\mathbf V}
\newcommand\bW {\mathbf W}
\newcommand\bX {\mathbf X}

\newcommand\bZ {\mathbf Z}

\newcommand\wbc {\widehat{\bc}}

\newcommand\wbk {\widehat{\bk}}

\newcommand\wtbc {\widetilde{\bc}}

\newcommand\itG {{\mathcal{G}}}

\newcommand\itJ {{\mathcal{J}}}

\newcommand\itU {{\mathcal{U}}}

\newcommand{\eps}{\varepsilon}

\newcommand\bbe {\mbox{\boldmath $\beta$}}

\newcommand\bla {\mbox{\boldmath $\lambda$}}
\newcommand\wtbla {\widetilde{\bla}}
\newcommand\blach {\mbox{\footnotesize$\bla$}}
\newcommand\wtblach {\mbox{\footnotesize$\wtbla$}}

\newcommand\bthe {\mbox{\boldmath $\theta$}}

\newcommand\bSi {\mbox{\boldmath $\Sigma$}}

\newcommand\wbeta {\widehat{\beta}}
\newcommand\wbbe {\widehat{\bbe}}

\newcommand\weta {\widehat{\eta}}

\newcommand\wmu {\widehat{\mu}}

\newcommand\wsigma {\widehat{\sigma}}

\newcommand\wtbeta {\widetilde{\beta}}

\newcommand\wtlam {\widetilde{\lambda}}

\def\real{\mathbb{R}}

\newcommand{\esp}{\mathbb{E}}

\newcommand{\trasp}{^{\mbox{\footnotesize \sc t}}}

\def\argmin{\mathop{\mbox{argmin}}}

\newcommand\noi{\noindent}

\def\square{\ifmmode\sqr\else{$\sqr$}\fi}
\def\sqr{\vcenter{
         \hrule height.1mm
         \hbox{\vrule width.1mm height2.2mm\kern2.18mm
\vrule width.1mm}
         \hrule height.1mm}}

\begin{document}
  
%\title{Simultaneous robust estimation and adaptive variable selection for partially linear additive models}
\title{A robust estimation and variable selection approach for sparse partially linear additive models}
\author{Alejandra Mercedes Mart\'{\i}nez\\ 
\small CONICET and Universidad Nacional de Luj\'an,   Argentina} 
  
\date{}
\maketitle
 
%%%%%%%%%%%%%%%%%% ABSTRACT%%%%%%%%%%%%%%%%%%%%%%%%%%%%%%%%
\begin{abstract}
In partially linear additive models the response variable is modelled with a linear component on a subset of covariates and an additive component in which the rest of the covariates enter to the model as a sum of univariate unknown functions. This structure is more flexible than the usual full linear or full nonparametric regression models, avoids the \lq\lq curse of dimensionality\rq\rq, is easily interpretable and
allows the user to include discrete or categorical variables in the linear part. On the other hand, in practice, the user incorporates all the available variables in the model no matter how they would impact on the response variable. For this reason, variable selection plays an important role since including covariates that has a null impact on the responses will reduce the prediction capability of the model. As in other settings, outliers in the data may harm estimations based on strong assumptions, such as normality of the response variable, leading to conclusions that are not representative of the data set. 

In this work, we propose a family of robust estimators that estimate and select variables from both the linear and the additive part of the model simultaneously. This family considers an adaptive procedure 
on a general class of penalties
%based on the SCAD penalty 
in the regularization part of the objetive function that defines the estimators. We study the behaviour of the proposal againts its least-squares counterpart under simulations and show the advantages of its use on a real data set.

\end{abstract} 

\vskip0.2in

\section{Introduction}{\label{intro}}

%\vspace{0.7cm}
Partially linear additive models (\textsc{plam}) assume that $(Y_i,\bZ_i\trasp,\bX_i\trasp)\trasp\in\real^{1+q+p}$, $1\leq i\leq n$, are independent and identically distributed random vectors with the same distribution as $(Y,\bZ\trasp,\bX\trasp)\trasp$ such that
\begin{equation}\label{plam}
Y=m(\bZ\trasp,\bX\trasp)+u=\mu+\bbe\trasp\bZ+\sum_{j=1}^p \eta_j(X_j)+\sigma\varepsilon
\end{equation}
where the constant $\mu\in\real$, vector $\bbe\in\real^q$, the univariate functions $\eta_j:\real\to\real$, $1\leq j\leq p$, and the scale parameter $\sigma>0$ are the objects to be estimated. In order to ensure identifiability of the additive components $\eta_j$, it is assumed that $\int\, \eta_j(x)\,dx=0$ and that $\eta_j:\mathcal{I}_j\to\real$, then $\int_{\mathcal{I}}\, \eta_j(x)\,dx=0$, for $1\leq j\leq p$. The errors $\varepsilon$ are assumed to be independent from the vector of covariates $(\bZ\trasp,\bX\trasp)\trasp$.  In the classical context, the errors $\varepsilon$ are considered to have $0$ mean and variance equals $1$, while in the context of robustness they are considered to have a symmetric distribution $F(\cdot)$ (that is, if they have a density, this density function has to be an even one) with scale parameter equals $1$.

In \citet{boente:martinez:2023} a robust approach to estimate all the parameters in the model is developed. A $B-$spline basis is used to approximate each additive function $\eta_j$ in (\ref{plam}), transforming the original model into a linear regression model. A robust regression approach is then applied to obtain the final estimators.
 In addition to having developed asymptotic theory for the proposal, a generalization of a BIC criteria was defined in order to robustly select the regularization parameters in the approximation procedure. However, this approach has its limitation. When many covariates are included, particularly in the additive part of the model, the estimation procedure becomes impractical, as it is either computationally very expensive or fails to produce unique solutions.

%En colaboraci\'on con la Dra. Graciela Boente de UBA y CONICET, bajo condiciones de regularidad, se obtuvo la consistencia fuerte uniforme de los estimadores propuestos, tasas de convergencia y la distribuci\'on asint\'otica de los estimadores del par\'ametro de regresi\'on lineal $\bbe$. Adem\'as, se prob\'o la consistencia fuerte del estimador preliminar $\wsigma$ de la escala de los residuos, que era una de las hip\'otesis necesarias para obtener dichos resultados. Finalmente, se propusieron diferentes estrategias para estimar la matriz de covarianzas asint\'otica de forma robusta.

Usually, in practice and in a first step of modelling, researchers introduce all possible variables into the model based on their own experience. In this sense, some variables that have a small impact on the response variable will reduce the prediction capability of the model. For this reason, under a sparse model, variable selection plays an important role. 
Even though there exists different methods for selecting variables such as \textit{stepwise} or \textit{best subset}, these used to be computational expensive,  their theoretical properties are hard to establish and, the most severe which is the lack of stability analyzed, for instance, by \citep{breiman:1996}. An alternative is to include a penalty term in the optimization problem that defines the estimators \citep[see][for an overview of penalization methods]{efron:hastie:2016}. 
An example of these is the least absolute shrinkage and selection operator (LASSO) estimator, related to the $\ell_1$ penalty, for linear regression models introduced by
 \citet{tibshirani:1996}. However, it has the disadvantages of selecting too many covariates and giving rise to biased solutions. To avoid the selection problem, \citet{zou:hastie:2005} considered a combination of the $L_1$ and $L_2$ penalties, named \textit{Elastic Net}, while \citet{fan:li:2001} introduced the SCAD penalty (see Section \ref{sec:penalty} for more details). Other penalties are the adaptive  LASSO, also called ADALASSO \citep{zou:2006}, and the Minimax Concave Penalty (MCP) introduced by \citet{zhang:2010}.

Similar to what happens with estimation procedures, simultaneous variable estimation and selection methods can be seriously affected if no robust procedures are used and there are even a few atypical observations presence in the data.
In order to solve this problem, \citet{smucler:yohai:2017} considered sparse robust estimators for linear regression models. In generalized linear models, \citet{Avella:ronchetti:2018} gave a robust proposal based on controlling the quasilikelihood for the variable selection procedure, and in \citet{Bianco:Boente:Chebi:2022} considered penalized robust estimators based on the deviance for the logistic model. On the other hand, \citet{Avella:2016} consideres robust approaches for generalized additive models. 

As we have already mentioned, many researchers have studied the problem of estimating and selecting variables in linear and also in partly linear models by considering that the regression parameter vector is sparse \citep[see, for instance,][for a review]{desboulets:2018}.
Specifically with respect to the simultaneous estimation and variable selection in \textsc{plam}, sparse models for the linear component were considered in \citet{liu:etal:2011}, \citet{du:etal:2012} and \citet{lian:2012} who developed a variable selection procedure based on the least squares regression estimator, spline approximation and penalty functions SCAD or ADALASSO for the regression parameters $\bbe$. More resistant approaches to atypical data have been studied in \citet{Koenker:2011}, \citet{Guo:etal:2013} and \citet{sherwood:wang:2016} who proposed methods based on quantile regression.  However, as it is mentioned in \citet{boente:martinez:2023}, the quantile estimators are related to an unbounded loss functions and, for this reason and as it happens in linear regression, it can be affected by high leverage points.

However, as far as we know, not many have studied the problem of estimating and simultaneously selecting variables from both the linear and the additive components of a \textsc{plam}. Among them, we can mention \citet{kasemi:etal:2018} who proposed a two-step estimation and variable selection method in a high dimensional setting, used the SCAD penalty for regularizing and considered an optimization procedure that minimizes the sum the squared residuals. Besides, they proposed a BIC-type criteria for selecting the penalty parameters and a \textit{2-fold cross-validation} mechanism for determining the tuning parameters of the $B-$splines approximation. \citet{hu:etal:2015} considered a hierarchical Bayesian estimation and model selection approach with quantile regression, while 
\citet{Banerjee:Ghosal:2014} proposed a Bayesian method for generalized partially linear additive models that uses group LASSO \citep[see][]{yuan:lin:2006}. When the response variable is a time series, \citet{feng:etal:2018} studied an estimation and variable selection procedure using the adaptive group LASSO. In all these papers, the additive functions have been approximated by splines, except in \citet{feng:etal:2018} where they used orthogonal series. However, all these approaches are not resistant to atypical data.
A more robust approach has been given in \citet{lv:yang:guo:2017} by combining modal estimators based on $B-$splines with an adaptive SCAD penalty. Even though this proposal is more robust that the ones previously mentioned since can afford heavy-tailed distributions, these estimators are highly sensitive to high leverage points in the linear component of the model associated to large residuals.

%Una propuesta m\'as robusta fue dada en \citet{lv:yang:guo:2017} a partir de estimadores modales basados en splines combinados con la penalidad SCAD.   Si bien esta propuesta es m\'as robusta que las anteriormente mencionadas pues tolera  errores con \textit{colas pesadas sim\'etricas}, los estimadores son altamente sensibles a la presencia de  datos at\'{\i}picos de alta palanca  en las covariables de la parte lineal asociados a residuos grandes.

The rest of the paper is organized as follows. In Section \ref{sec:proposal} the robust approach that estimates and selects variables from both the linear and the additive components of a \textsc{plam} is proposed. In Section \ref{sec:bicpenalty}, a robust BIC criterion is defined to select the penalty parameters, while in Section \ref{sec:bicregularization} it is mentioned a robust BIC criterion for selecting the parameters related to the $B-$splines approximation. Section \ref{sec:penalty} contains a brief review of the differences among some commonly used penalty functions. In Section \ref{sec:generalalgo}, a general algorithm to obtain the penalized robust estimators is derived. The results of a simulation study carried out to compare the performance of the robust proposal against its least squares counterpart is shown in Section \ref{sec:simu}. Section \ref{sec:realdata} contains the analysis of the well-known plasma beta-carotene level data set where the advantages of using the penalized robust approach to simultaneously estimate and select variables from both linear and additive part of the \textsc{plam} are highlighted. Finally, some final comments can be found in Section \ref{sec:comments}.

\section{The proposal}\label{sec:proposal}

In order to estimate and select variables from both components of a \textsc{plam}, a $B-$spline basis is used to approximate each additive function $\eta_j$ of (\ref{plam}). Similar to what is done in \citet{boente:martinez:2023}, $\eta_j$ is approximated as a  $\sum_{s=1}^{k_j-1}c_s^{(j)}B_s^{(j)}(t)$ with $B_s^{(j)}$, for $1\leq s\leq k_j-1$ and $1\leq j\leq p$, the $s$th element of the basis of centered $B-$splines of order $\ell_j$, that is, $B_s^{(j)}$ is such that $B_s^{(j)}(t)=\widetilde{B}_s^{(j)}(t)-\int_{\mathcal{I}_j}\widetilde{B}_s^{(j)}(t)\,dt$ with $\{\widetilde{B}_s^{(j)}\,:\, 1\leq s\leq k_j\}$ a spline basis of order $\ell_j$. Besides, as it is explained in that paper, only $k_j-1$ elements of the basis are considered instead of the $k_j$ total in order to allow the paramerer $\mu$ to be identifiable. Then, the \textsc{plam} in (\ref{plam}) is approximated by the following linear regression model 
\begin{equation}\label{mo-approx}
Y=\mu+\bbe\trasp\bZ+\sum_{j=1}^p \sum_{s=1}^{k_j-1} c_s^{(j)} B_s^{(j)}(X_j)+\sigma\varepsilon\,.
\end{equation}

%In order to define the robust estimators in Boente and Mart\'{\i}nez (2023), it is first computed a preliminary robust $S-$estimator of the scale parameter $\wsigma$ and then it is used an $M-$estimator to solve the following optimization problem
%\begin{equation}\label{rplam}
%(\wmu,\wbbe,\wbd)=\argmin_{a\in\real,\bb\in\real^q,\bd\in\real^K}\sum_{i=1}^n\rho\left(\frac{r_i(a,\bb,\bd)}{\wsigma}\right)\,,
%\end{equation}
%with $\wbd=(\wbc^{(1)},\dots,\wbc^{(p)})\trasp$, $\bc^{(j)}=(c_1^{(j)},\dots,c_{k_j-1}^{(j)})\trasp\in\real^{k_j-1}$, $K=\sum_{j=1}^p k_j-p$, where $\rho$ is a bounded $\rho-$function as defined in Maronna \textit{et al.} (2019) and $r_i(a,\bb,\bd)=r_i(a,\bb,\bc^{(1)\mbox{\footnotesize{\sc t}}},\dots,\bc^{(p)\mbox{\footnotesize{\sc t}}})$ $=Y_i-a-\bb\trasp\bZ_i-\sum_{j=1}^p\sum_{s=1}^{k_j-1}c_s^{(j)}B_s^{(j)}(X_{ij})$ are the residuals. Finally, the estimators of the additive functions $\eta_j$ are defined as $$\weta_j(x)=\sum_{s=1}^{k_j-1}\wc_s^{(j)}B_s^{(j)}(x)\,,$$ and the estimator of the multivariate regression function as $$\wm(\bz\trasp,\bx\trasp)=\wmu+\wbbe\trasp\bz+\sum_{j=1}^p \weta_j(x_j)$$ for any $\bz\in\real^q$ and $\bx=(x_1,\dots,x_p)\trasp$.

Let $\bbe=(\beta_1,\dots, \beta_q)\trasp\in\real^q$, $\bc^{(j)}=(c_1^{(j)},\dots,c_{k_j-1}^{(j)})\trasp\in\real^{k_j-1}$, $\bc=( \bc^{(1)\mbox{\footnotesize{\sc t}}},\dots,\bc^{(p)\mbox{\footnotesize{\sc t}}})\in\real^K$, $K=\sum_{j=1}^p k_j-p$, and $\bV=(\bV^{(1)}(X_{1})\trasp,\dots,\bV^{(p)}(X_{p})\trasp)\trasp\in\real^K$ where $\bV^{(j)}(t)=(B_1^{(j)}(t),\dots,B_{k_j-1}^{(j)}(t))\trasp\in\real^{k_j-1}$. Then, (\ref{mo-approx}) can be written in a simplified form as
$$Y=\mu+\bbe\trasp\bZ+\bc\trasp\bV_i +\sigma\varepsilon.$$ 

Let $\bb=(b_1,\dots,b_q)\trasp\in\real^q$ and $\bd^{(j)}=(d_1^{(j)},\dots,d_{k_j-1}^{(j)})\trasp\in\real^{k_j-1}$ two general vectors to approximate $\bbe$ and $\bc^{(j)}$, for $1\leq j\leq p$, and $\bd=( \bd^{(1)\mbox{\footnotesize{\sc t}}},\dots,\bd^{(p)\mbox{\footnotesize{\sc t}}})\in\real^K$. Let $\wmu$ be an estimator of $\mu$ and $\wsigma$ a scale estimator of $\sigma$, then we define the penalized robust estimators for the partially linear additive models as
\begin{equation}\label{opt-plam}(\wbbe,\wbc)=\argmin_{\bb\in\real^q,\bd\in\real^K}PL_{n,\blach,\bk}(\bb,\bd)\,,
\end{equation}
where
\begin{equation}\label{pl}
PL_{n,\blach,\bk}(\bb,\bd)= L_n(\wmu,\wsigma,\bb,\bd) + \itJ_{\blach,\bk}(\bb,\bd)\,,
\end{equation} %p_{\blach,\bk}(\bb,\bc)
with 
$$ L_n(a,\varsigma,\bb,\bd)= \frac{1}{n}\sum_{i=1}^n\rho\left(\frac{r_i(a,\bb,\bd)}{\varsigma}\right)\,, \qquad \qquad r_i(a,\bb,\bd)  = Y_i-a-\bb\trasp\bZ_i- \bd\trasp\bV_i\,,
$$
and $\itJ_{\blach,\bk}(\bb,\bd)$ is an arbitrary penalty function (selected by the user) depending on a regularization parameter $\bla=(\lambda_{1,1},\dots, \lambda_{1,q},\lambda_{2,1},\dots,\lambda_{2,p})\trasp\in\real^{q+p}$ that determines the model complexity, that is, $\lambda_{1,s}$, $1\leq s\leq q$, and $\lambda_{2,j}$, $1\leq j\leq p$, are the penalty parameters that control the sparsity of the parametric and nonparametric components, respectively, and $\bk=(k_1,\dots,k_p)\trasp$ is the parameter vector that controls the $B-$spline approximation of the additive parte of the model. For instance, $\itJ_{\blach,\bk}(\bb,\bd)$ can be defined as
$$\itJ_{\blach,\bk}(\bb,\bd)=\sum_{s=1}^q p_{\lambda_{1,s}}(|b_s|)+\sum_{j=1}^p p_{\lambda_{2,j}}(\|\bd^{(j)}\|_{H_j})$$
with $p_\lambda$ a univariate penalty function such as the SCAD or the MCP penalties. Additionally, the proposal includes, but it is not mandatory, the adaptive setting, as proposed in \citet{zhao:etal:2014}, as follows
\begin{equation}\label{cond}
\lambda_{1,s}=\frac{\wtlam_1}{|\wtbeta_{s}|}\qquad \mbox{ and } \qquad \lambda_{2,j}=\frac{\wtlam_2}{\|\wtbc^{(j)}\|_{H_j}}\,,
\end{equation}
%\textcolor{blue}{
%$$PL_{\bla}(a,\bb,\bd)=\frac{1}{n}\sum_{i=1}^n r_i^2(a,\bb,\bd) +p_{\bla}(\bb,\bd)
%$$}
where $\wtlam_1$ and $\wtlam_2$ are auxiliary penalty parameters. In addition,  $\|\bd^{(j)}\|_{H_j}=(\bd^{(j)\mbox{\footnotesize\sc t}}H_j\bd^{(j)})^{1/2}$ with $H_j$ of dimension $(k_j-1)\times (k_j-1)$ with its $(s,s')$ element given by $\int_0^1 B_s^{(j)}(t)B_{s'}^{(j)}(t)\,dt$, $\rho$ is a bounded $\rho-$function as defined in \citet{maronna:etal:2019libro}. Besides, $\wtbeta_s$, for $1\leq s\leq q$, and $\wtbc^{(j)}$, for $1\leq j\leq p$, are unpenalized estimators of $\beta_s$ and $\bc^{(j)}$, respectively. A possible choice is to compute the robust estimators obtained with the non-penalized procedure proposed in \citet{boente:martinez:2023} from where it can also be obtained $\wmu$ and $\wsigma$. It is worth noting that when the adaptive setting is considered, the penalty parameters $\lambda_{1,s}$ and $\lambda_{2,j}$, for $1\leq s\leq q$, $1\leq j\leq p$, associated to components $b_s$ and $\bd^{(j)}$, respectively, are given in (\ref{cond}) through the auxiliary penalty parameters $\wtlam_1$ and $\wtlam_2$. In this context, the search for the \lq\lq best\rq\rq\, subset of parameters is reduced from dimension $q+p$ to $2$, as it is explained in Section \ref{sec:penalty}.
%Let denote $\eta_{j,c^{(j)}}=\sum_{s=1}^{k_j-1}c_s^{(j)}B_s^{(j)}(x)$, then $\|\bc^{(j)}\|_{H_j}=\|\eta_{j,c^{(j)}}\|_2$.

\noindent\textit{Remark}. When the objetive function $PL_{n,\blach,\bk}$ is defined through auxiliary coefficients $\wtlam_1$ and $\wtlam_2$ as in (\ref{cond}), to determine the penality parameters for each component of the regression parameter vectors, say $\bb$ and $\bd$, the resulting procedure is called \textit{adaptive}. An example of this is the adaptive LASSO, called ADALASSO, defined in \citet{zou:2006}.

\section{Selection of penalty parameters}\label{sec:bicpenalty}

Since any robust estimation and variable selection approach also needs robust methods for selecting the regularization and penalty parameters, we will consider a robust procedure for selecting the penalty parameters when the the number of terms to approximate the additive functions, that is, $\bk$, is fixed.  Following the ideas in \citet{zhao:etal:2014} and \citet{lv:yang:guo:2017} for modal regression, for $\bk$ fixed, we propose the following robust BIC criterion
\begin{equation}\label{rbic-lambdas-gen}
\mbox{RBIC}_{\blach}(\bla)=\log\left(\wsigma^2 \sum_{i=1}^n  \rho\left(\frac{r_i({\wbeta}_{\blach},\wbc_{\blach})}{\wsigma}\right)\right) +df_c\frac{\log(n)}{n}+df_n \frac{\log(n/K)}{n/K}
\end{equation}
%\textcolor{blue}{
%$$BIC_{\bla}(\wtlam_1,\wtlam_2)=\log\left(\sum_{i=1}^n  r_i^2(\wmu_{\bla},{\wbeta}_{\bla},\wbd_{\bla})\right) +df_c\frac{\log(n)}{n}+df_n \frac{\log(n/K)}{n/K}$$
%}
where $K=\sum_{j=1}^p k_j-p$, $(\wbeta_{\blach},\wbc_{\blach})$ is the solution of (\ref{opt-plam}) with penalty vector $\bla=(\lambda_{1,1},\dots,\lambda_{1,q},\lambda_{2,1},\linebreak \dots,\lambda_{2,p})\trasp$, $df_c$ is the number of nonzero parametric components, $df_n$ is the number of nonzero nonparametric components and $\wsigma$ is the preliminary $S-$estimator.

However, this proposal needs to select $\bla$ over a grid of dimension $q+p$ which is very expensive if this number is very large. For this reason and taking into account the relation between $\lambda_{1,s}$ and $\wtlam_1$ and between $\lambda_{2,j}$ and $\wtlam_2$ as stated in (\ref{cond}), we will now propose a method for selecting only the penalty parameters $\wtlam_1$ and $\wtlam_2$ which implies a selection over a grid of dimension $2$. The modified robust BIC criterion is defined in the following way:
\begin{equation}\label{rbic-lambdas}
\mbox{RBIC}_{\blach}(\wtbla)=\log\left(\wsigma^2 \sum_{i=1}^n  \rho\left(\frac{r_i({\wbeta}_{\wtblach},\wbc_{\wtblach})}{\wsigma}\right)\right) +df_c\frac{\log(n)}{n}+df_n \frac{\log(n/K)}{n/K}
\end{equation}
where $K=\sum_{j=1}^p k_j-p$, $df_c$, $df_n$ and $\wsigma$ are defined as before, and where $(\wbeta_{\wtblach},\wbc_{\wtblach})$ is now the solution of (\ref{opt-plam}) with $\bla$ given through  $\wtbla=(\wtlam_1,\wtlam_2)$ as in (\ref{cond}).

%We will now propose a method for selecting the penalty parameters $\wtlam_1$ and $\wtlam_2$ when $\bk=(k_1,\dots,k_p)\trasp$, the vector containing the information of the number of terms to approximate the additive functions, is fixed. The proposed variable selection method avoids the problem of selecting simultaneously $q+p$ parameters when each element of the regression vector is penalized, that is, when condition (\ref{cond}) is not used and so the vector of unknown penalization parameteres is $\bla=(\lambda_{1,1},\dots,\lambda_{1,q},\lambda_{2,1},\dots,\lambda_{2,p})\trasp$. When no adaptive procedure is considered, the following RBIC criterion can be used but will depend directly on the vector $\bla$ so the minimization will be on a grid of points in $\real^{q+p}$.

\section{Selection of regularization parameters}\label{sec:bicregularization}

Any $B-$spline approach requires the selection of a set of knots to define the basis functions. In this sense, since each additive function $\eta_j$, $1\leq j\leq p$, is approximated by a $B-$spline curve, $p$ sequences of knots are necessary. \citet{stone:1985} mentions that the number of knots is more important than their locations. For this reason, we will consider a generalization of a BIC criterion for the number of knots considered for each additive component, that is, for $k_1,\dots,k_p$. However, regarding their locations, two possible choices are: the equally spaced knots and the quantile knots. Uniform knots are used when the function $\eta_j$ does not present dramatic changes in its derivates, while non-uniform knots are preferred when the function shows very different behaviours in different regions. When the latter is the case, quantile knots are used, that is, the knots are obtained as the quantiles of the observed explanatory variable with uniform percentile ranks.

Since, once again, a robust criterion should be used to obtain final robust estimators, in this paper, in order to determine $\bk=(k_1,\dots,k_p)\trasp$, we will use the robust BIC criterion defined in \citet{boente:martinez:2023} which is the following:
\begin{equation}\label{rbic-k}
\mbox{RBIC}_{\bk}(\bk)=\log\left(\wsigma^2\sum_{i=1}^n \rho\left(\frac{r_i(\wbeta_{\blach,\bk},\wbc_{\blach,\bk})}{\wsigma}\right)\right)+\frac{\log(n)}{2n}\sum_{j=1}^p k_j
\end{equation}
where $\bk=(k_1,\dots,k_p)\trasp$, $r_i(\wbeta_{\blach,\bk},\wbc_{\blach,\bk})$, for $1\leq i\leq n$, are the residuals obtained using a basis of dimension $k_j$ to approximate the additive function $\eta_j$ and $\bla$ is given, $\rho$ is the $\rho-$function used to compute the $M-$estimator and $\wsigma$ is a preliminary scale estimator of $\sigma$. When $p=1$, that is, when it is considered a partially linear model with one covariate entering in the nonparametric part of the model, and $\rho(t)=t^2$, this criterion is reduced to the criterion proposed in \citet{he:etal:2002}. When $\bla$ in defined through $\wtbla$ as in (\ref{cond}), the dependence of the estimators can be hightlighted as $\wbeta_{\wtblach,\bk}$ and $\wbc_{\wtblach,\bk}$.

Besides, when the number of elements to approximate each additive function is the same, that is, when $k_j=k$ for $1\leq j\leq p$, the robust BIC criterion reduces to
\begin{equation}\label{rbic}
\mbox{RBIC}_{\bk}(k)=\log\left(\wsigma^2\sum_{i=1}^n \rho\left(\frac{r_i(\wbeta_{\blach,\bk},\wbc_{\blach,\bk})}{\wsigma}\right)\right)+\frac{\log(n)}{2n}pk\,.
\end{equation}

As it is mentioned in \citet{he:etal:2002} and also in \citet{boente:martinez:2023}, since for cubic splines the smallest possible number of knots is $4$, the $j$th component of $\bk$, that is, $k_j$, in (\ref{rbic-k}) can be chosen in the interval $[\max(n^{1/5}/2,4);8+2n^{1/5}]$ when cubic splines are considered. The numerical computations are even more reduced when the size of the basis is the same for all $j$. In this case, it is enough to look for the first local minimum of $\mbox{RBIC}_\bk(k)$ defined in (\ref{rbic}) for $\max(n^{1/5}/2,4)\leq k \leq 8+2n^{1/5}$.

%\textcolor{red}{Mencionar lo de la grilla si usamos la misma cantidad de elementos para aproximar cada componente.}

\section{The penalty functions}\label{sec:penalty} 

There are many univariate penalty functions $p_\lambda$ that can be considered in the regularization component of (\ref{pl}), such as the $L_1$ or, in general, the $L_{q}$ penalties \citep[see][]{tibshirani:1996, tibshirani:1997, Frank:friedman:1993, fu:1998}, the hard thresholding \citep[see][for an overview and a comparison of penalty functions]{fan:li:2001, potscher:leeb:2009}, the Minimax Concave Penalty (MCP) introduced by \citet{zhang:2010} or the  Smoothly Clipped Absolute Deviation Penalty (SCAD) proposed by \citet{fan:li:2001}.

\begin{figure}[H]
 \begin{center}
\begin{tabular}{c}
\includegraphics[scale=0.5]{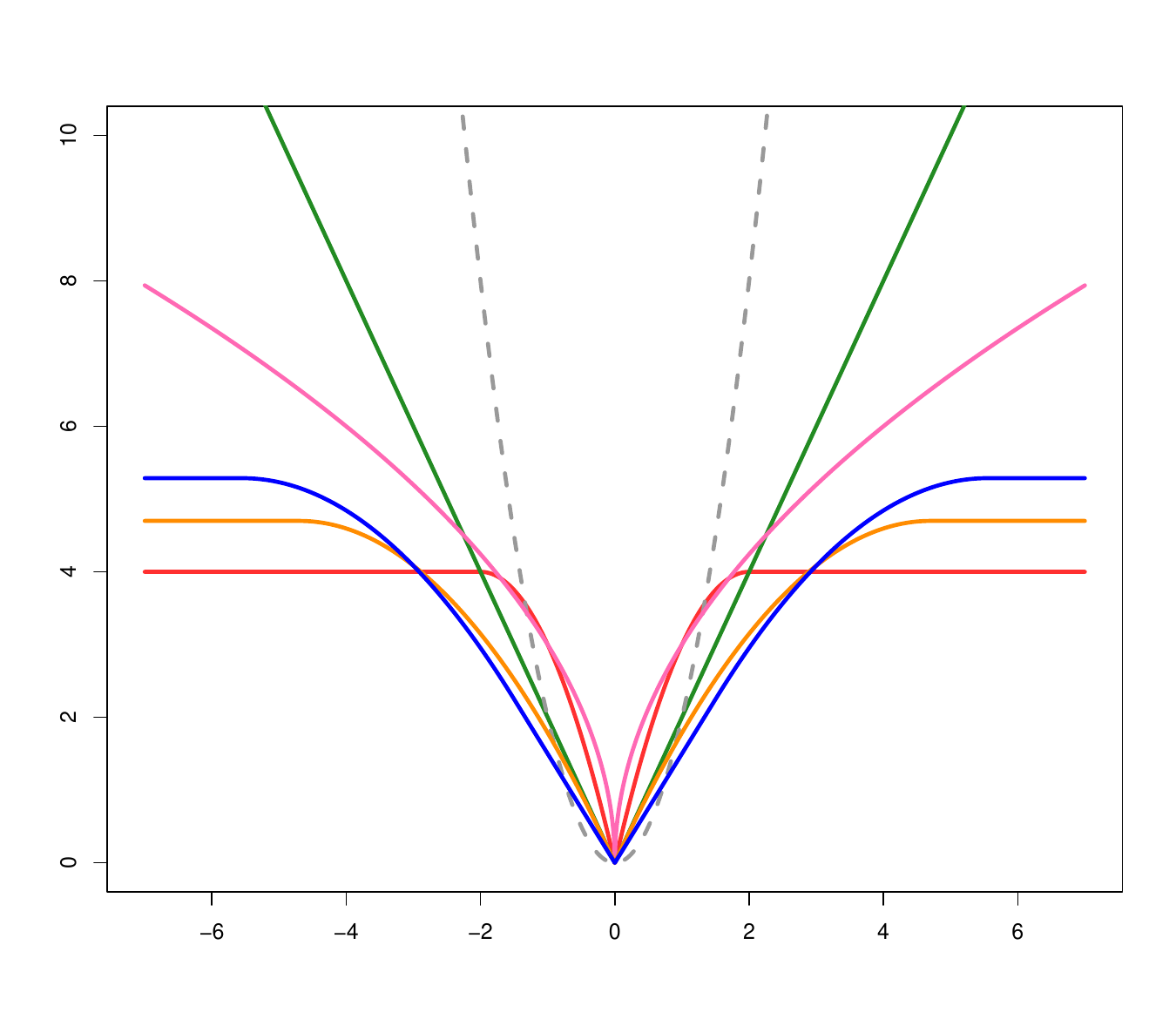}
\end{tabular}
\caption{\label{fig:penalidades} Plot of penalty functions: the hard penalty in red, the $L_1$ in green, the $L_{0.5}$ in pink, the MCP in orange, the SCAD penalty in blue and the $L_2$ in dashed gray line.}
\end{center}
\end{figure}

The hard thresholding penalty is defined as $p_{\lambda}(\theta)=\lambda^2-(|\theta|-\lambda)^2I\{|\theta|<\lambda\}$, while the $L_q$ penalty is defined as $p_{\lambda}(\theta)=\lambda|\theta|^q$. When $q=1$, we get the $L_1$ penalty defined as $p_\lambda(\theta)=\lambda|\theta|$ which is associated to the LASSO estimators. Additionally, the MCP is defined as 
$$p_{\lambda}(\theta)=\left\{\begin{array}{cl}
 \lambda|\theta|-\frac{\theta^2}{2\gamma} & |\theta|\leq \lambda\gamma\\
\frac{\gamma\lambda^2}{2} &  |\theta|> \lambda\gamma
\end{array}\right.,$$
where $\gamma>1$ is a tuning parameter, and the SCAD penalty is defined, for $\lambda>0$ (and some $a>2$), as
$$p_\lambda(\theta)=\left\{\begin{array}{cl}
\lambda|\theta|,  & |\theta|\leq \lambda\\
-(\theta^2-2a\lambda|\theta|+\lambda^2)/(2(a-1)), & \lambda<|\theta|\leq a\lambda\\
(a+1)\lambda^2/2, & |\theta|>a\lambda
\end{array}\right..$$
The SCAD penalty has advantages over the $L_q$ and the hard thresholding penalties such as its simultaneous unbiased, sparsity and continuity properties of the resulting estimators. For instace, the $L_q$ penalties give rise to continuos solutions when $q\geq 1$ but for $q>1$ it does not produce sparse solutions, and when $q=1$ the resulting estimator is shifted by a constant and so it is not unbiased; while the hard thresholding does not produce continuous estimatos \citep[see][for more details]{fan:li:2001}. It can also be appreciated that the penalties hard thresholding, SCAD and MCP are bounded functions while the others are unbounded. Even though the similarities between the SCAD and the MCP penalties, as it is mentioned in \citet{Bianco:Boente:Chebi:2022}, in some regression settings when considering classical estimadors, the first one outperforms the latter. In allow the reader some visual comparison, Figure \ref{fig:penalidades} shows the plots of the penalty functions previously mentioned. In red solid line it is plotted the hard thresholding, in green line the $L_1$ penalty, in pink the $L_{0.5}$ penalty, in orange the MCP and in blue line the SCAD penalty. As previously mentioned, the $L_2$ penalty function does not yield sparse solutions due to its differentiability at the origin, so its curve is represented by a dashed gray line.
%Since, as it has already been mentioned, the penalty function $L_2$ does not give rise to sparse solutions due to its differentiability property at the origin, its curve has been drawn on dashed gray line.

%For these reasons, for all the computational studies carried out throughout this article, we consider the Smoothly Clipped Absolute Deviation Penalty (SCAD). From now on, as it is usual, the value of $a$ in the definition of the SCAD penalty is set to $3.7$ so it is not a parameter to be selected.

For all computational studies conducted in this article, we use the Smoothly Clipped Absolute Deviation (SCAD) penalty. As is customary, the parameter $a$ in the SCAD penalty definition is fixed at $3.7$, so it does not require selection.

\section{General algorithm}\label{sec:generalalgo}

In this section, we derive an algorithm to compute the penalized estimators for the general setting in which the penalized vector is given by $\bla=(\lambda_{1,1},\dots,\lambda_{1,q},\lambda_{2,1},\dots, \lambda_{2,p})\trasp$ and the RBIC criterion for selecting the penalty parameters is defined in (\ref{rbic-lambdas-gen}). However, when an adaptive procedure is used, that is, when the components of the penalty vector $\bla$ are defined according to (\ref{cond}), the proposed algorithm can be easily modified by considering a grid for $\wtbla=(\wtlam_1,\wtlam_2)\trasp$ and minimizing the RBIC criterion defined in (\ref{rbic-lambdas}). Similarly, although we describe the general framework for selecting the regularization parameters $\bk=(k_1,\dots,k_p)\trasp$ where the RBIC criterion is defined in (\ref{rbic}), a straightforward modification of the algorithm allows us to minimize the RBIC defined in (\ref{rbic-k}) under the assumption that $k_j=k$ for all $j=1,\dots,p$.
%in order to include the general setting where the the penalization parameter does not have an set form.
In this sense, Algorithm \ref{alg1} presents a general procedure for obtaining the proposed estimators. % when considering a grid for the penalty parameters $\bla$, a grid for the regularization parameters $\bk$ and both robust BIC criteria defined in Sections and \ref{sec:bicpenalty} and \ref{sec:bicregularization}. %In Section \ref{subproblem}, a method to solve the optimization problem in (\ref{subproblema}) is developed for a certain type of penalty functions such as the SCAD.  %In order to do this, we will first mention the general algorithm that uses the $RBIC$. A similar algorithm but for the $BIC$ when considering the least squares estimator can be used. 

\begin{algorithm}[ht!]
\caption{General algorithm}
\label{alg1}
\begin{algorithmic}[1]
\STATE Obtain preliminary robust estimators $\wmu$, $\wbbe$, $\weta_j$, for $1\leq j\leq p$, and a robust scale estimator $\wsigma$. A possible choice is to compute the robust estimators obtained with the non-penalized procedure proposed in \citet{boente:martinez:2023}.% and as scale estimator of $\sigma$ it can be used a robust scale estimator, such as the MAD, over the initial residuals $r_{\ini,i}=Y_i-\wmu-\wbbe\trasp\bZ_i-\sum_{j=1}^p \weta_j(X_{ij})$. % Another one is to use initial estimates of $\mu$, $\weta_j$ and $\wbbe$ described in Section \ref{sec:ini} and then to consider a robust scale estimator, such as the MAD, over the initial residuals $r_{\ini,i}=Y_i-\wmu-\wbbe\trasp\bZ_i-\sum_{j=1}^p \weta_j(X_{ij})$ to obtain an estimator $\wsigma$ of $\sigma$. \textcolor{red}{Se me ocurre que lo har\'{\i}a as\'{\i} para calcular un estimador de $\wsigma$ usando el estimador preliminar propuesto en la secci\'on anterior. Sin embargo, el nombre de la secci\'on anterior, en este momento, s\'olo menciona que se tienen estimadores preliminares de $\mu$ y de $\eta_j$ cuando ac\'a, por lo menos con lo que pens\'e yo, tambi\'en necesitar\'{\i}a el estimador (preliminar) que se obtiene de $\bbe$.}
\STATE For $j=1,\dots, p$, fix the spline order $\ell_j$.
\STATE Consider a grid for $\bk=(k_1,\dots,k_p)\trasp$ of $M$ elements: $\bk_1,\dots,\bk_M$, and consider a grid for $\bla$ of $N$ elements: $\bla_1,\dots,\bla_N$. 
\FOR{$\ell = 1$ \TO $M$}
\FOR{$j = 1$ \TO $N$} 
\STATE Solve the optimization problem
\begin{equation}\label{subproblema}
(\wbeta_{\blach_j,\bk_\ell},\wbc_{\blach_j,\bk_\ell})=\argmin_{\bb\in\real^q,\bd\in\real^K}PL_{n,\blach_j,\bk_\ell}(\bb,\bd)\,.
\end{equation}
Note that $(\wbeta_{\blach_j,\bk_\ell},\wbc_{\blach_j,\bk_\ell})$ is the argument that minimizes the optimization problem when $\bk=\bk_\ell$ and $\bla=\bla_j$ are fixed.
\STATE Compute and save $\mbox{RBIC}_{\blach}(\bla_j)$. %This procedure uses $(\wmu_{j,\ell},\wbeta_{j,\ell},\wbc_{j,\ell})$.
\ENDFOR
\STATE Solve
$$\bla_{\bk_\ell}=\argmin_{j}\,\{\mbox{RBIC}_{\blach}(\bla_j)\}\,.$$
That is, $\bla_{\bk_\ell}$ is the optimal penalty parameter vector for $\bk=\bk_\ell$ fixed. 
\STATE Compute and save $\mbox{RBIC}_{\bk}(\bk_\ell)$. 
\ENDFOR
\STATE Obtain $\wbk$ as 
$$\wbk=\argmin_\ell\,\{\mbox{RBIC}_\bk(\bk_\ell)\}\,.$$
\STATE The penalized estimators are $(\wbeta_{\blach_{\wbk},\wbk},\wbc_{\blach_{\wbk},\wbk})$.
\end{algorithmic}
\end{algorithm}

As it can be appreciated, Algorithm \ref{alg1} does not include how to solve the optimization problem (\ref{subproblema}), which we still need to describe. In Subsection \ref{sec:subproblem}, a method to solve the minimization task, for $\bk$ and $\wtbla$ fixed, is developed. The procedure considers certain type of penalty functions, such as the SCAD and MCP penalties.

\subsection{Solving the optimization problem for $\bk$ and $\bla$ fixed}\label{sec:subproblem}

In what follows, we consider univariate penalties $p_\lambda$ that are twice continuously differentiable on $(0,M)$, with $M>0$, and satisfy $p_\lambda(0)=0$. These conditions allow for local approximation by quadratic functions near in a neighbourhood of the origin and include some common penalty functions such as the $L_1$, SCAD and MCP penalties. 

%In order to implement the proposed estimators, we need to use the first derivative of $p_\lambda(\theta)$ for $\theta\neq 0$. When the penalty function is the SCAD penalty, if $\theta\neq 0$, then $$p_\lambda^\prime(|\theta|)=\lambda I_{\{|\theta|\leq \lambda\}}+\frac{(a\lambda-|\theta|)_+}{(a-1)}I_{\{|\theta|>\lambda\}}$$ with $(t)_+=tI_{\{t>0\}}$.

For the sake of simplicity, let assume that $\mu=0$. Then, for $\bla_j$ and $\bk_\ell$ fixed, we have \begin{equation}\label{pl-sina}
PL_{n,\blach_j,\bk_\ell}(\bb,\bd)=\frac{1}{n}\sum_{i=1}^n\rho\left(\frac{r_i(\bb,\bd)}{\wsigma}\right)+\itJ_{\blach_j,\bk_\ell}(\bb,\bd)\,.
\end{equation}
with $r(\bb,\bd)=Y_i-\bb\trasp\bZ_i-\bd\trasp\bV_i$.
If $\mu\neq 0$, obtain $\widehat{\mu}$ an estimator of $\mu$, for instance using the unpenalized robust estimator defined in \citet{boente:martinez:2023} or using the $S-$ or the $MM-$Ridge estimators defined in \citet{maronna:2011}, and take the response variables $Y_i-\widehat{\mu}$ instead of $Y_i$.

%Remark. An estimator of $\mu$ is the one obtained by the unpenalized approach.

%Now, the optimization problem (\ref{subproblema}) in Algorithm \ref{alg1} is rewritten as 
%\begin{equation}\label{subproblema2}
%(\wbeta_{\wtblach_j,\bk_\ell},\wbc_{\wtblach_j,\bk_\ell})=\argmin_{\bb\in\real^q,\bd\in\real^K}PL_{\wtblach_j,\bk_\ell}(\bb,\bd)
%\end{equation}
%with $PL_{\wtblach_j,\bk_\ell}(\bb,\bd)$ defined in (\ref{pl-sina}). 
Assume that the penalty function used $p_\lambda$ is the same for both the linear and the additive parts of the model.
%, such as the SCAD penalty, is singular at the origin and does not have second order derivatives. 
To avoid burden notation, let $p=p_{\lambda_{1,s}}$ and $p=p_{\lambda_{2,j}}$ when it corresponds. 
The assumptions on $p$ previously mentioned guarantee that, if $t_0$ is a initial point close to $0$, then there exists a quadratic function $q(t)$ that is symmetric around $0$, $q(t_0)=p(|t_0|)$ and $q'(t_0)=p^\prime(|t_0|)\,\mbox{sgn}(t_0)$.
Using the symmetry, we have that $q(t)=a+bt^2$. Then, the conditions on $q(t_0)$ and $q^\prime(t_0)$ imply that
$$\left\{
\begin{array}{rcl}
a+b|t_0|^2&=&p(|t_0|)\\
2bt_0&=&p^\prime(|t_0|)\,\mbox{sgn}(t_0)
\end{array}
\right.\Leftrightarrow 
\left\{
\begin{array}{rcl}
a+b|t_0|^2&=&p(|t_0|)\\
2b|t_0|&=&p^\prime(|t_0|)
\end{array}
\right..$$
Then, from the second equation we have that $b=p^\prime(|t_0|)/(2|t_0|)$ and replacing in the first equation we get that $a=p(|t_0|)-p^\prime(|t_0|)/(2|t_0|)t_0^2$. Finally, the quadratic approximation is
\begin{eqnarray*}
q(t)&=&p(|t_0|)-\frac{p^\prime(|t_0|)}{2|t_0|}t_0^2+\frac{p^\prime(|t_0|)}{2|t_0|}t^2\\
&=&p(|t_0|)+\frac{p^\prime(|t_0|)}{2|t_0|}(t^2-t_0^2),
\end{eqnarray*}
that is, for $t$ close to $t_0$, we have that
$$p(|t|)\approx p(|t_0|)+\frac{p^\prime(|t_0|)}{2|t_0|}(t^2-t_0^2)\,.$$

In this way, given $(\bb_0,\bd_0)$ close to the minimizer of (\ref{subproblema}) and $(\bb,\bd)$ close to $(\bb_0,\bd_0)$ with $|b_{0s}|>0$ and $\|\bd_0^{(j)}\|_{H_j}>0$, for $s=1,...,q$ and $j=1,...,p$, we have that
$$p(|b_s|)\approx p(|b_{0s}|)+\frac{p^\prime(|b_{0s}|)}{2|b_{0s}|}(b_s^2-b_{0s}^2)$$
%Since $p(\|\bc^{(j)}\|_{H_j})=p( (\bc^{(j)\mbox{\footnotesize\sc t}}H_j\bc^{(j)})^{1/2} )$ and so 
%$$\frac{\partial}{\partial \bc^{(j)}}p^\prime(\|\bc^{(j)}\|_{H_j})=p^\prime(\|\bc^{(j)}\|_{H_j})\frac{\partial}{\partial \bc^{(j)}}(\|\bc^{(j)}\|_{H_j})=p^\prime(\|\bc^{(j)}\|_{H_j}) \frac{1}{2}(\bc^{(j)\mbox{\footnotesize\sc t}} H_j\bc^{(j)})^{1/2} 2H_j^{1/2}\bc H_j^{1/2}=\frac{p^\prime(\|\bc^{(j)}\|_{H_j})}{\|\bc^{(j)}\|_{H_j}} H_j\bc^{(j)}$$
and
$$p(\|\bd^{(j)}\|_{H_j})\approx p(\|\bd_0^{(j)}\|_{H_j})+\frac{p^\prime(\|\bd_0^{(j)}\|_{H_j})}{2\|\bd^{(j)}\|_{H_j}}(\|\bd^{(j)}\|_{H_j}^2-\|\bd_0^{(j)}\|_{H_j}^2)\,.$$

Then, recalling that $\|\bd^{(j)}\|_{H_j}^2=\bd^{(j)\mbox{\footnotesize\sc t}}H_j\bd^{(j)}$, if we define 
$$\bSi(\bb_0,\bd_0)=\mbox{diag}\left\{\frac{p^\prime(|b_{01}|)}{2|b_{01}|},\dots,\frac{p^\prime(|b_{0q}|)}{2|b_{0q}|},\frac{p^\prime(\|\bd_0^{(1)}\|_{H_1})}{2\|\bd^{(1)}\|_{H_1}} H_1,\dots, \frac{p^\prime(\|\bd_0^{(p)}\|_{H_p})}{2\|\bd^{(p)}\|_{H_p}}H_p \right\}\in\real^{(q+K-p)\times (q+K-p)}$$
we get that the objective function given in (\ref{pl-sina}) can be approximated by
\begin{equation}\label{quadratic-pl}
\frac{1}{n}\sum_{i=1}^n\rho\left(\frac{r_i(\bb,\bd)}{\wsigma}\right)+(\bb\trasp,\bd\trasp) \bSi(\bb_0,\bd_0) (\bb\trasp,\bd\trasp)\trasp
\end{equation}
and so the estimators obtained by minimizing (\ref{quadratic-pl}) will be close to the minimizers of (\ref{subproblema}) with the penalized function given in (\ref{pl-sina}).

Denote $\bthe=(\bb\trasp,\bd\trasp)\trasp$ and $\bW_i=(\bZ_i\trasp,\bV_i\trasp)\trasp$. The minimization of (\ref{quadratic-pl}) can be obtained by a reweighted procedure as it is usual for $M-$estimators. 

\textbf{Deduction of the iteration procedure}. Suppose now we have $\bthe^{(m)}=(\bb^{(m) \mbox{\footnotesize\sc t}}, \bd^{(m) \mbox{\footnotesize\sc t}})\trasp$ the estimators obtained in the $m-$step and $\wsigma$ is a preliminary robust estimator of $\sigma$. By first derivating (\ref{quadratic-pl}) with respect to $\bthe$ and then multiplying and dividing by $(Y_i-\bW_i\trasp\bthe)/\wsigma$, we obtain
\begin{eqnarray*}
\frac{1}{n}\sum_{i=1}^n \psi\left(\frac{Y_i-\bW_i\trasp\bthe}{\wsigma}\right)\left(-\frac{\bW_i}{\wsigma}\right)+2\bSi(\bthe^{(m)}) \bthe &=&\textbf{0}\\
\frac{1}{n}\sum_{i=1}^n \psi\left(\frac{Y_i-\bW_i\trasp\bthe}{\wsigma}\right)\frac{\wsigma}{Y_i-\bW_i\trasp\bthe}\frac{Y_i-\bW_i\trasp\bthe}{\wsigma}\left(-\frac{\bW_i}{\wsigma}\right)+2\bSi(\bthe^{(m)}) \bthe &=&\textbf{0}\\
\frac{1}{n}\sum_{i=1}^n w\left(\frac{r_i(\bb,\bd)}{\wsigma}\right)\frac{Y_i-\bW_i\trasp\bthe}{\wsigma}\left(-\frac{\bW_i}{\wsigma}\right)+2\bSi(\bthe^{(m)}) \bthe &=&\textbf{0}
\end{eqnarray*}
where $w(t)=\psi(t)/t$ and $r_i(\bb,\bd)=Y_i-\bW_i\trasp\bthe$. Then, the estimator  $\bthe^{(m+1)}=(\bb^{(m+1) \mbox{\footnotesize\sc t}}, \bd^{(m+1) \mbox{\footnotesize\sc t}})\trasp$ of the $(m+1)-$step is defined as the solution of
$$\frac{1}{n}\sum_{i=1}^n w_{i,m}\frac{Y_i-\bW_i\trasp\bthe}{\wsigma}\left(-\frac{\bW_i}{\wsigma}\right)+2\bSi(\bthe^{(m)}) \bthe =\textbf{0}$$
where $w_{i,m}=W(r_i(\bthe^{(m)})/\wsigma)$ are the weights obtained by replacing the estimator calculated in the previous step. Finally, the latter equation is equivalent to
$$-\frac{1}{n}\sum_{i=1}^n \frac{w_{i,m}}{\wsigma^2}\bW_i Y_i+\left(\frac{1}{n}\sum_{i=1}^n \frac{w_{i,m}}{\wsigma^2}\bW_i\bW_i\trasp+2\bSi(\bthe^{(m)})\right)\bthe=\textbf{0}$$
and so the iteration step is 
\begin{equation}\label{theta-m}
\bthe^{(m+1)}=\left(\frac{1}{n}\sum_{i=1}^n \frac{w_{i,m}}{\wsigma^2}\bW_i\bW_i\trasp+2\bSi(\bthe^{(m)})\right)^{-1}\frac{1}{n}\sum_{i=1}^n \frac{w_{i,m}}{\wsigma^2}\bW_i Y_i\,.
\end{equation}

Finally, we obtain the following algorith to minimize (\ref{quadratic-pl}).

\textbf{The iteratively reweighted algorithm}. Consider $\epsilon>0$ the stopping rule, then
\begin{enumerate}
\item Compute $\wsigma$ and an initial $\bthe^{(0)}$ such as the one mentioned in Step 1 in Algorithm \ref{alg1}.
\item For $m=0, 1, 2,\dots$ compute the weights $w_{i,m}$, for $i=1,\dots,n$, and then $\bthe^{(m+1)}$ in (\ref{theta-m}).
\item Stop when the incrementation is small enough, for instance, if $\|\bthe^{(m+1)}-\bthe^{(m)}\|/\|\bthe^{(m)}\|<\epsilon$. 
\end{enumerate}

%\textcolor{blue}{En el caso cl\'asico, queremos minimizar $$\frac{1}{n}\sum_{i=1}^n (Y_i-\bS_i\trasp\bthe)^2+\bthe\trasp\Sigma\bthe$$ en lugar de (\ref{quadratic-pl}) y derivando y viendo que en los dos t\'erminos queda el 2, te termina quedando que  $$\bthe^{(m+1)}=\left(\frac{1}{n}\sum_{i=1}^n \bS_i\bS_i\trasp+\Sigma\left(\bthe^{(m)}\right)\right)^{-1}\frac{1}{n}\sum_{i=1}^n \bS_i Y_i$$  Copiar los pasos anteriores para que se vea todo lo que depende de $m$.}

\section{Monte Carlo Study}\label{sec:simu}

This section contains the results of a simulation study in which it is compared the robust proposal with the corresponding estimator based on least squares. %, that it, the one that is computed using $\rho(t)=t^2$.
All computations were carried out in {\tt{R}}. The code used is available at {\url{https://github.com/alemermartinez/rplam-vs}}. The LS$-$estimator corresponds to use $\rho(t)=t^2$ in (\ref{pl}). Observe that, in this case, no preliminary scale estimator is needed. For the robust estimator,the loss function considered is the Tukey's bisquare loss which is of the form $\rho_c(t)=\min\{1-(1-(t/c)^2)^3,1\}$. The tuning constant $c>0$ balances the robustness and efficiency properties of the associated estimators. For the simulation study, the tuning constant was fixed at $c=4.685$, which is the constant necessary to obtain a 95\% efficiency for linear regression estimators. Besides, an $S-$estimator is used for the preliminary scale estimator. From now on, the robust procedure proposed in this paper will be denoted as {\textsc{rob}} and as {\textsc{ls}} the approach based on least squares.

The samples $\{(Y_i,\bZ_i\trasp,\bX_i\trasp)\trasp\}_{i=1}^n$ are generated with the same distribution as $(Y,\bZ\trasp,\bX\trasp)\trasp$, \linebreak $\bZ=(Z_1,\dots,Z_q)\in\real^q$, $\bX=(X_1,\dots,X_p)\trasp\in\real^p$, with $q=10$ and $p=10$ and three different sample sizes $n=200, 400$ and $600$. The covariates $\bZ_i=(Z_{i1},\dots,Z_{iq})\trasp$ are generated from a multivariate normal distribution with zero mean and variance-covariance matrix $\Sigma$ defined as $(\Sigma)_{i,\ell}=\mbox{Corr}(Z_{ik},Z_{i\ell})=0.5^{|k-\ell|}$, for $1\leq k,\ell\leq q$, and $X_{ij}$, for $j=1,\dots,p$, are uniformly distributed on the interval $[0;1]$ and are independent from each other and from the vector $\bZ_i$.

In all cases, the response and the covariates satisfy the partially linear additive model
$$Y=\mu+\bbe\trasp\bZ+\sum_{j=1}^p \eta_j(X_j)+u$$
where $\bbe=(3,1.5,2,-1.5,0,0,0,0,0,0)\in\real^{10}$, the additive functions are
$$\eta_1(x)=5x-\frac{5}{2}\,,\qquad\qquad \eta_2(x)=3(2x-1)^2-1\,,$$
$$\eta_3(x)= 60x^3-90x^2+30x\,, \qquad \eta_4(x)= 2\sin(\pi x)-\frac{4}{\pi}\,,$$
$\eta_j(x)\equiv 0$ for $j=5,\dots,p$, $u=\sigma\eps$, $\sigma=1$ and $\mu=0$. In all cases the additive functions are such that $\esp \eta_j(X_j)=0$ for all $j=1,\dots,10$. For clean samples, the error's distribution is $\epsilon\sim N(0,1)$, denoted from now on as $C_0$.  

In order to study the effect of atypical data on the estimators, seven different contamination schemes were considered:
\begin{itemize}
\item $C_1$\,:\, $u \sim t_3$
\item $C_2$\,:\, $u \sim 0.9 N(0,\sigma^2)+0.1 N(0,25\sigma^2)$
\item $C_3$\,:\, $u \sim 0.95 N(0,\sigma^2)+0.05 N(0,100\sigma^2)$
\item $C_4$\,:\, $u \sim 0.95 N(0,\sigma^2)+0.05 N(15,\sigma^2)$% \textcolor{red}{Este orden es mejor (en la simulación la C5 es la C4 y viceversa)}
\item $C_5$\,:\, $U \sim 0.85 N(0,\sigma^2)+0.15 N(15,\sigma^2)$
\item $C_6$\,:\, A 5\% of covariates $\bZ_i$ were replaced by $(20,\dots,20)\in\real^{10}$.
\item $C_7$\,:\, A 10\% of covariates $\bZ_i$ were replaced by $(20,\dots,20)\in\real^{10}$.
\end{itemize}
Contamination $C_1$ is a heavy-tailed contamination setting where the errors are distributed as a Student's t$-$dis\-tri\-bu\-tion with 3 degrees of freedom. $C_2$ and $C_3$ are variance contamination settings where a 10\% and a 5\%, respectively, of the errors are normaly distributed with enlarged variances. $C_4$ and $C_5$ are bias contamination schemes where 5\% and 15\% of the errors, respectively, have their location shifted to 15. These contamination schemes give rise to what are called \textit{vertical outliers}. $C_6$ and $C_7$ are high-leverage contaminations where a 5\% and 10\%, respectively, of the covariates of the linear part were randomly chosen to take another value. In this case, the response variables and the errors are not modified.
It is interesting to note that none of the contamination settings considered introduces outliers in the additive component of the model or in the response variable depending on a particular set of the additive components. The main reason for not considering outliers in the covariates of the additive part is that this would only generate high computational costs since the estimators would be defined using a larger number of terms to approximate the additive functions. With respect to contamination of the errors depending on the values taken by the covariates belonging to the additive part, similarly as what was done in \citep{boente:martinez:SB:2017}, it would be needed a much larger number of sample sizes. For instance, if we want to contaminate errors with their 10 covariates in the interval $[0,0.1]$, since all the covariates are distributed as $\itU[0,1]$, this would lead to $10^{-10}$\% of each sample with contaminated data, and even with the highest sample size of $n=600$, this would represent an amount of $6.10 ^{-8}$ observations.

With respect to the selection of the penalty and regularization parameters, both generalizations of the BIC criteria stated in Sections \ref{sec:bicpenalty} and \ref{sec:bicregularization}, respectively, were used. For the regularization parameters and for both robust and least squares estimators, equally spaced knots are used and the same number of terms to approximate each additive function were used, that is, $k_j=k$ for all $1\leq j\leq 10$, and so the selection problem consists of minimizing the simplified $\mbox{RBIC}_{\bk}$ defined in (\ref{rbic}). As it was mentioned in Section \ref{sec:bicregularization}, for $n=200$, the grid for $k$ was $\{4;5;\dots;13\}$, for $n=400$, tit was $\{4;5;\dots;14\}$ and, for $n=600$, it was $\{4;5;\dots;15\}$. For selecting the penalty parameters, an adaptive procedure was considered. The robust estimators proposed in \citep{boente:martinez:2023} were used as initial estimators for the robust proposal and the least-squares version for the LS$-$approach. Then, the $\mbox{RBIC}_{\blach}$ was minimized over the grid $\{(\wtlam_1,\wtlam_2)\,:\, \wtlam_1\in\{0.05;0.1;\dots;0.35\}, \wtlam_2\in\{0.20;0.25;\dots;0.50\}\}$ for $n=200$ and both robust and LS$-$estimators, for $n=400$ and $n=600$, the robust estimator was minimized over $\{(\wtlam_1,\wtlam_2)\,:\, \wtlam_1\in\{0.15;0.2;\dots;0.35\}, \wtlam_2\in\{0.05;0.1;\dots;0.25\}\}$, while for the estimator based on least squares the grid used was $\{(\wtlam_1,\wtlam_2)\,:\, \wtlam_1\in\{0.05;0.1;\dots;0.35\}, \wtlam_2\in\{0.05;0.1;\dots;0.40\}\}$.

Different measures were used for determining the variable selection results. For each sample, it is computed
\begin{itemize}
\item[] C: the number of zero components correctly estimated to be zero,
\item[] IC: the number of zero components incorrectly estimated to be zero, and
\item[] CF: 1 if the true model is selected  and 0 otherwise.
\end{itemize}
Then, they are averaged over the 500 replications to obtain the final values. All these three measures were calculated for both parametric and nonparametric components separately and also for the complete model, and for both estimators. For instance, for both the parametric and nonparametric parts of the models, the closer it is the C measure to 6, the better. However, when the C measure is considered for the whole model, the closer to 12, the better. In all cases, the closer the IC measure is to 0 and the CF measure is to 1, the better. 

%the proportion of times the correct variables are selected.

To evaluate the performance of the parametric components, for each replication, we calculated the generalized mean square error (GMSE) defined as
$$\mbox{GMSE}=(\wbbe-\bbe)\trasp \Sigma (\wbbe-\bbe)$$
%\textcolor{red}{Ac\'a hac\'es este comentario: Esto esta bien para LS  porque  Z es indepnediente de X y porque la covarianza de $\hat{\beta}$ es $\Sigma^{-1}$ pero en el robusto deberia aparecer otro termino que es como en nuestro trabajo y depende de $\psi$. Deberias mencionarlo.\\
%No entiendo. Estamos intentando estimar algo ac\'a? Mir\'a que la $\Sigma$ es la del modelo (no est\'a estimada). Lo que puedo hacer, es cambiar la medida. Yo us\'e esta medida porque la usaban en otros trabajos (como medida \textit{global} para los betas). Nosotras no hacemos esto en el paper (mostramos el comportamiento de los betas por separado, digamos).}
where $\Sigma\in\real^{q\times q}$ is the true variance-covariance matrix of $\bZ$. While for evaluating the performance of the estimator of the nonparametric components, we used the square root of the average square error (RASE) defined as
%$$\mbox{RASE}(\eta_j)=\sqrt{\frac{1}{m}\sum_{k=1}^m \|\weta_j(t_k)-\eta_j(t_k)\|^2}$$
$$\mbox{RASE}=\sqrt{\frac{1}{m}\sum_{j=1}^p \sum_{k=1}^m \|\weta_j(t_k)-\eta_j(t_k)\|^2}$$
where $\{t_k\}_{k=1}^m$ is a grid of $m=1000$ equidistant points in the interval $[0,1]$. Then, once again, they are averaged over replications to obtain the final values and also, in order to visualize the variability, it is calculated the standard deviation fot the GMSE and the RASE measures.

Both estimation measures, the GMSE for the parametric part and the RASE for the nonparametric components, were considered for the oracle estimators, that is, for the estimators obtained when using only the non-zero components.% The corresponding notations are Oracle(GMSE) and Oracle(RASE).

Table \ref{tab:Selection} shows the measures obtained for the least squares and robust variable selection procedures for the parametric and nonparametric parts and for the complete model, under all contamination cases and the three sample sizes. When no contaminated data and for all sample sizes considered, both estimators throw similar values of the C, IC and CF measures. It can be appreciated that, in some cases, the robust approach performs slightly better than the LS$-$estimator, probably due to the more stable algorithm used to compute the robust estimator. For $C_1$ to $C_7$, the least squares estimator throws much smaller values of the C and CF measures. By analizing the C measure, for instance, for the linear part of the model and under $C_4$ and $C_5$, the C measure of the least squares estimator is reduced in more than a 42\% with respect ot the $C_0$ for all sample sizes.
A significant reduction in the C measure can also be observed for both the additive part and the complete model for the LS$-$estimator. For instance, for the complete model, under contamination $C_4$, the C values have been reduced to less than a 35\% of the values obtained under the noncontaminated setting.
 In a more extreme way, for contamination $C_7$, the least squares estimator throws values of the C measure that are less than the 29\% of the values obtained for noncontaminated data. On the other hand, across all the contamination scenarios and for all sample sizes, the robust approach shows a more stable behaviour, throwing values of the C measure close to the ones obtain under $C_0$. A similar analysis can be done for the CF measure for the three parts of the model. In this case, it is also even more notorious the bad performance of the LS$-$estimator across all the scenarios where, in some cases, the correct variables are never or almost never selected. With respect to the linear part of the model, when the sample size is $200$, the correct variables are almost never selected under schemes $C_4$ to $C_7$, and never or almost never under $C_5$ to $C_7$ for sample sizes $400$ and $600$. For the additive part, when $n=200$  and under $C_4$ to $C_7$, the values of the CF measure of the least squares estimator are close to $0$. Moreover, under contaminations $C_5$ to $C_7$, for the complete model, the values of the CF obtained by the LS$-$estimator under all the sample sizes considered are zero. This means that the components of whole model are never selected all at once. In contrast, the robust approach performs quite similarly across all the sample sizes and all contamination settings, yielding C measure values close to 6 for the linear and additive parts and to 12 for the complete model, and values of the CF measure close to 1, especially when the sample size increases. With respect to the IC measure, these values are 0 or close to 0 for both estimators and under all the different scenarios probably meaning that it is hard for the procedures to indicate as zero variables that are important to the model.

\begin{table}[ht!]
\begin{center}
\scriptsize
\begin{tabular}{|ccc|ccc|ccc|ccc|}
  \hline
\multicolumn{2}{|c}{} & \multicolumn{1}{c|}{} & \multicolumn{3}{c|}{Linear part} & \multicolumn{3}{c|}{Additive part}  & \multicolumn{3}{c|}{Complete model}\\
  \hline
$n$ & \textsc{Cont} & \textsc{Method} & \textsc{C} & \textsc{IC} & \textsc{CF} & \textsc{C} & \textsc{IC} & \textsc{CF} & \textsc{C} & \textsc{IC} & \textsc{CF} \\ 
  \hline
200 & $C_0$ & \textsc{ls} & 5.12 & 0.00 & 0.76 & 5.95 & 0.00 & 0.95 & 11.07 & 0.00 & 0.73 \\ 
  200 & $C_0$ & \textsc{rob} & 5.55 & 0.00 & 0.88 & 5.92 & 0.00 & 0.93 & 11.47 & 0.00 & 0.84 \\ 
  200 & $C_1$ & \textsc{ls} & 3.78 & 0.00 & 0.39 & 5.02 & 0.01 & 0.39 & 8.80 & 0.01 & 0.16 \\ 
  200 & $C_1$ & \textsc{rob} & 5.38 & 0.00 & 0.87 & 5.94 & 0.03 & 0.91 & 11.32 & 0.03 & 0.81 \\ 
  200 & $C_2$ & \textsc{ls} & 3.31 & 0.00 & 0.29 & 4.82 & 0.01 & 0.29 & 8.13 & 0.01 & 0.10 \\ 
  200 & $C_2$ & \textsc{rob} & 5.51 & 0.00 & 0.90 & 5.94 & 0.01 & 0.95 & 11.45 & 0.01 & 0.86 \\ 
  200 & $C_3$ & \textsc{ls} & 2.98 & 0.00 & 0.19 & 3.80 & 0.03 & 0.11 & 6.78 & 0.03 & 0.03 \\ 
  200 & $C_3$ & \textsc{rob} & 5.65 & 0.00 & 0.90 & 5.93 & 0.00 & 0.94 & 11.58 & 0.00 & 0.86 \\ 
  200 & $C_4$ & \textsc{ls} & 2.96 & 0.01 & 0.07 & 2.11 & 0.06 & 0.01 & 5.07 & 0.07 & 0.00 \\ 
  200 & $C_4$ & \textsc{rob} & 5.79 & 0.00 & 0.94 & 5.95 & 0.00 & 0.96 & 11.74 & 0.00 & 0.90 \\ 
  200 & $C_5$ & \textsc{ls} & 2.91 & 0.04 & 0.01 & 0.72 & 0.06 & 0.00 & 3.64 & 0.10 & 0.00 \\ 
  200 & $C_5$ & \textsc{rob} & 5.75 & 0.00 & 0.95 & 5.99 & 0.02 & 0.97 & 11.74 & 0.02 & 0.93 \\ 
  200 & $C_6$ & \textsc{ls} & 1.37 & 0.01 & 0.00 & 2.95 & 0.03 & 0.01 & 4.32 & 0.04 & 0.00 \\ 
  200 & $C_6$ & \textsc{rob} & 5.59 & 0.00 & 0.90 & 5.91 & 0.00 & 0.93 & 11.50 & 0.00 & 0.86 \\ 
  200 & $C_7$ & \textsc{ls} & 1.44 & 0.01 & 0.00 & 2.62 & 0.03 & 0.00 & 4.06 & 0.04 & 0.00 \\ 
  200 & $C_7$ & \textsc{rob} & 5.60 & 0.02 & 0.85 & 5.89 & 0.01 & 0.94 & 11.49 & 0.03 & 0.82 \\ \hline
  400 & $C_0$ & \textsc{ls} & 5.64 & 0.00 & 0.92 & 6.00 & 0.00 & 1.00 & 11.64 & 0.00 & 0.92 \\ 
  400 & $C_0$ & \textsc{rob} & 5.79 & 0.00 & 0.95 & 5.99 & 0.00 & 1.00 & 11.79 & 0.00 & 0.95 \\ 
  400 & $C_1$ & \textsc{ls} & 4.64 & 0.00 & 0.66 & 5.77 & 0.00 & 0.80 & 10.41 & 0.00 & 0.54 \\ 
  400 & $C_1$ & \textsc{rob} & 5.87 & 0.00 & 0.98 & 6.00 & 0.00 & 1.00 & 11.87 & 0.00 & 0.97 \\ 
  400 & $C_2$ & \textsc{ls} & 4.37 & 0.00 & 0.56 & 5.62 & 0.00 & 0.65 & 9.99 & 0.00 & 0.37 \\ 
  400 & $C_2$ & \textsc{rob} & 5.94 & 0.00 & 0.99 & 6.00 & 0.00 & 1.00 & 11.94 & 0.00 & 0.99 \\ 
  400 & $C_3$ & \textsc{ls} & 3.50 & 0.00 & 0.37 & 5.10 & 0.00 & 0.36 & 8.59 & 0.00 & 0.17 \\ 
  400 & $C_3$ & \textsc{rob} & 5.85 & 0.00 & 0.97 & 5.99 & 0.00 & 1.00 & 11.84 & 0.00 & 0.97 \\ 
  400 & $C_4$ & \textsc{ls} & 2.87 & 0.00 & 0.17 & 3.86 & 0.02 & 0.04 & 6.73 & 0.02 & 0.01 \\ 
  400 & $C_4$ & \textsc{rob} & 5.95 & 0.00 & 0.99 & 5.99 & 0.00 & 1.00 & 11.94 & 0.00 & 0.99 \\ 
  400 & $C_5$ & \textsc{ls} & 2.95 & 0.00 & 0.05 & 1.68 & 0.06 & 0.00 & 4.63 & 0.06 & 0.00 \\ 
  400 & $C_5$ & \textsc{rob} & 5.96 & 0.00 & 0.99 & 6.00 & 0.00 & 1.00 & 11.96 & 0.00 & 0.99 \\ 
  400 & $C_6$ & \textsc{ls} & 0.79 & 0.00 & 0.00 & 4.73 & 0.01 & 0.18 & 5.52 & 0.01 & 0.00 \\ 
  400 & $C_6$ & \textsc{rob} & 5.93 & 0.00 & 0.98 & 5.98 & 0.00 & 0.99 & 11.92 & 0.00 & 0.98 \\ 
  400 & $C_7$ & \textsc{ls} & 0.89 & 0.00 & 0.00 & 4.53 & 0.01 & 0.14 & 5.42 & 0.01 & 0.00 \\ 
  400 & $C_7$ & \textsc{rob} & 5.90 & 0.00 & 0.97 & 6.00 & 0.00 & 1.00 & 11.90 & 0.00 & 0.96 \\ \hline
  600 & $C_0$ & \textsc{ls} & 5.88 & 0.00 & 0.97 & 6.00 & 0.00 & 1.00 & 11.88 & 0.00 & 0.97 \\ 
  600 & $C_0$ & \textsc{rob} & 5.94 & 0.00 & 0.99 & 6.00 & 0.00 & 1.00 & 11.94 & 0.00 & 0.99 \\ 
  600 & $C_1$ & \textsc{ls} & 5.11 & 0.00 & 0.78 & 5.88 & 0.00 & 0.88 & 10.99 & 0.00 & 0.70 \\ 
  600 & $C_1$ & \textsc{rob} & 5.94 & 0.00 & 0.99 & 6.00 & 0.00 & 1.00 & 11.94 & 0.00 & 0.99 \\ 
  600 & $C_2$ & \textsc{ls} & 4.97 & 0.00 & 0.75 & 5.89 & 0.00 & 0.89 & 10.86 & 0.00 & 0.67 \\ 
  600 & $C_2$ & \textsc{rob} & 5.95 & 0.00 & 0.99 & 6.00 & 0.00 & 1.00 & 11.95 & 0.00 & 0.99 \\ 
  600 & $C_3$ & \textsc{ls} & 4.28 & 0.00 & 0.54 & 5.56 & 0.00 & 0.63 & 9.84 & 0.00 & 0.36 \\ 
  600 & $C_3$ & \textsc{rob} & 5.95 & 0.00 & 0.99 & 6.00 & 0.00 & 1.00 & 11.95 & 0.00 & 0.99 \\ 
  600 & $C_4$ & \textsc{ls} & 3.14 & 0.00 & 0.27 & 4.66 & 0.01 & 0.16 & 7.80 & 0.01 & 0.05 \\ 
  600 & $C_4$ & \textsc{rob} & 5.92 & 0.00 & 0.98 & 5.99 & 0.00 & 1.00 & 11.91 & 0.00 & 0.98 \\ 
  600 & $C_5$ & \textsc{ls} & 2.78 & 0.00 & 0.09 & 2.63 & 0.06 & 0.00 & 5.40 & 0.06 & 0.00 \\ 
  600 & $C_5$ & \textsc{rob} & 5.95 & 0.00 & 0.99 & 6.00 & 0.00 & 1.00 & 11.95 & 0.00 & 0.99 \\ 
  600 & $C_6$ & \textsc{ls} & 0.43 & 0.00 & 0.00 & 5.36 & 0.00 & 0.49 & 5.78 & 0.00 & 0.00 \\ 
  600 & $C_6$ & \textsc{rob} & 5.92 & 0.00 & 0.98 & 6.00 & 0.00 & 1.00 & 11.92 & 0.00 & 0.98 \\ 
  600 & $C_7$ & \textsc{ls} & 0.48 & 0.00 & 0.00 & 5.20 & 0.00 & 0.40 & 5.68 & 0.00 & 0.00 \\ 
  600 & $C_7$ & \textsc{rob} & 5.97 & 0.00 & 0.99 & 5.99 & 0.00 & 1.00 & 11.96 & 0.00 & 0.99 \\ 
   \hline
\end{tabular}
\caption{\label{tab:Selection}\footnotesize C, IC, CF obtained for the least squares and robust approaches, for different sample sizes.} 
\end{center}
\end{table}

\begin{figure}[ht!]
 \begin{center}
\small
\begin{tabular}{cc}
$C_0$ & $C_1$ \vspace{-0.5cm}\\
\includegraphics[scale=0.4]{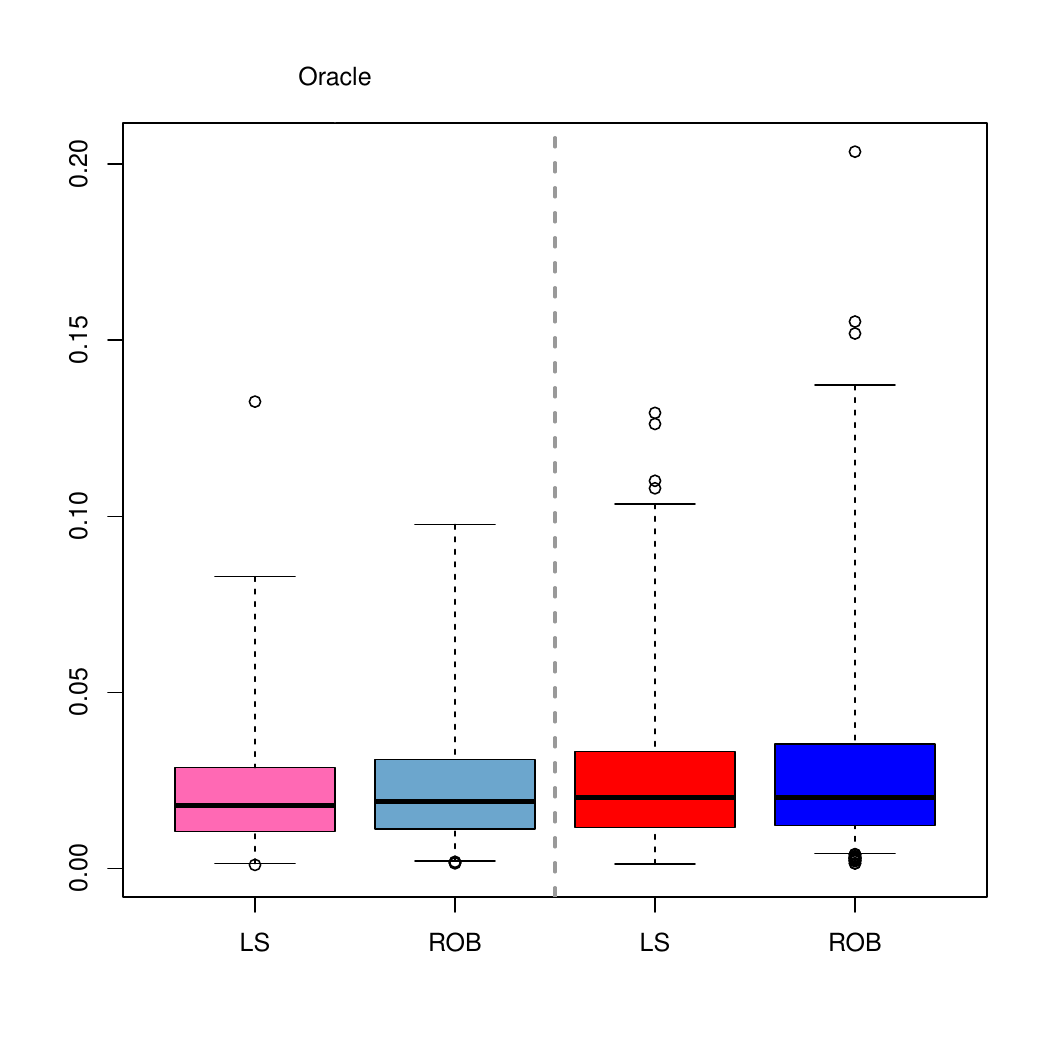} &
\includegraphics[scale=0.4]{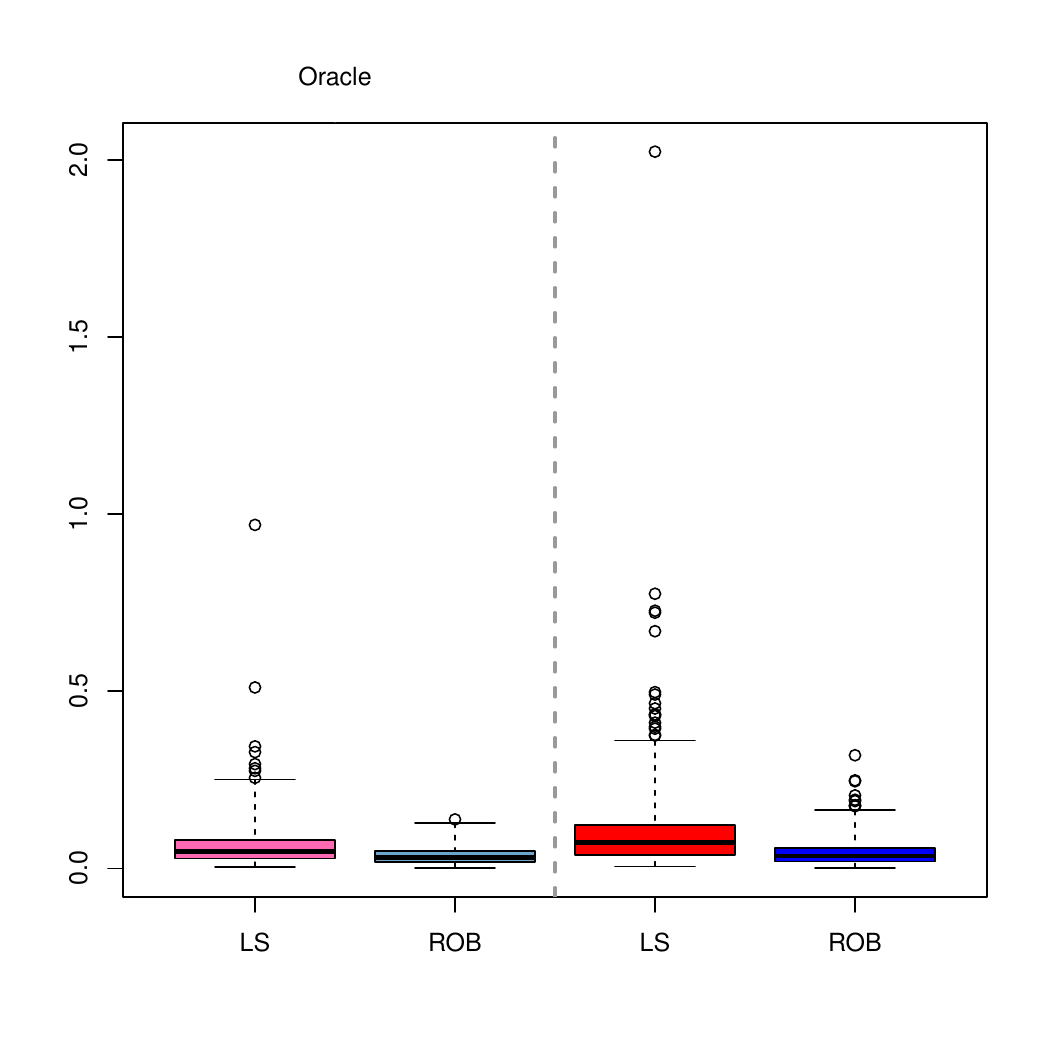} \\
& \\
 $C_2$ &  $C_4$ \vspace{-0.5cm}\\ 
\includegraphics[scale=0.4]{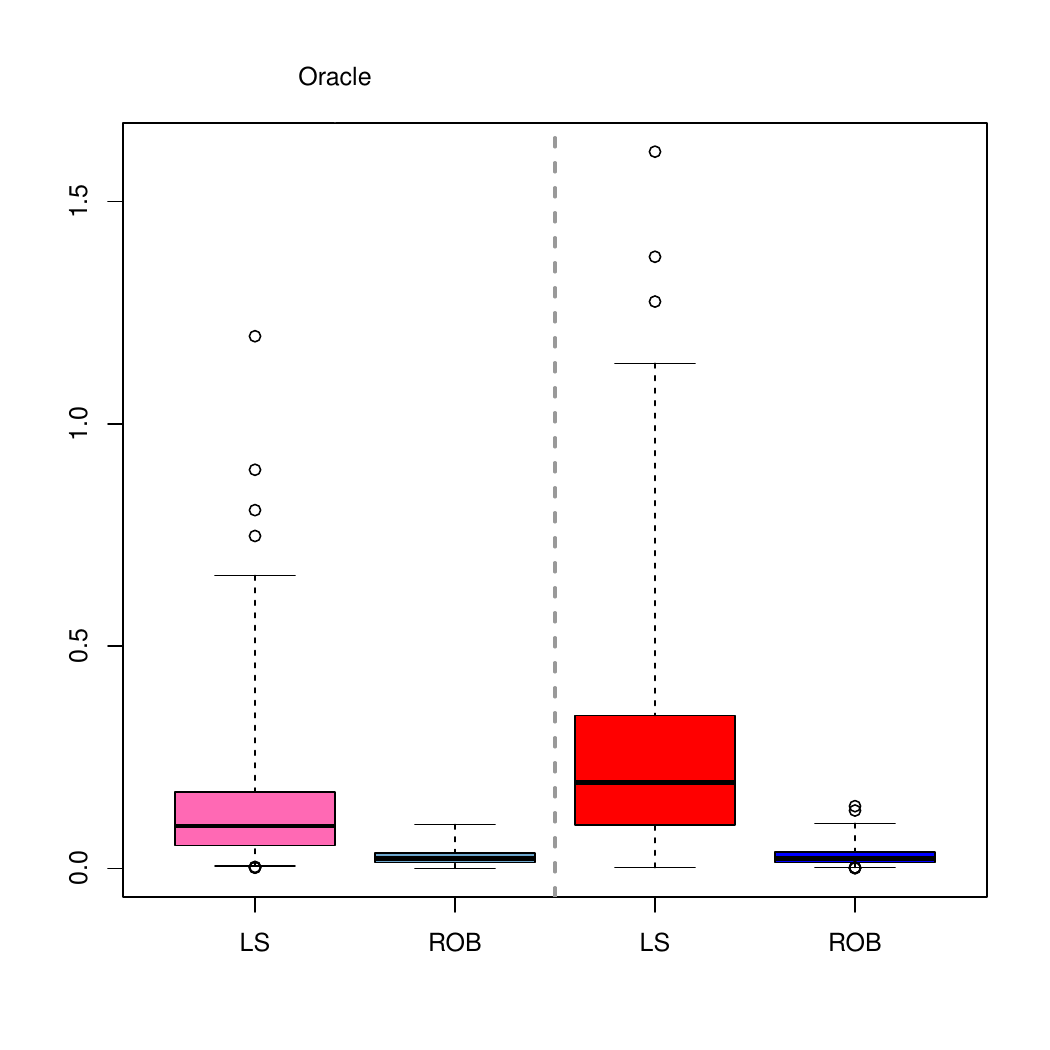}  &
\includegraphics[scale=0.4]{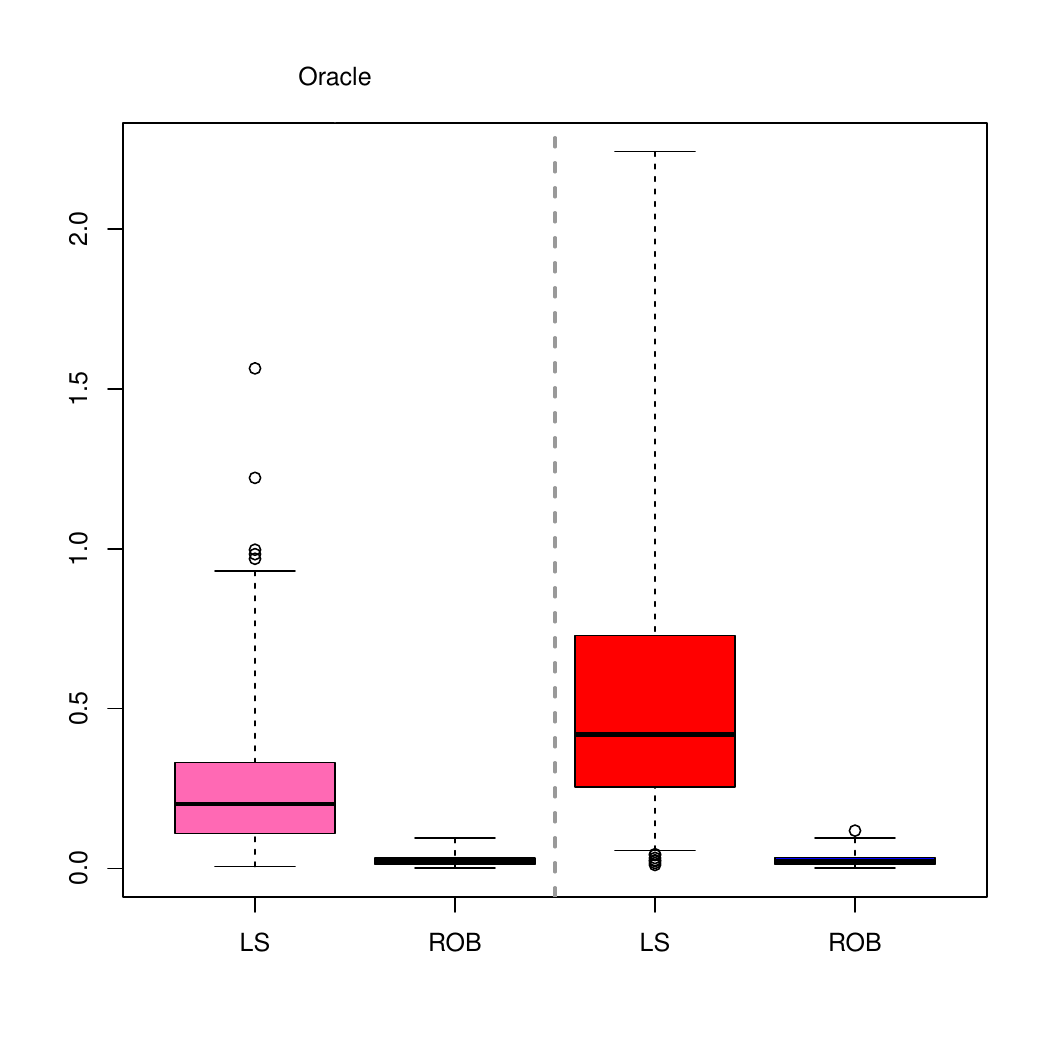}\\
& \\
 $C_7$ & \vspace{-0.5cm} \\
\includegraphics[scale=0.4]{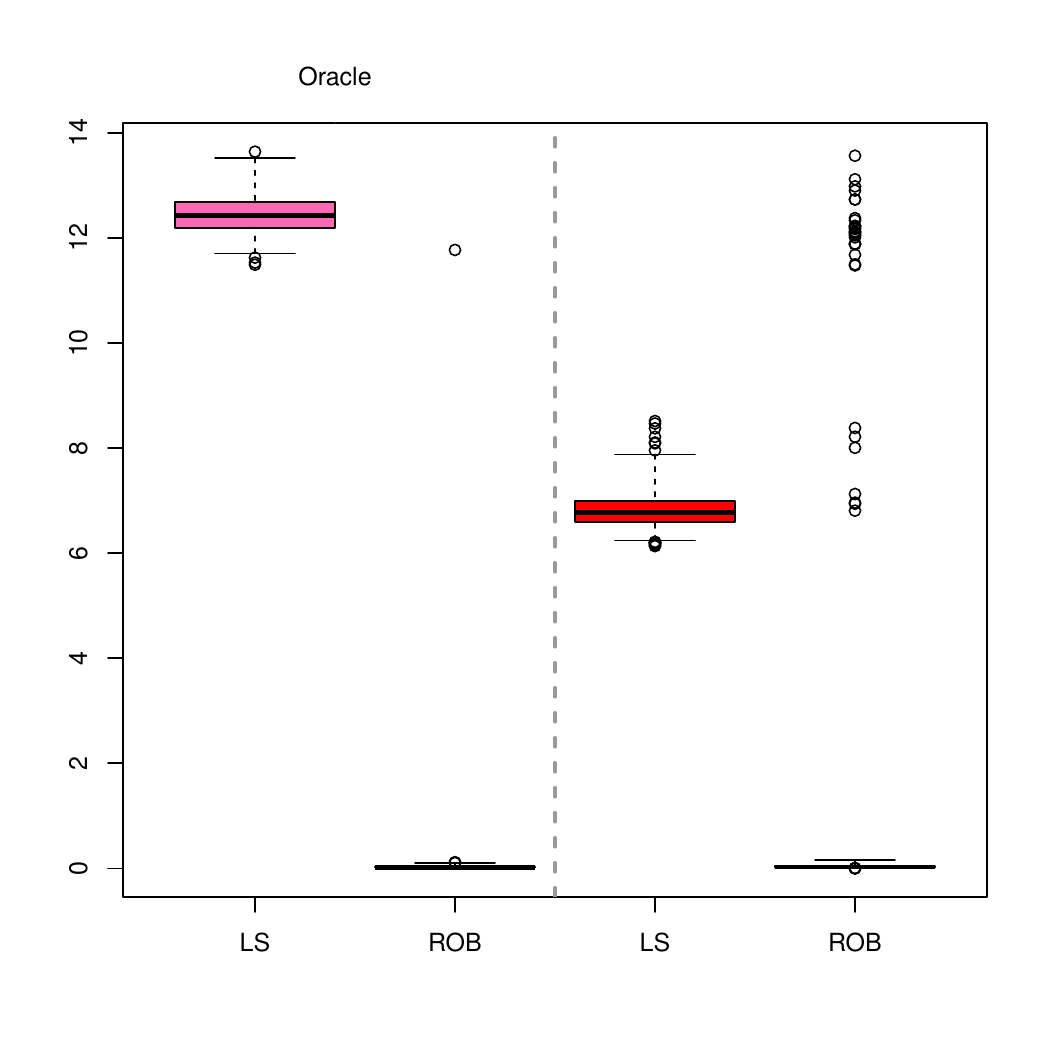} & 
\\ 
\end{tabular}
\vspace{-0.5cm} 
\caption{\label{fig:GMSE-n200}  Adjusted boxplots for the GMSE values under $C_0$, $C_1$, $C_3$, $C_4$ and $C_7$ for both oracle estimators and both penalized estimators for sample size $n=200$.}
\end{center}
\end{figure}

\begin{figure}[ht!]
 \begin{center}
\small
\begin{tabular}{cc}
$C_0$ & $C_1$ \vspace{-0.5cm} \\
\includegraphics[scale=0.4]{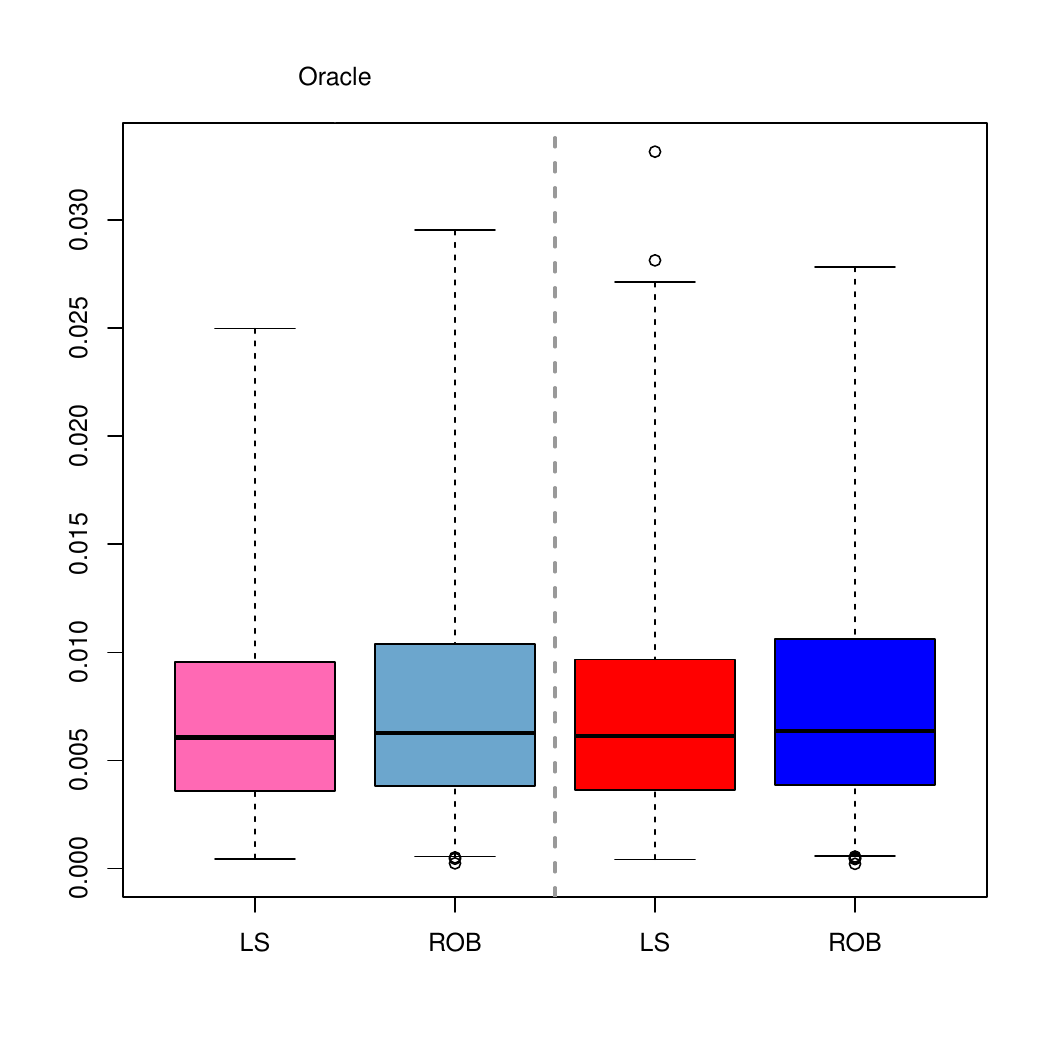} &
\includegraphics[scale=0.4]{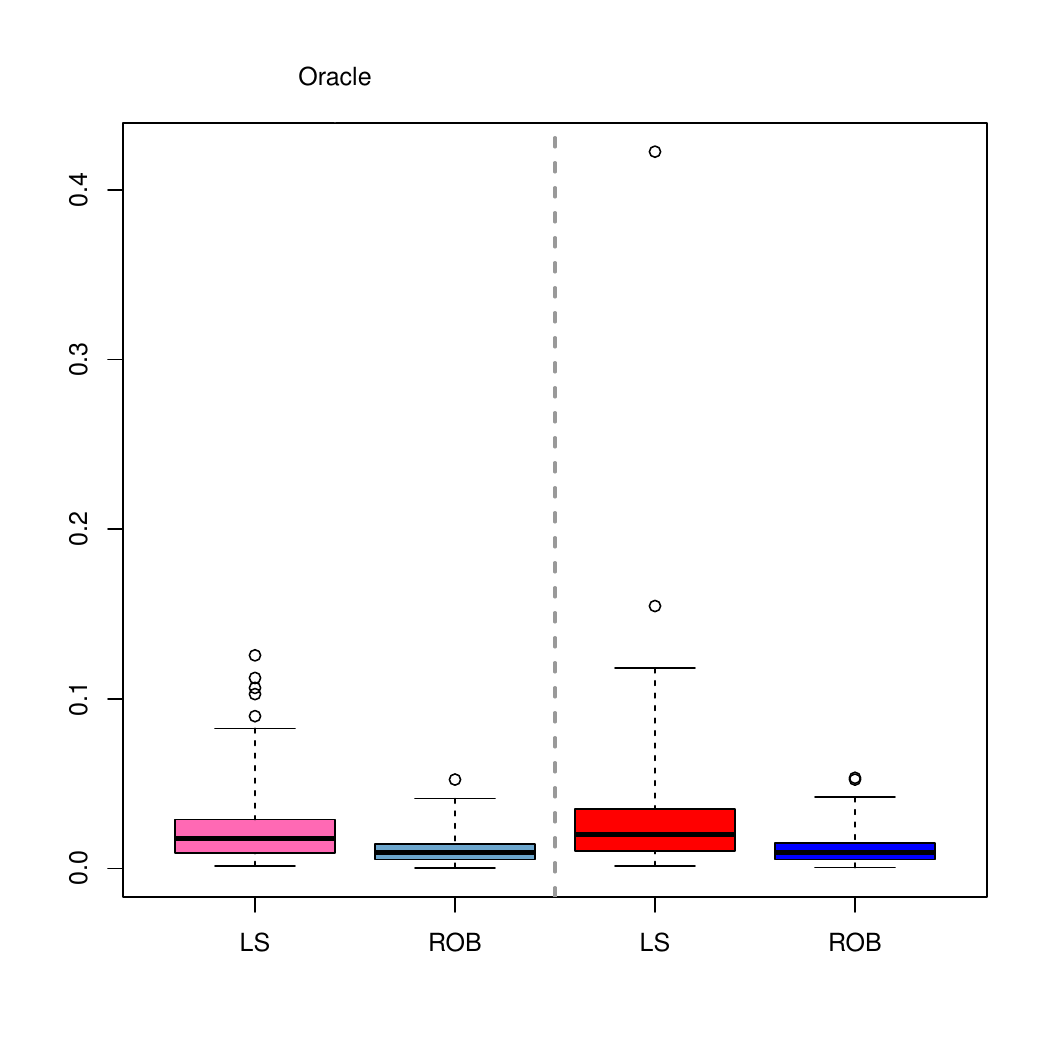} \\
$C_2$ & $C_4$ \vspace{-0.5cm}\\ 
\includegraphics[scale=0.4]{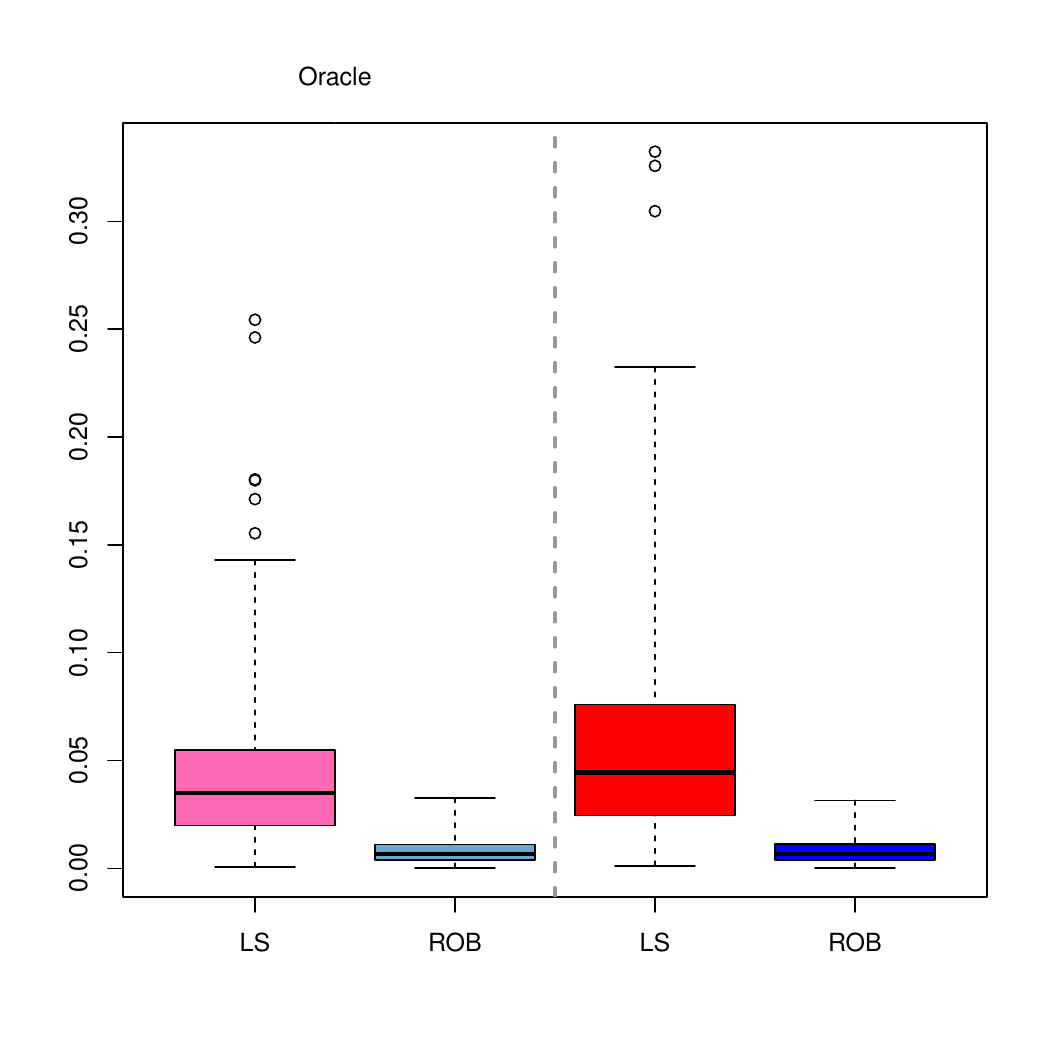} &
\includegraphics[scale=0.4]{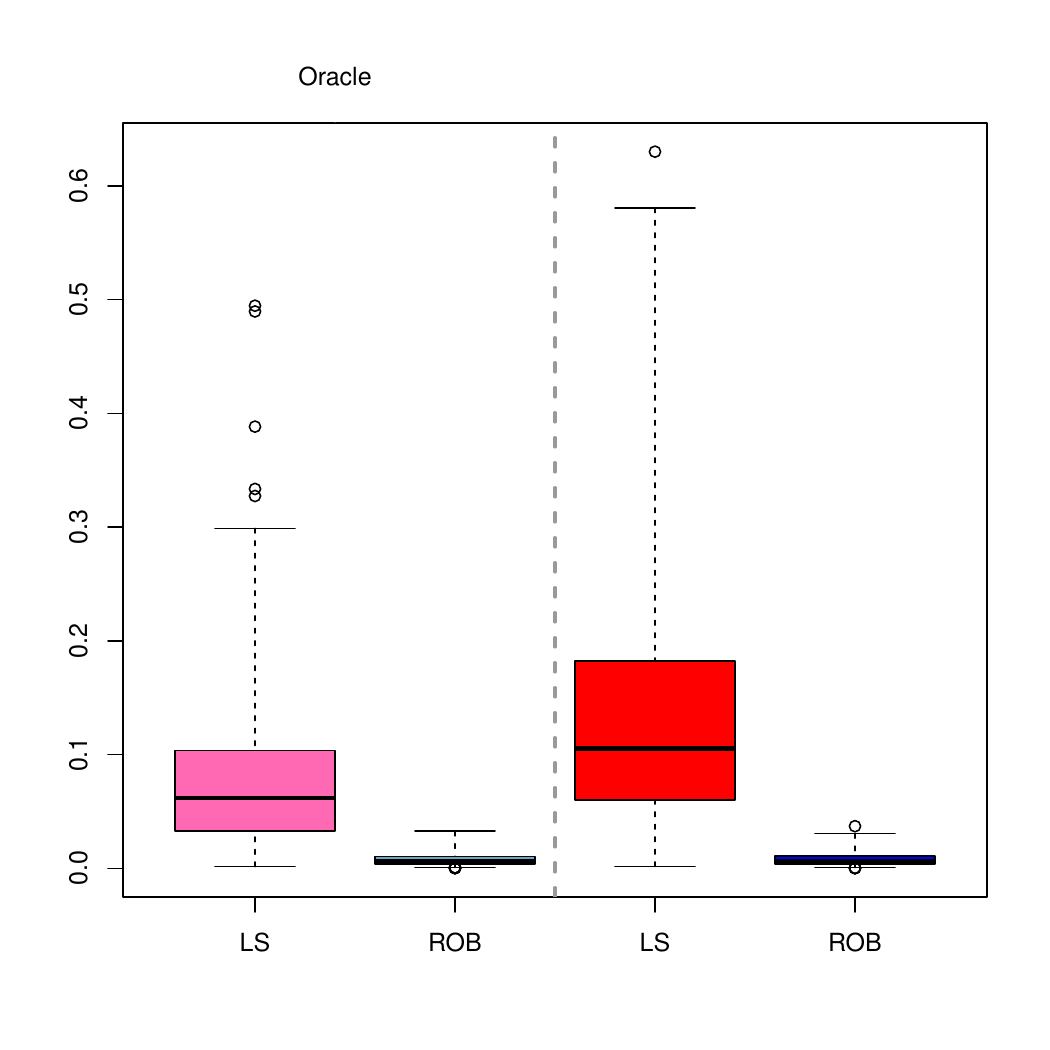} \\
$C_7$ & \vspace{-0.5cm} \\
\includegraphics[scale=0.4]{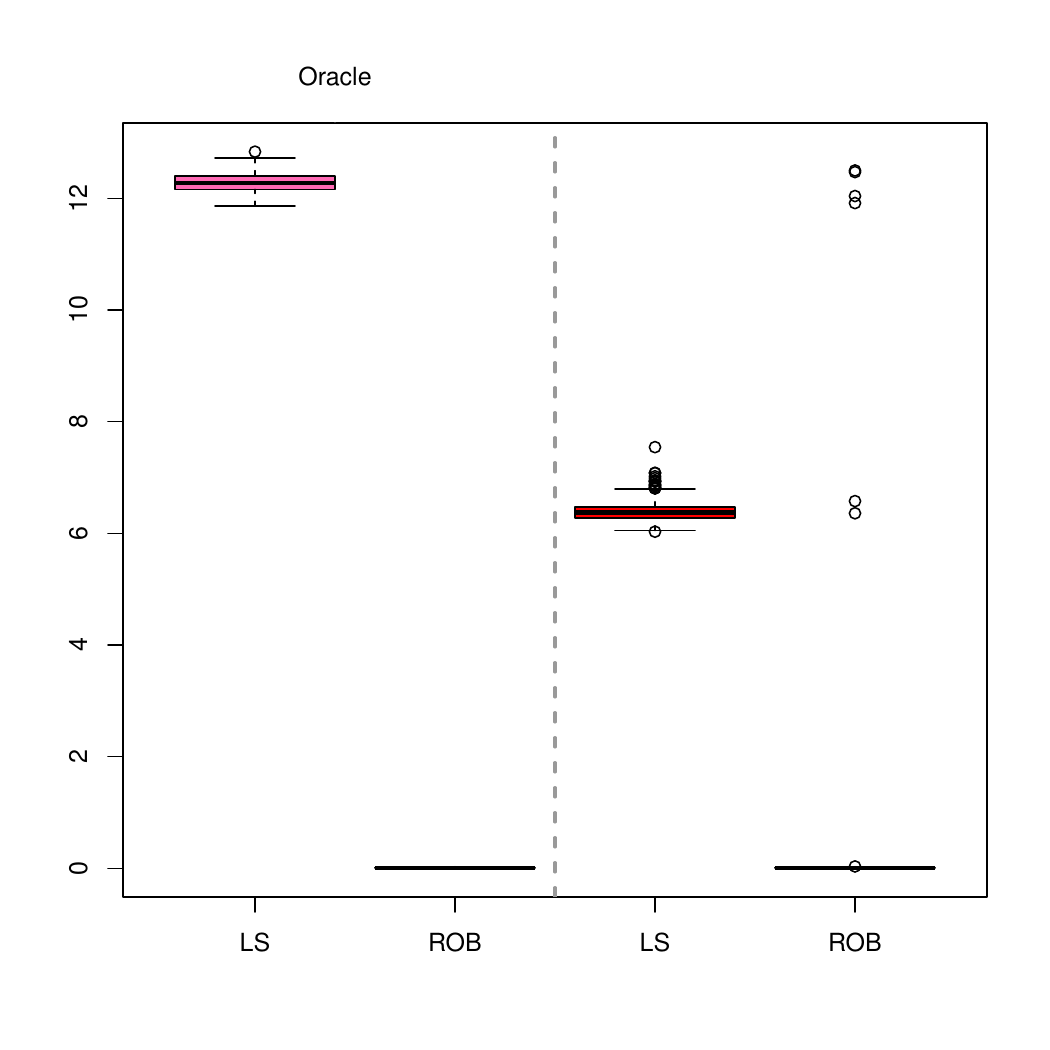} & \\
\end{tabular}
\vspace{-0.5cm} 
\caption{\label{fig:GMSE-n600} Adjusted boxplots for the GMSE values under $C_0$, $C_1$, $C_3$, $C_4$ and $C_7$ for both oracle estimators and both penalized estimators for sample size $n=600$.}
\end{center}
\end{figure}

\begin{figure}[ht!]
 \begin{center}
\small
\begin{tabular}{cc}
$C_0$ & $C_1$ \vspace{-0.5cm}  \\
\includegraphics[scale=0.4]{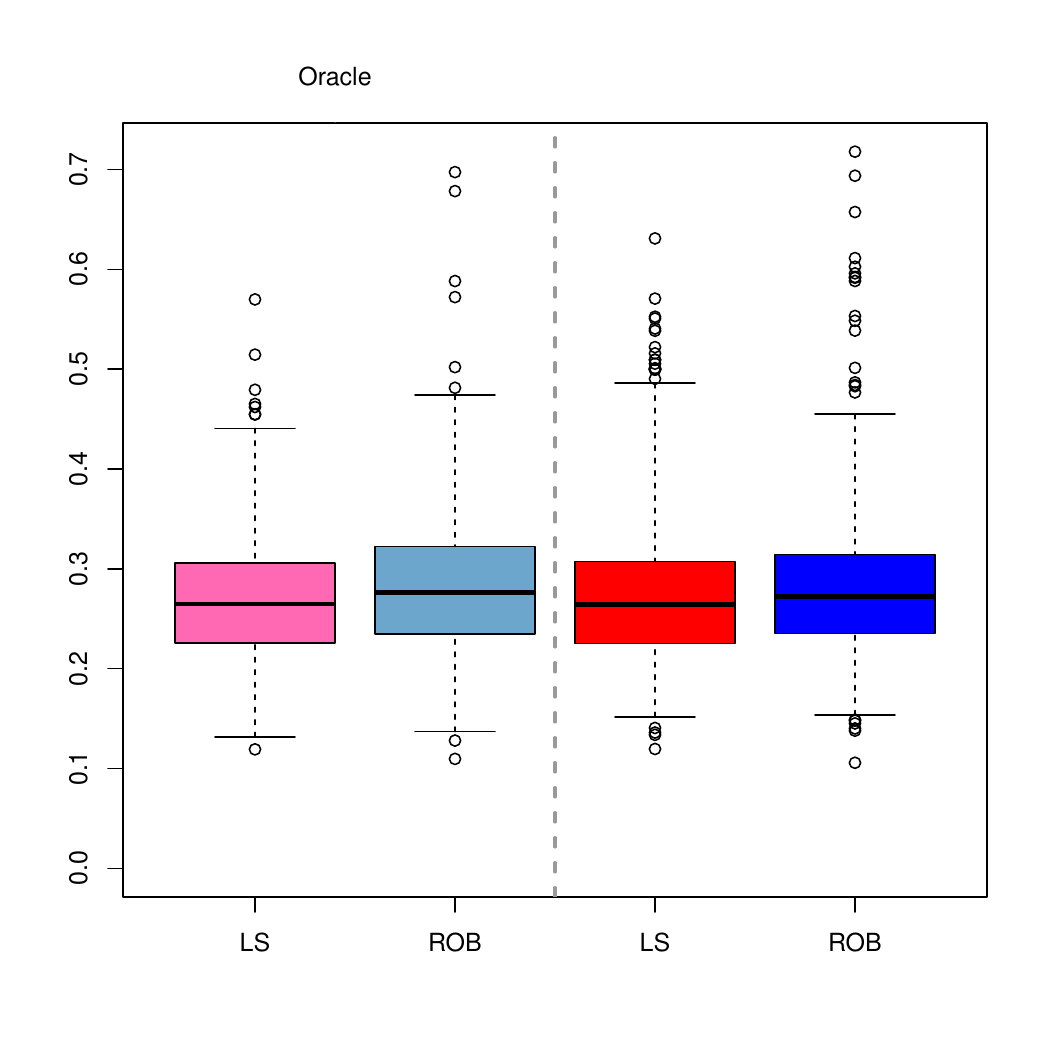} &
\includegraphics[scale=0.4]{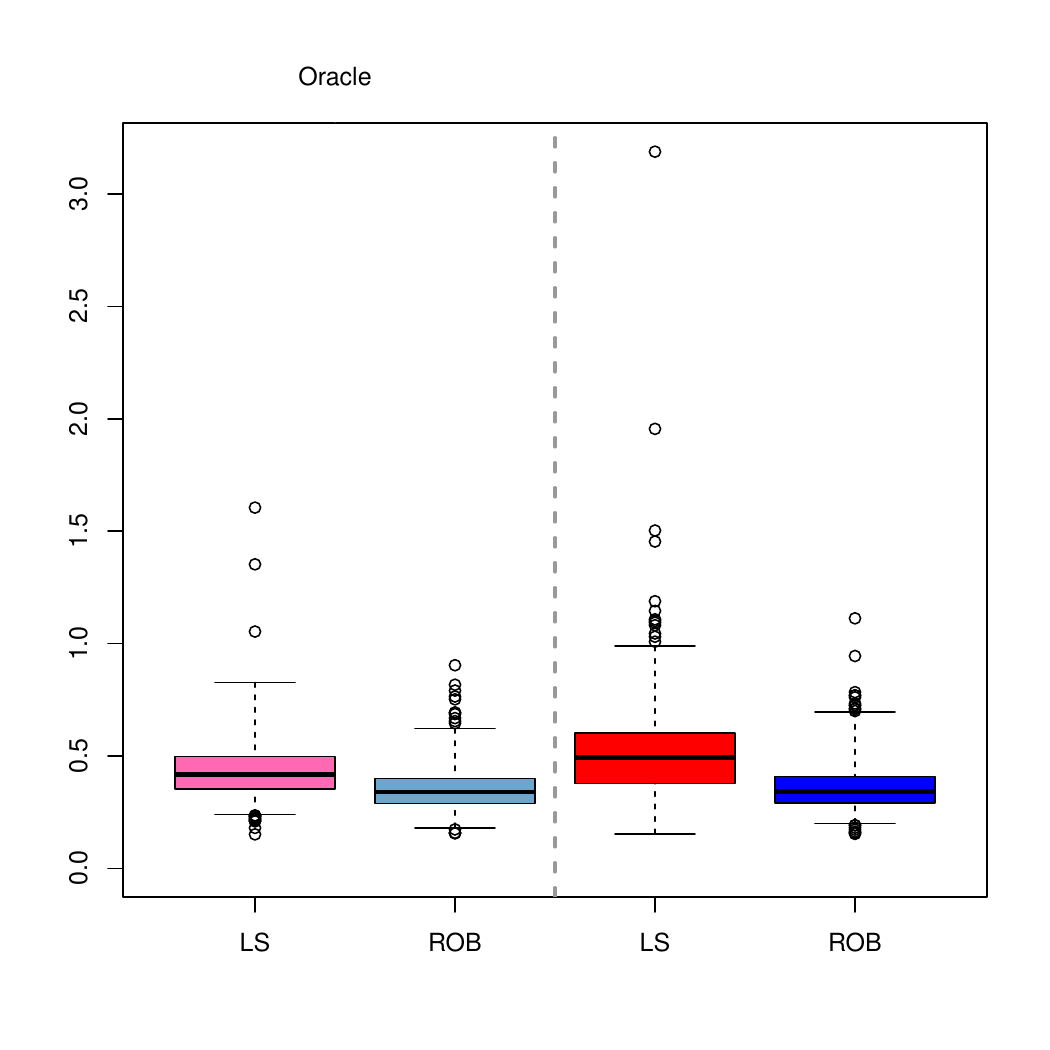} \\
$C_2$  & $C_4$ \vspace{-0.5cm}  \\ 
\includegraphics[scale=0.4]{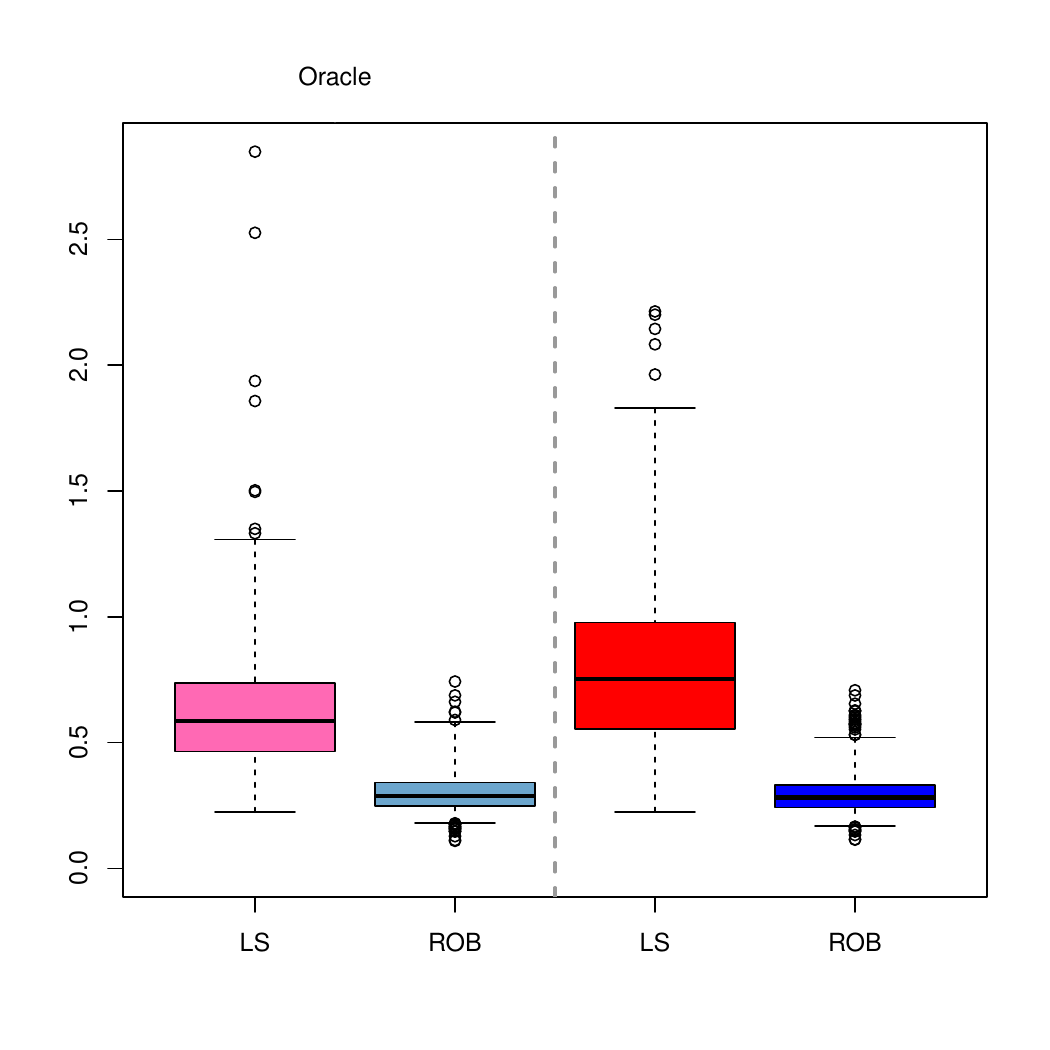}  &
\includegraphics[scale=0.4]{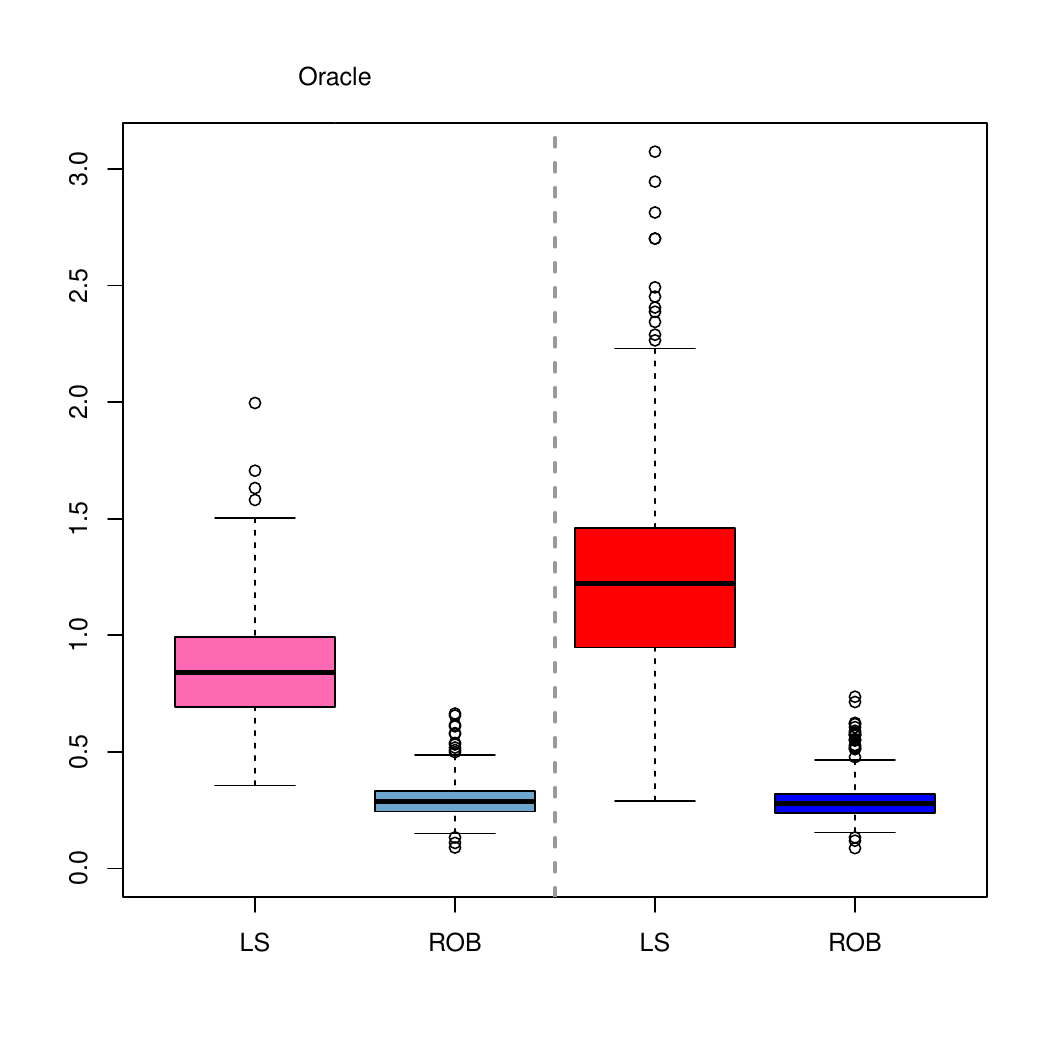} \\
 $C_7$ & \vspace{-0.5cm}  \\
\includegraphics[scale=0.4]{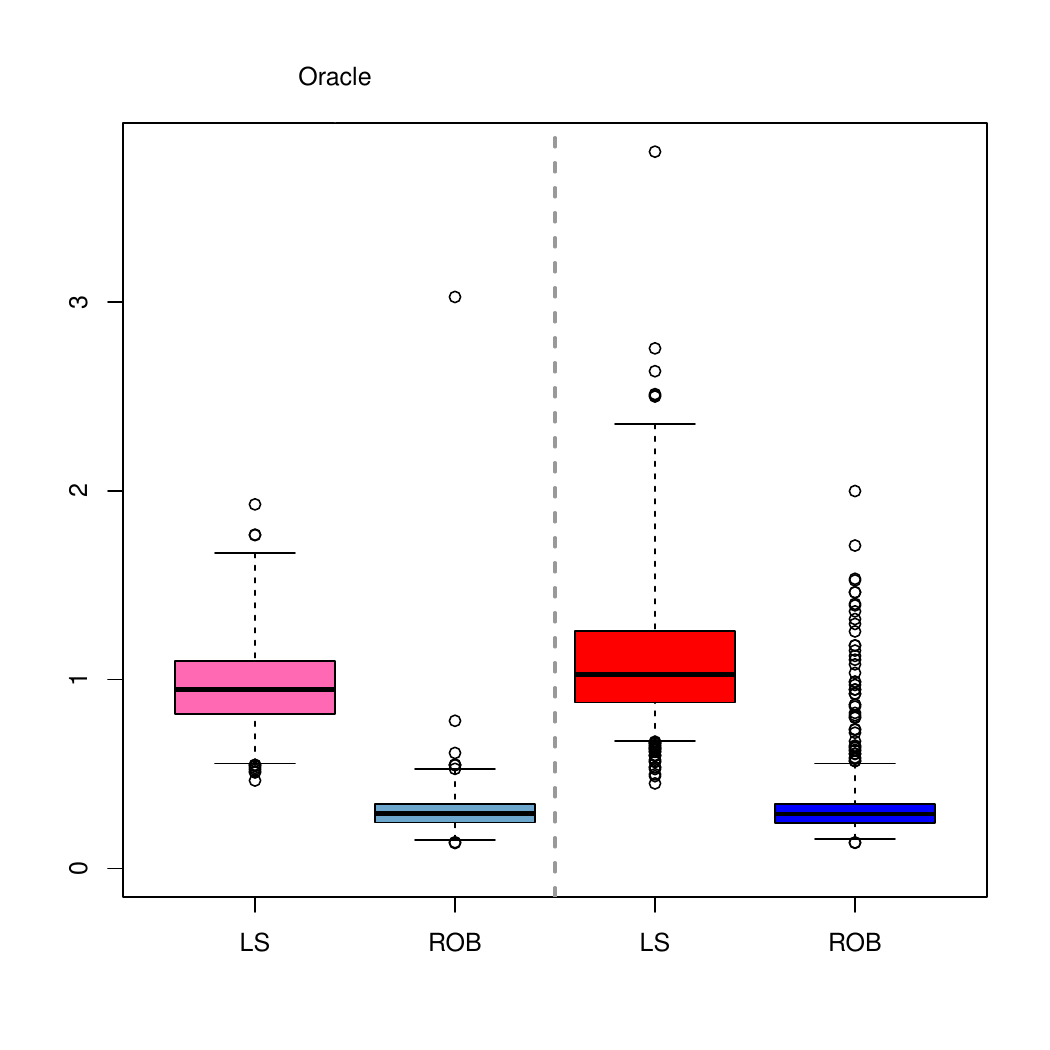} & \\
\end{tabular}
\vspace{-0.5cm} 
\caption{\label{fig:RASE-n200} Adjusted boxplots for the RASE values under $C_0$, $C_1$, $C_3$, $C_4$ and $C_7$ for both oracle estimators and both penalized estimators for sample size $n=200$.}
\end{center}
\end{figure}

\begin{figure}[ht!]
 \begin{center}
\small
\begin{tabular}{cc}
$C_0$ & $C_1$ \vspace{-0.5cm} \\
\includegraphics[scale=0.4]{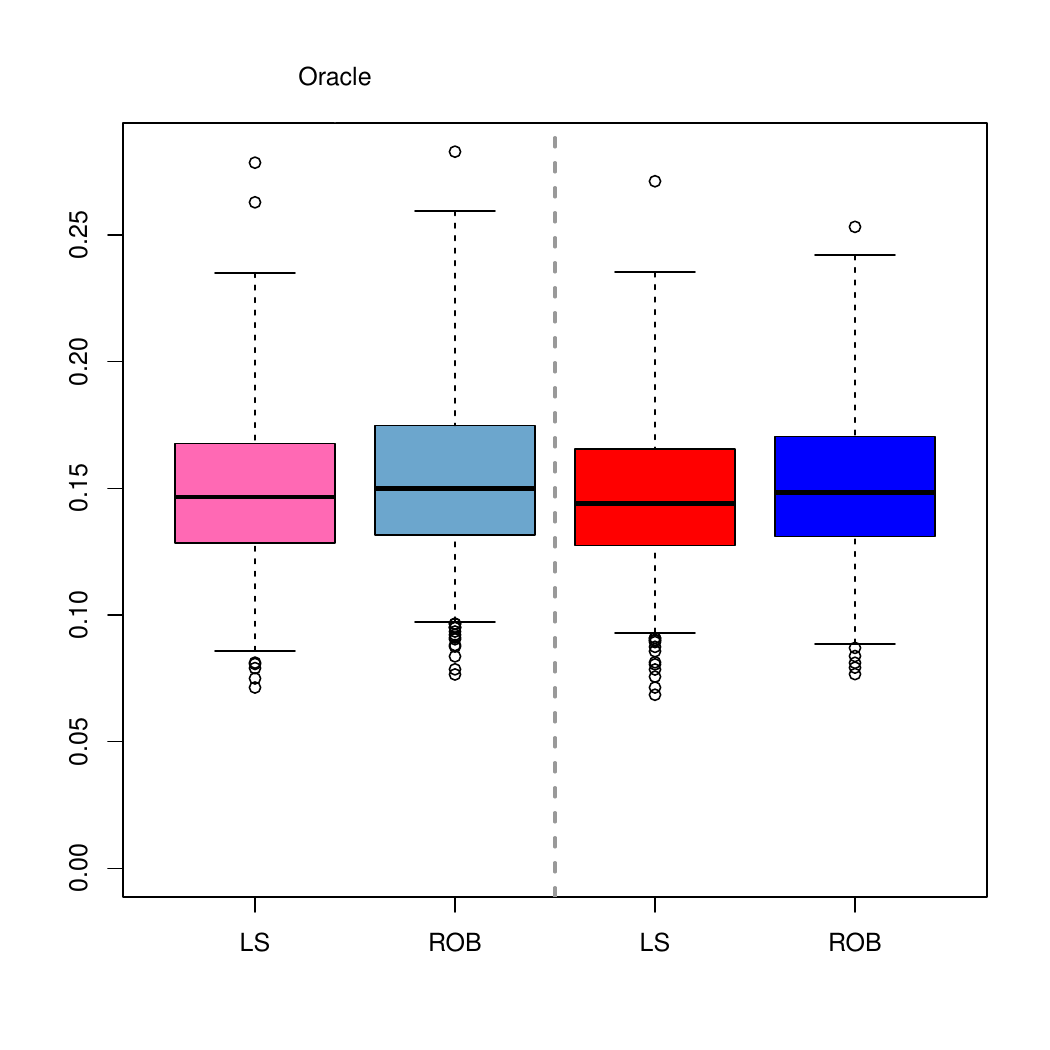} &
\includegraphics[scale=0.4]{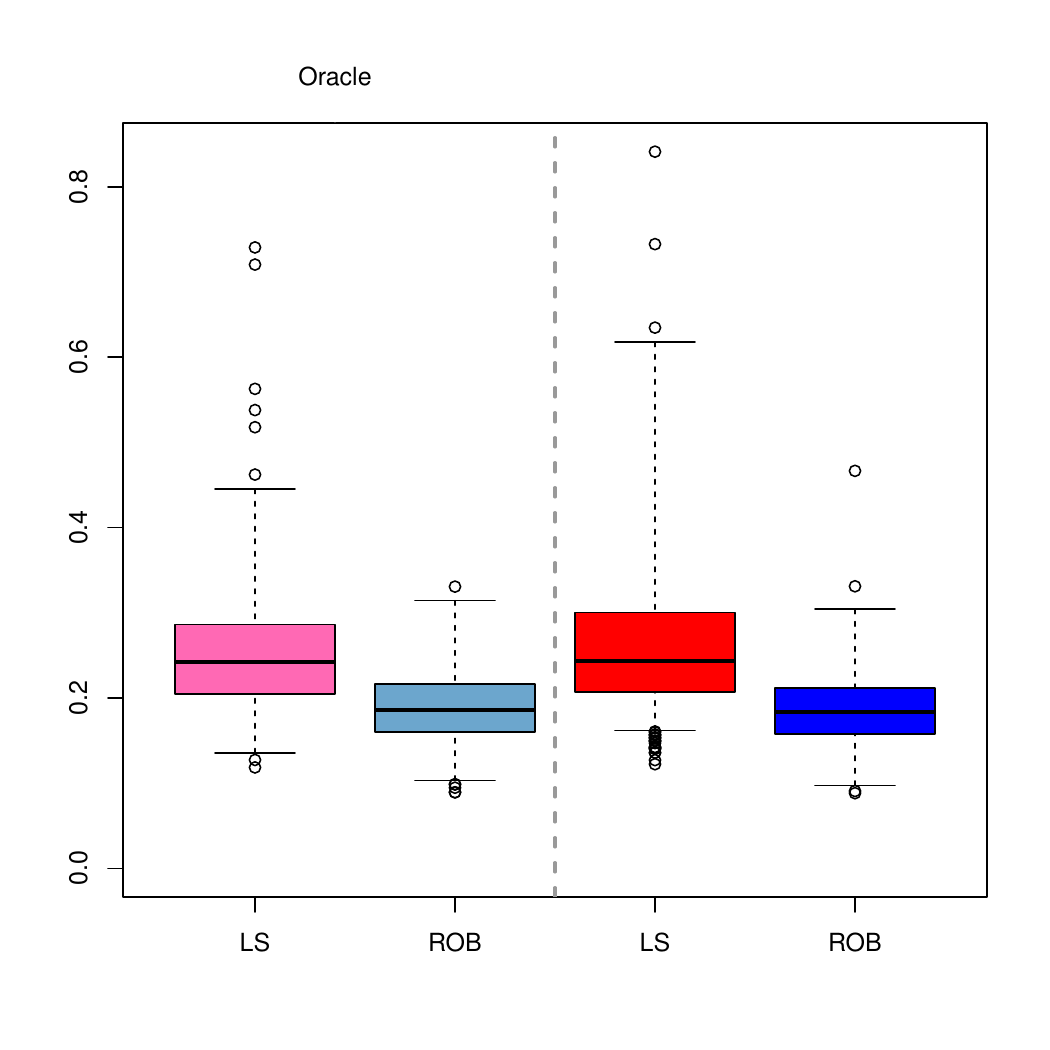} \\
$C_2$  & $C_4$ \vspace{-0.5cm} \\ 
\includegraphics[scale=0.4]{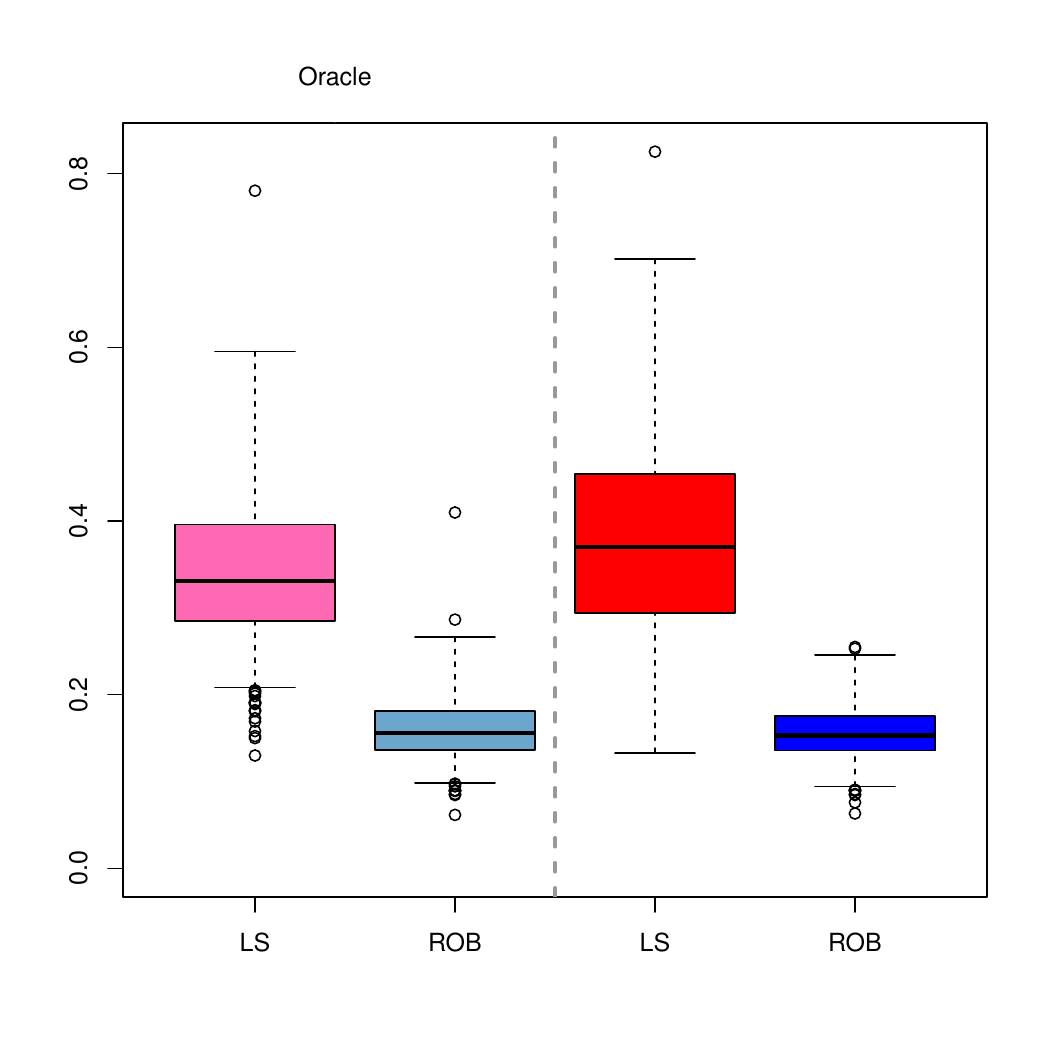} &
 \includegraphics[scale=0.4]{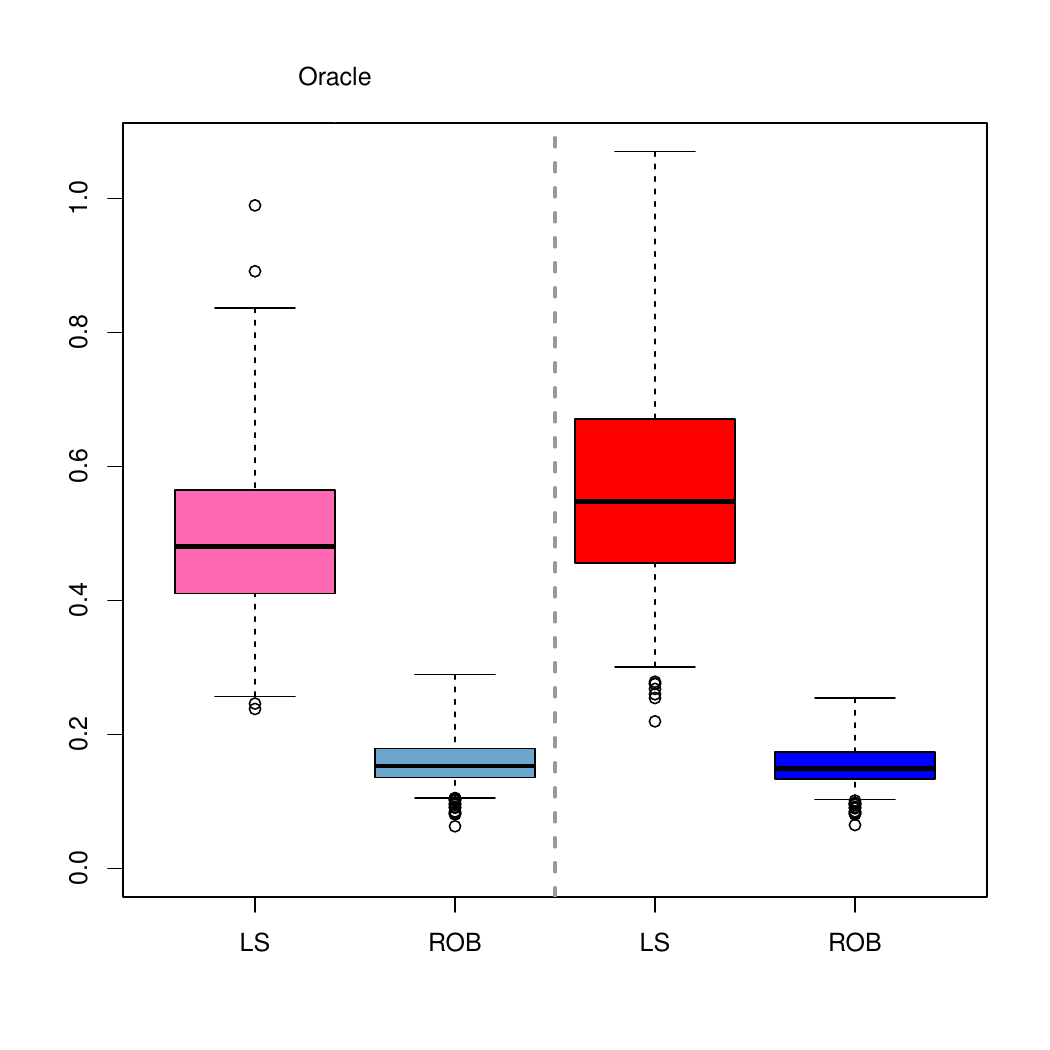} \\
$C_7$ & \vspace{-0.5cm} \\
\includegraphics[scale=0.4]{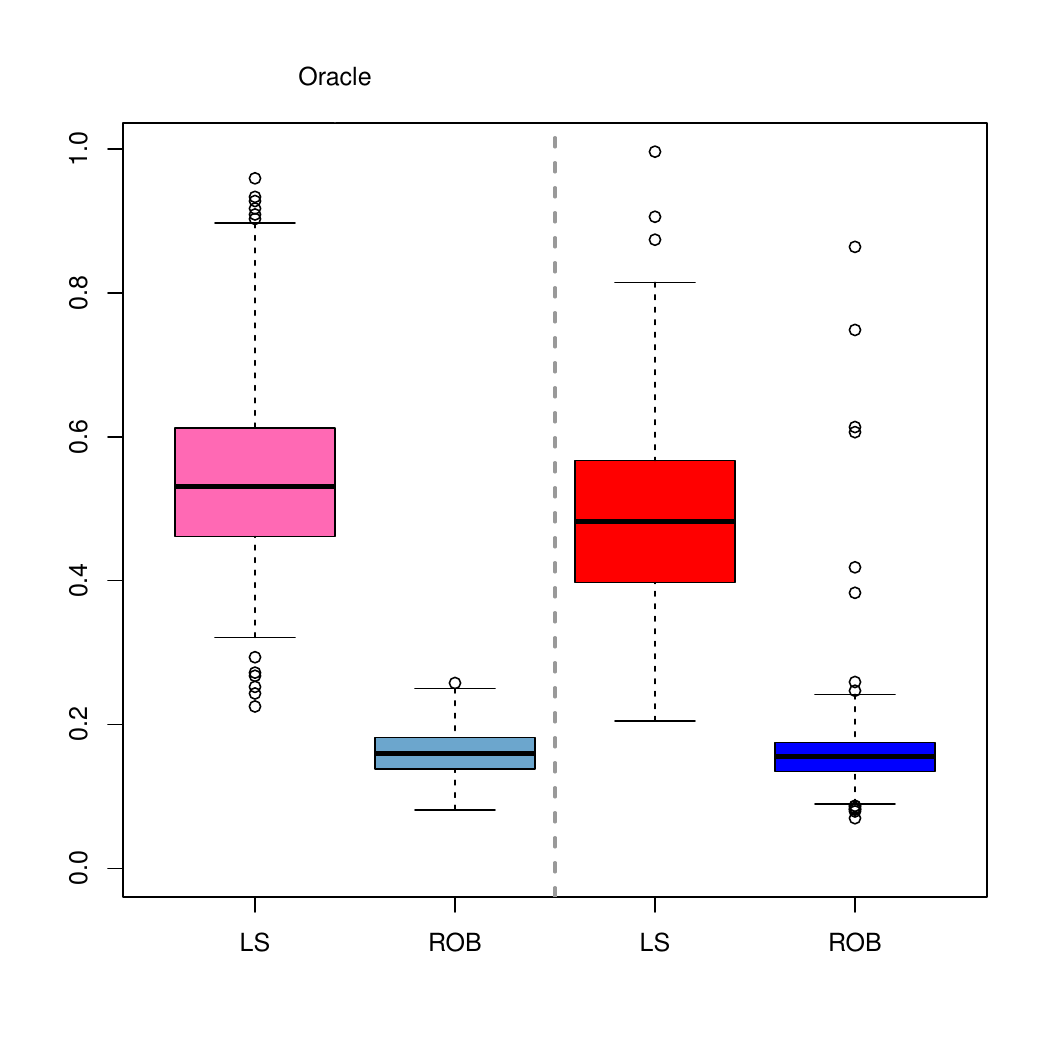}  & \\
\end{tabular}
\vspace{-0.5cm} 
\caption{\label{fig:RASE-n600} Adjusted boxplots for the RASE values under $C_0$, $C_1$, $C_3$, $C_4$ and $C_7$ for both oracle estimators and both penalized estimators for sample size $n=600$.}
\end{center}
\end{figure}

Table \ref{tab:Estimation} shows the results of the estimation measures GMSE and RASE for the linear part and the additive part, respectively, for the penalized and the oracle LS$-$ and robust estimators, under different contamination settings and different sample sizes. For each estimator, it was computed the mean of the estimation measure, denoted \textsc{mean}, and also the standard deviation, denoted \textsc{sd}.
As it is expected, under the noncontaminated scheme $C_0$, both the oracle and penalized  robust estimators throw slightly larger values of these measures than their least squares counterparts under the different sample sizes considered, caused by the lack of efficiency of the robust estimators usually have. Besides, under this noncontaminted setting, the penalized approaches throw similar values to the ones obtained by the oracle versions since, as it was already mentioned when analysing Table \ref{tab:Selection}, the correct model was highly selected. This behaviour can also be appreciated by observing the standard deviations obtained.
For the contamination schemes $C_1$ to $C_7$, similar values of the oracle estimator with respect to the penalized estimator can be still observe but only for the robust approach showing a more stable behaviour across all the contamination schemes and for both GMSE and RASE measures. However, despite of the greater stability of the robust approach, it can be appreciated that, for the $C_7$ case and the GMSE measure, the robust proposal yields considerably higher results than the oracle estimator for all sample sizes, along with much larger standard deviations.
% for all sample sizes, the results obtained by the robust proposal are much larger than the oracle ones, including also much higher standard deviations. 
 It can be checked that a few atypical observations appeared in the data under this contamination setting (see, for instance Figures \ref{fig:GMSE-n200} and \ref{fig:GMSE-n600}). For $n=200$, a 7.4\% of the GMSE values were detected as outliers by the adjusted boxplot, which is a boxplot adapted to asymmetric data. For $n=400$ and $n=600$, these percentages are reduced to 4\% and 1.4\%, respectively.
Table \ref{tab:Estimation-C7} shows the results of the $10\%-$trimmed mean of the GMSE values for the oracle and penalized LS$-$ and robust estimators. By computed the trimmed mean, the results obtained by the robust approach are quite similar to those obtained by its oracle version. 
Let us continue analysing the GMSE measure. 
When looking only the penalized estimators, the robust proposal shows a more stable behaviour across all the contamination settings than its least squares counterpart. 
%The LS--penalized estimator throws values that are much larger than the ones obtained under the noncontaminated case, going from 3.5 times up to 913 times larger under $C_7$. 
On the contrary, the LS$-$estimator shows a more erratic behaviour under all sample sizes and all contamination settings, especially under $C_5$ o $C_7$ where the mean of GMSE obtained are between 51 and 913 times larger than under the $C_0$ case. In order to visualize the results obtained for the GMSE even more, Figures \ref{fig:GMSE-n200} and \ref{fig:GMSE-n600} show the adjusted boxplots for contaminations $C_0$, $C_1$, $C_3$, $C_4$ and $C_7$ and sample sizes $n=200$ and $n=600$, respectively.
A similar analysis can be done for the RASE measure, that is, the measure used for the additive part. Under all sample sizes, when no contaminated data, both oracle estimators and both penalized estimators behave similarly between them, with also the robust approaches throwing slightly larger values than the least squares counterparts, as it was already mentioned. 
Across the contamination schemes $C_1$ to $C_7$, both estimators based on least squares throw larger values of the RASE than the ones obtained under $C_0$, especially under $C_4$ to $C_7$, achieving values up to 8 times larger. On the other hand, the robust proposal, and also its oracle version, shows a more stable behaviour across all the contamination scenarios. In order to visualize these behaviours, Figures \ref{fig:RASE-n200} and \ref{fig:RASE-n600} show the adjusted boxplots of the RASE values for contaminations $C_0$, $C_1$, $C_3$, $C_4$ and $C_7$, for the oracle and penalized estimators, and sample sizes $n=200$ and $n=600$, respectively.

\begin{table}[ht!]
\begin{center}
\scriptsize
\begin{tabular}{|ccc|cc|cc|cc|cc|}
  \hline
  \multicolumn{3}{|c|}{} & \multicolumn{4}{c|}{GMSE} & \multicolumn{4}{c|}{RASE}\\\hline
    \multicolumn{3}{|c|}{} & \multicolumn{2}{c|}{Oracle} & \multicolumn{2}{c|}{Penalized} & \multicolumn{2}{c|}{Oracle} & \multicolumn{2}{c|}{Penalized}\\\hline
$n$ & \textsc{Cont} & \textsc{Method} & \textsc{mean} & \textsc{sd} & \textsc{mean} & \textsc{sd} & \textsc{mean} & \textsc{sd} & \textsc{mean} & \textsc{sd} \\ 
  \hline
  200 & $C_0$ & \textsc{ls} & 0.022 & 0.016 & 0.026 & 0.021 & 0.270 & 0.064 & 0.278 & 0.082 \\ 
  200 & $C_0$ & \textsc{rob} & 0.023 & 0.016 & 0.027 & 0.023 & 0.284 & 0.074 & 0.283 & 0.080 \\ 
  200 & $C_1$ & \textsc{ls} & 0.064 & 0.068 & 0.105 & 0.135 & 0.436 & 0.131 & 0.522 & 0.227 \\ 
  200 & $C_1$ & \textsc{rob} & 0.036 & 0.025 & 0.044 & 0.037 & 0.355 & 0.103 & 0.369 & 0.128 \\ 
  200 & $C_2$ & \textsc{ls} & 0.076 & 0.062 & 0.129 & 0.102 & 0.481 & 0.140 & 0.557 & 0.179 \\ 
  200 & $C_2$ & \textsc{rob} & 0.029 & 0.020 & 0.034 & 0.028 & 0.313 & 0.083 & 0.314 & 0.101 \\ 
  200 & $C_3$ & \textsc{ls} & 0.132 & 0.127 & 0.261 & 0.234 & 0.625 & 0.258 & 0.805 & 0.344 \\ 
  200 & $C_3$ & \textsc{rob} & 0.026 & 0.018 & 0.028 & 0.021 & 0.302 & 0.086 & 0.298 & 0.092 \\ 
  200 & $C_4$ & \textsc{ls} & 0.249 & 0.198 & 0.527 & 0.384 & 0.859 & 0.226 & 1.252 & 0.418 \\ 
  200 & $C_4$ & \textsc{rob} & 0.025 & 0.017 & 0.026 & 0.019 & 0.297 & 0.081 & 0.289 & 0.084 \\ 
  200 & $C_5$ & \textsc{ls} & 0.654 & 0.494 & 1.522 & 0.959 & 1.370 & 0.331 & 2.222 & 0.519 \\ 
  200 & $C_5$ & \textsc{rob} & 0.027 & 0.018 & 0.029 & 0.023 & 0.299 & 0.099 & 0.307 & 0.106 \\ 
  200 & $C_6$ & \textsc{ls} & 12.172 & 0.445 & 6.708 & 0.368 & 0.962 & 0.236 & 1.027 & 0.354 \\ 
  200 & $C_6$ & \textsc{rob} & 0.025 & 0.018 & 0.027 & 0.021 & 0.299 & 0.086 & 0.295 & 0.094 \\ 
  200 & $C_7$ & \textsc{ls} & 12.450 & 0.359 & 6.826 & 0.373 & 0.973 & 0.232 & 1.112 & 0.375 \\ 
  200 & $C_7$ & \textsc{rob} & 0.050 & 0.526 & 0.745 & 2.790 & 0.304 & 0.144 & 0.353 & 0.243 \\ \hline
  400 & $C_0$ & \textsc{ls} & 0.011 & 0.008 & 0.012 & 0.009 & 0.186 & 0.041 & 0.184 & 0.041 \\ 
  400 & $C_0$ & \textsc{rob} & 0.012 & 0.008 & 0.012 & 0.009 & 0.192 & 0.044 & 0.188 & 0.040 \\ 
  400 & $C_1$ & \textsc{ls} & 0.031 & 0.027 & 0.042 & 0.040 & 0.309 & 0.084 & 0.335 & 0.112 \\ 
  400 & $C_1$ & \textsc{rob} & 0.016 & 0.012 & 0.017 & 0.012 & 0.234 & 0.057 & 0.229 & 0.057 \\ 
  400 & $C_2$ & \textsc{ls} & 0.036 & 0.026 & 0.048 & 0.036 & 0.325 & 0.077 & 0.362 & 0.104 \\ 
  400 & $C_2$ & \textsc{rob} & 0.014 & 0.010 & 0.014 & 0.010 & 0.211 & 0.050 & 0.205 & 0.046 \\ 
  400 & $C_3$ & \textsc{ls} & 0.064 & 0.052 & 0.101 & 0.079 & 0.421 & 0.117 & 0.484 & 0.155 \\ 
  400 & $C_3$ & \textsc{rob} & 0.012 & 0.008 & 0.013 & 0.009 & 0.201 & 0.048 & 0.194 & 0.041 \\ 
  400 & $C_4$ & \textsc{ls} & 0.125 & 0.089 & 0.223 & 0.159 & 0.598 & 0.155 & 0.720 & 0.204 \\ 
  400 & $C_4$ & \textsc{rob} & 0.012 & 0.008 & 0.012 & 0.009 & 0.198 & 0.048 & 0.192 & 0.043 \\ 
  400 & $C_5$ & \textsc{ls} & 0.314 & 0.230 & 0.620 & 0.382 & 0.949 & 0.218 & 1.374 & 0.287 \\ 
  400 & $C_5$ & \textsc{rob} & 0.013 & 0.009 & 0.013 & 0.009 & 0.206 & 0.049 & 0.199 & 0.042 \\ 
  400 & $C_6$ & \textsc{ls} & 12.005 & 0.273 & 6.402 & 0.230 & 0.648 & 0.135 & 0.590 & 0.153 \\ 
  400 & $C_6$ & \textsc{rob} & 0.012 & 0.008 & 0.012 & 0.009 & 0.198 & 0.047 & 0.191 & 0.040 \\ 
  400 & $C_7$ & \textsc{ls} & 12.293 & 0.210 & 6.491 & 0.225 & 0.665 & 0.141 & 0.621 & 0.154 \\ 
  400 & $C_7$ & \textsc{rob} & 0.012 & 0.009 & 0.319 & 1.890 & 0.201 & 0.046 & 0.209 & 0.101 \\ \hline
  600 & $C_0$ & \textsc{ls} & 0.007 & 0.005 & 0.007 & 0.005 & 0.149 & 0.031 & 0.147 & 0.029 \\ 
  600 & $C_0$ & \textsc{rob} & 0.008 & 0.005 & 0.008 & 0.005 & 0.154 & 0.032 & 0.151 & 0.030 \\ 
  600 & $C_1$ & \textsc{ls} & 0.022 & 0.018 & 0.026 & 0.028 & 0.253 & 0.072 & 0.267 & 0.093 \\ 
  600 & $C_1$ & \textsc{rob} & 0.011 & 0.008 & 0.011 & 0.008 & 0.190 & 0.042 & 0.186 & 0.041 \\ 
  600 & $C_2$ & \textsc{ls} & 0.023 & 0.017 & 0.028 & 0.020 & 0.265 & 0.061 & 0.285 & 0.086 \\ 
  600 & $C_2$ & \textsc{rob} & 0.009 & 0.006 & 0.009 & 0.006 & 0.169 & 0.036 & 0.164 & 0.032 \\ 
  600 & $C_3$ & \textsc{ls} & 0.042 & 0.032 & 0.057 & 0.046 & 0.344 & 0.088 & 0.381 & 0.114 \\ 
  600 & $C_3$ & \textsc{rob} & 0.008 & 0.006 & 0.008 & 0.006 & 0.160 & 0.036 & 0.155 & 0.031 \\ 
  600 & $C_4$ & \textsc{ls} & 0.077 & 0.063 & 0.129 & 0.096 & 0.494 & 0.118 & 0.568 & 0.150 \\ 
  600 & $C_4$ & \textsc{rob} & 0.008 & 0.005 & 0.008 & 0.006 & 0.158 & 0.033 & 0.153 & 0.031 \\ 
  600 & $C_5$ & \textsc{ls} & 0.201 & 0.154 & 0.374 & 0.229 & 0.775 & 0.171 & 1.023 & 0.235 \\ 
  600 & $C_5$ & \textsc{rob} & 0.009 & 0.006 & 0.009 & 0.006 & 0.164 & 0.035 & 0.161 & 0.036 \\ 
  600 & $C_6$ & \textsc{ls} & 11.970 & 0.221 & 6.300 & 0.177 & 0.525 & 0.108 & 0.462 & 0.116 \\ 
  600 & $C_6$ & \textsc{rob} & 0.008 & 0.005 & 0.008 & 0.005 & 0.158 & 0.033 & 0.153 & 0.030 \\ 
  600 & $C_7$ & \textsc{ls} & 12.280 & 0.174 & 6.391 & 0.177 & 0.540 & 0.116 & 0.491 & 0.124 \\ 
  600 & $C_7$ & \textsc{rob} & 0.008 & 0.006 & 0.132 & 1.162 & 0.161 & 0.034 & 0.162 & 0.061 \\
   \hline
\end{tabular}
\caption{\label{tab:Estimation}\footnotesize GMSE and RASE obtained for the least squares and robust approaches and both oracle estimators, for different sample sizes.} 
\end{center}
\end{table}

\begin{table}[ht!]
\begin{center}
\scriptsize
\begin{tabular}{|cc|c|c|}
  \hline
  \multicolumn{2}{|c|}{} & \multicolumn{2}{c|}{GMSE}\\\hline
$n$ & \textsc{Method} & {Oracle} & {Penalized} \\\hline
200 & \textsc{ls} & 12.437 & 6.790 \\ 
  200 & \textsc{rob} & 0.024 & 0.028 \\\hline
400 & \textsc{ls} & 12.292 & 6.471 \\
400 & \textsc{rob} & 0.011 & 0.012 \\\hline
600 & \textsc{ls} & 12.277 & 6.375 \\
600 & \textsc{rob} & 0.007 & 0.008 \\ 
   \hline
\end{tabular}
\caption{\label{tab:Estimation-C7}\footnotesize 10\%$-$trimmed mean of the GMSE values for the LS and robust approaches and both oracle estimators, for different sample sizes and under the $C_7$ contamination setting.} 
\end{center}
\end{table}

In order to see how the penalized approaches estimates the additive functions, Figures \ref{fig:curvas-cl-rob-C0} to  \ref{fig:curvas-cl-rob-C7} show the functional boxplots obtained for $n=200$ under the contamination settings $C_0$, $C_1$, $C_2$, $C_4$ and $C_7$, respectively, for the least squares and robust estimators. Similar plots can be obtained for sample sizes $n=400$ and $n=600$. Functional boxplots were introduced by \citet{sun:genton:2011} and are a useful tool to visualize a collection of curves. The area in purple represents the 50\% inner band of curves, and the solid blue and the dotted red lines correspond to the whiskers and outlying curves, respectively. Moreover, the black line indicates the deepest function. The true functions are also plotted in solid green line. All $y-$limits were fixed across all the contamination scenarios for the ease of comparison.
 %Since it is well-known the bad behaviour the spline approximation has on the bounderies, the plots were considered only in the inner interval $[0.05 , 0.95]$.
 Recalling that Figure \ref{fig:curvas-cl-rob-C0} shows the results obtained under $C_0$, it can be appreciated that both methods show similar estimates, for instance by noting that the black curves, corresponding to the deepest curve, and the purple boxes, corresponding to the 50\% percent of the inner curves, are quite similar. Taking into account that only the first four additive functions are nonidentically zero functions and the last six are null functions, both estimators seem to be estimating the correct true functions (in green). %This same behaviour can be seen in the last six covariates where the object to be estimated is the constant function zero. 
 In the last six functional boxplots, except for the constant curves in green that are clearly visible, the boxes are almost indistinguishable and only the few outlying curves can be seen. It is also worth noting, especially for the first four additive functions, that near the boundaries, for both robust and least squares estimators, the boxes tend to be wider. The reason of this is the well-known bad behaviour of spline approximations in the boundaries due to the reduced number of data points. This can also be observed for the rest of the figures. To study the effect of the atypical data, Figures \ref{fig:curvas-cl-rob-C1} to \ref{fig:curvas-cl-rob-C7} show the same functional boxplots but for contamination squemes $C_1$, $C_2$, $C_4$ and $C_7$. Recall that none of the contamination settings considered introduce outliers in the additive components of the model and, for this reason, the black curves representing the deepest curves are always close to the true one (in green). However, these functional boxplots allow us to see the general behaviour of the estimators by observing the width of the boxes and whiskers and the outlying curves. For instance, Figure \ref{fig:curvas-cl-rob-C1}, which shows the functional boxplots when considering the contamination scheme $C_1$, for the LS--estimator, boxes of the first four covariates are slightly larger than under $C_0$ and present a few outlying curves. For the last six additive functions, the presence of outlying curves is quite remarkable. %On the other hand, the robust estimator seems to behave similarly to case $C_0$. 
 On the other hand, the robust estimator seems to show a behavior similar to that of $C_0$.
 An analogous analysis can be done for the contamination scenario $C_2$. However, for contamination scenarios $C_4$ and $C_7$, %the estimations obtained by the LS--estimator show a worse behaviour. T
the notorious poor performance of the least squares proposal can be observed in all additive functions.
 Furthermore and in contrast to the other settings, an increase of the size of the boxes and longer whiskers can be appreciated when the approach based on LS estimates the null functions. On the other hand, the robust approach seems to consistently estimate the ten additive functions very well, leading to estimations similar to those obtained under $C_0$. Interestingly and as it was already mentioned, under the most extreme contamination setting considered, that is, under $C_7$ (see Figure \ref{fig:curvas-cl-rob-C7}), the robust approach presents some extra outlying curves that did not appear in the other plots. However, this unexpected behavior is not so bad when compared to the damage caused to the least squares-based estimator.

%Across all the contamination settings, the robust estimator presents a more stable behaviour than the LS$-$estimator. 

%The functional boxplots of the estimators based on least squares have larger boxes for all the ten covariates. Besides, even though the medians of these estimators, represented by the black curves inside the boxes, are close to the true function, as it can be appreciated in the first two rows where the first 20 curves are diplayed, the estimated curves don't behave as the true one.

\begin{landscape}

\begin{figure}[htbp]
 \begin{center}
\small
\begin{tabular}{ccccc}
\includegraphics[scale=0.22]{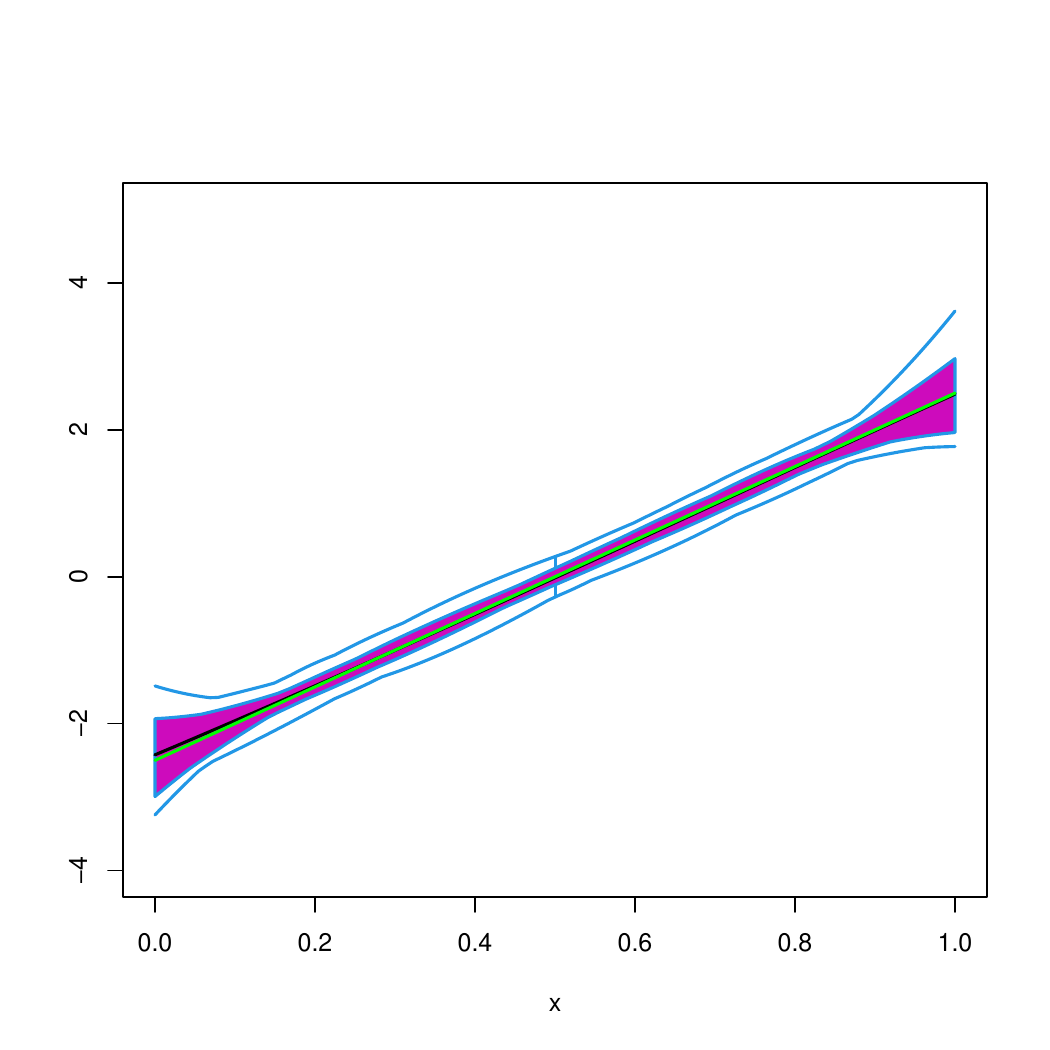} &
 \includegraphics[scale=0.22]{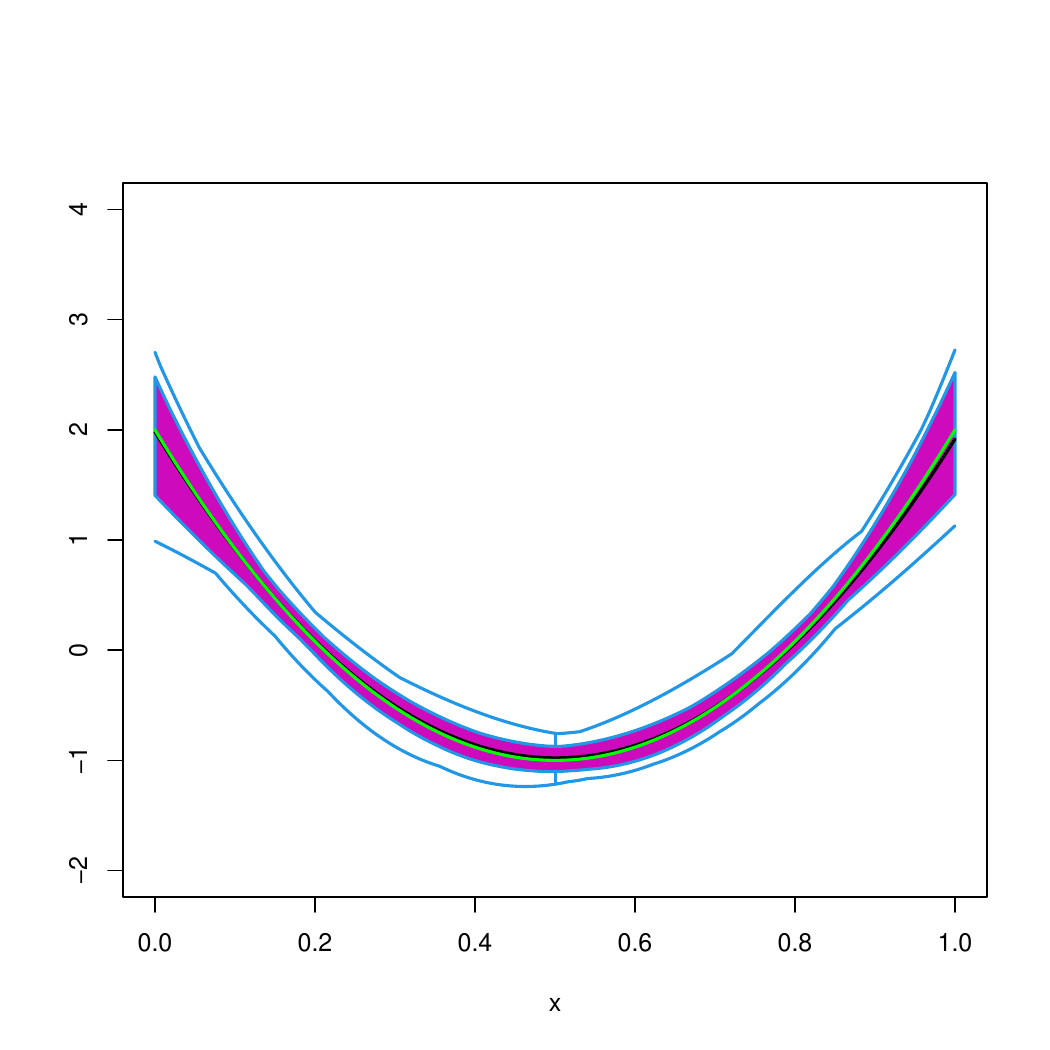} &
\includegraphics[scale=0.22]{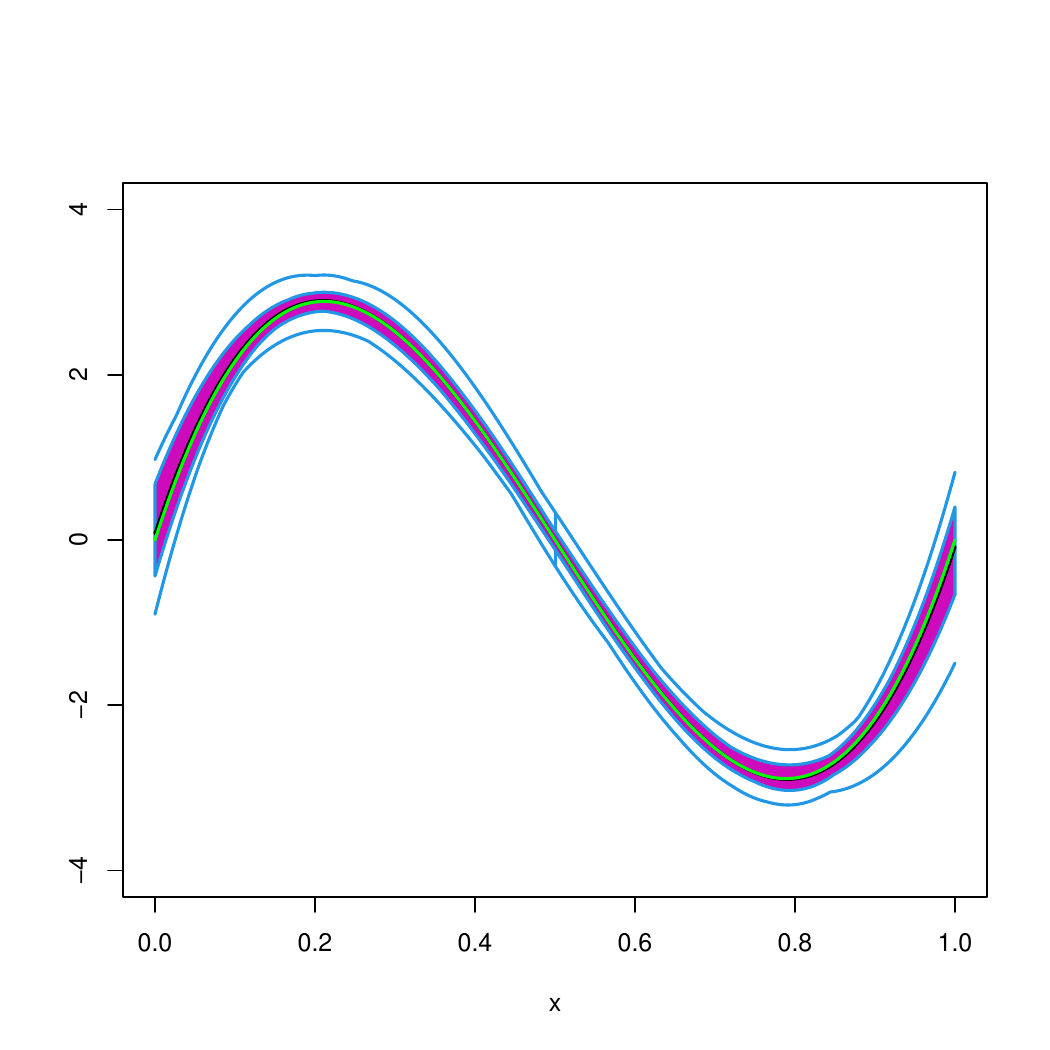}&
\includegraphics[scale=0.22]{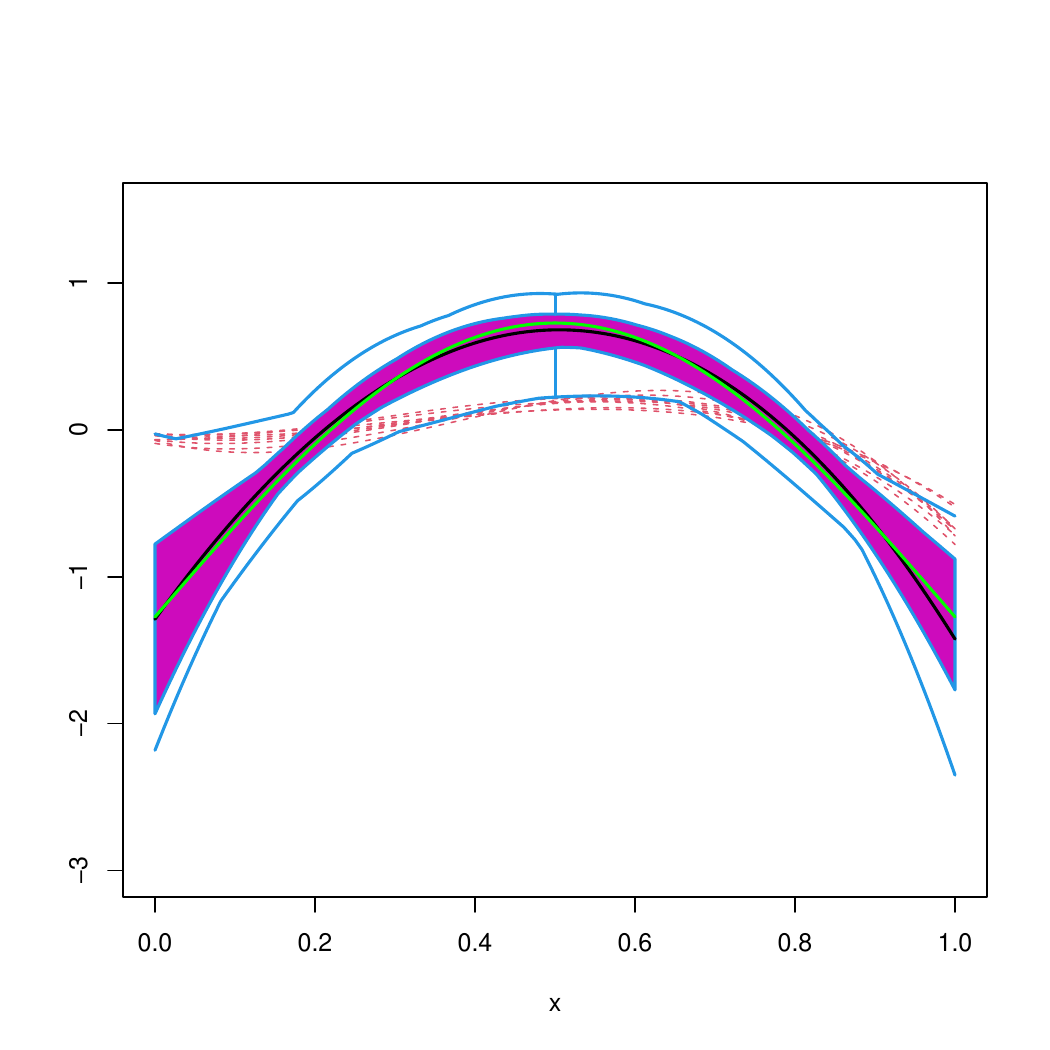}&
\includegraphics[scale=0.22]{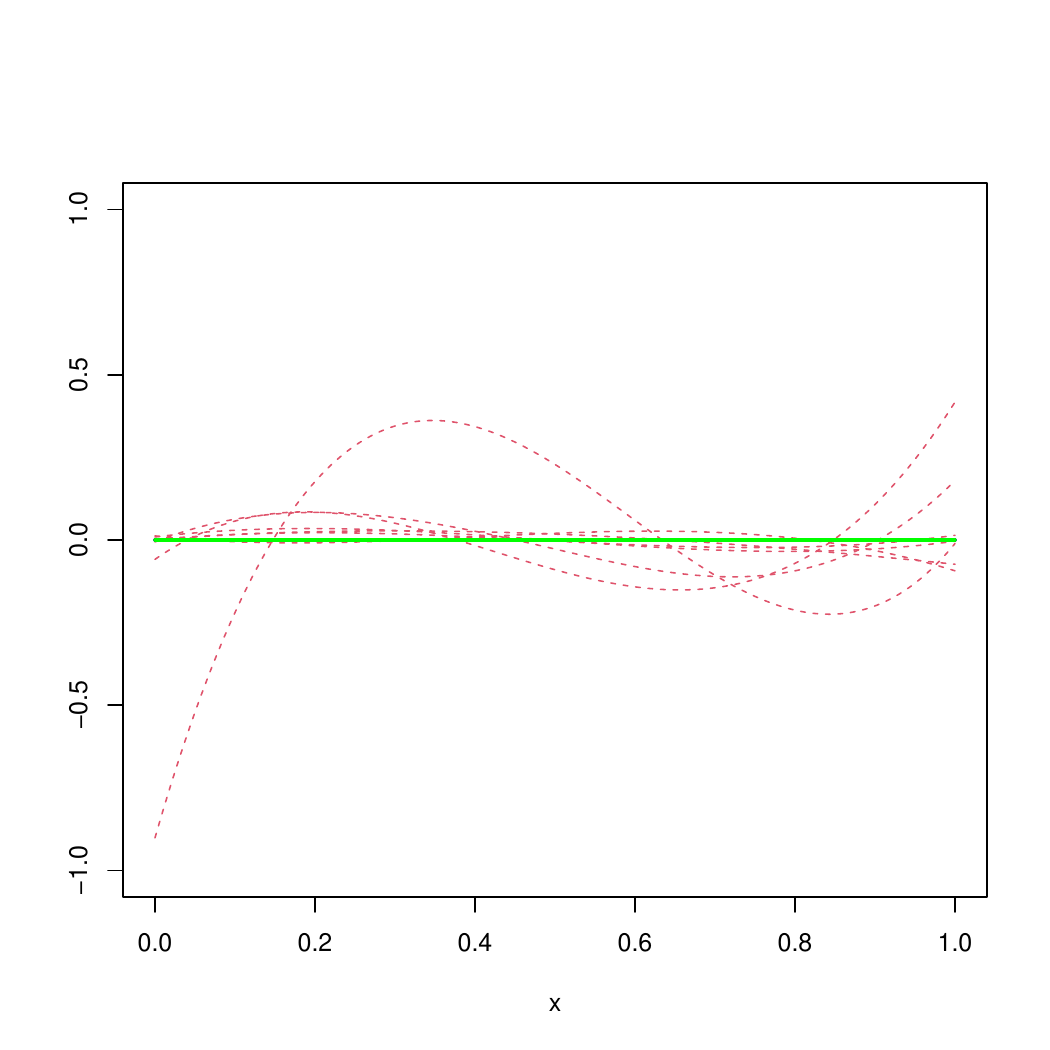}\\
\includegraphics[scale=0.22]{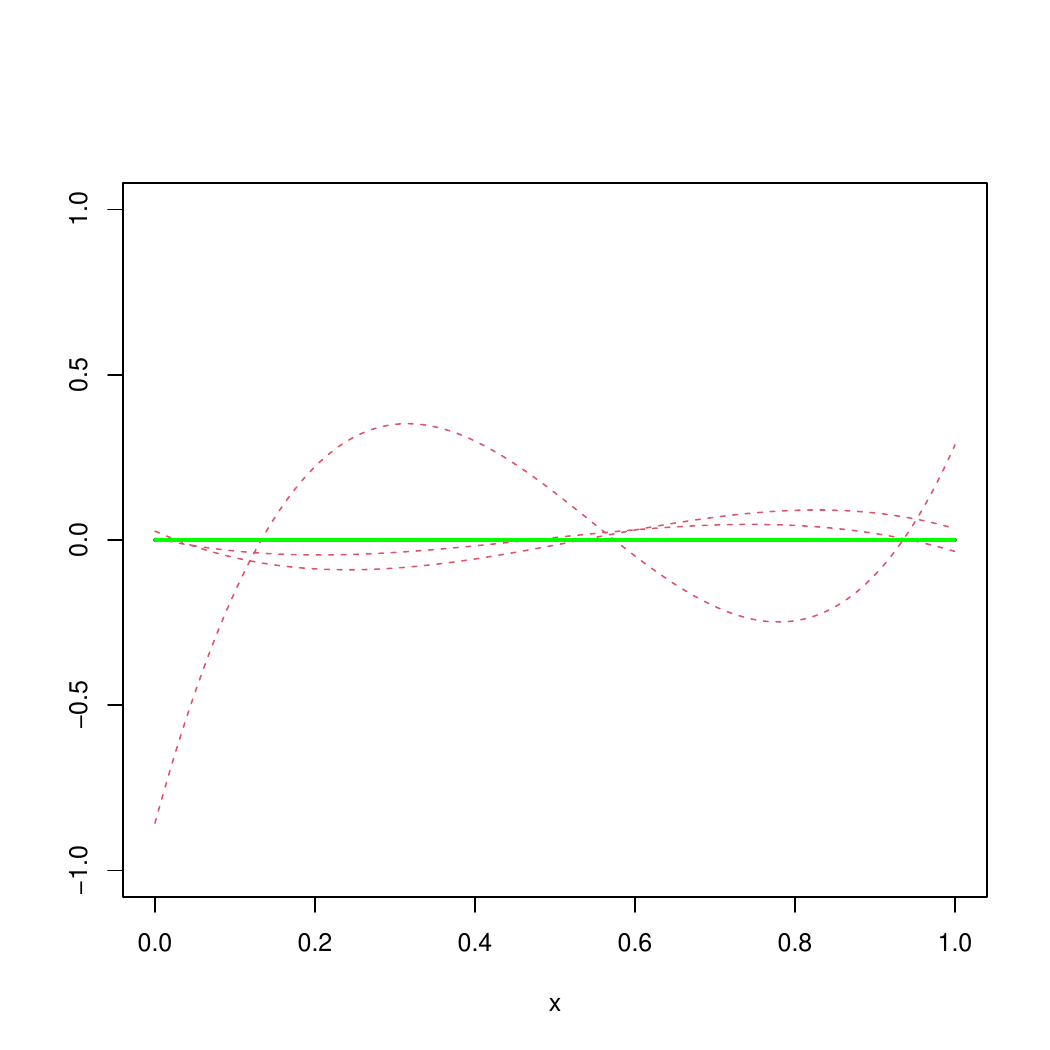} &
 \includegraphics[scale=0.22]{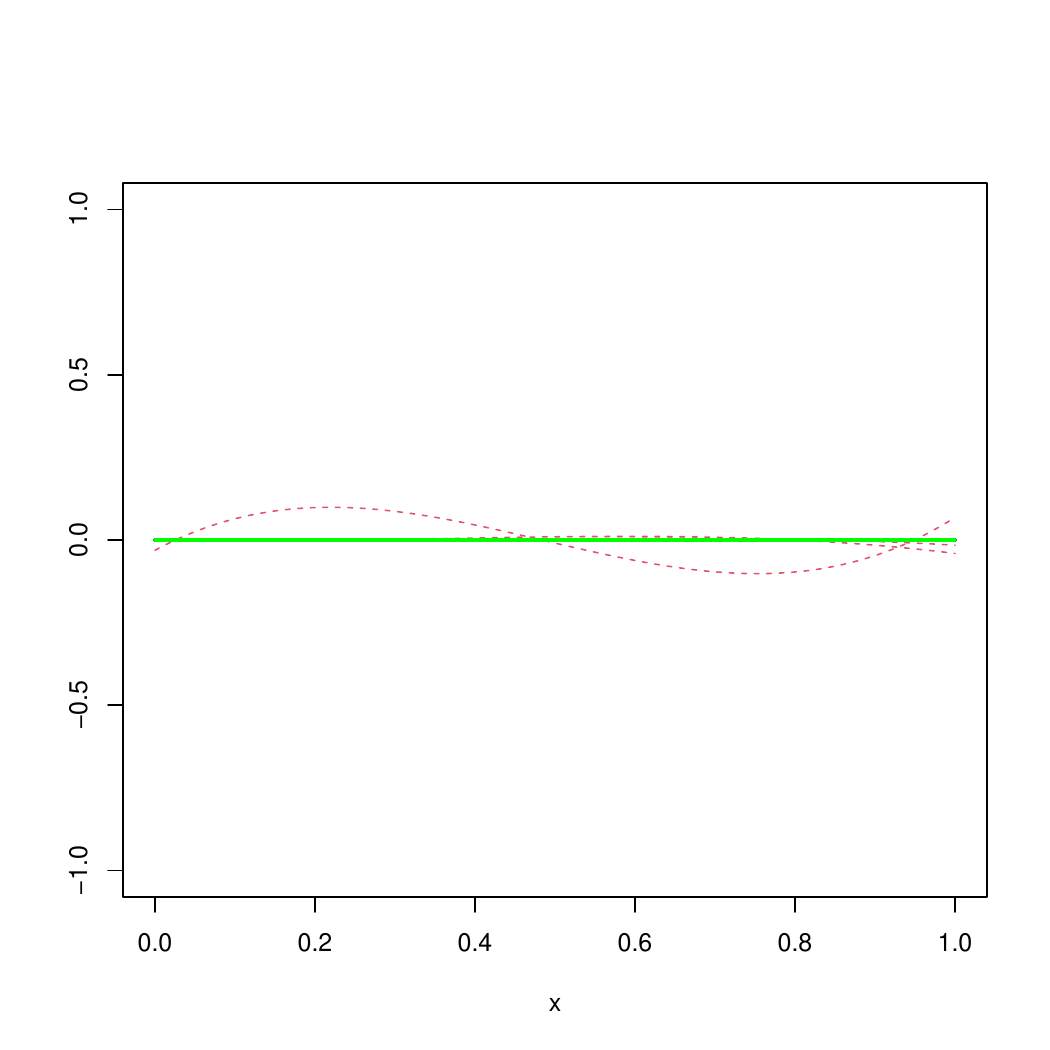}&
 \includegraphics[scale=0.22]{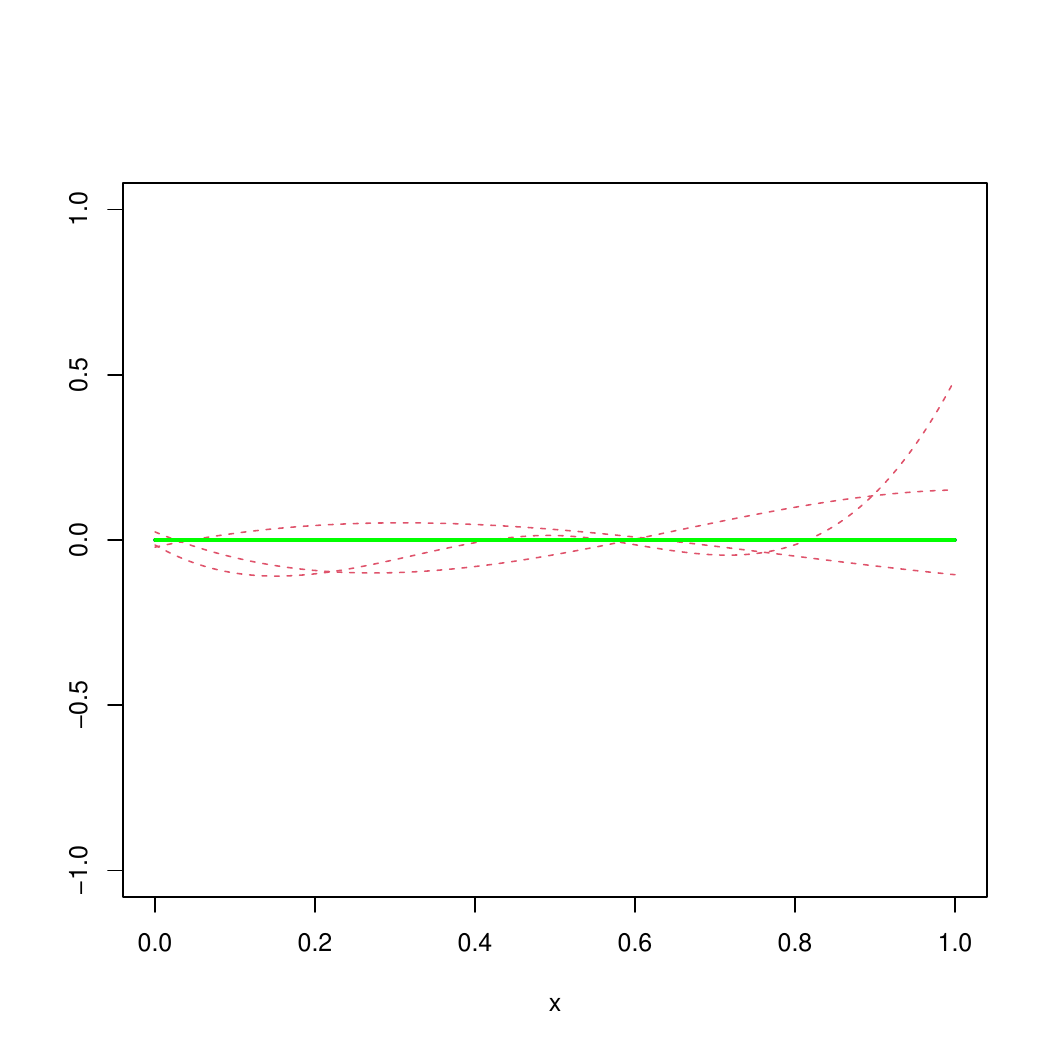}&
  \includegraphics[scale=0.22]{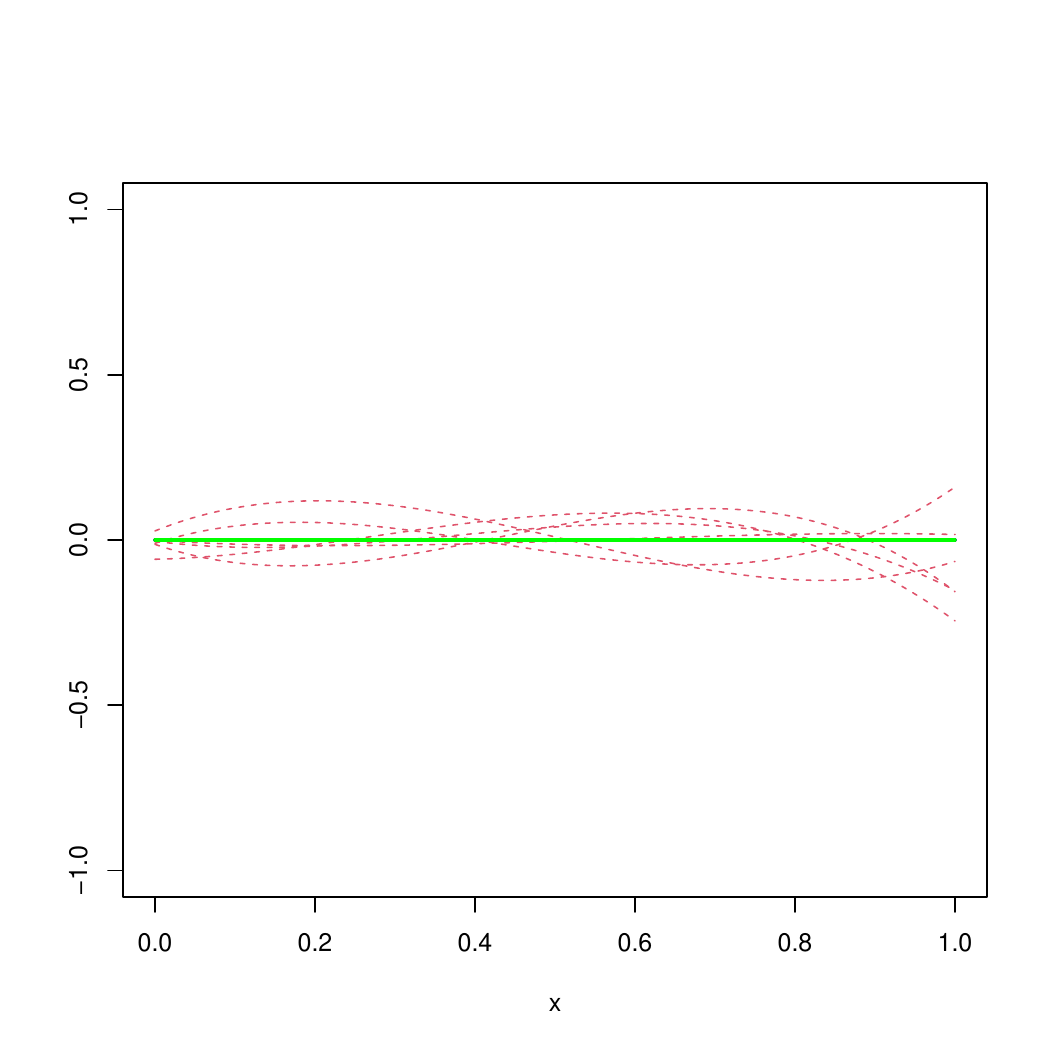}&
   \includegraphics[scale=0.22]{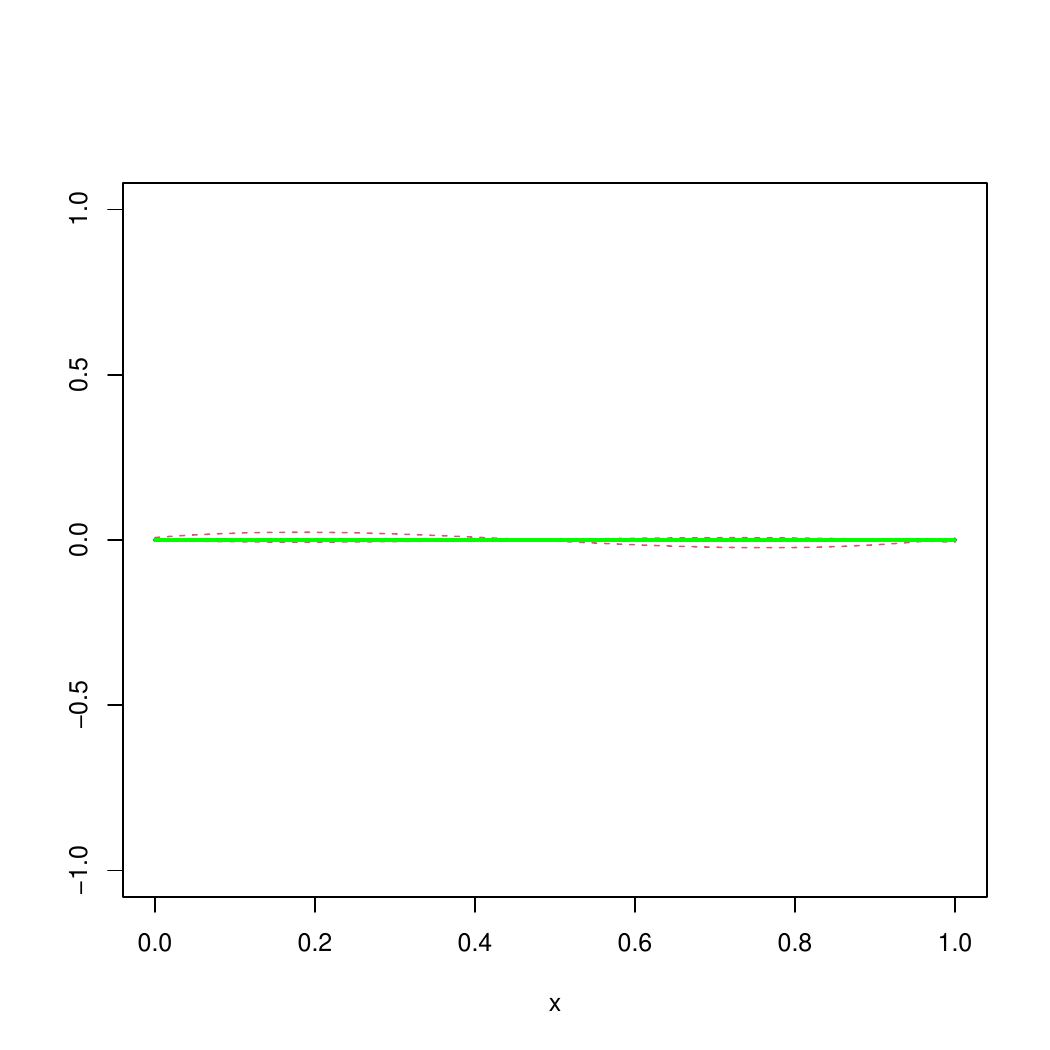} \\
   \includegraphics[scale=0.22]{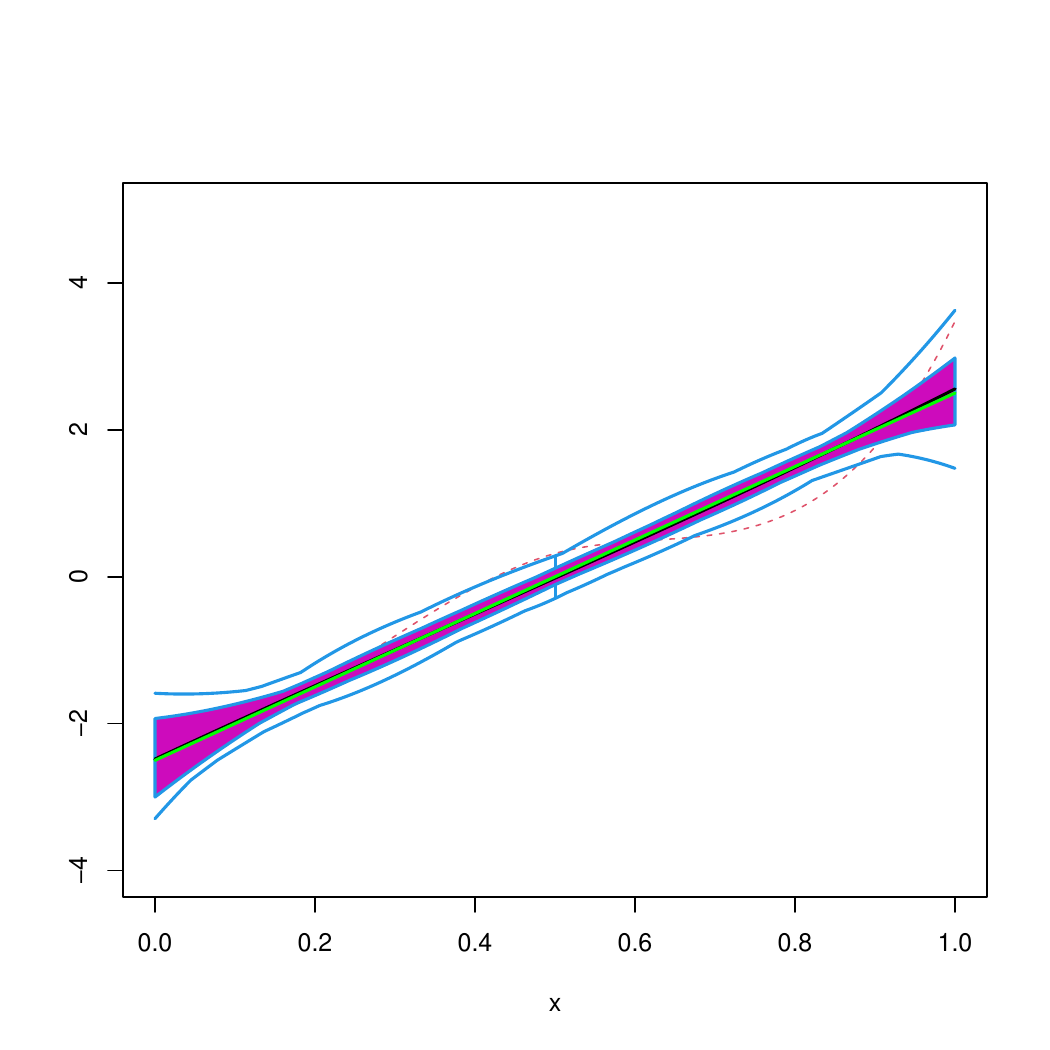} &
 \includegraphics[scale=0.22]{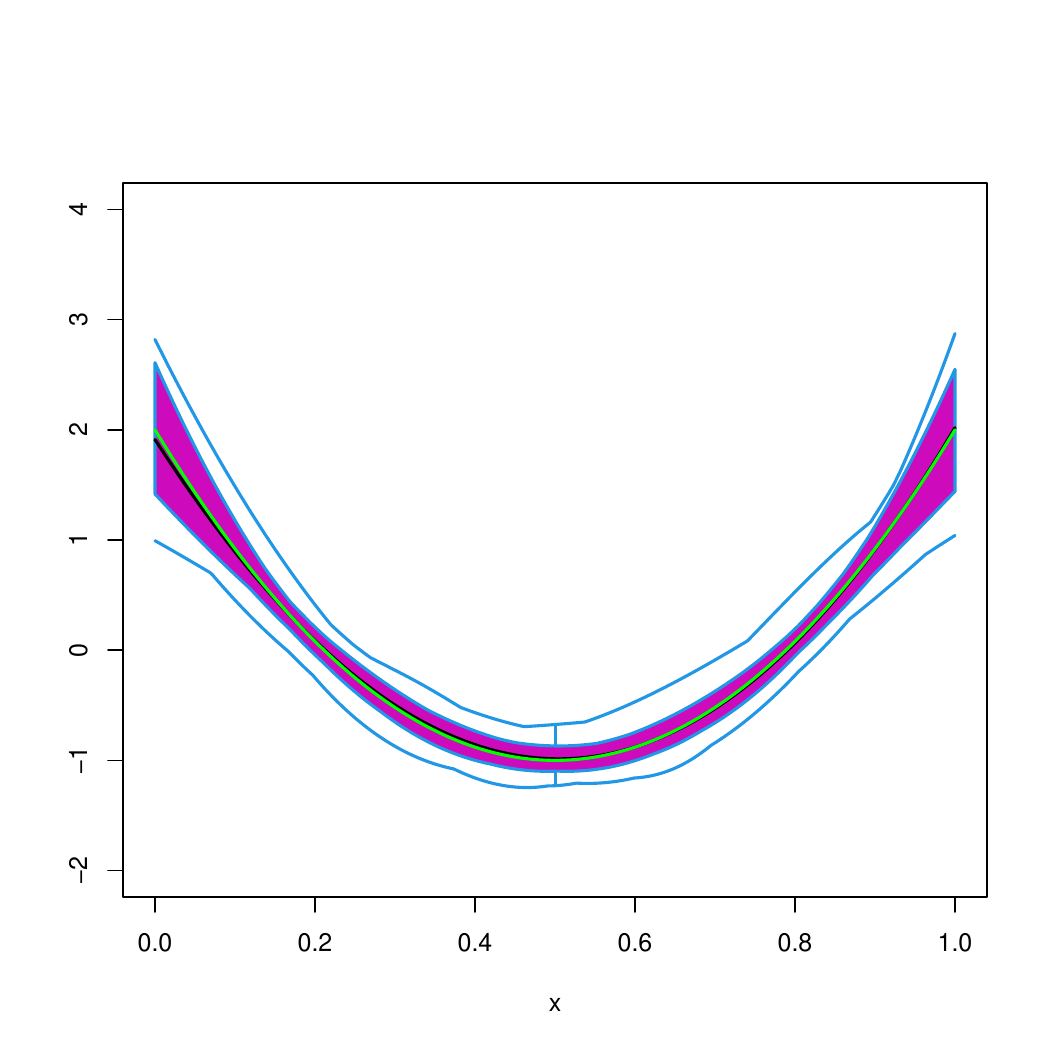} &
\includegraphics[scale=0.22]{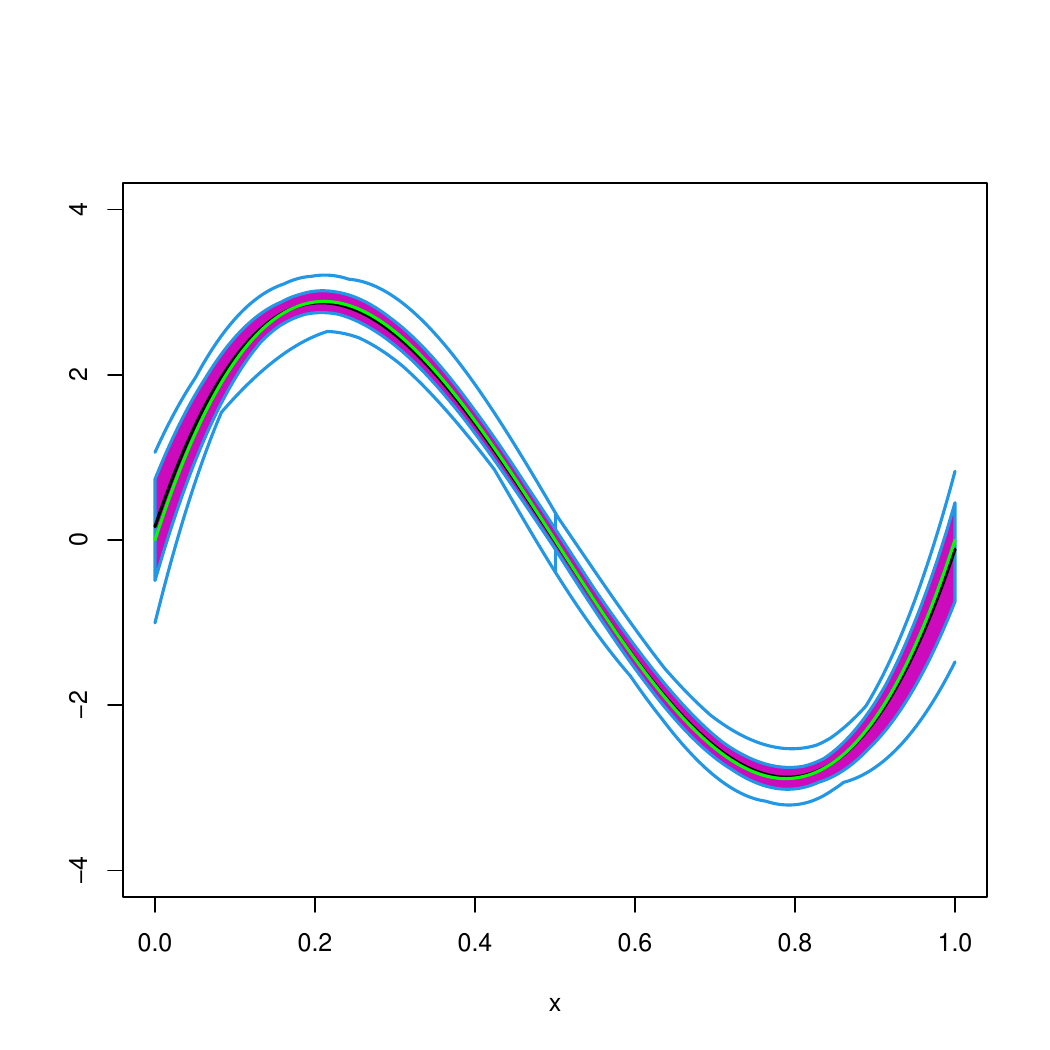}&
\includegraphics[scale=0.22]{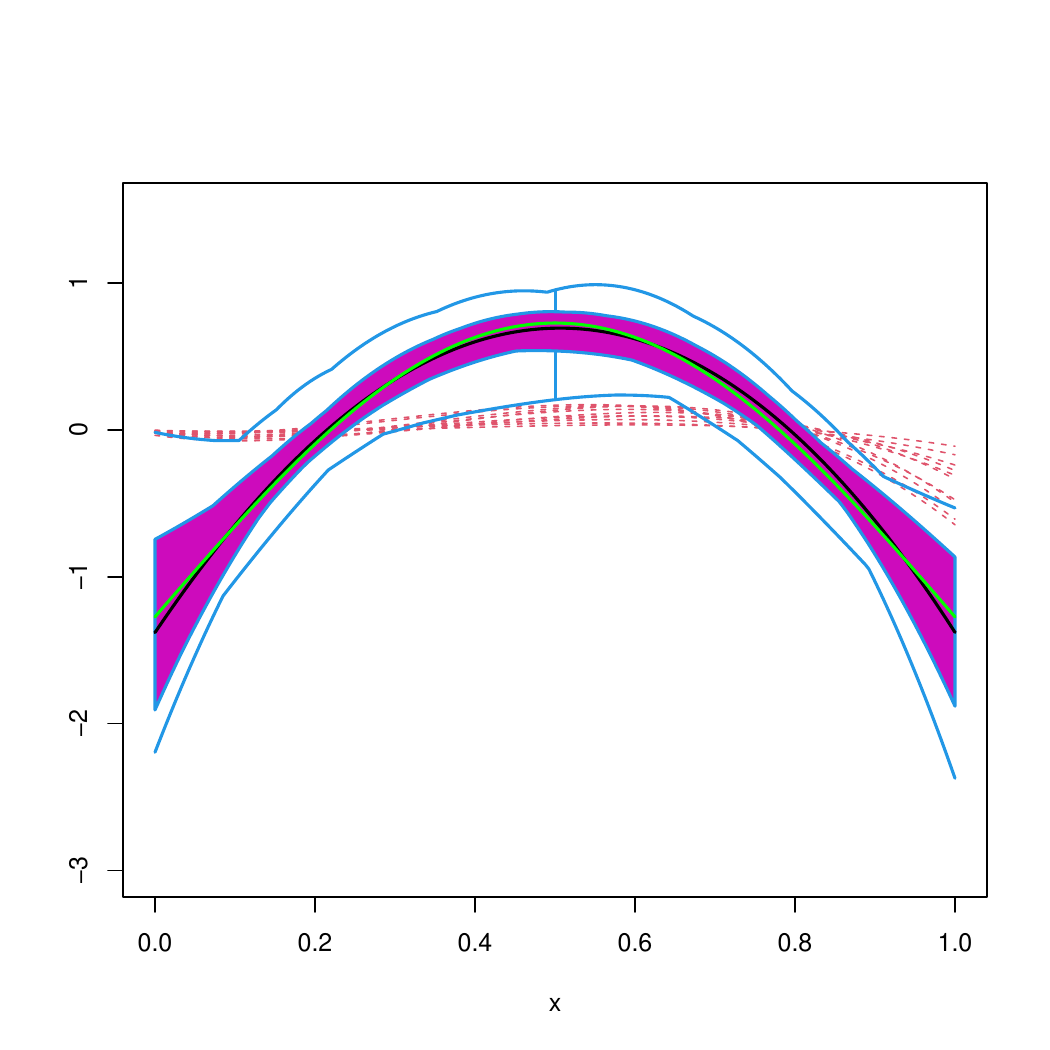}&
\includegraphics[scale=0.22]{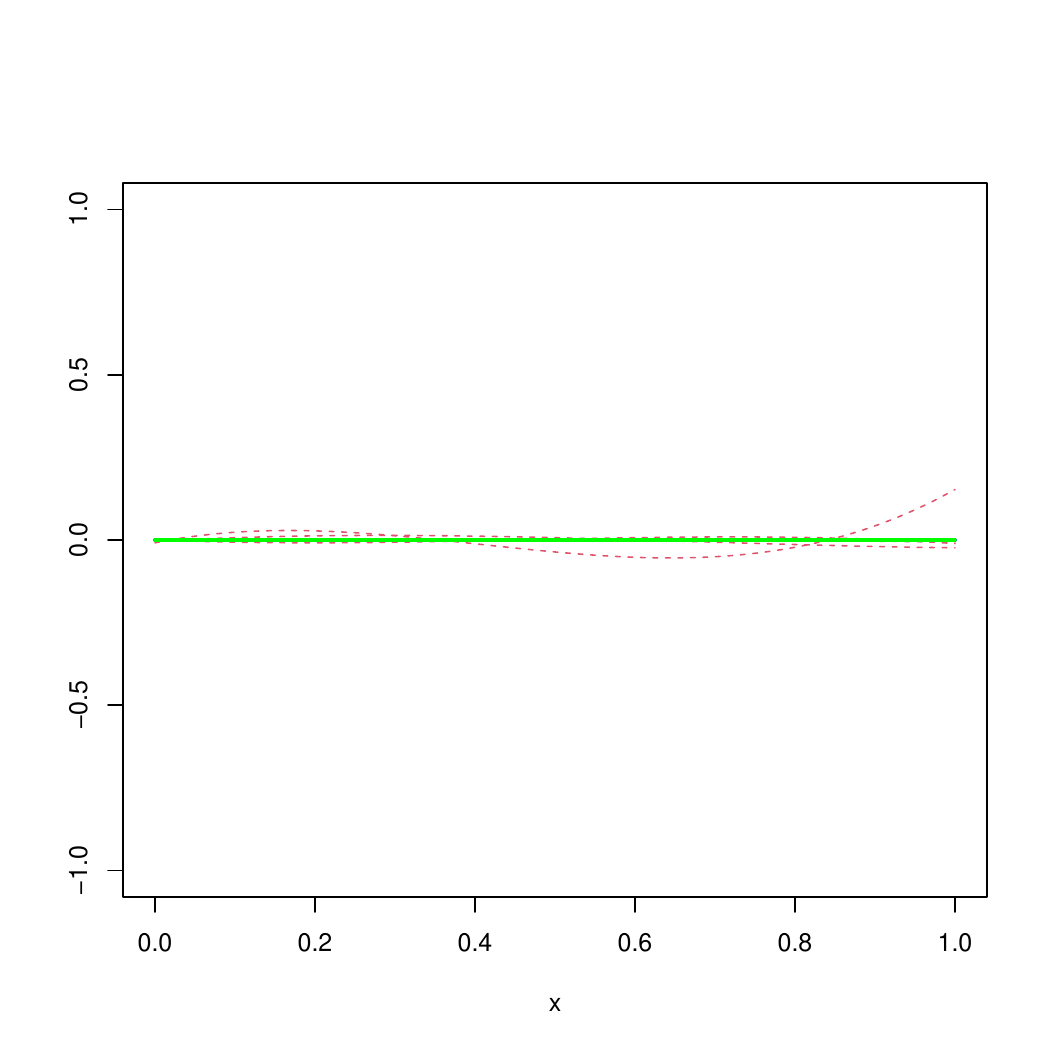}\\
\includegraphics[scale=0.22]{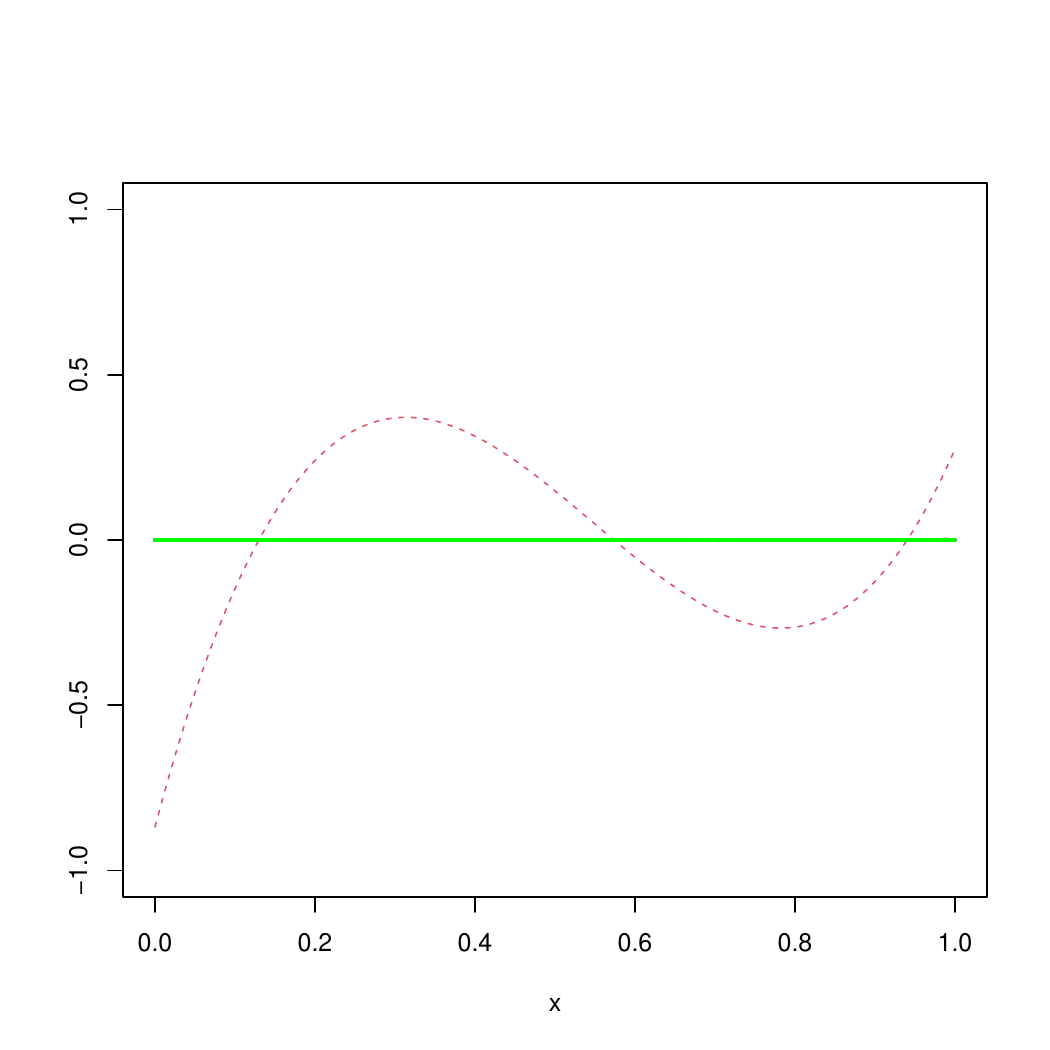} &
 \includegraphics[scale=0.22]{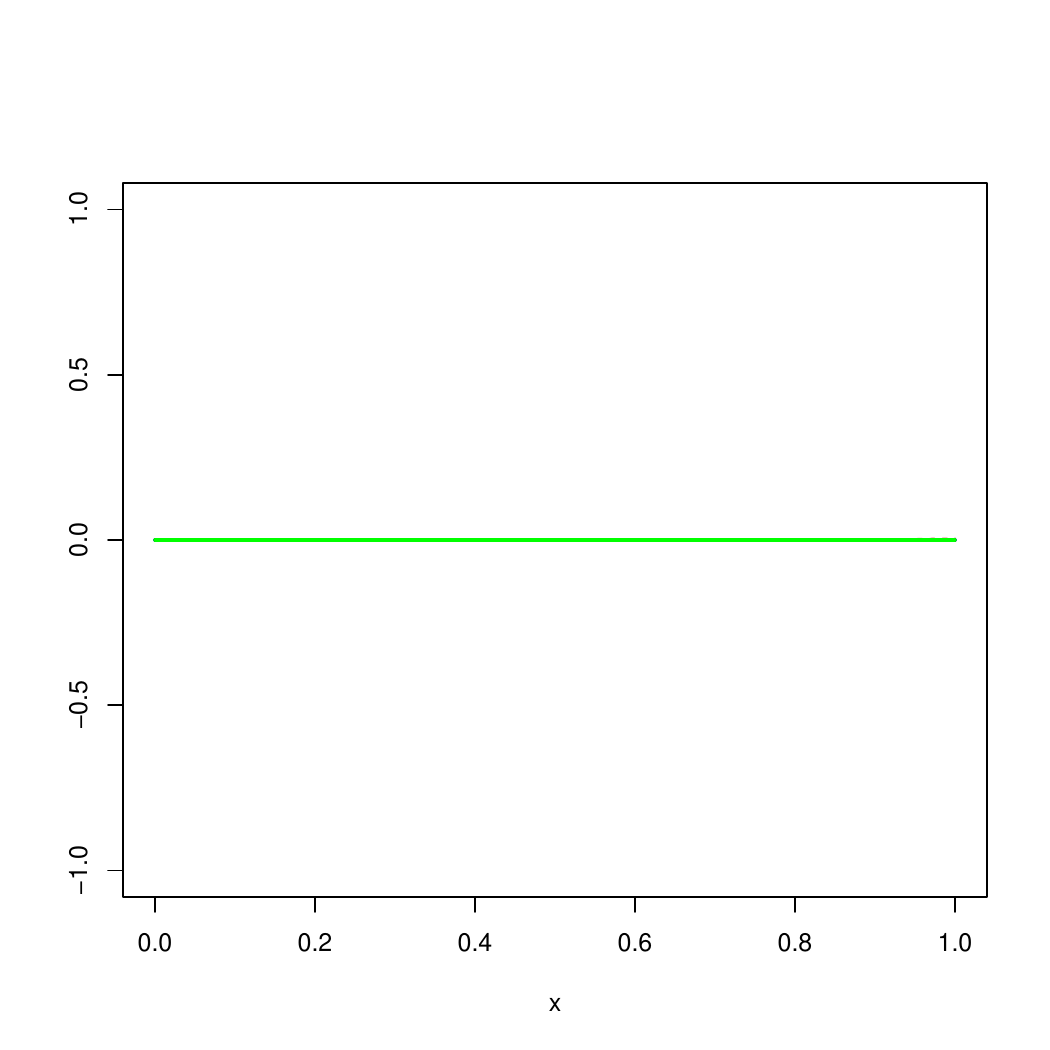}&
 \includegraphics[scale=0.22]{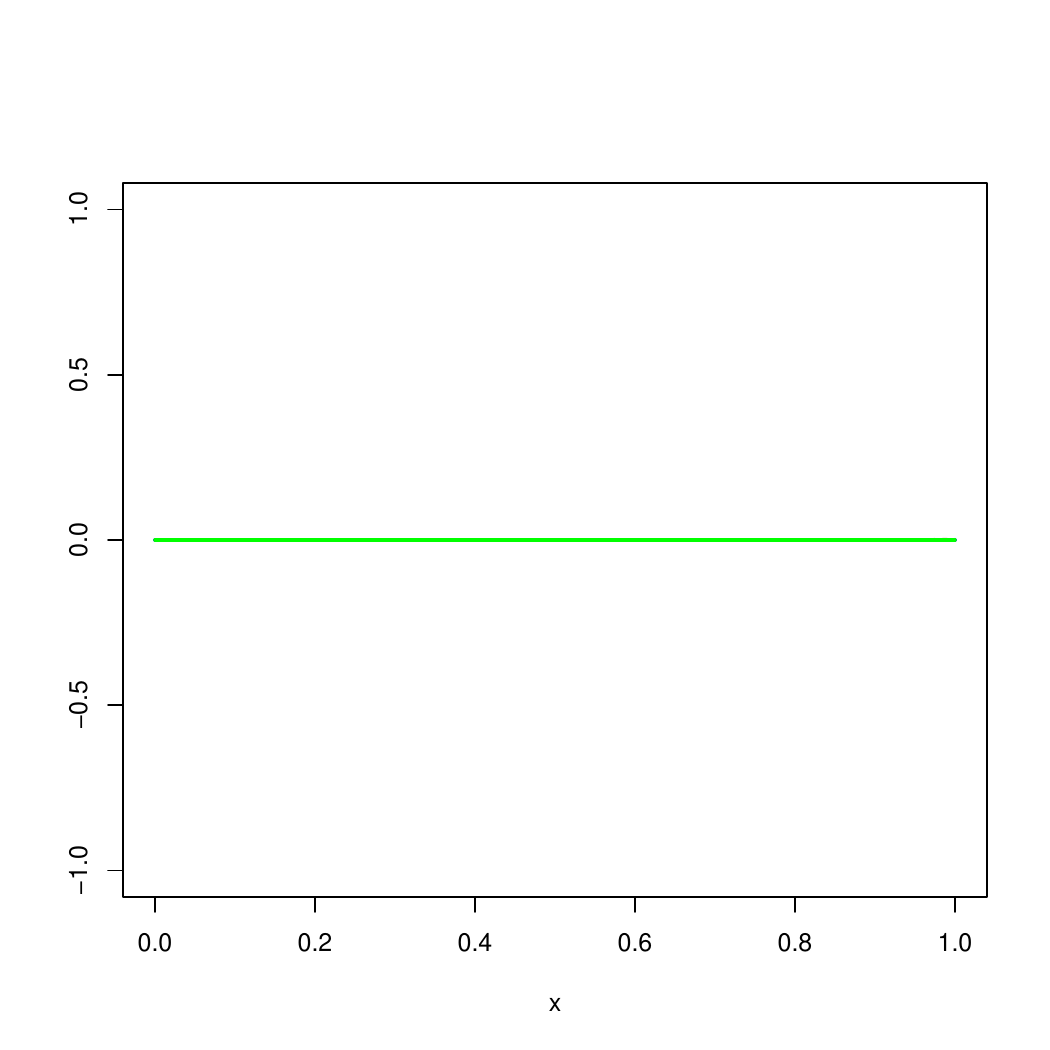}&
  \includegraphics[scale=0.22]{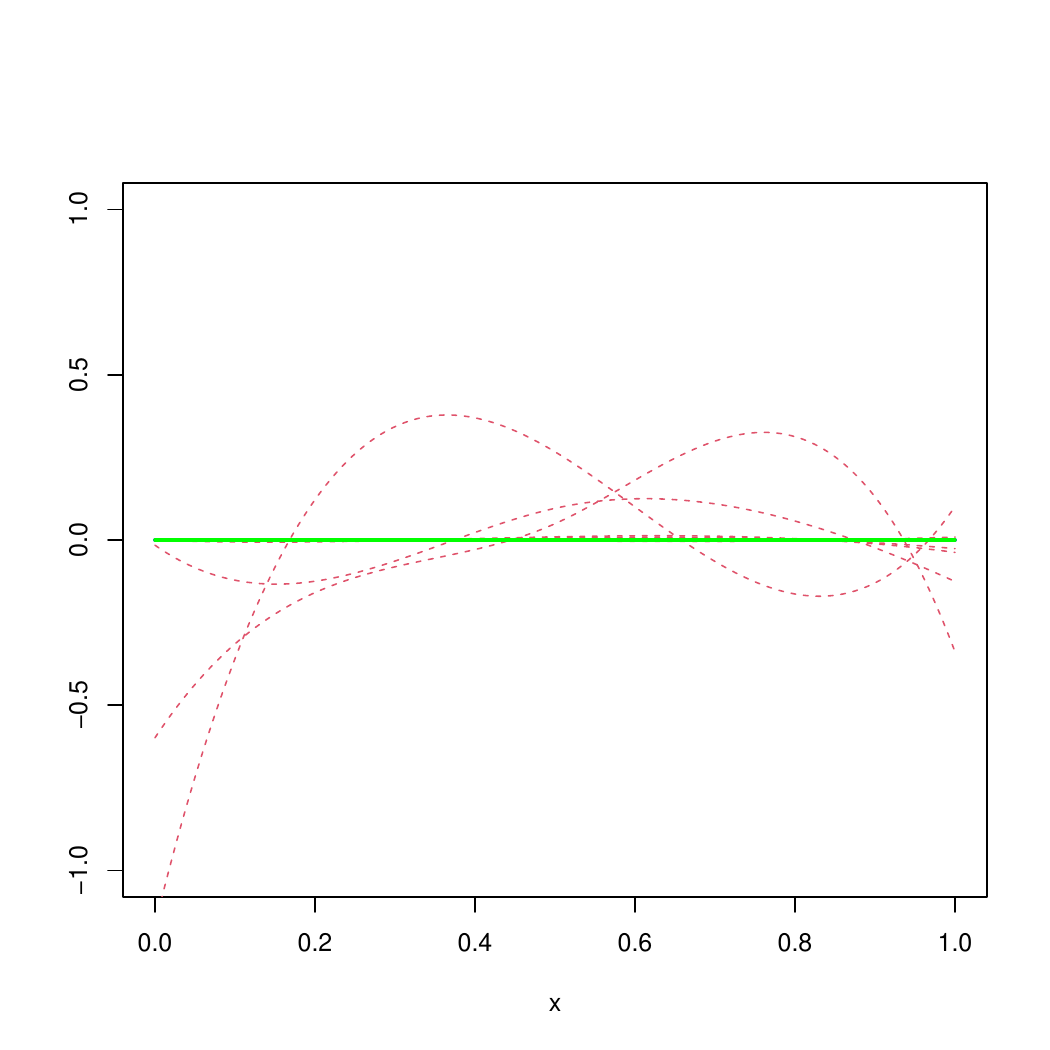}&
   \includegraphics[scale=0.22]{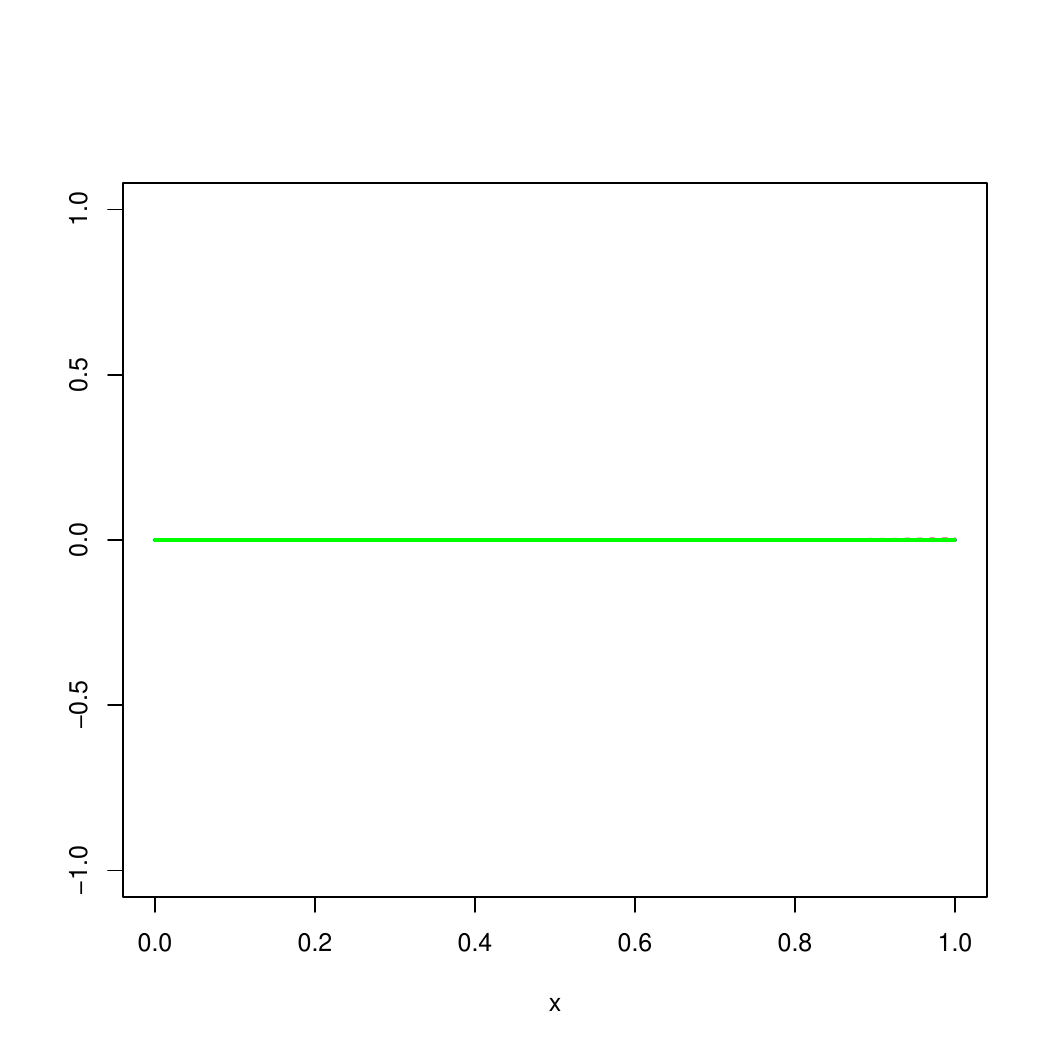}
   \end{tabular}
\caption{\small \label{fig:curvas-cl-rob-C0} The first two rows contain the functional boxplots of the estimated additive functions using the least square-based estimator while the last two rows contain the functional boxplots of the estimated additive functions by the robust approach, for $n=200$ when no contaminated data.} 
\end{center}
\end{figure}

\begin{figure}[htbp]
 \begin{center}
\small
\begin{tabular}{ccccc}
\includegraphics[scale=0.22]{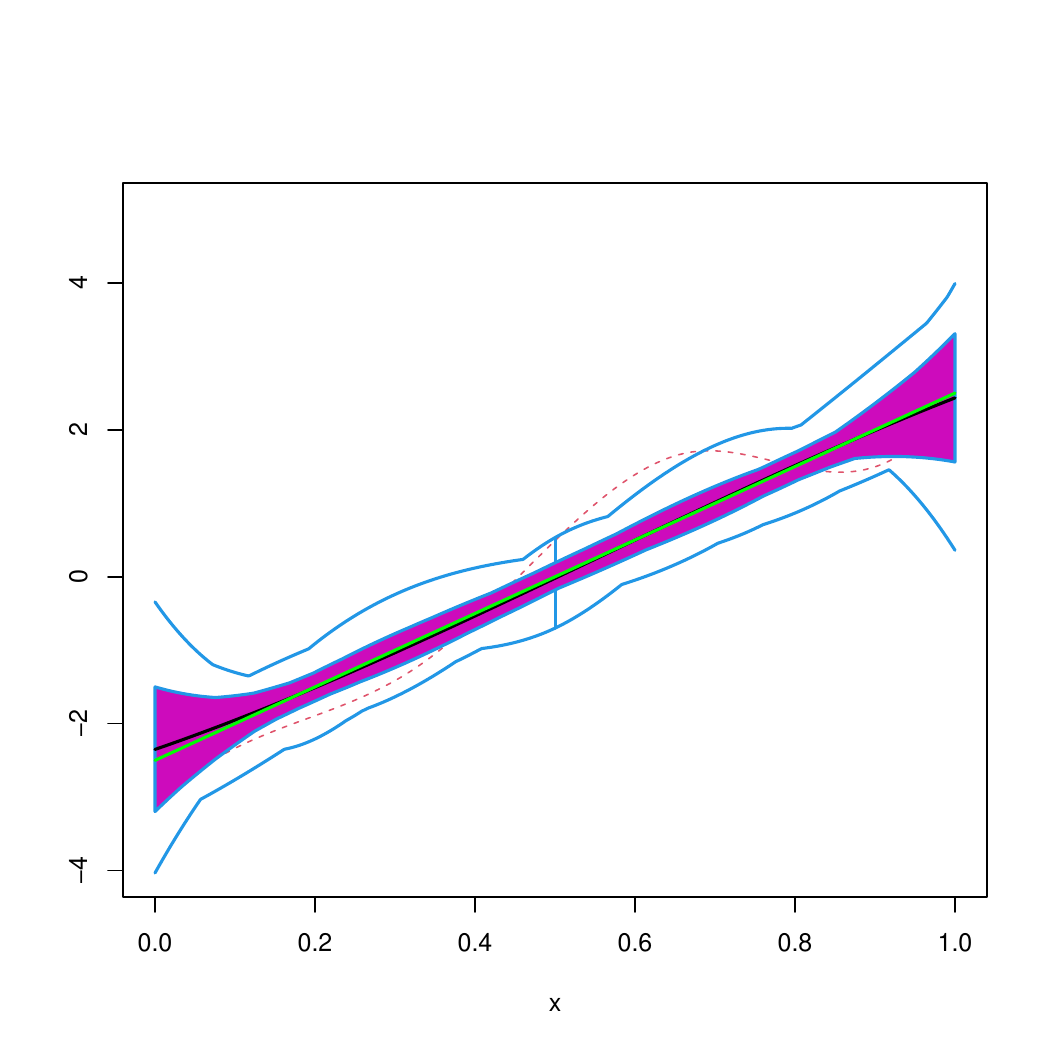} &
 \includegraphics[scale=0.22]{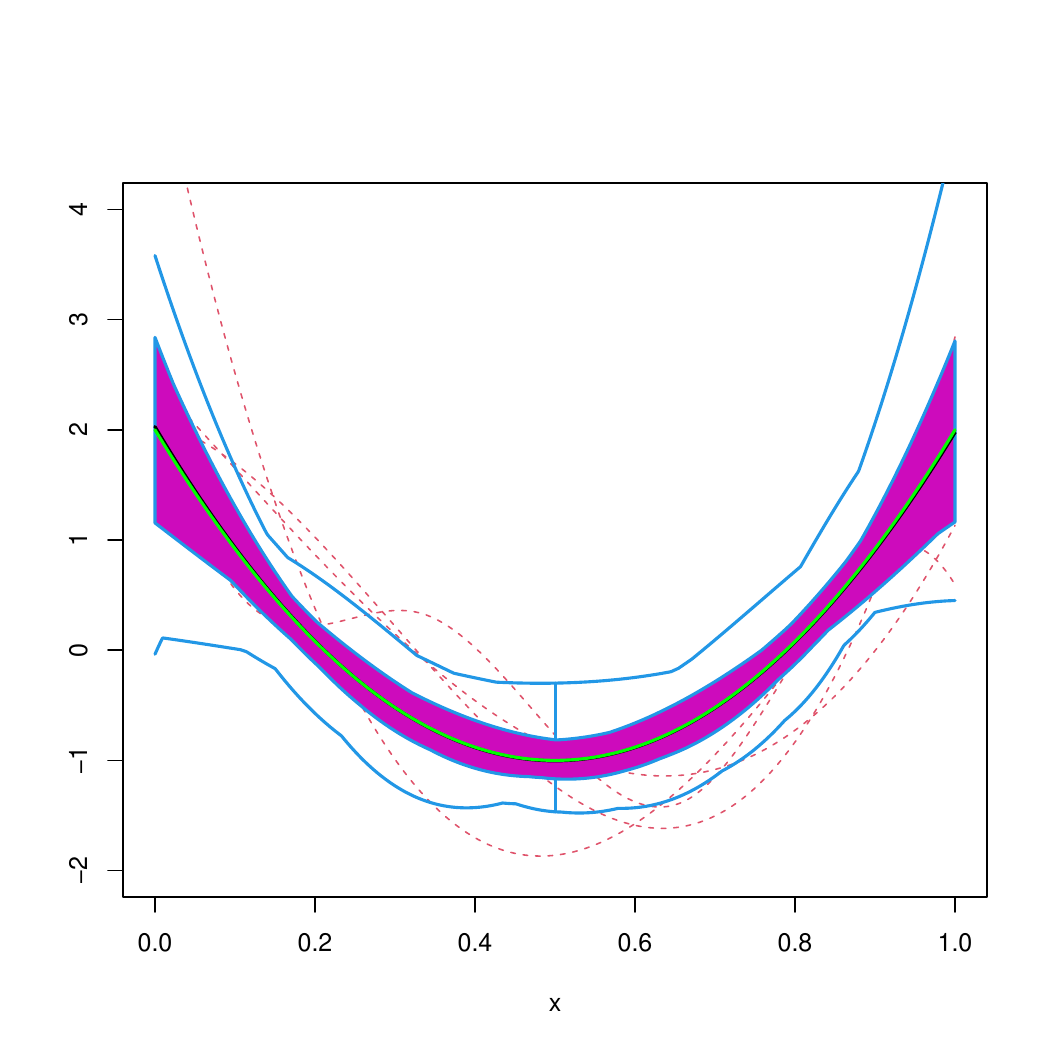} &
\includegraphics[scale=0.22]{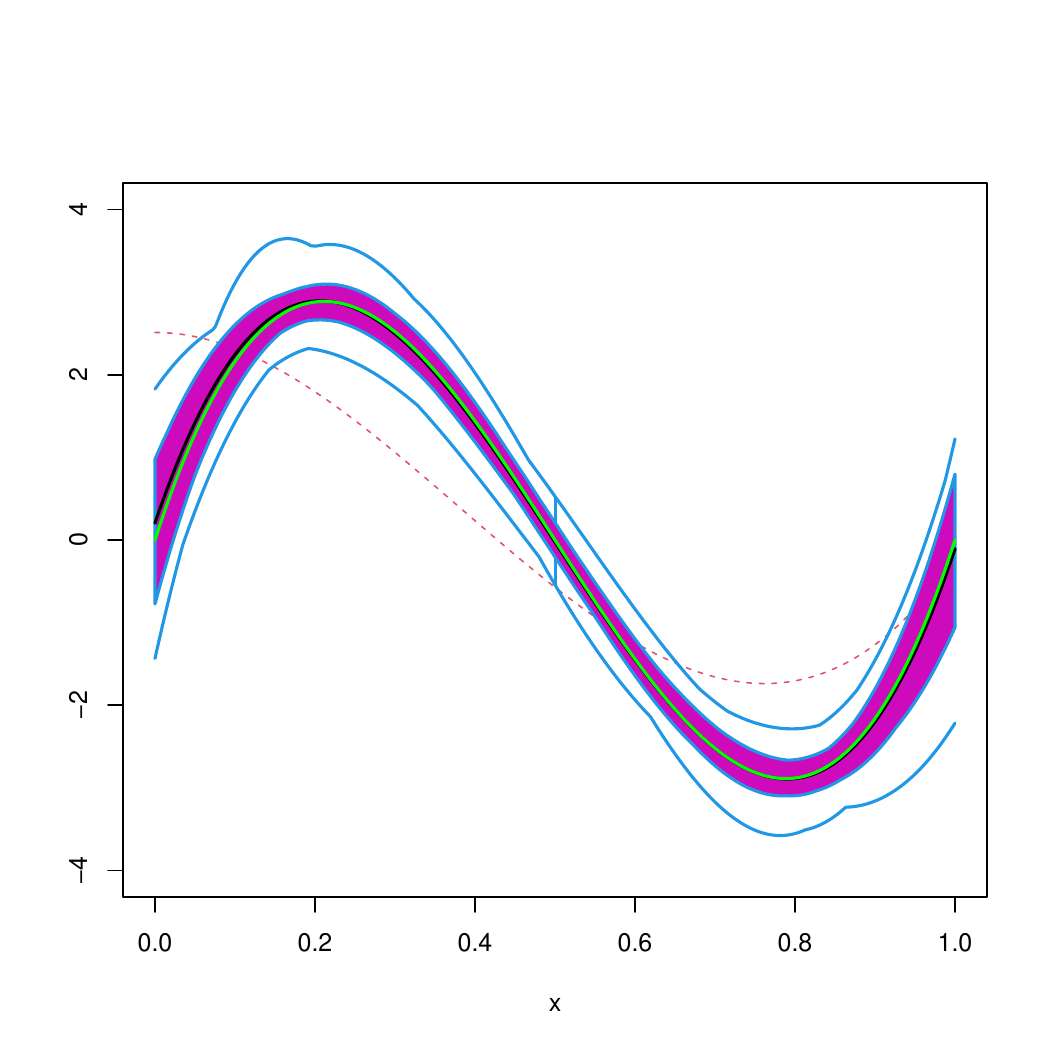}&
\includegraphics[scale=0.22]{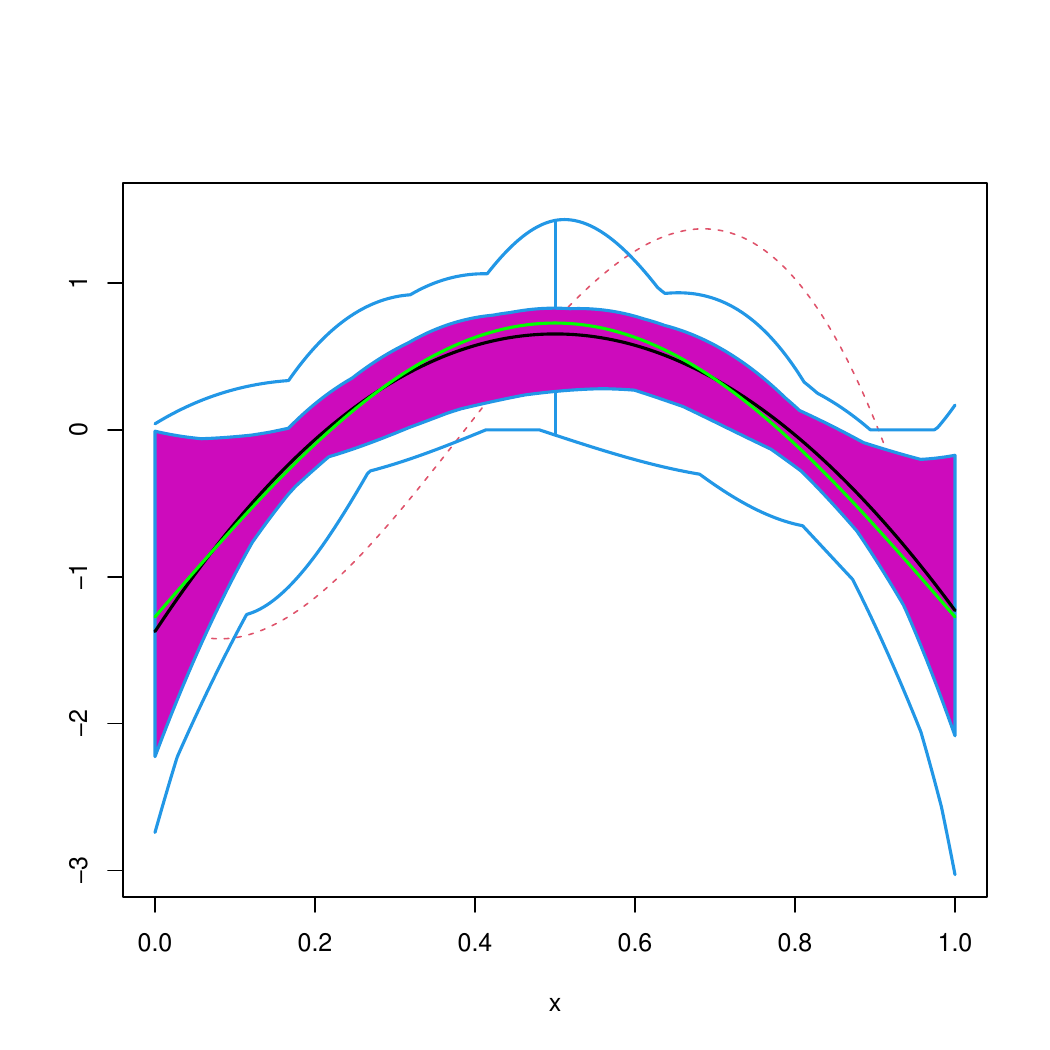}&
\includegraphics[scale=0.22]{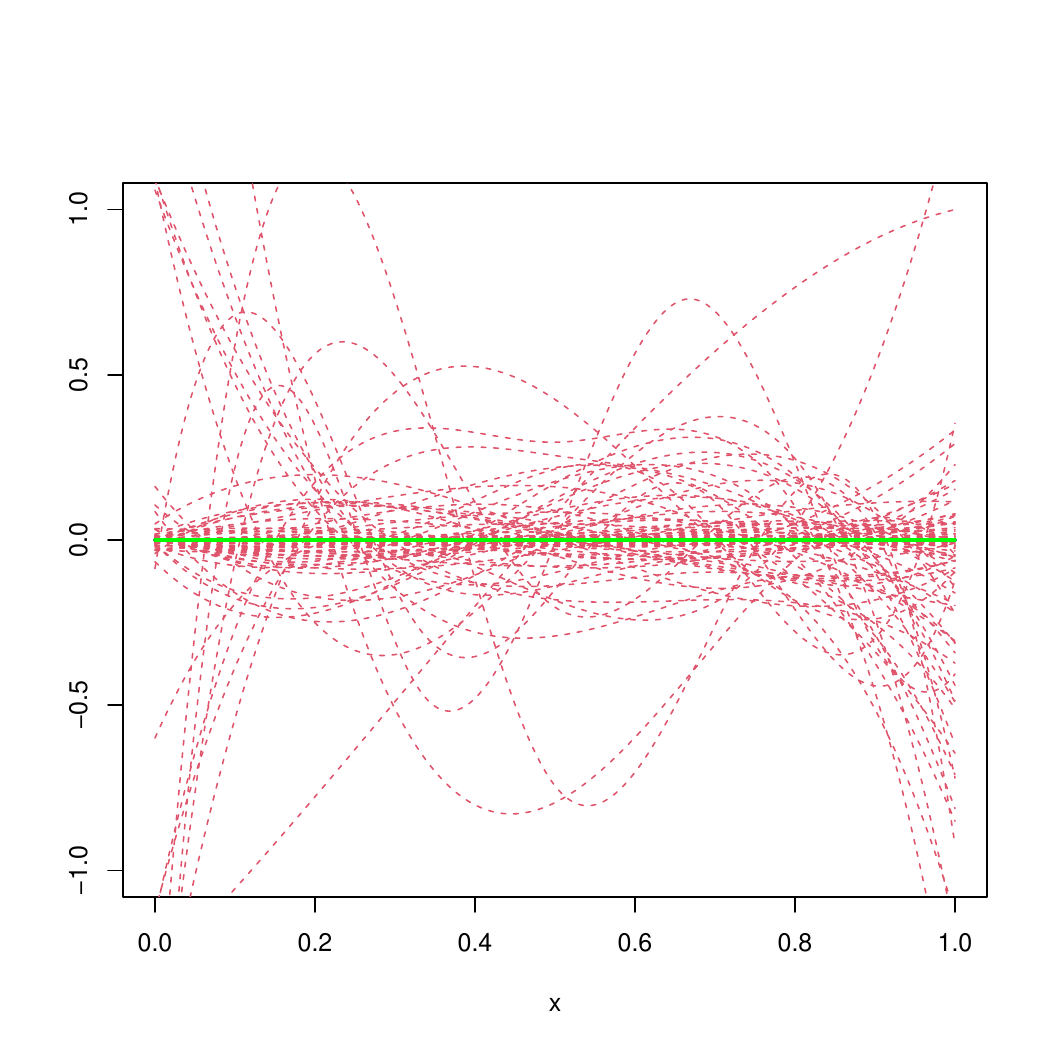}\\
\includegraphics[scale=0.22]{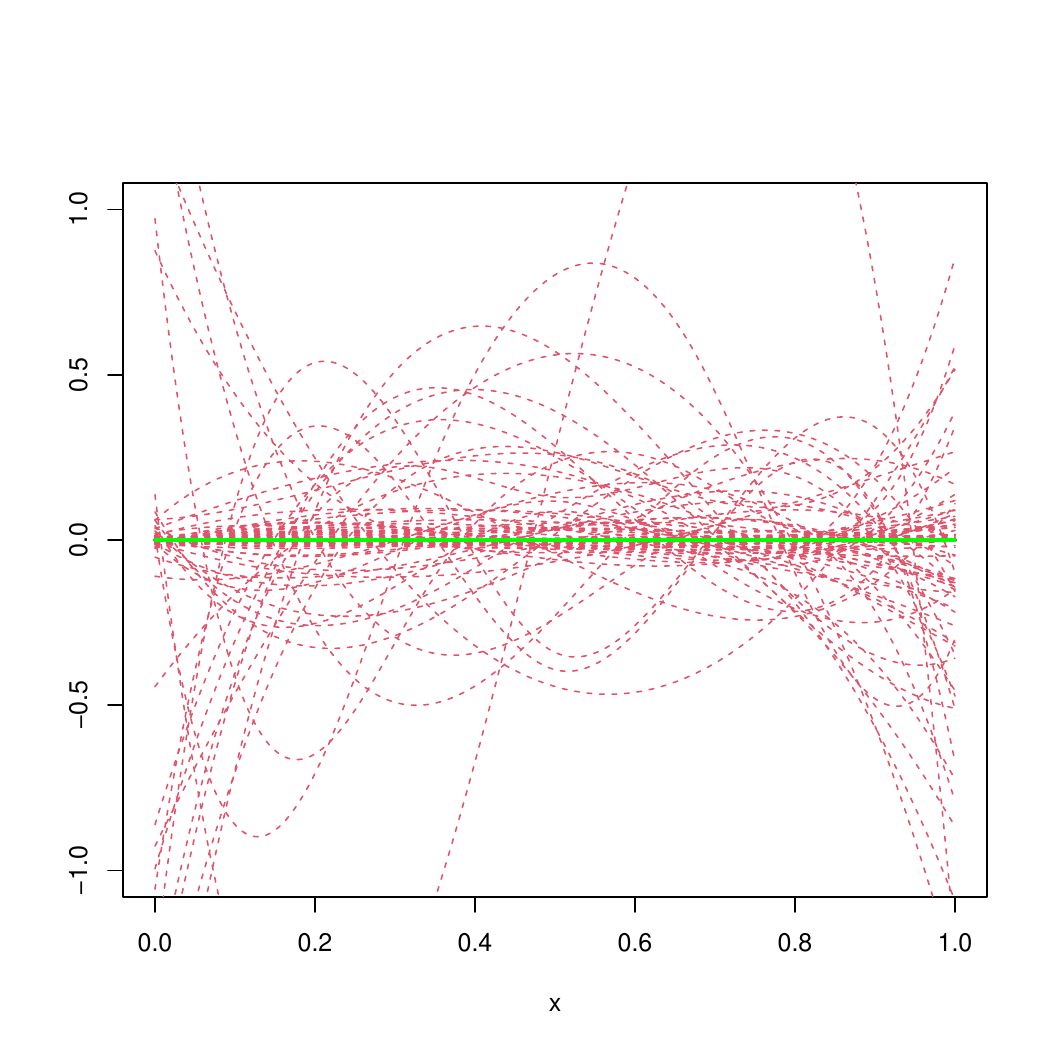} &
 \includegraphics[scale=0.22]{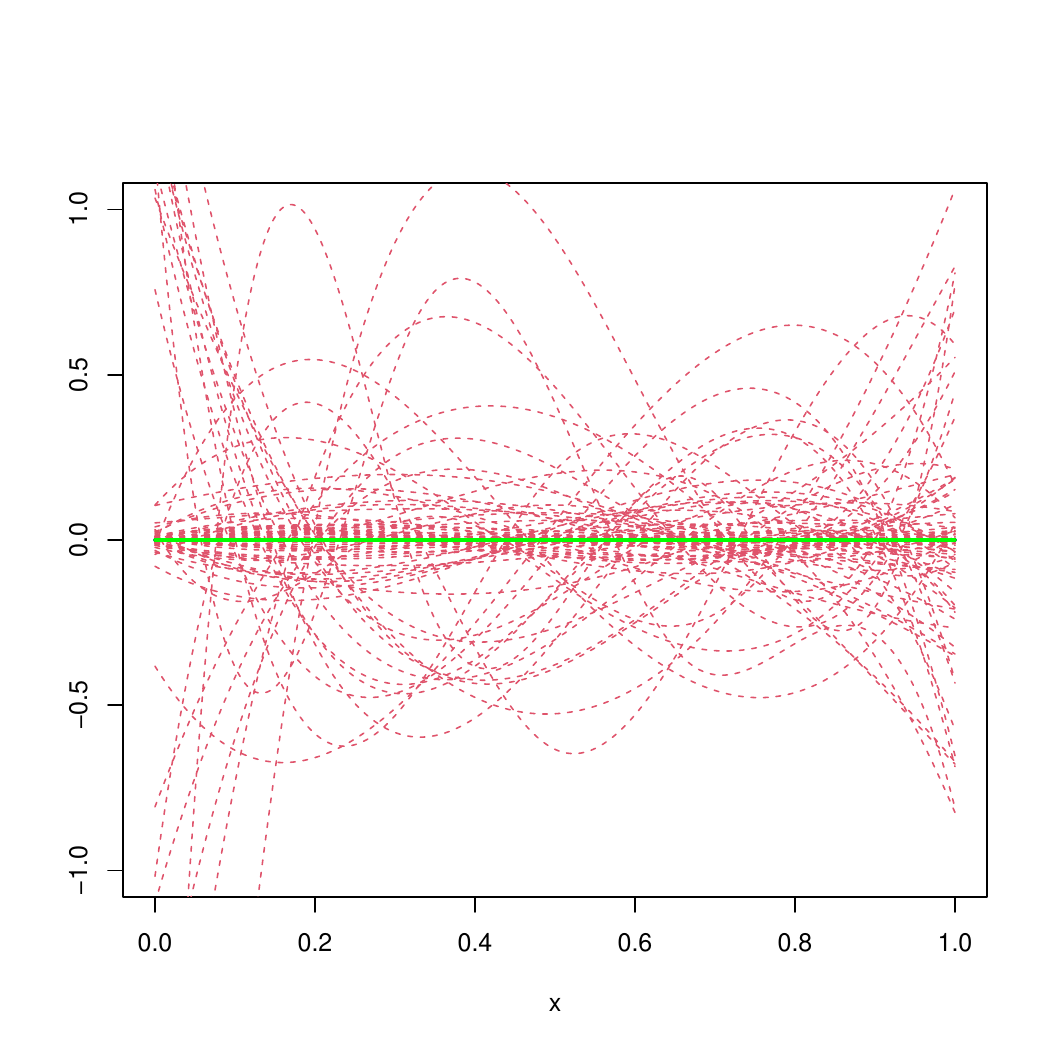}&
 \includegraphics[scale=0.22]{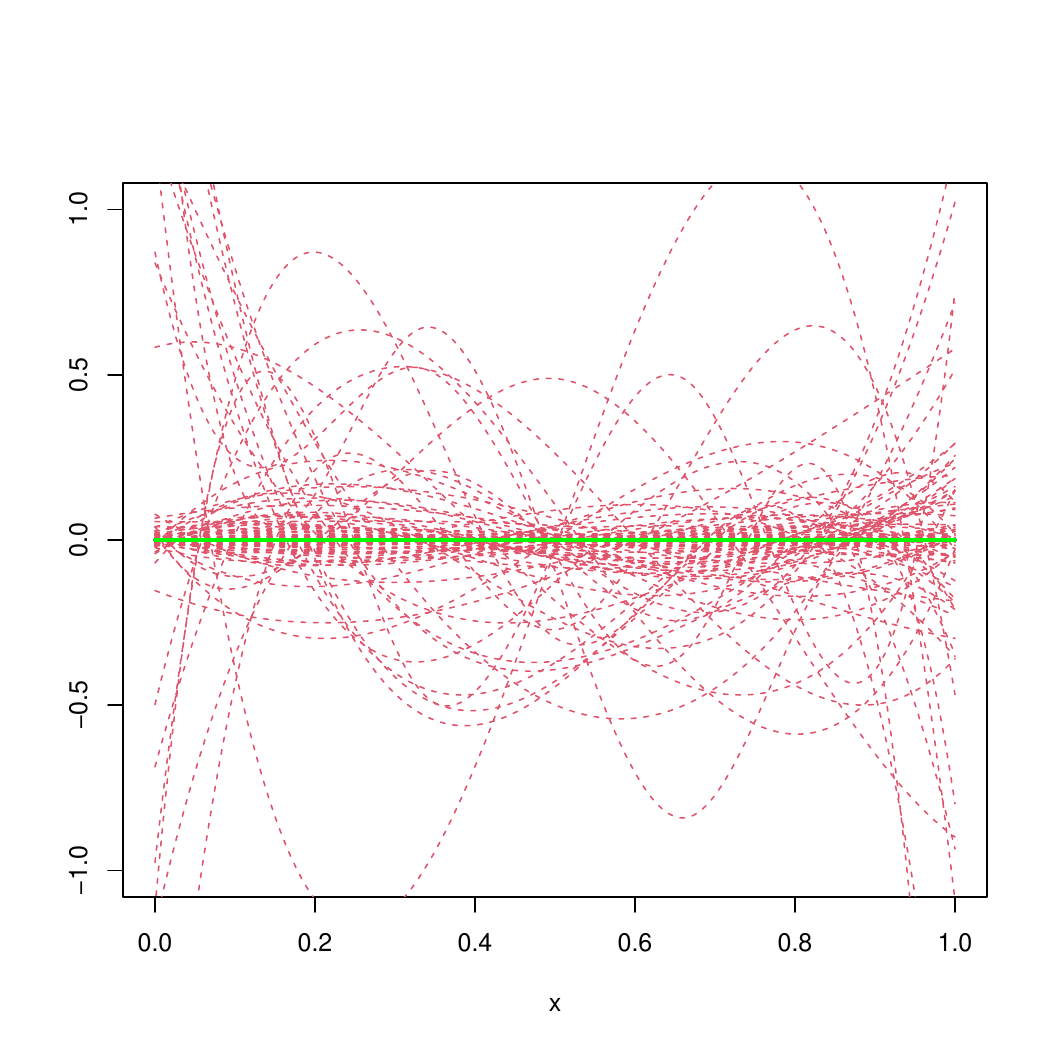}&
  \includegraphics[scale=0.22]{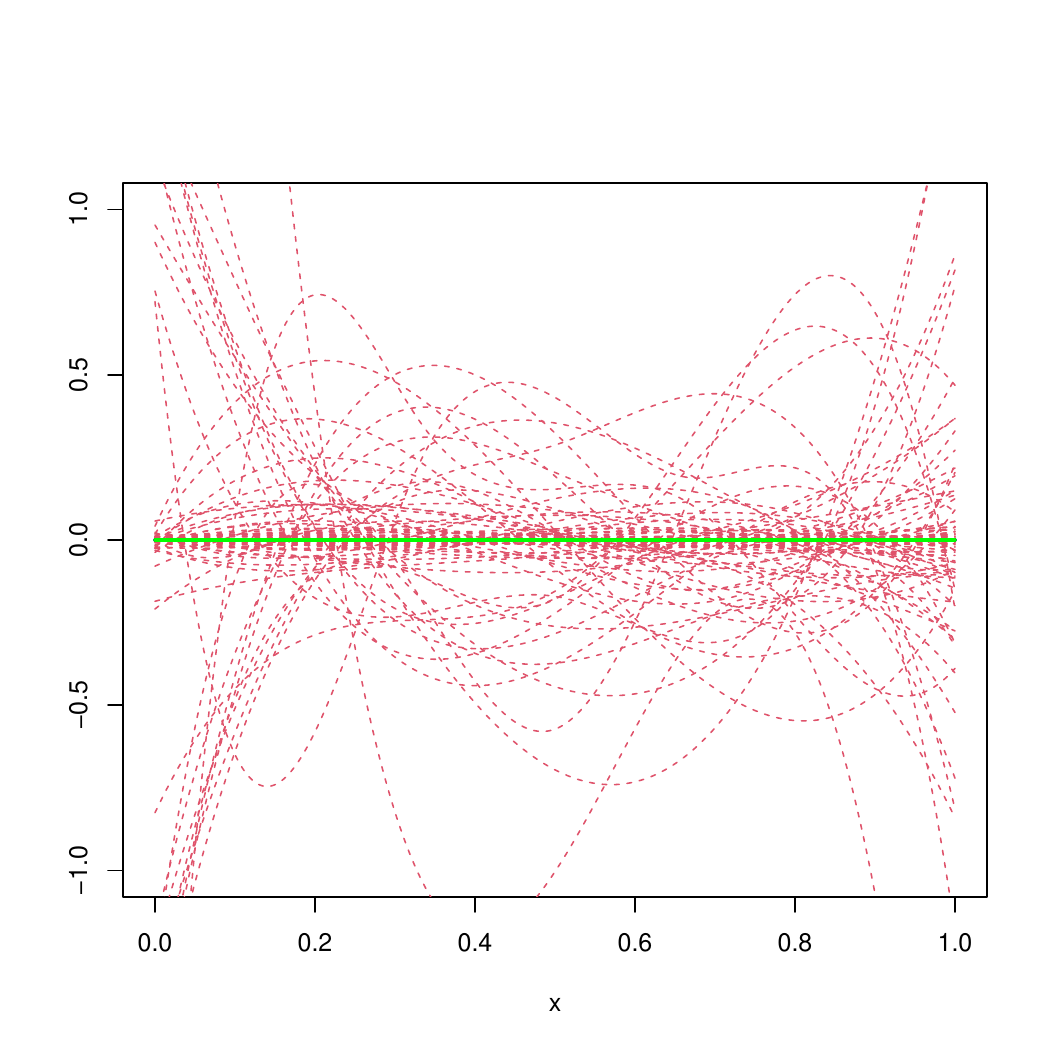}&
   \includegraphics[scale=0.22]{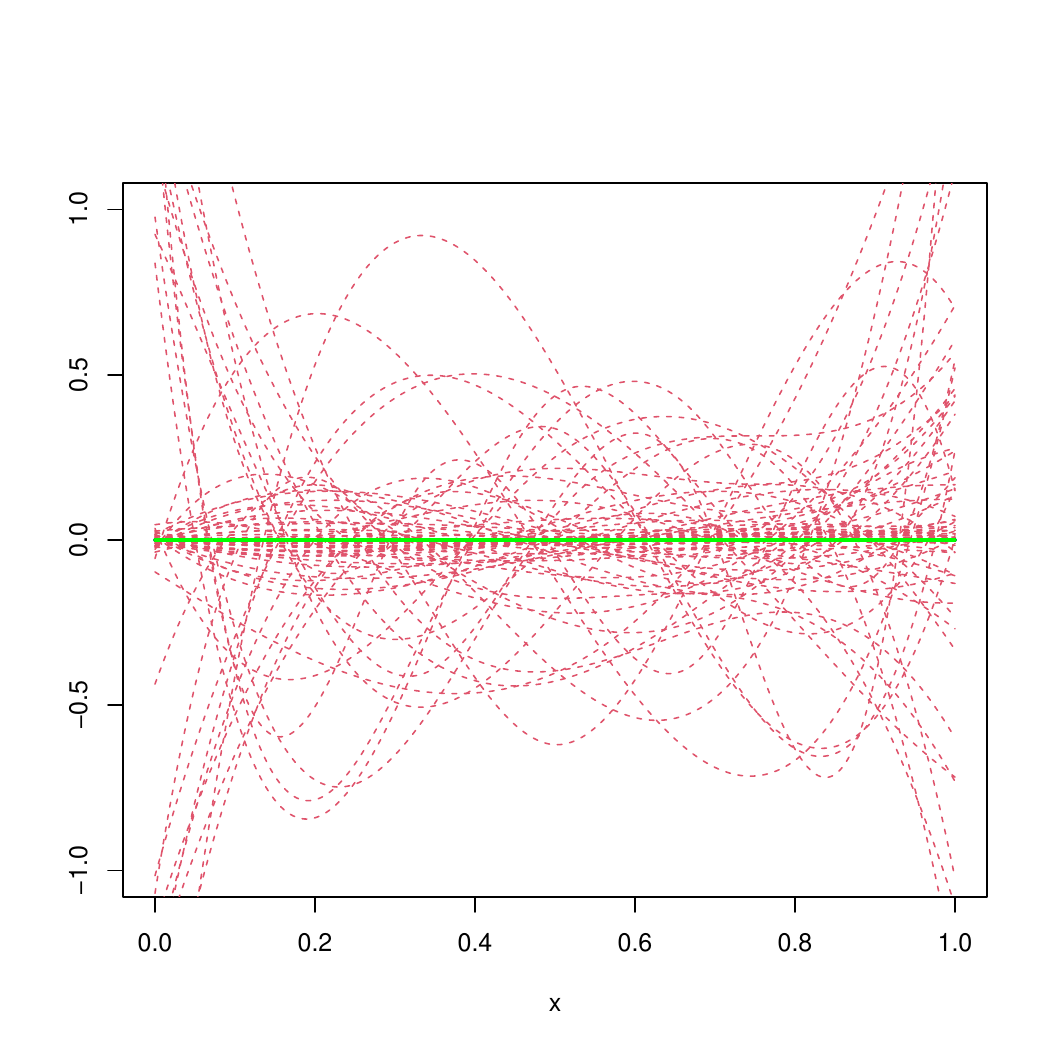} \\
   \includegraphics[scale=0.22]{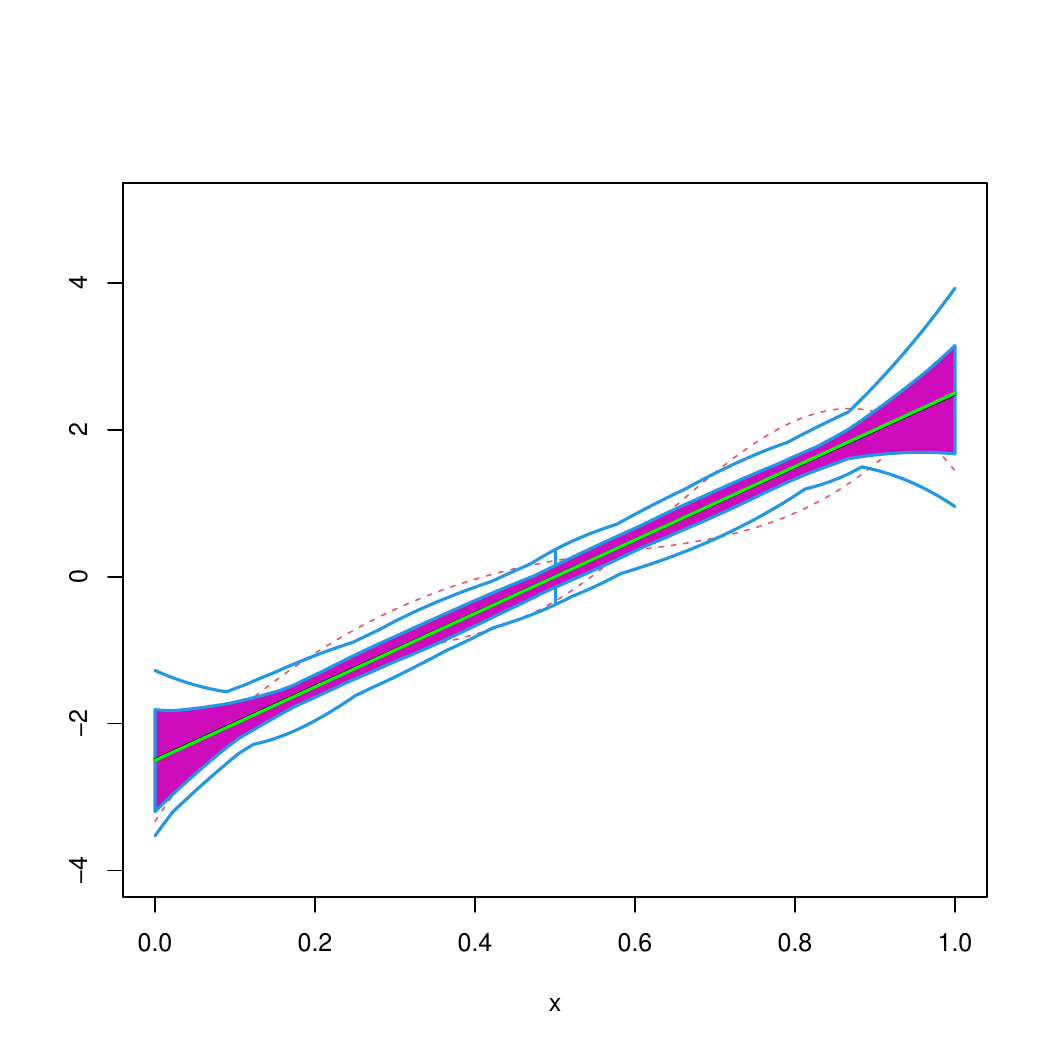} &
 \includegraphics[scale=0.22]{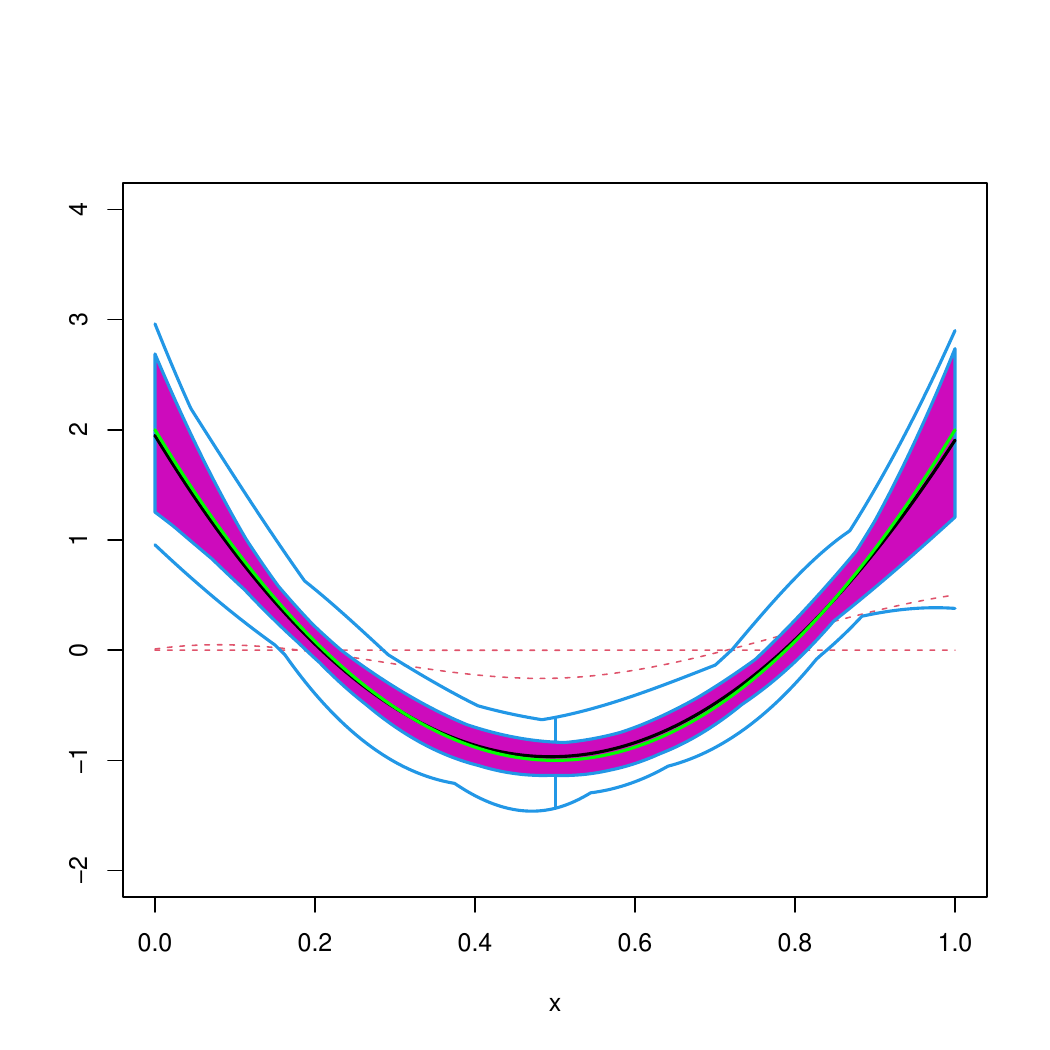} &
\includegraphics[scale=0.22]{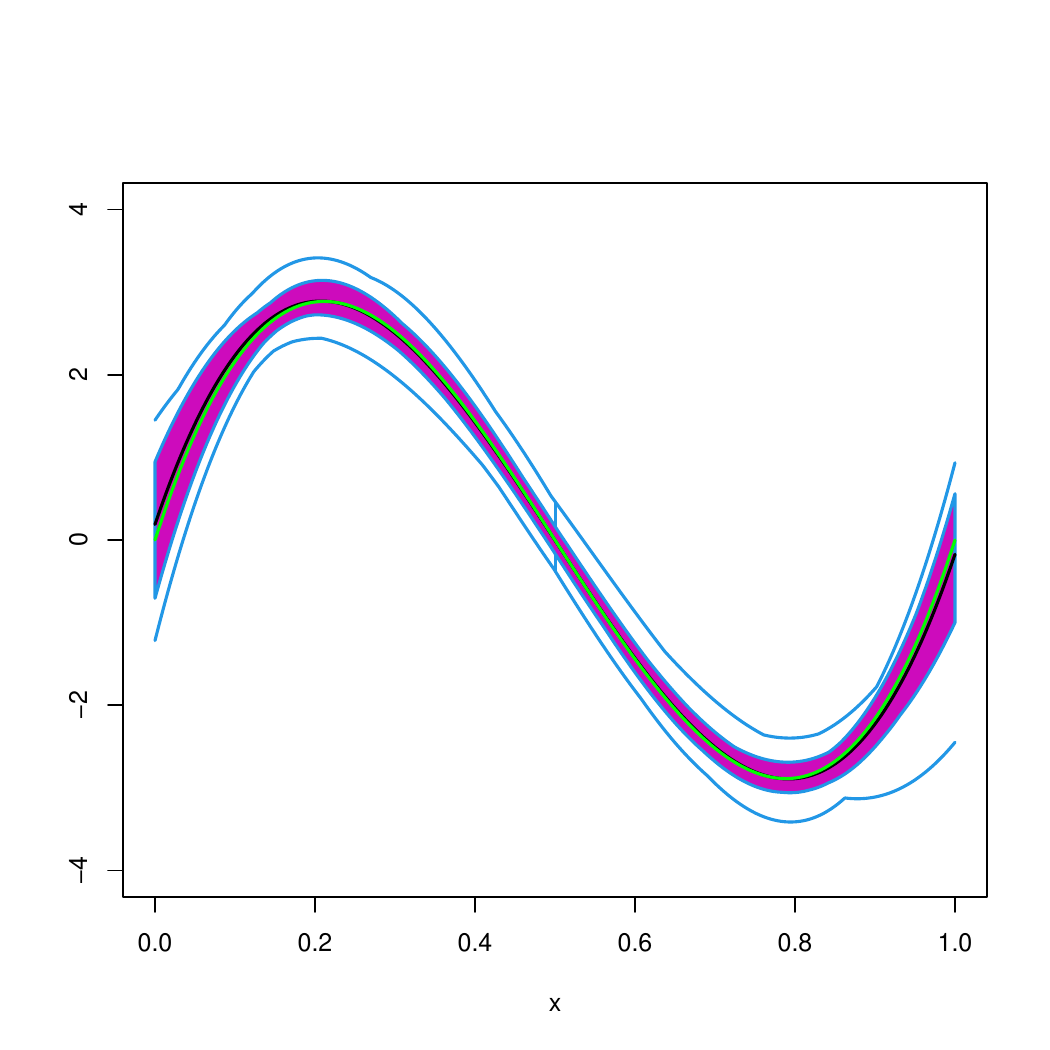}&
\includegraphics[scale=0.22]{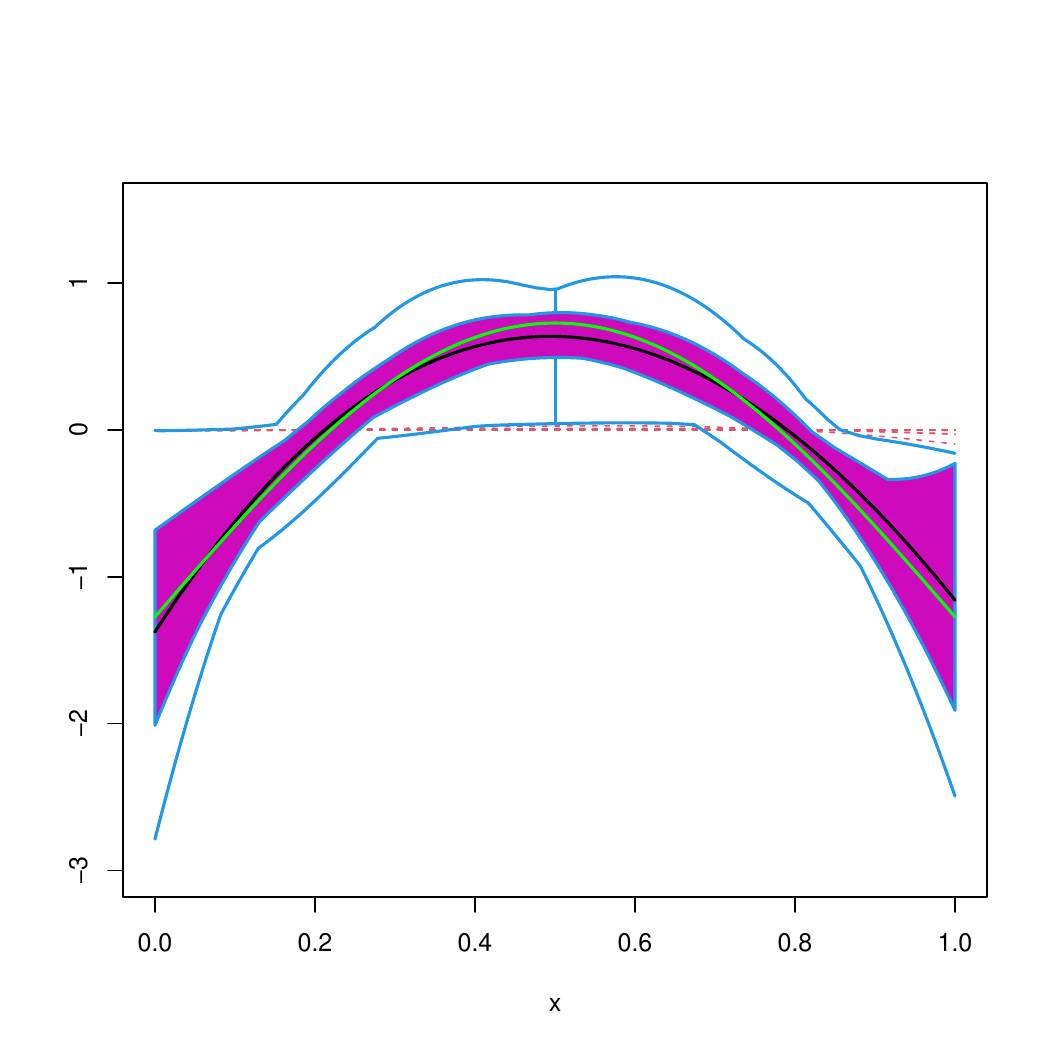}&
\includegraphics[scale=0.22]{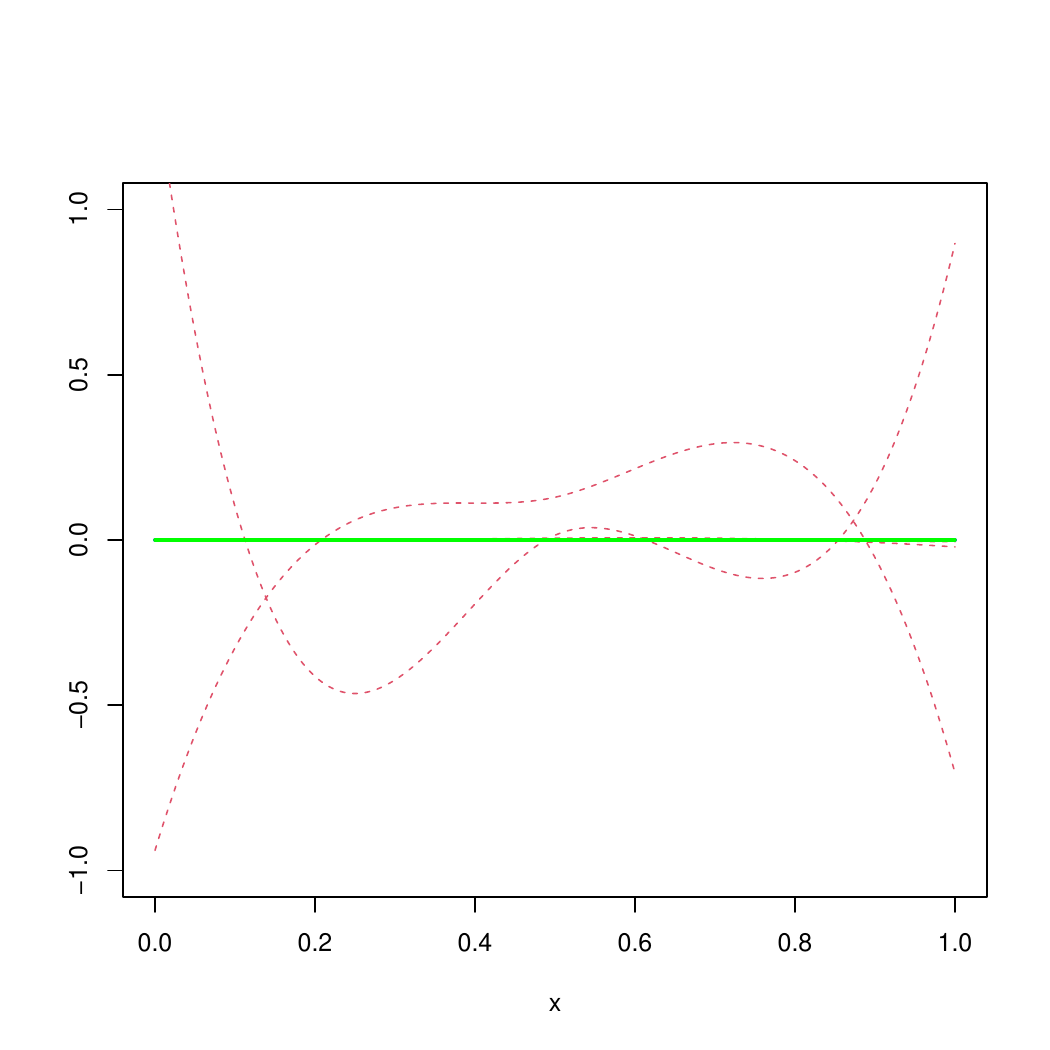}\\
\includegraphics[scale=0.22]{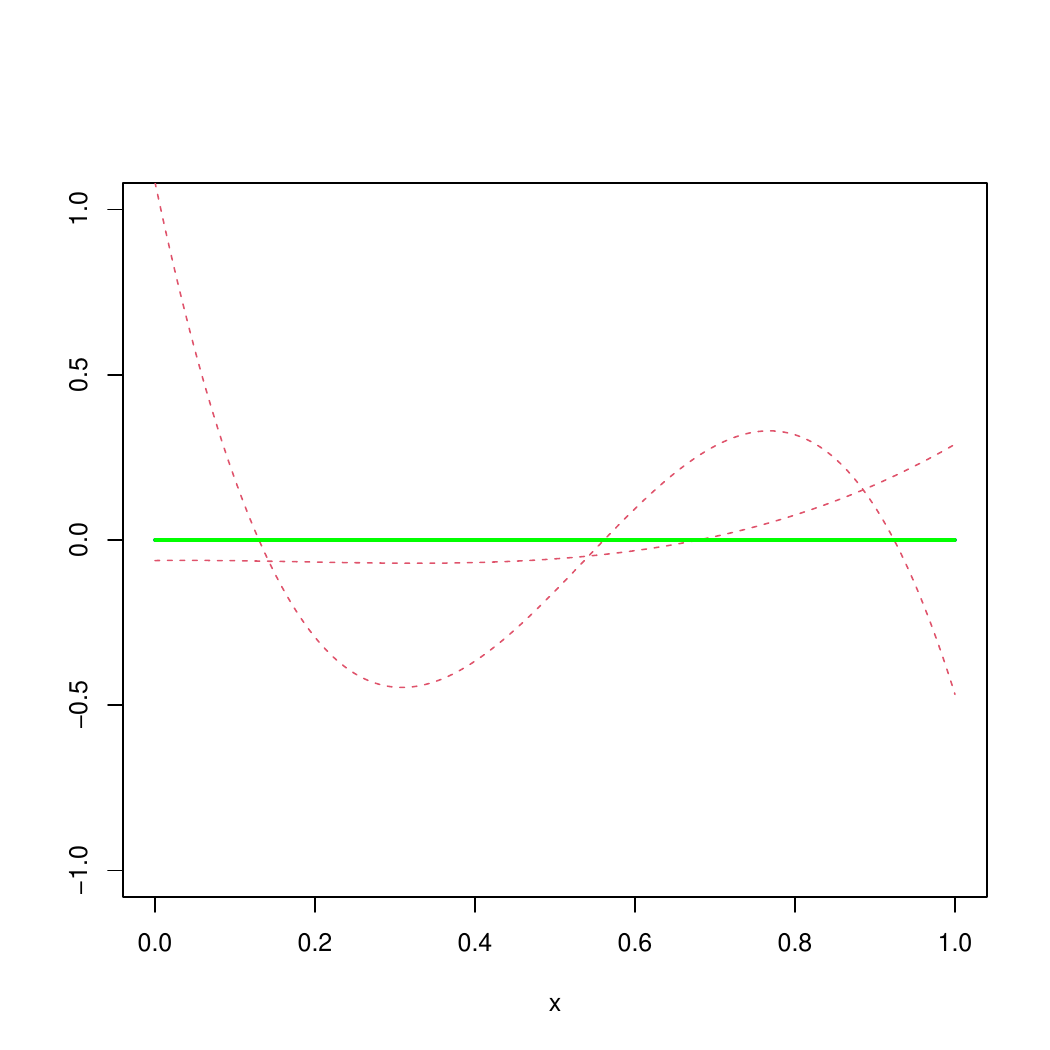} &
 \includegraphics[scale=0.22]{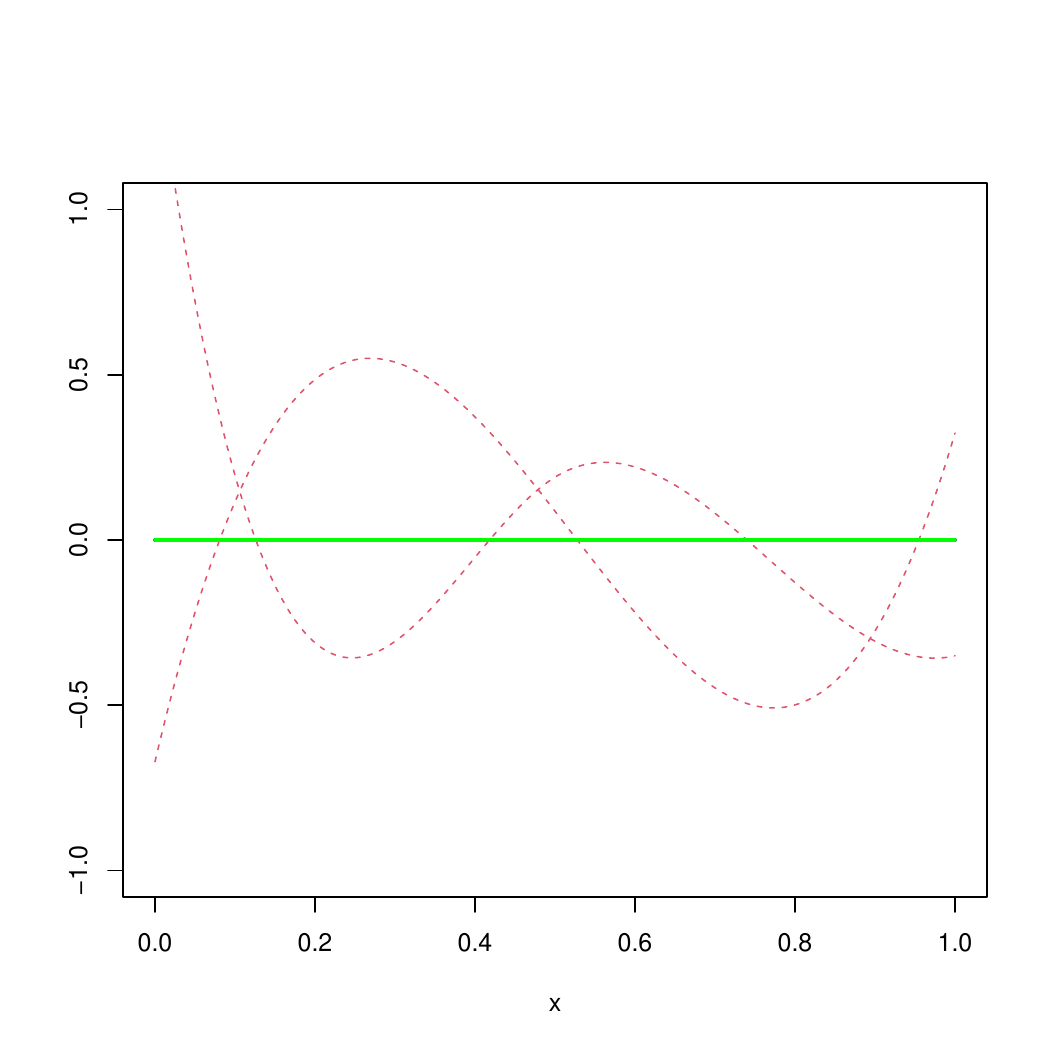}&
 \includegraphics[scale=0.22]{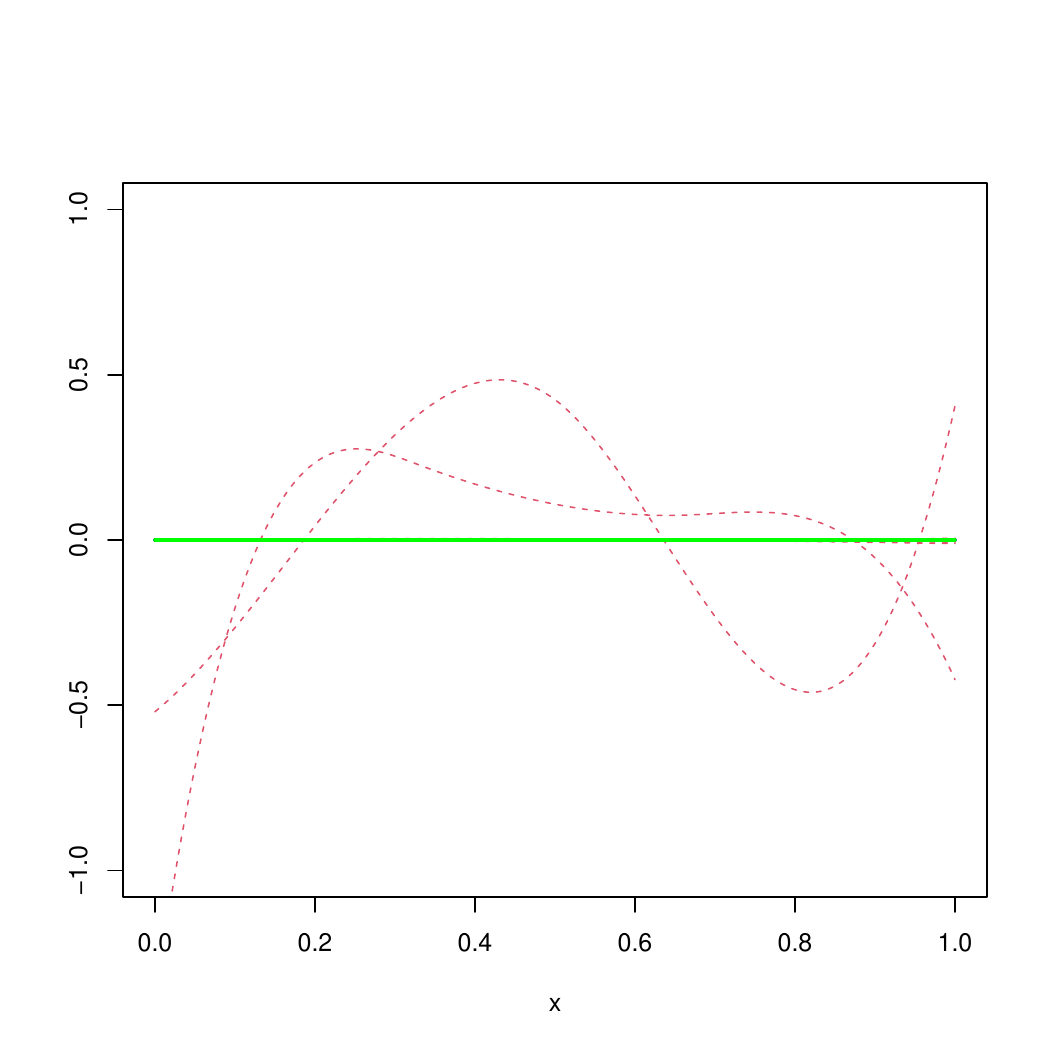}&
  \includegraphics[scale=0.22]{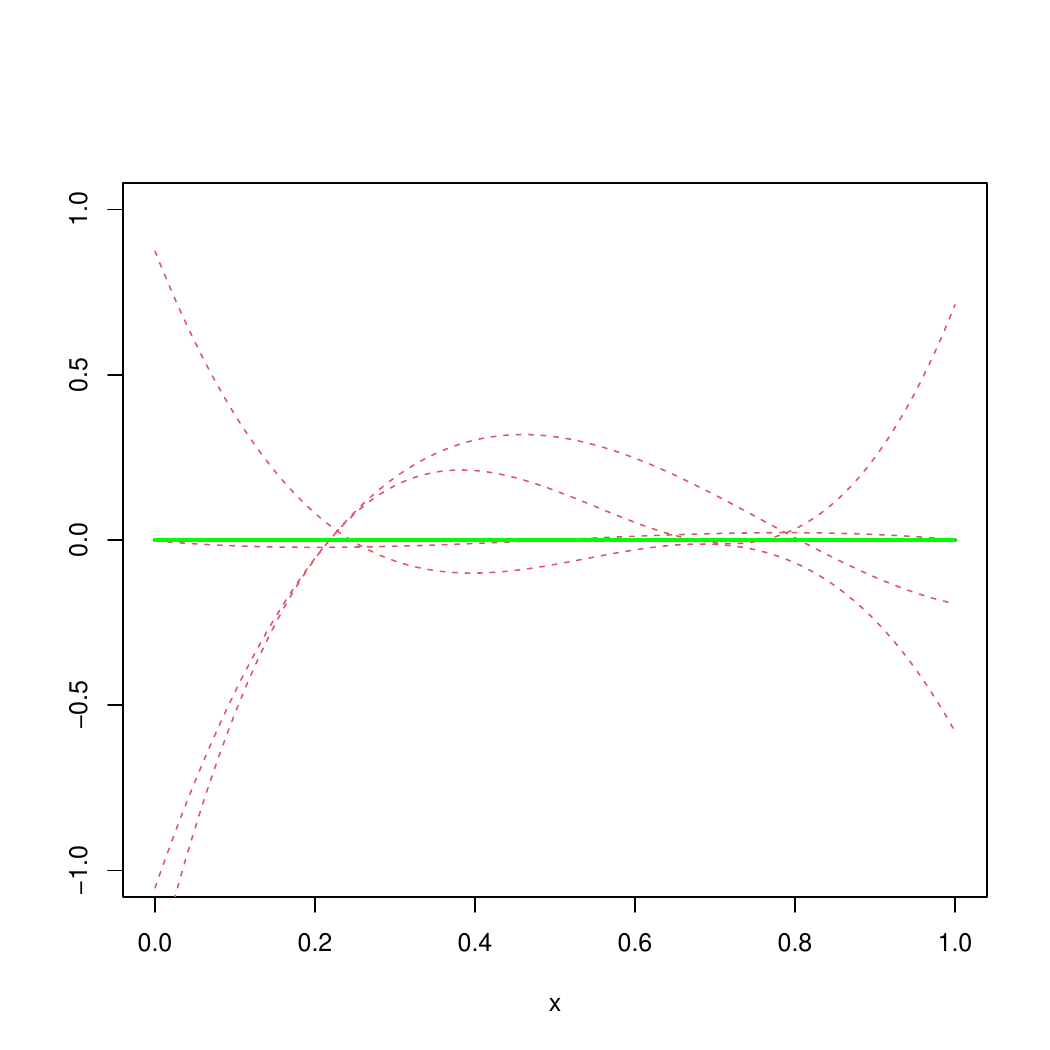}&
   \includegraphics[scale=0.22]{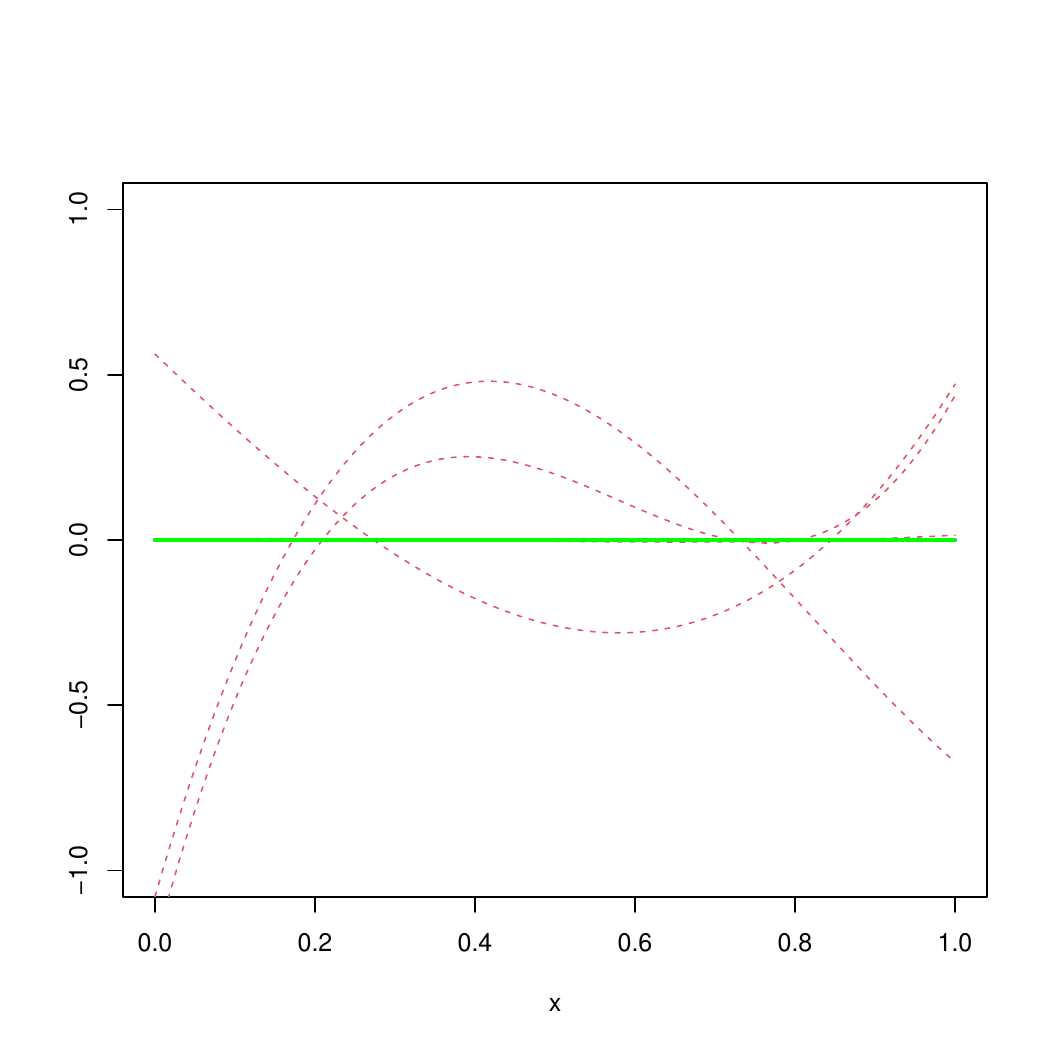}
   \end{tabular}
\caption{\small \label{fig:curvas-cl-rob-C1} The first two rows contain the functional boxplots of the estimated additive functions using the least square-based estimator while the last two rows contain the functional boxplots of the estimated additive functions by the robust approach, for $n=200$ and the contamination setting $C_1$.} 
\end{center}
\end{figure}

\begin{figure}[htbp]
 \begin{center}
\small
\begin{tabular}{ccccc}
\includegraphics[scale=0.22]{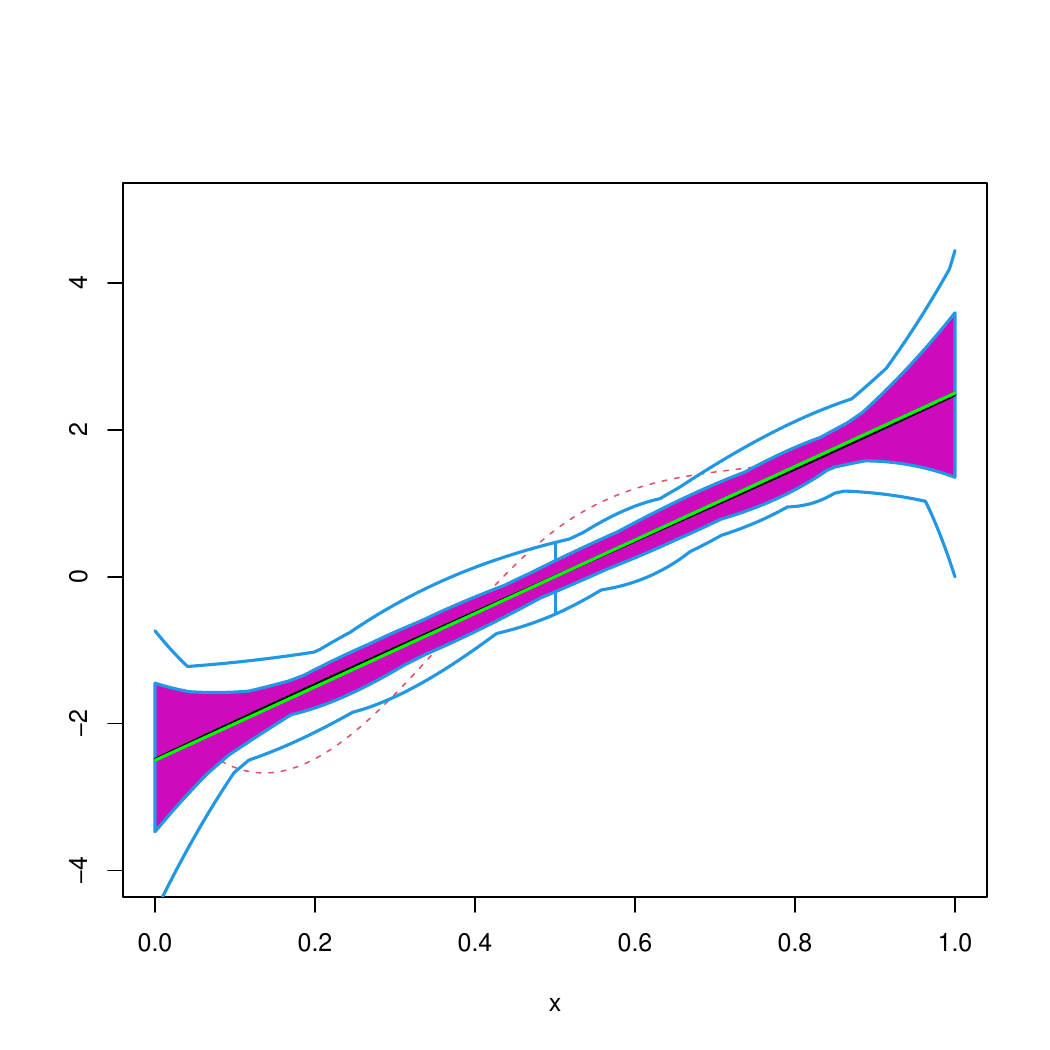} &
 \includegraphics[scale=0.22]{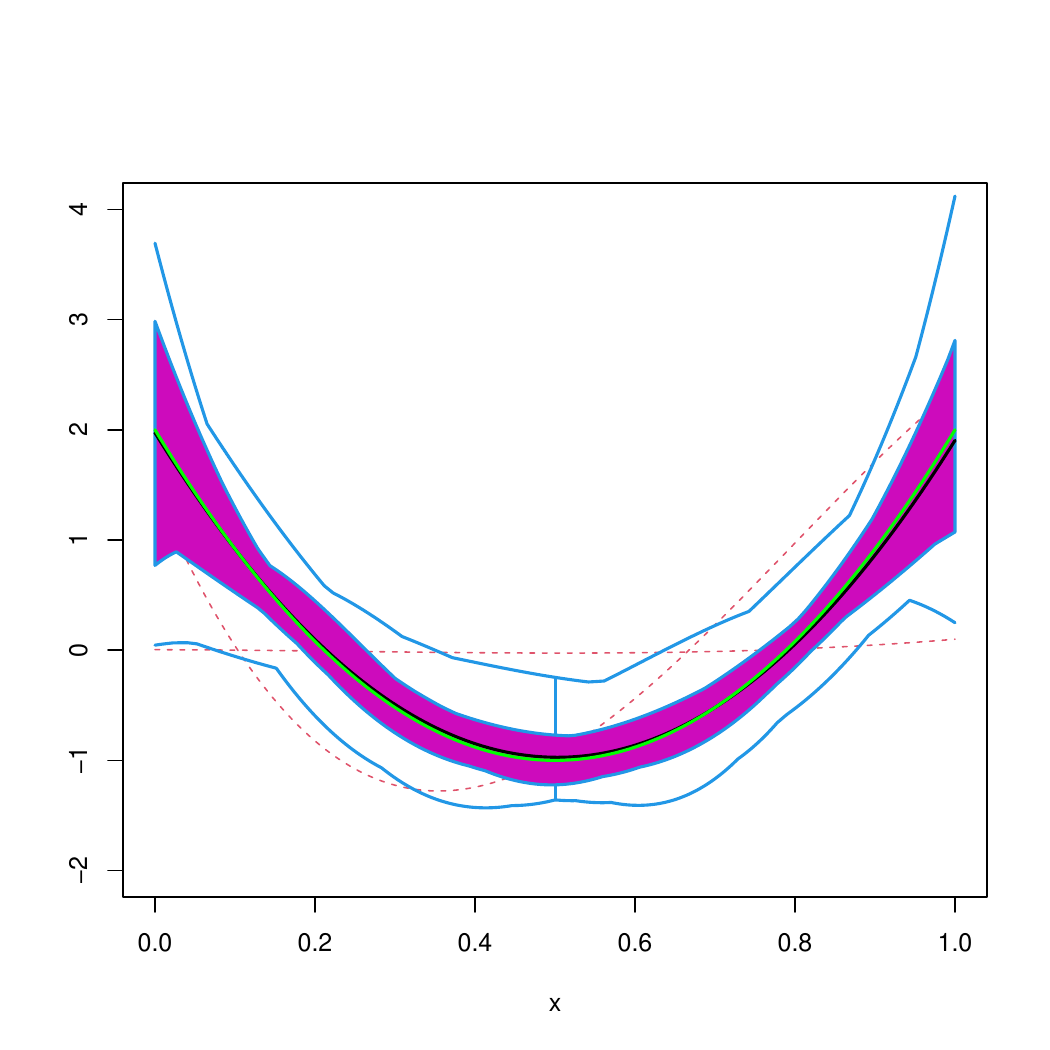} &
\includegraphics[scale=0.22]{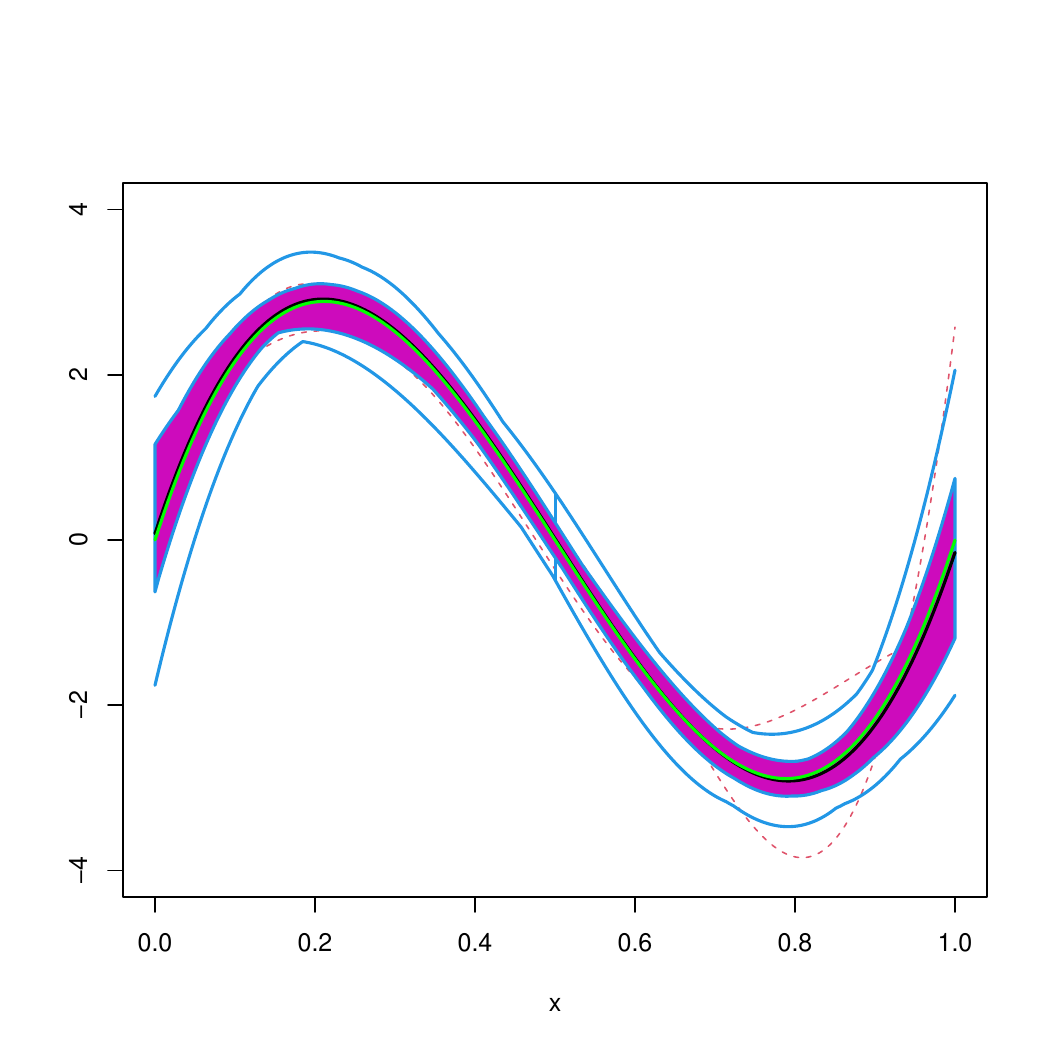}&
\includegraphics[scale=0.22]{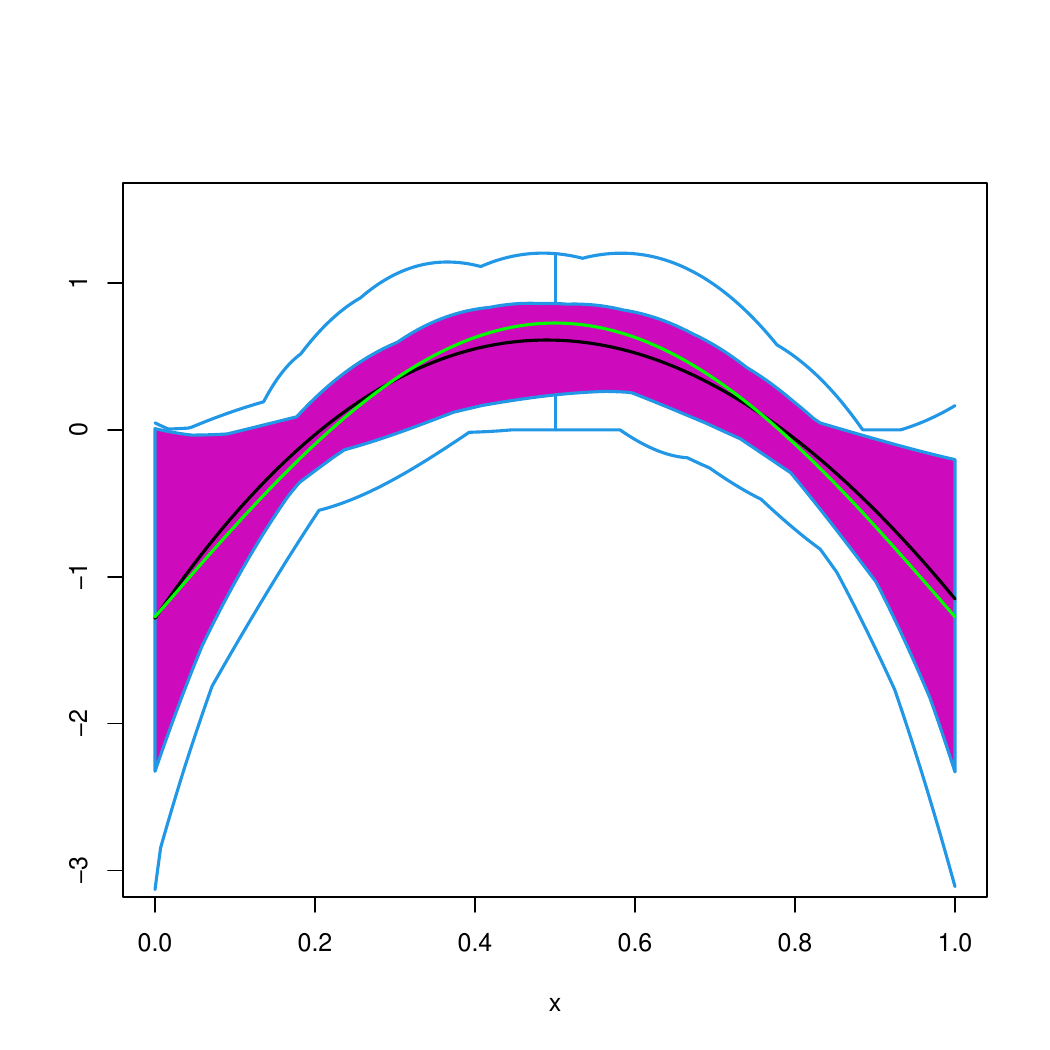}&
\includegraphics[scale=0.22]{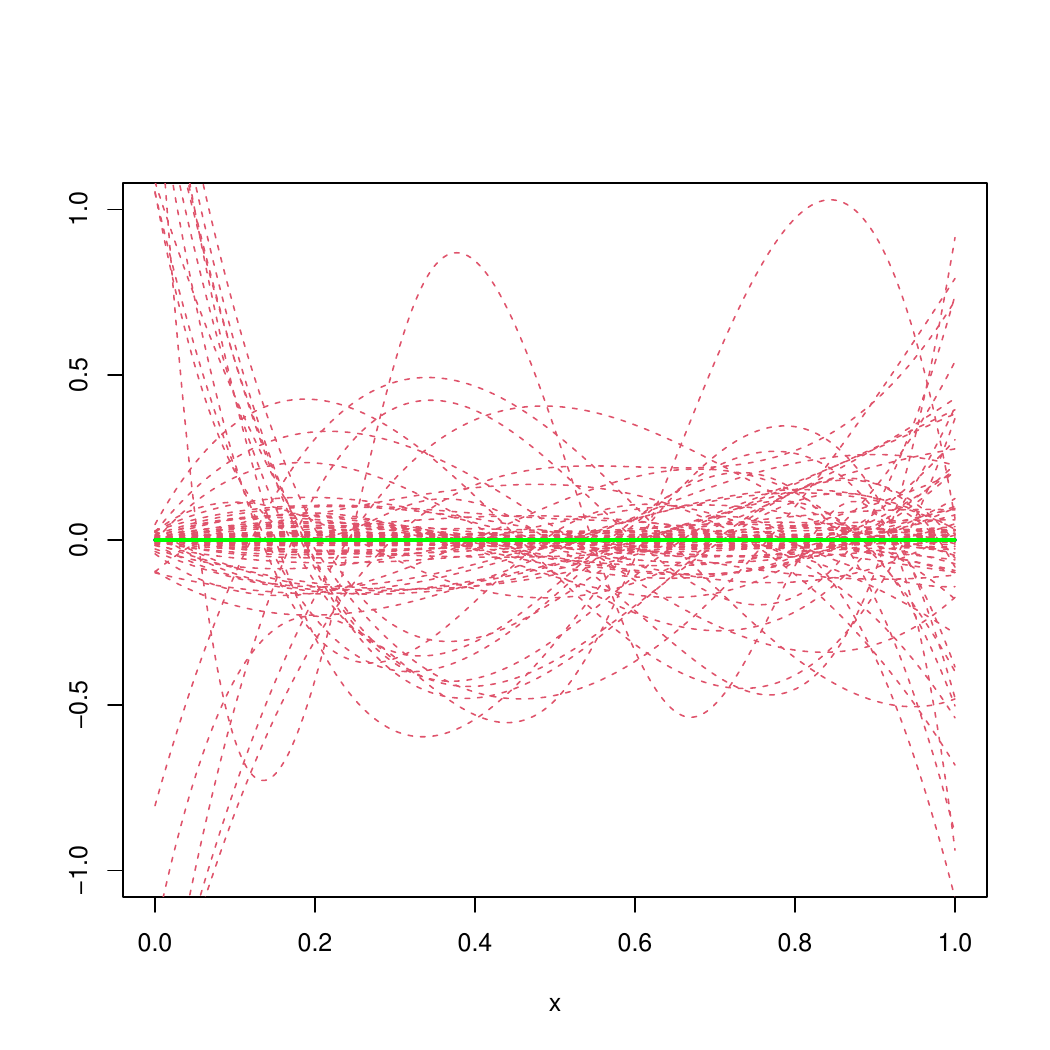}\\
\includegraphics[scale=0.22]{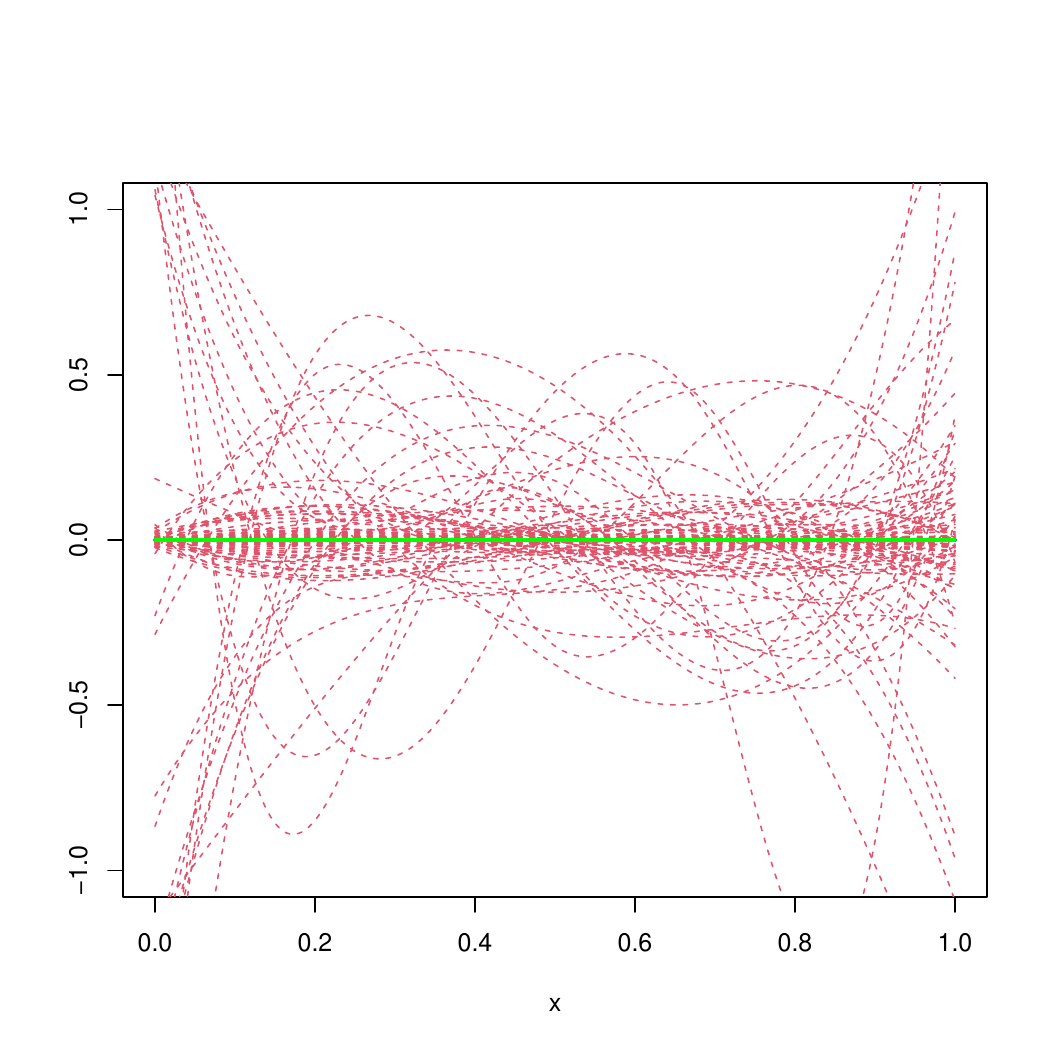} &
 \includegraphics[scale=0.22]{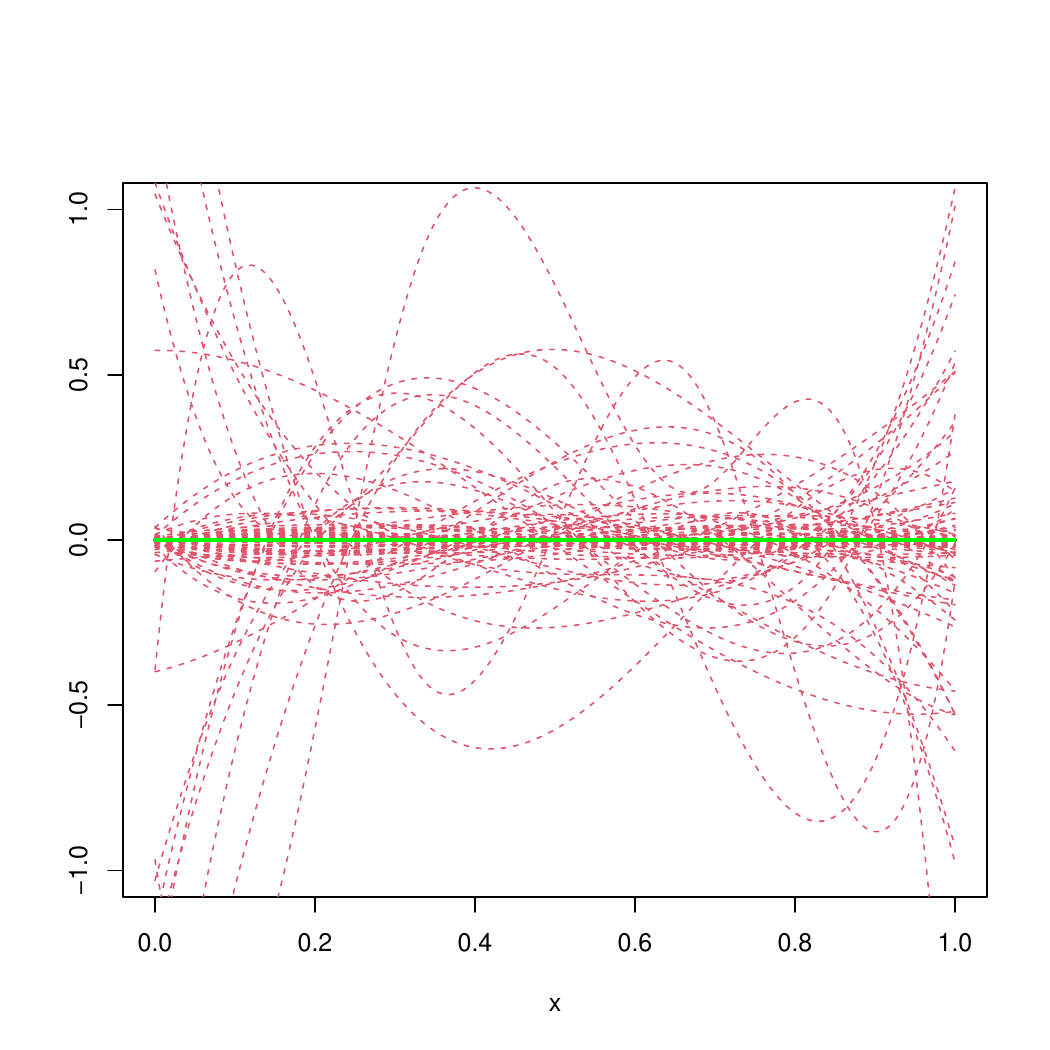}&
 \includegraphics[scale=0.22]{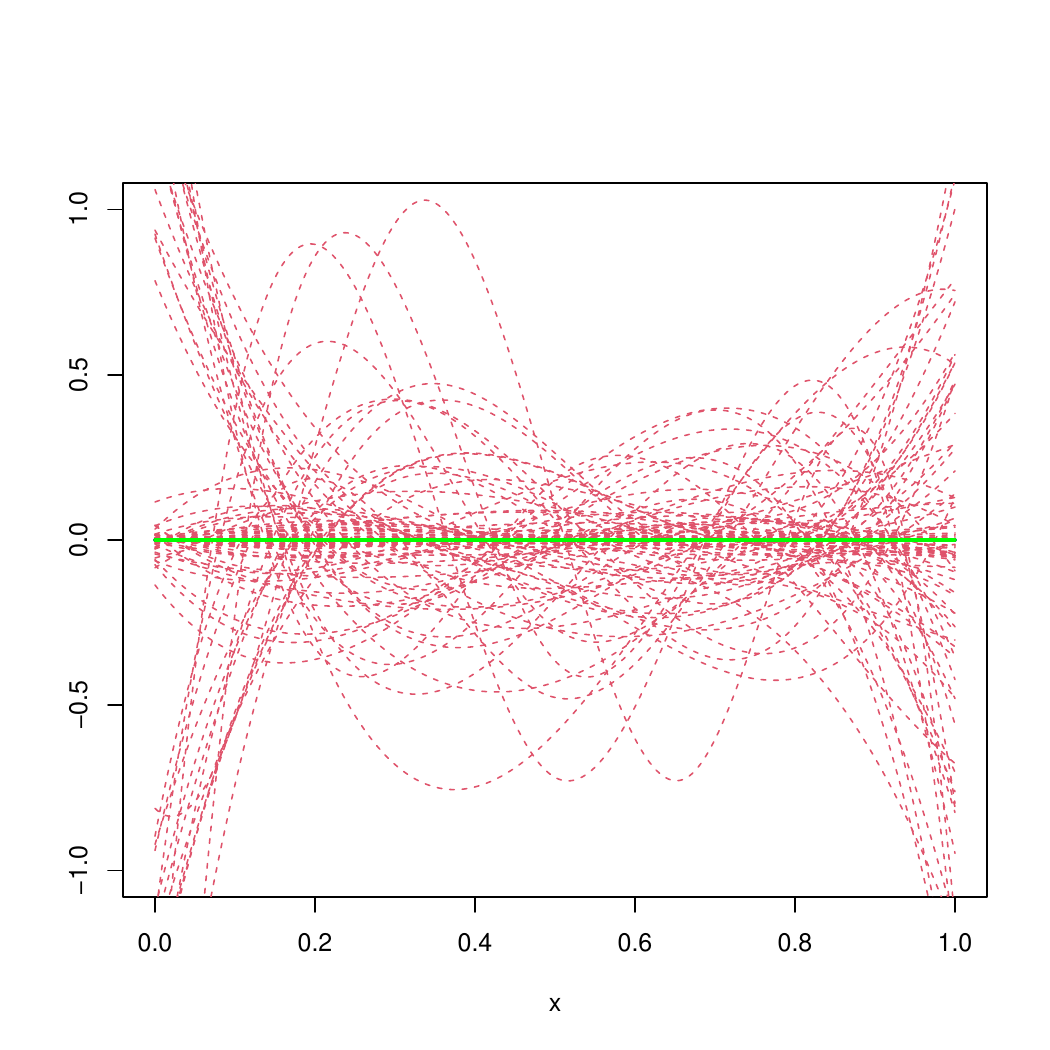}&
  \includegraphics[scale=0.22]{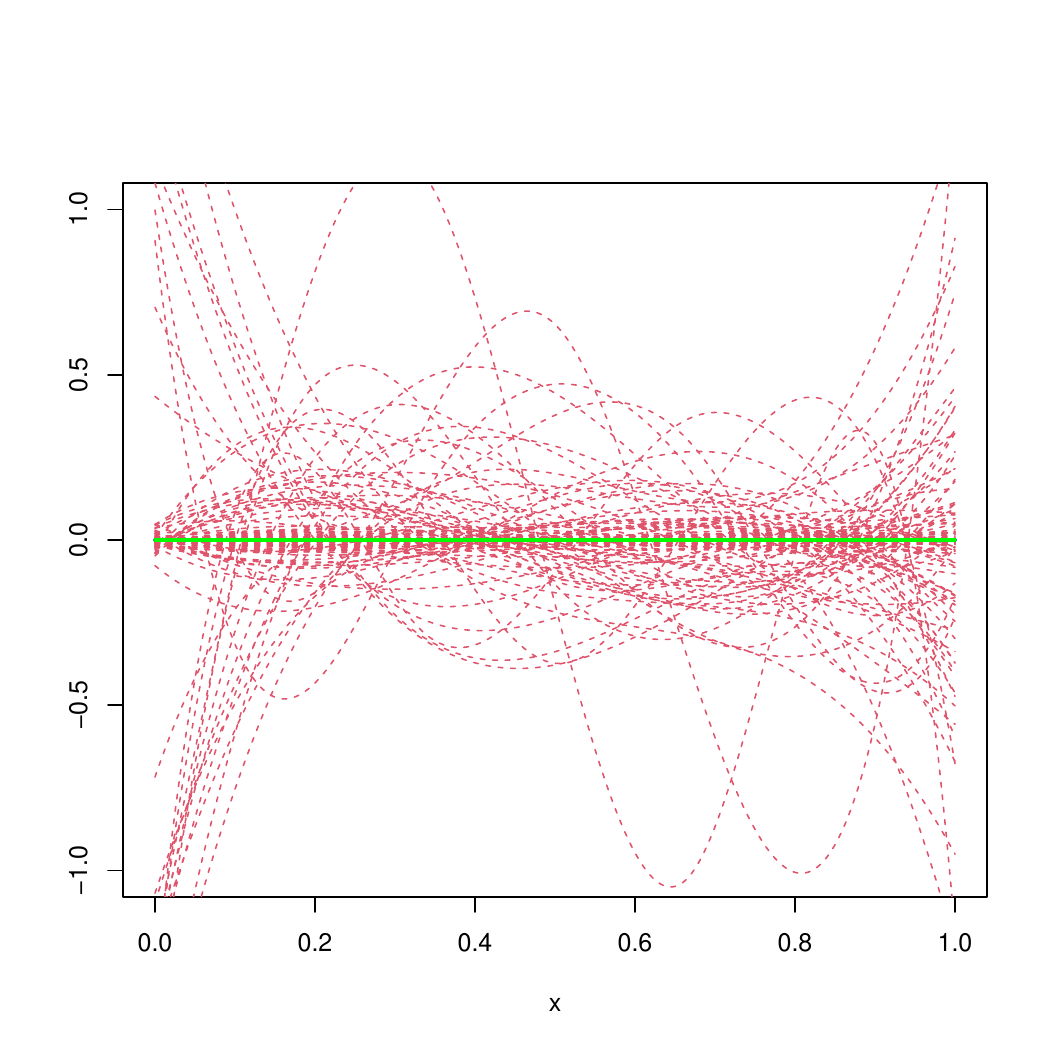}&
   \includegraphics[scale=0.22]{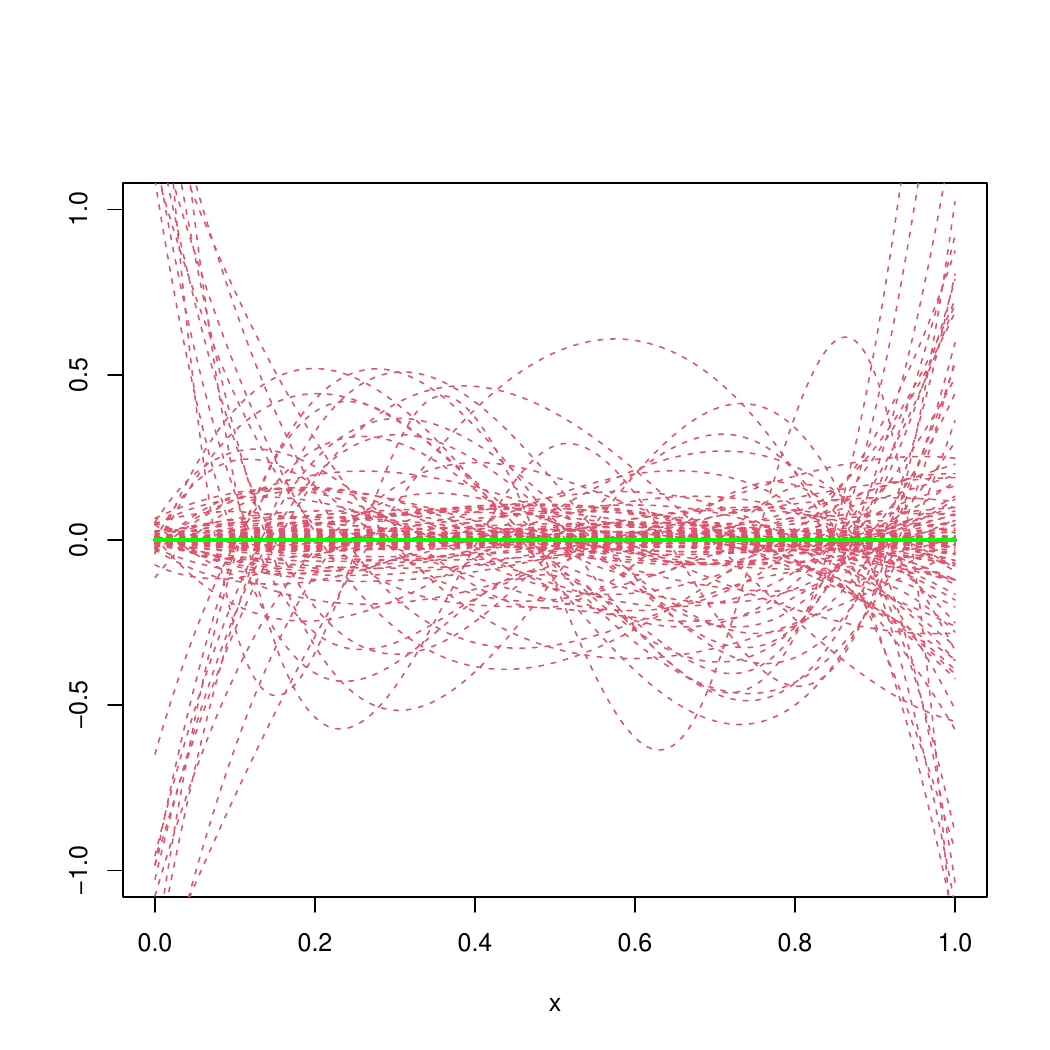} \\
   \includegraphics[scale=0.22]{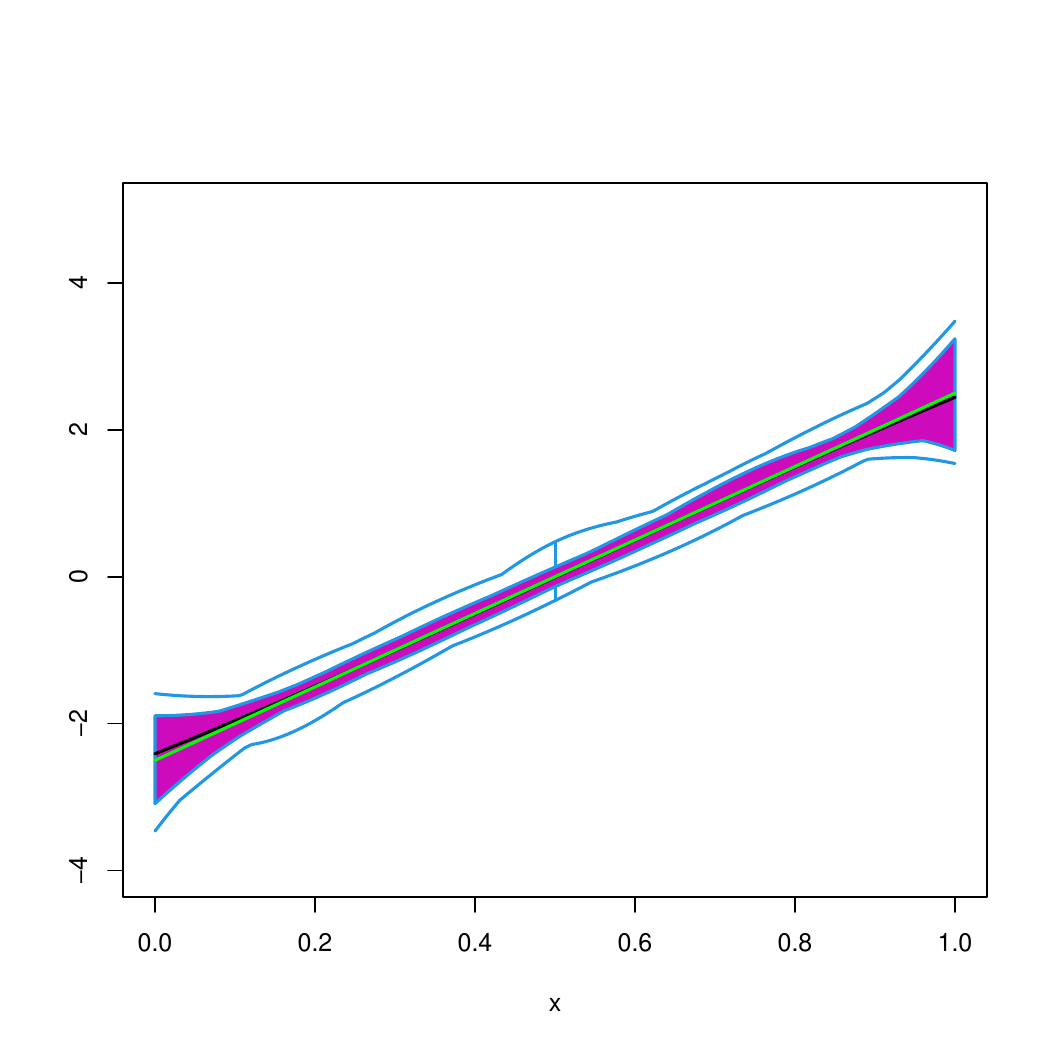} &
 \includegraphics[scale=0.22]{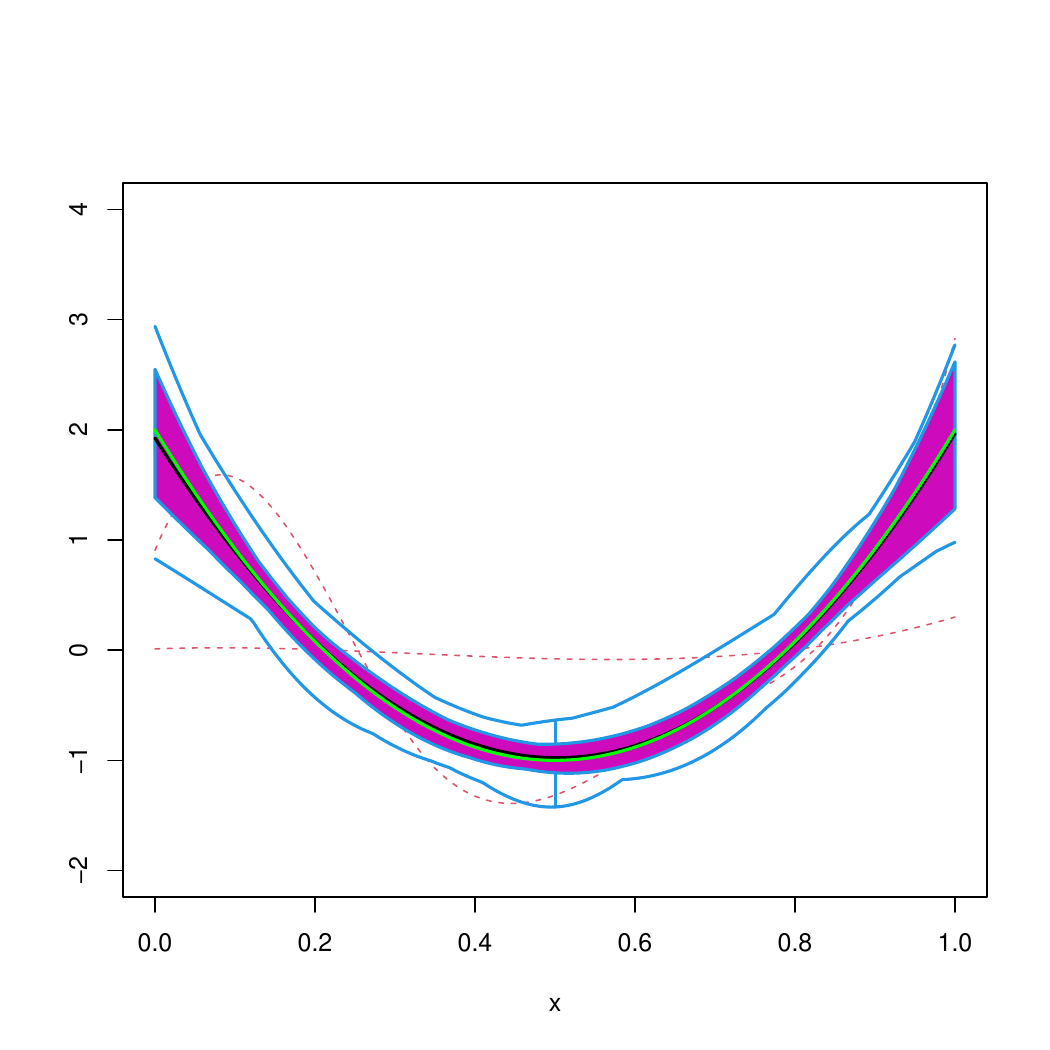} &
\includegraphics[scale=0.22]{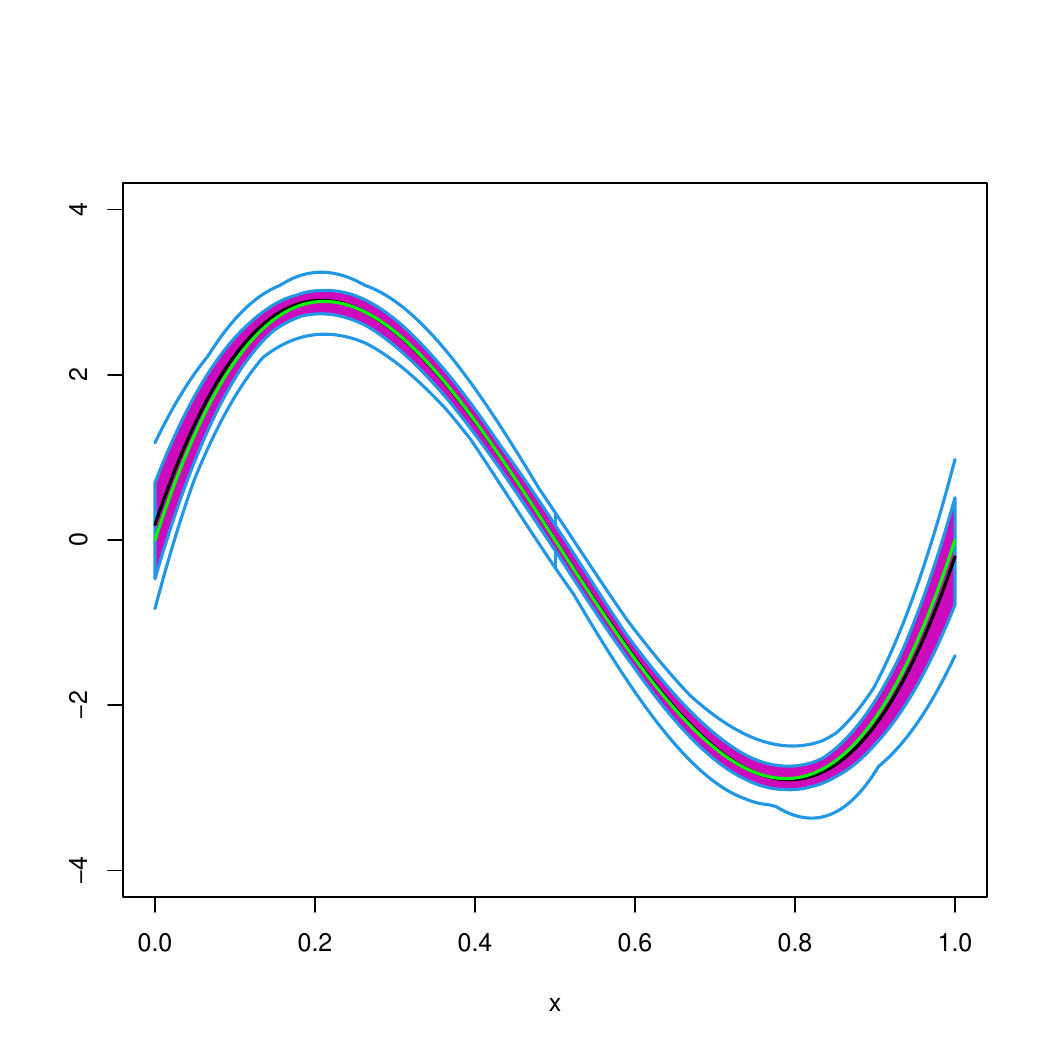}&
\includegraphics[scale=0.22]{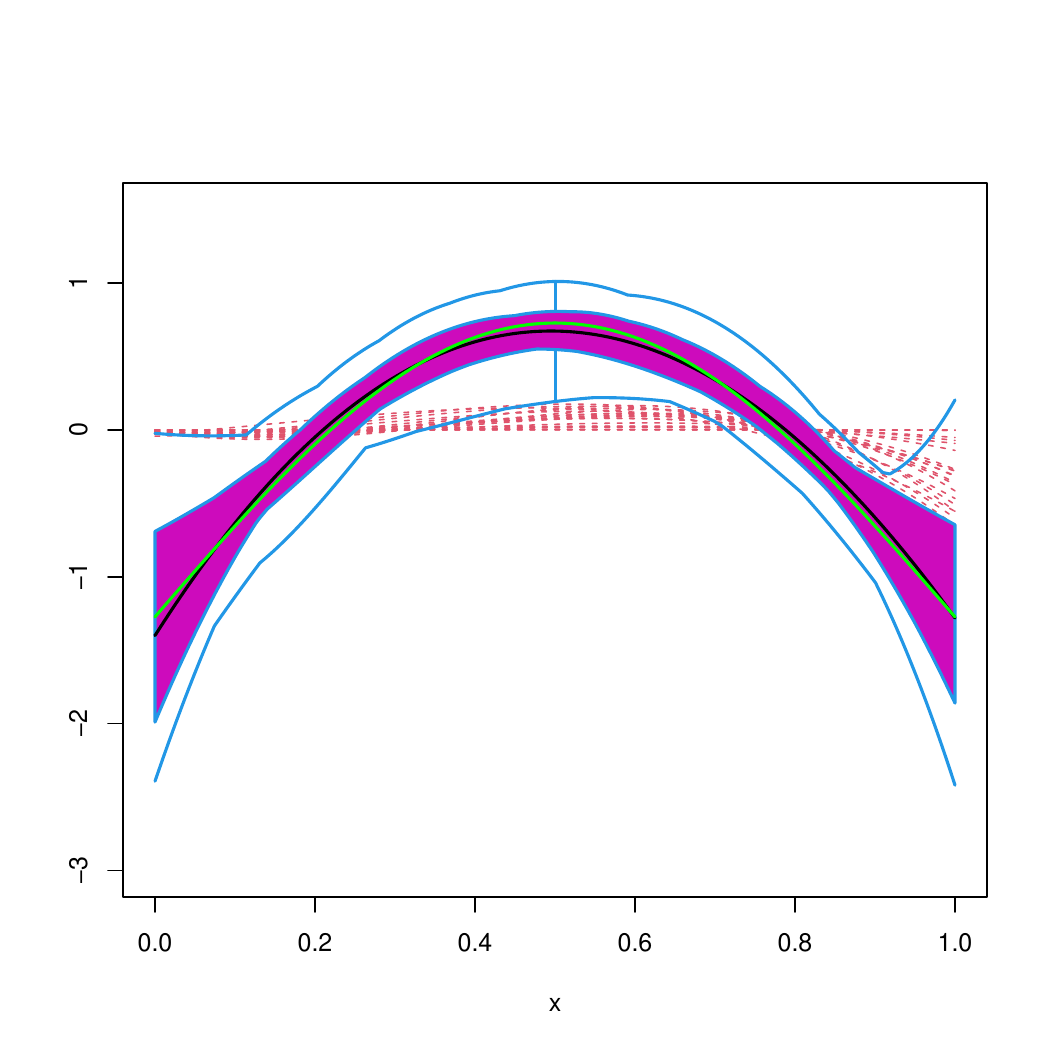}&
\includegraphics[scale=0.22]{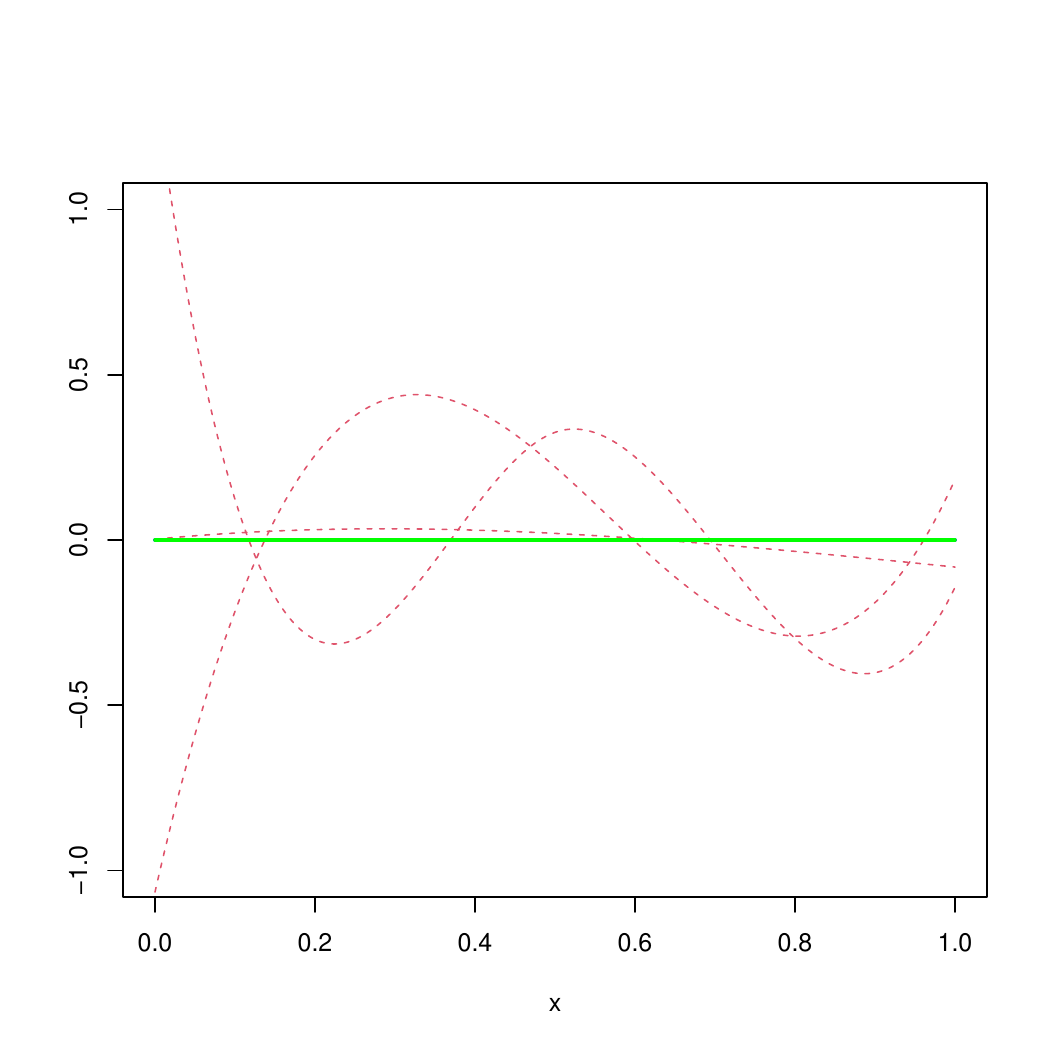}\\
\includegraphics[scale=0.22]{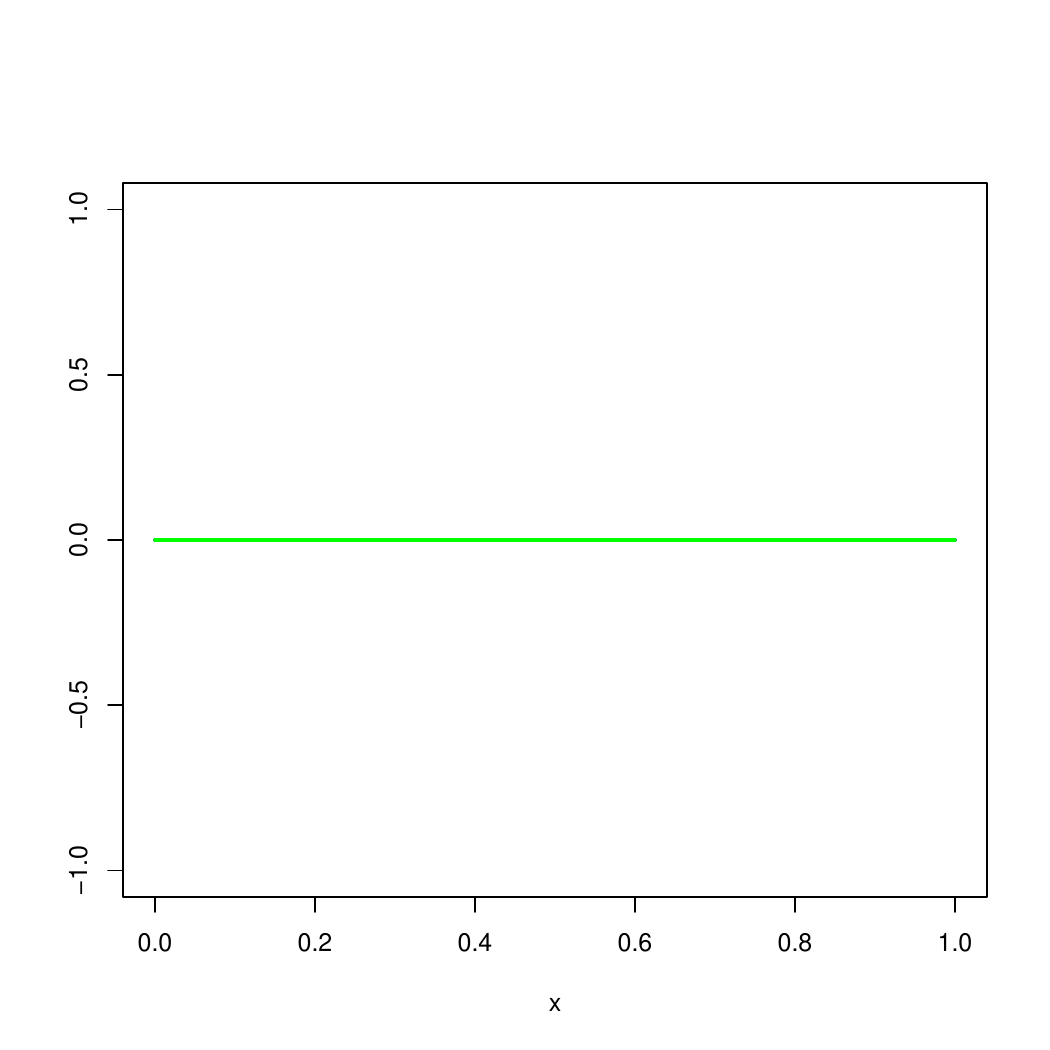} &
 \includegraphics[scale=0.22]{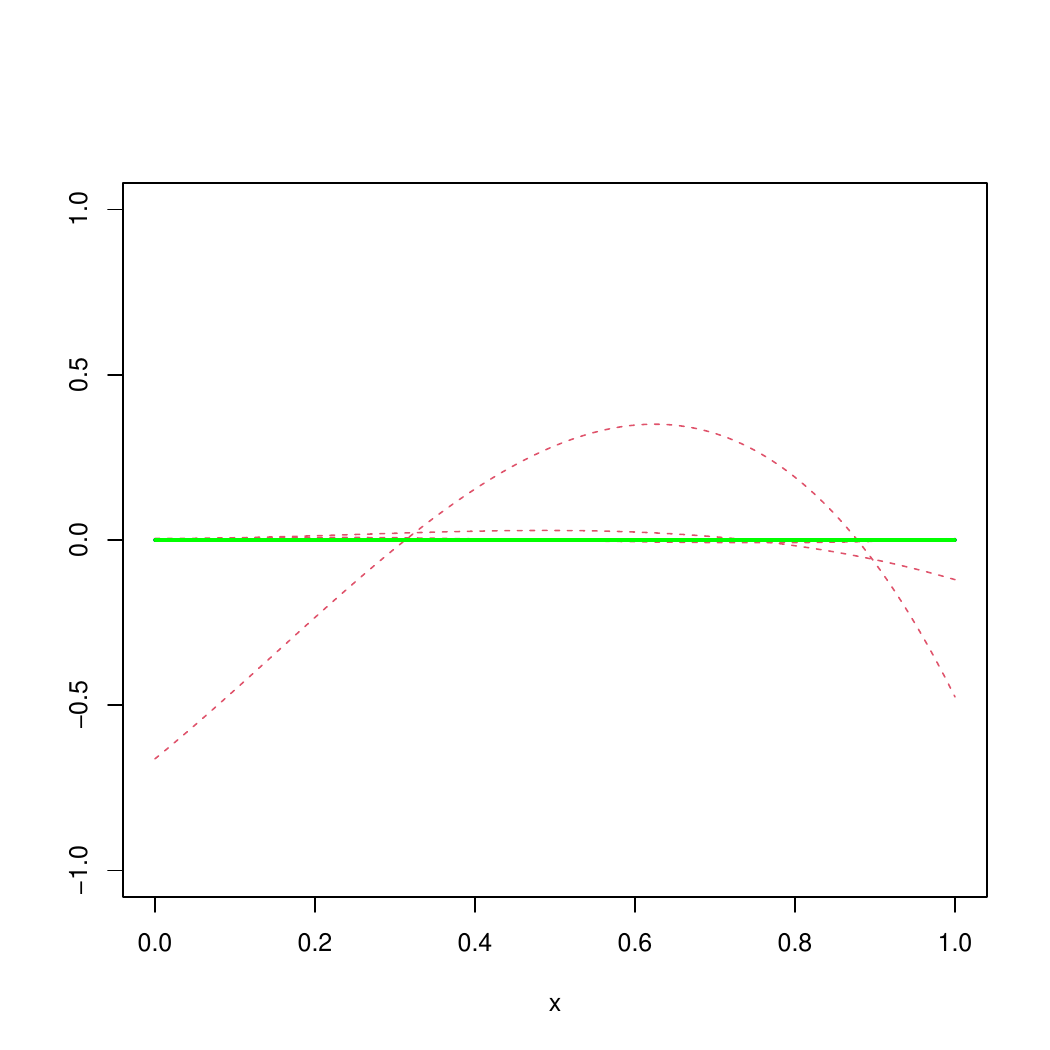}&
 \includegraphics[scale=0.22]{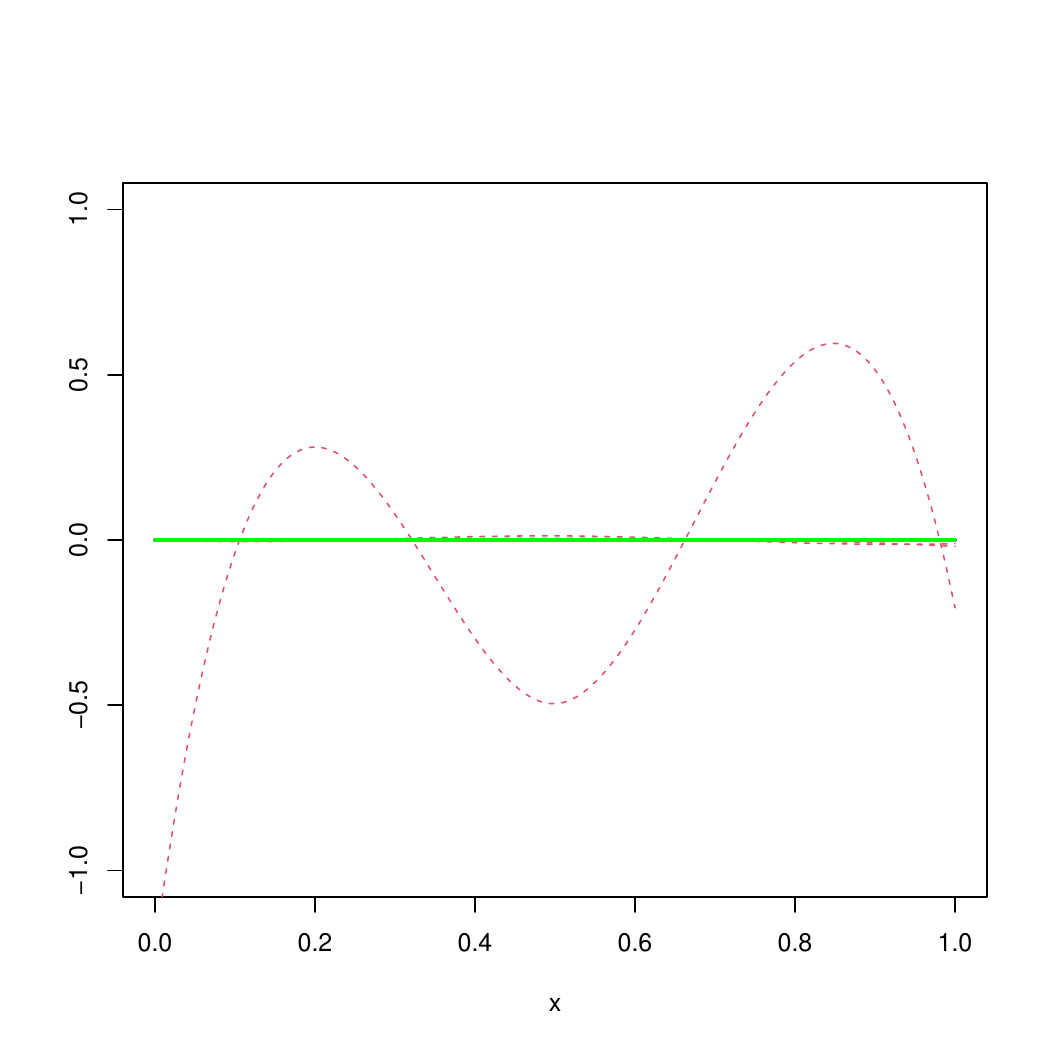}&
  \includegraphics[scale=0.22]{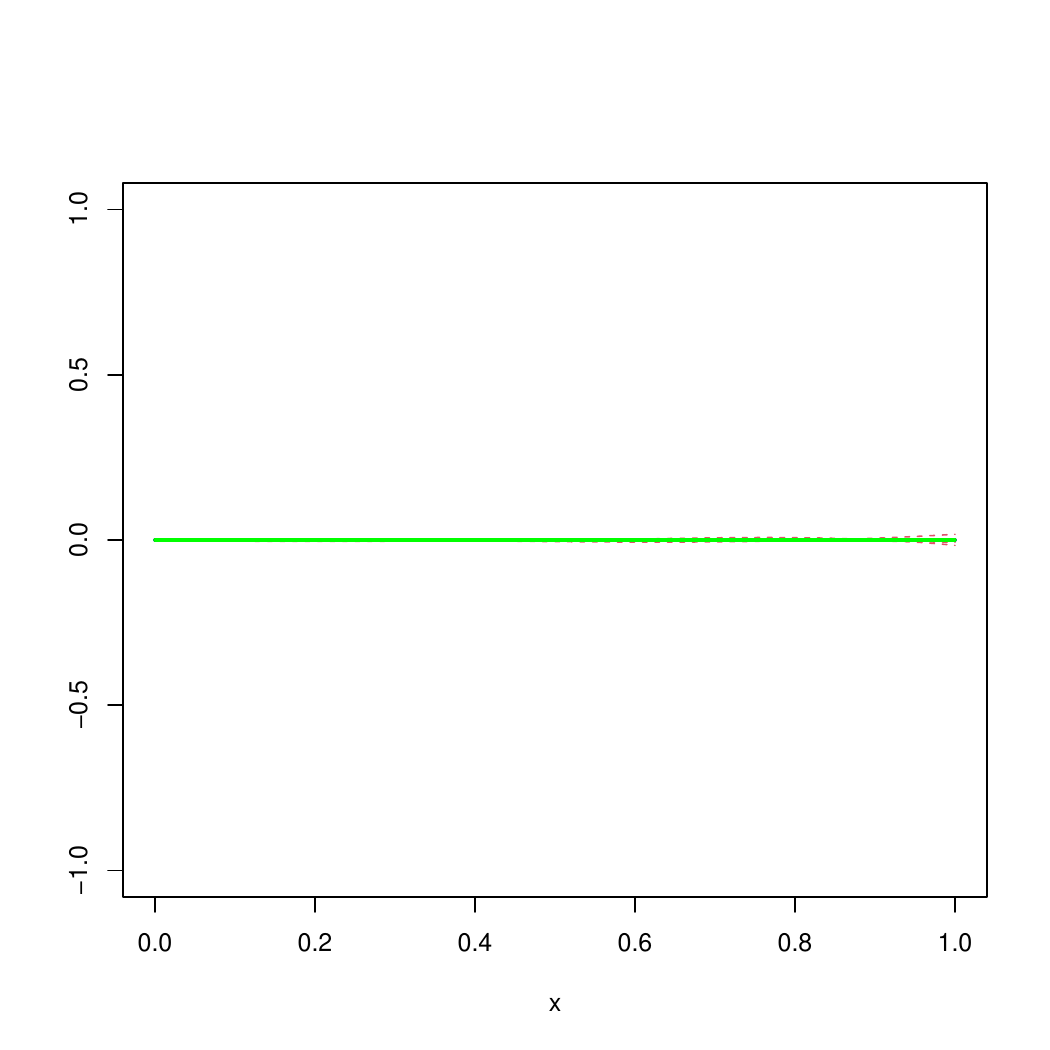}&
   \includegraphics[scale=0.22]{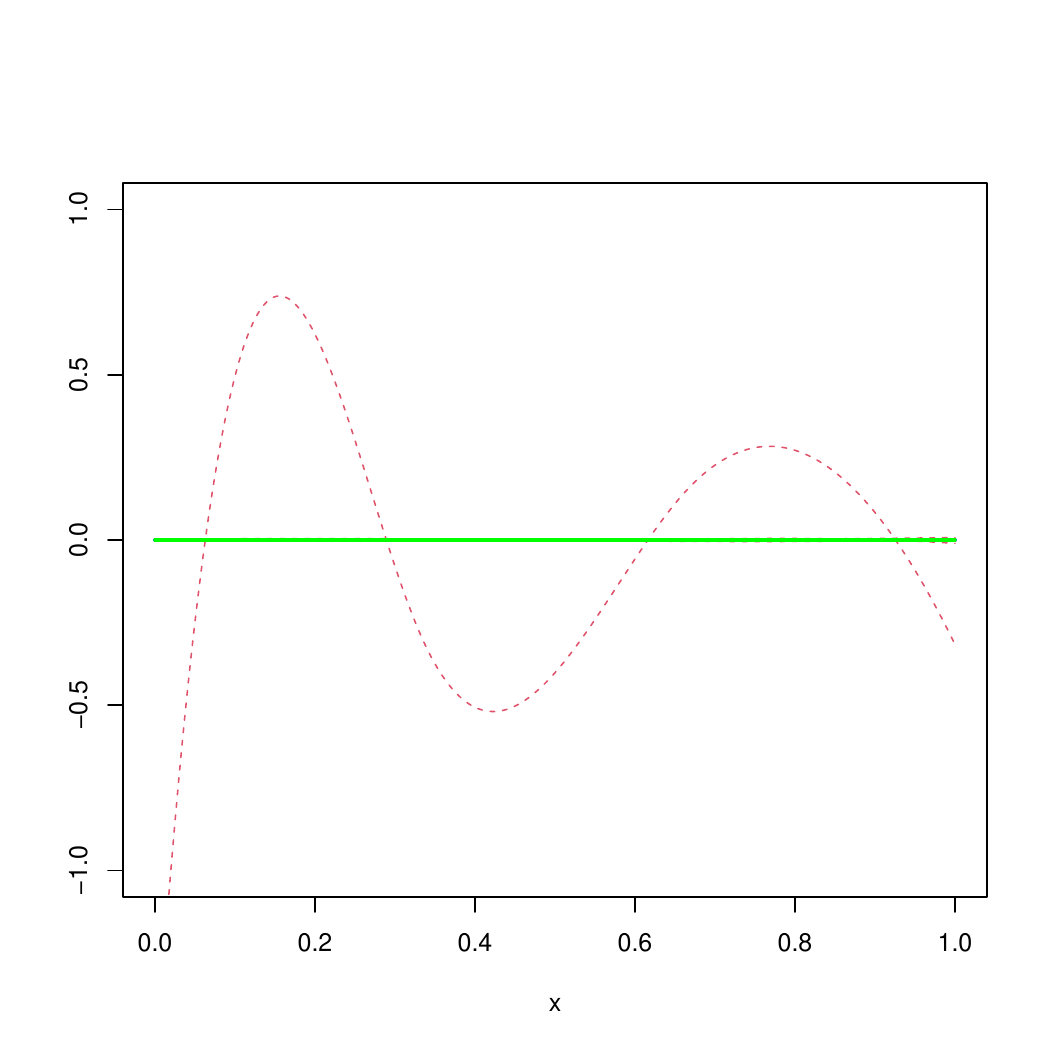}
   \end{tabular}
\caption{\small \label{fig:curvas-cl-rob-C2} The first two rows contain the functional boxplots of the estimated additive functions using the least square-based estimator while the last two rows contain the functional boxplots of the estimated additive functions by the robust approach, for $n=200$ and the contamination setting $C_2$.} 
\end{center}
\end{figure}

\begin{figure}[htbp]
 \begin{center}
\small
\begin{tabular}{ccccc}
\includegraphics[scale=0.22]{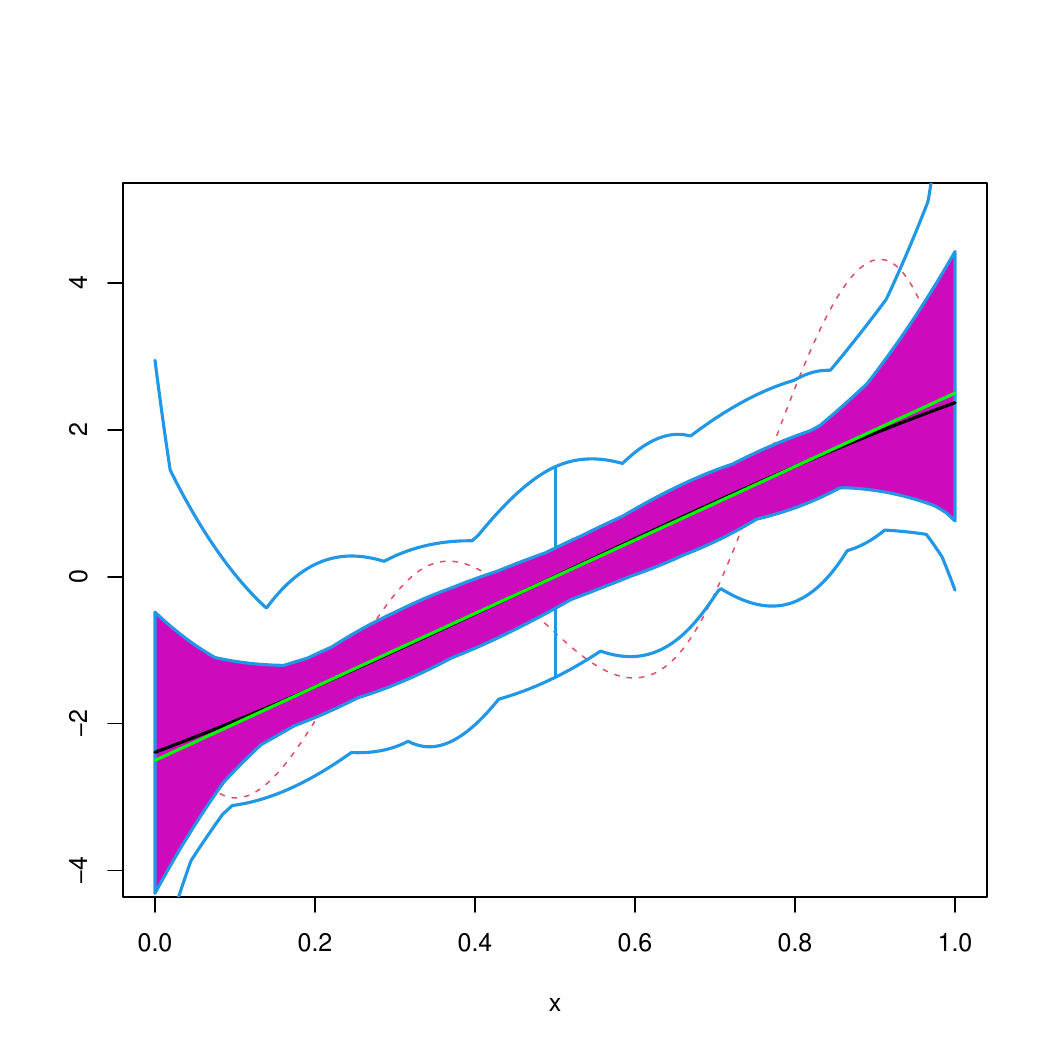} &
 \includegraphics[scale=0.22]{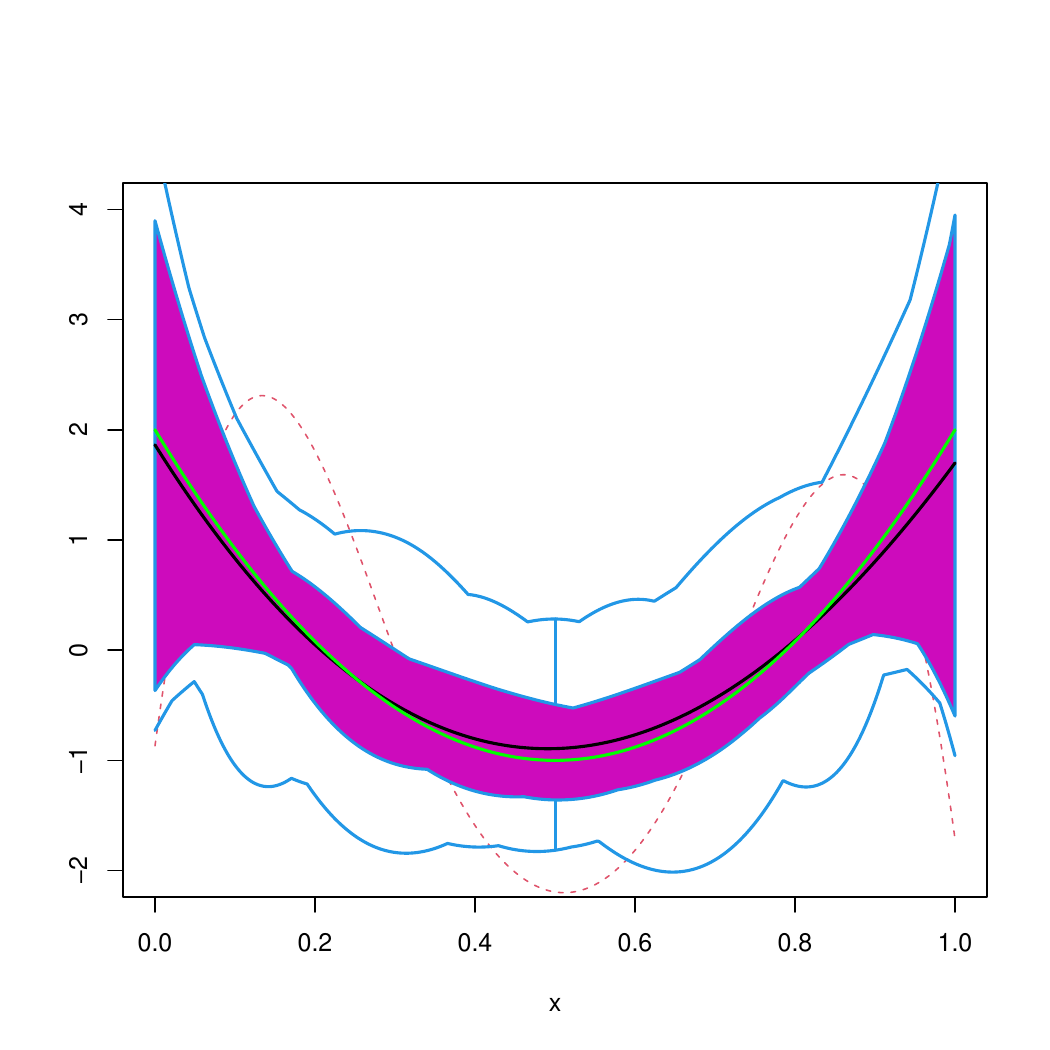} &
\includegraphics[scale=0.22]{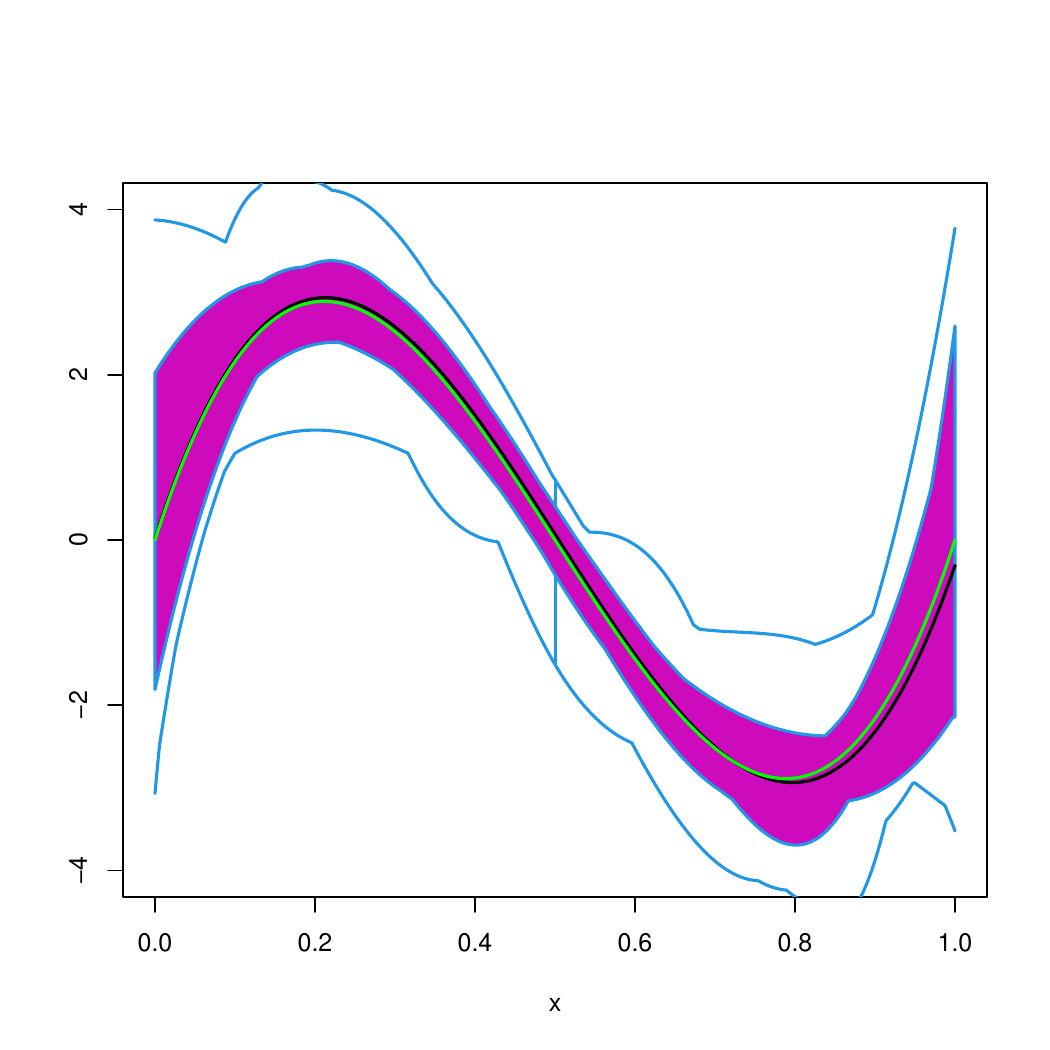}&
\includegraphics[scale=0.22]{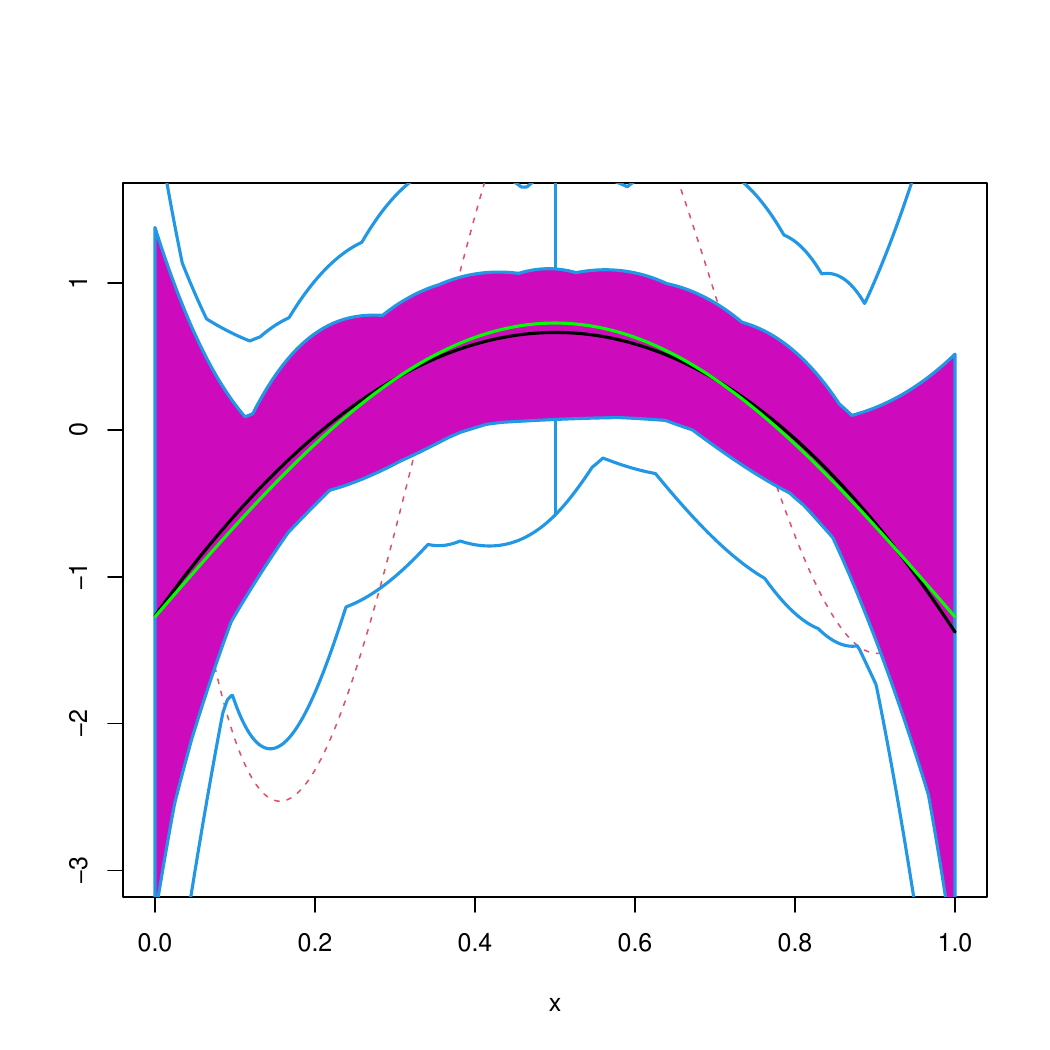}&
\includegraphics[scale=0.22]{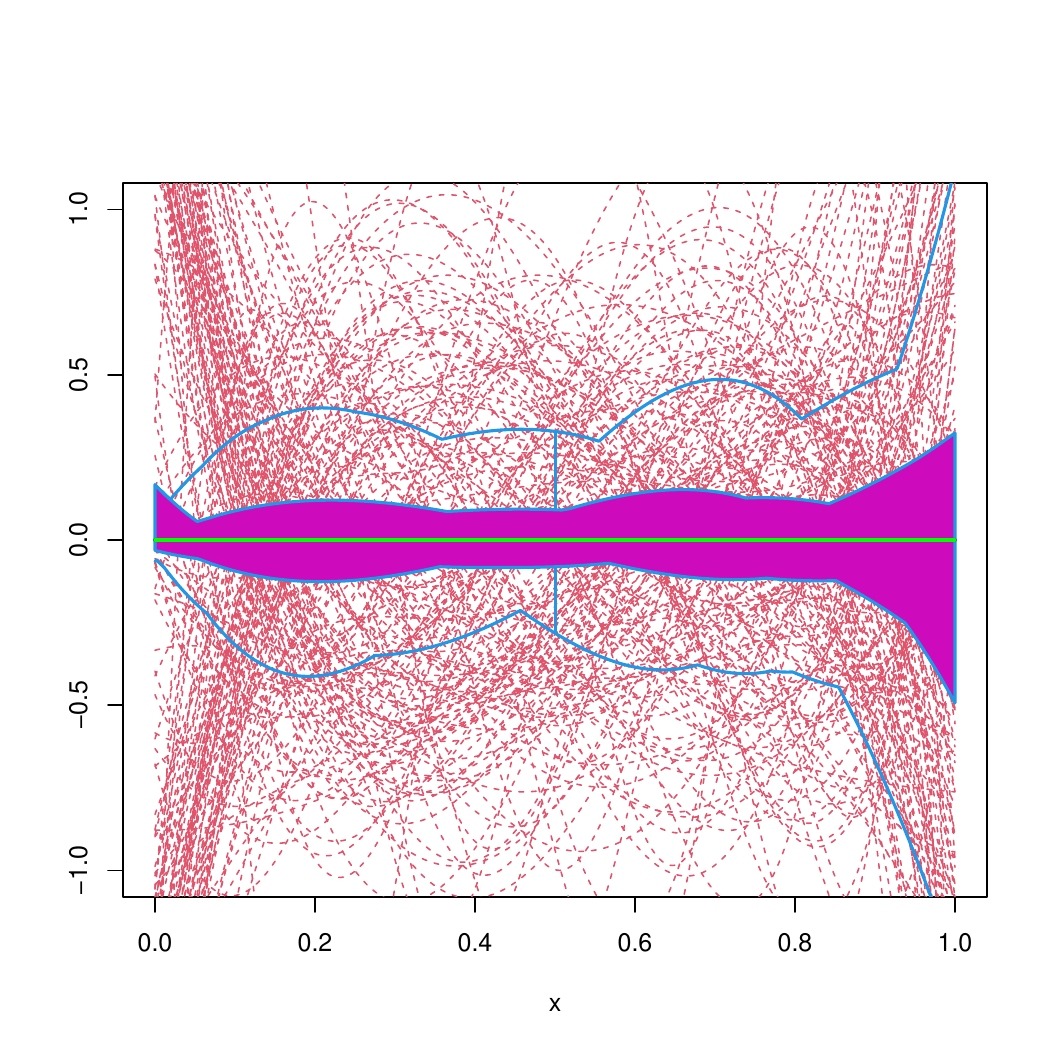}\\
\includegraphics[scale=0.22]{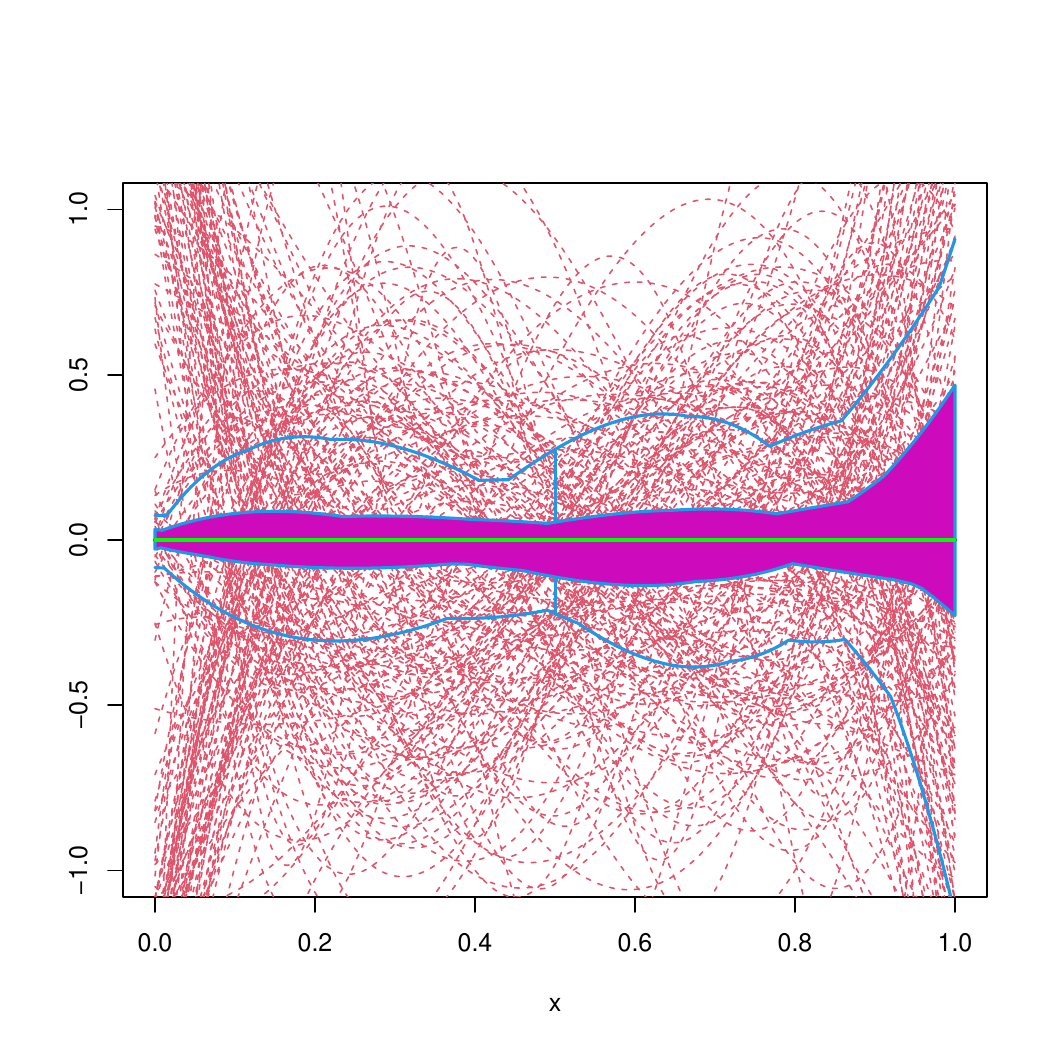} &
 \includegraphics[scale=0.22]{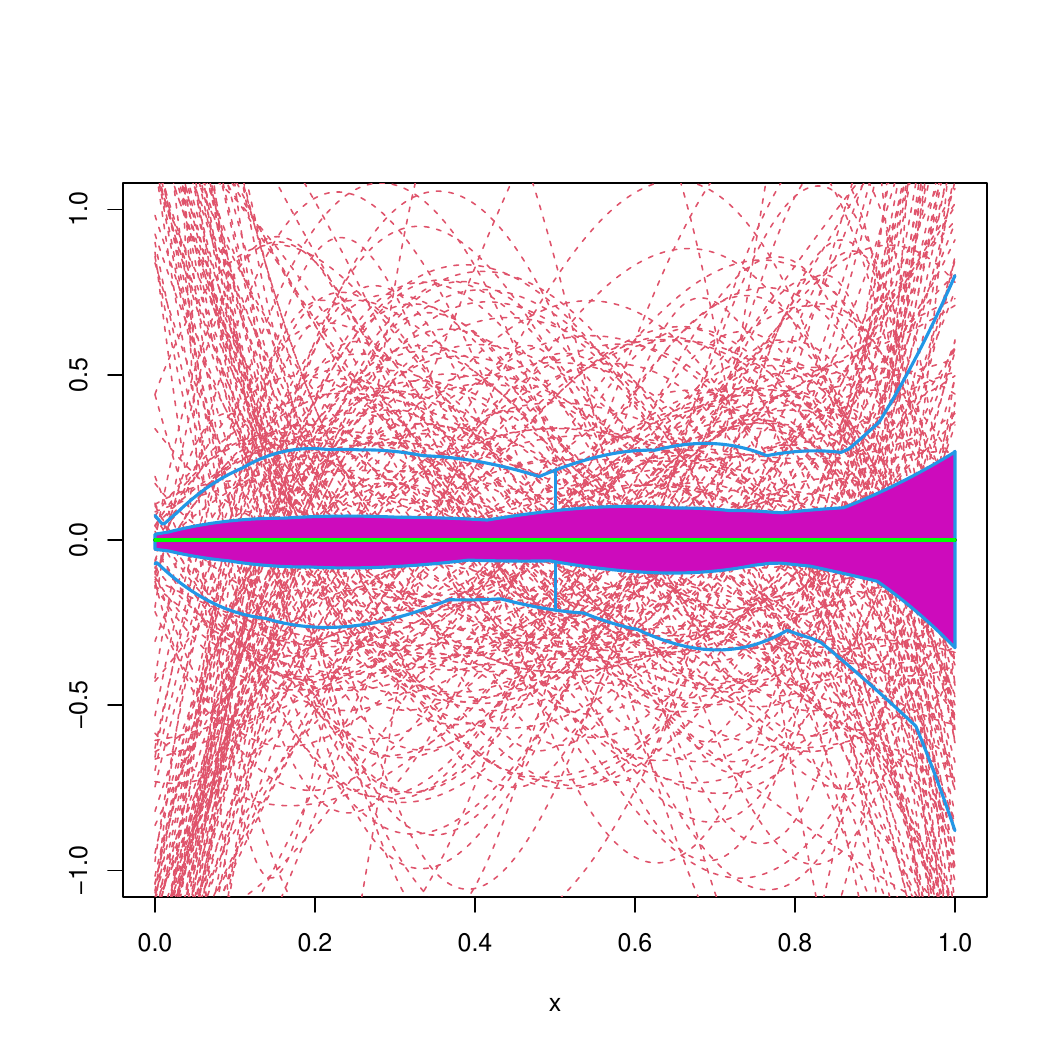}&
 \includegraphics[scale=0.22]{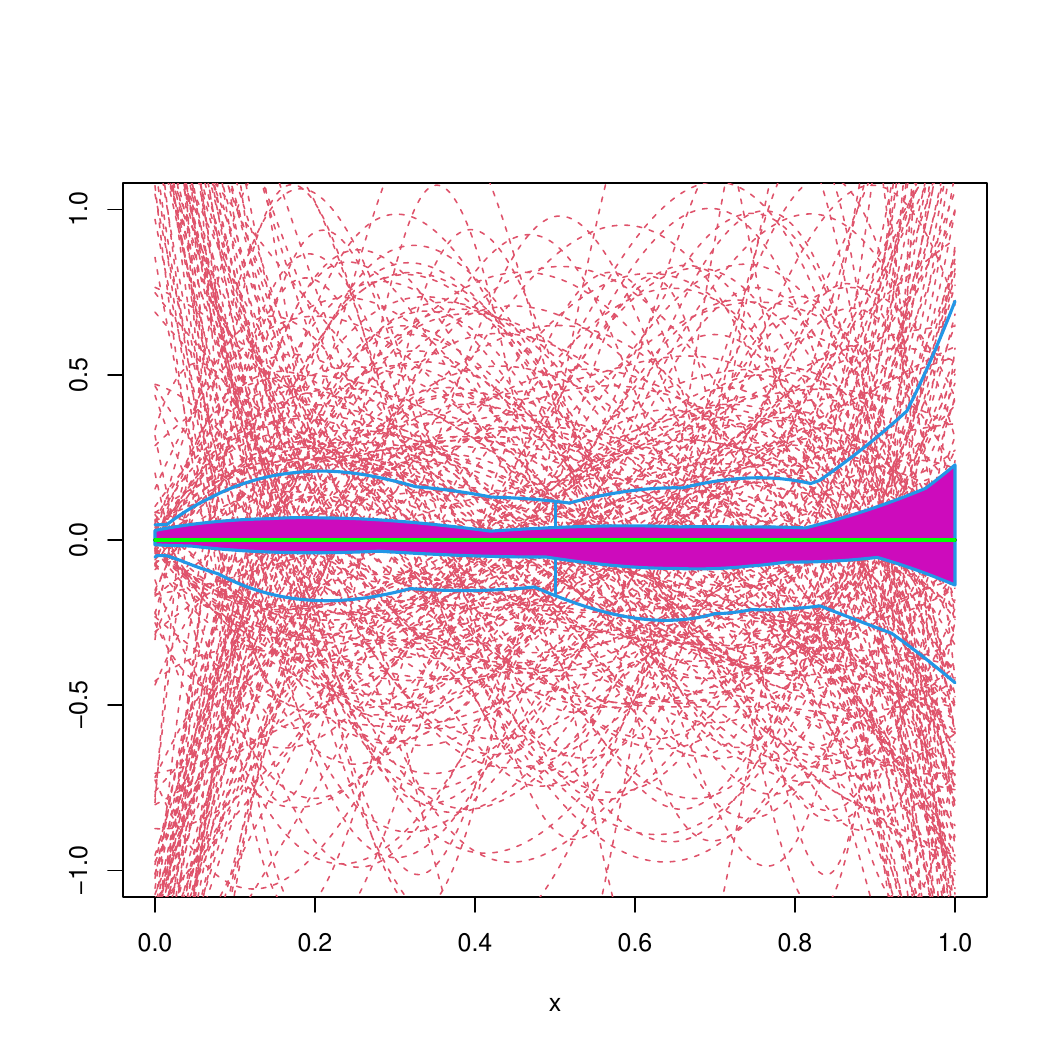}&
  \includegraphics[scale=0.22]{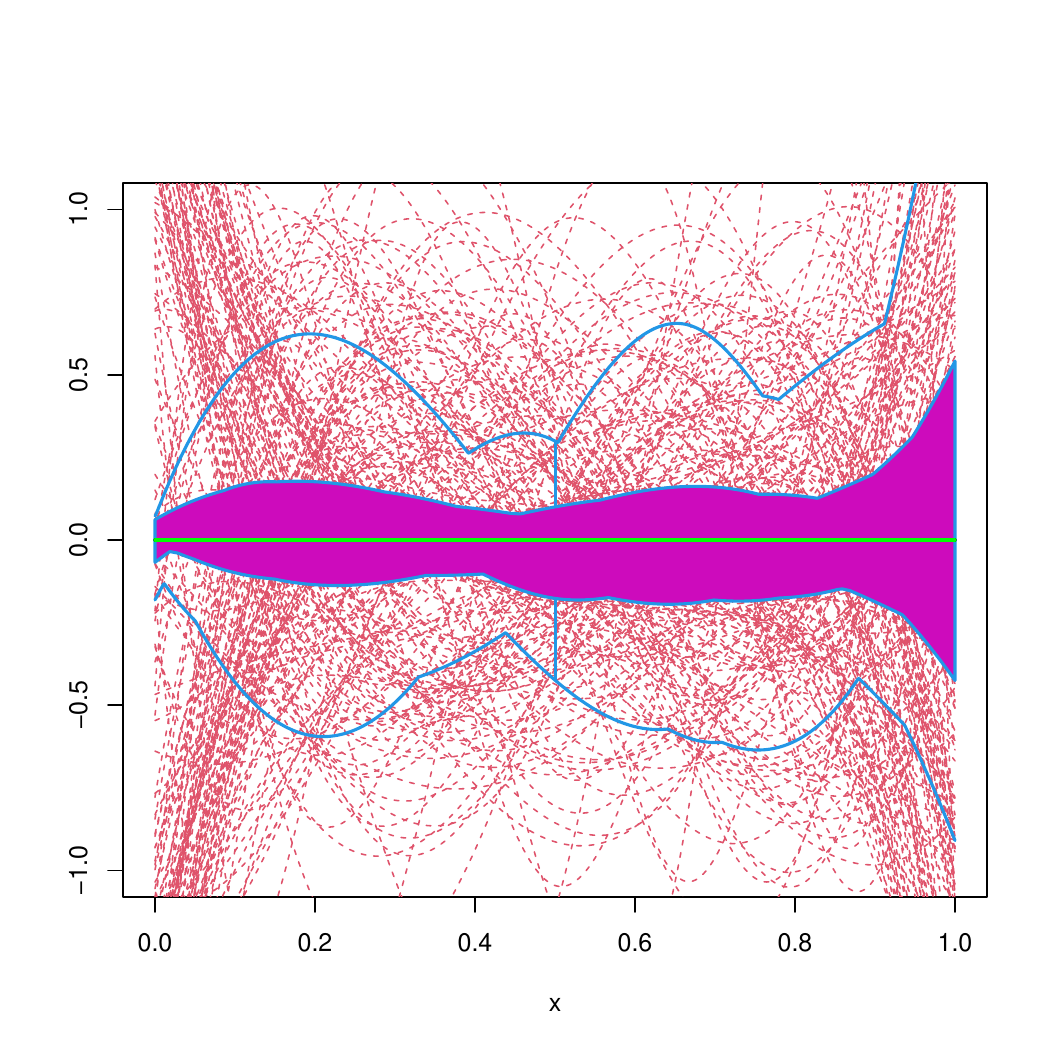}&
   \includegraphics[scale=0.22]{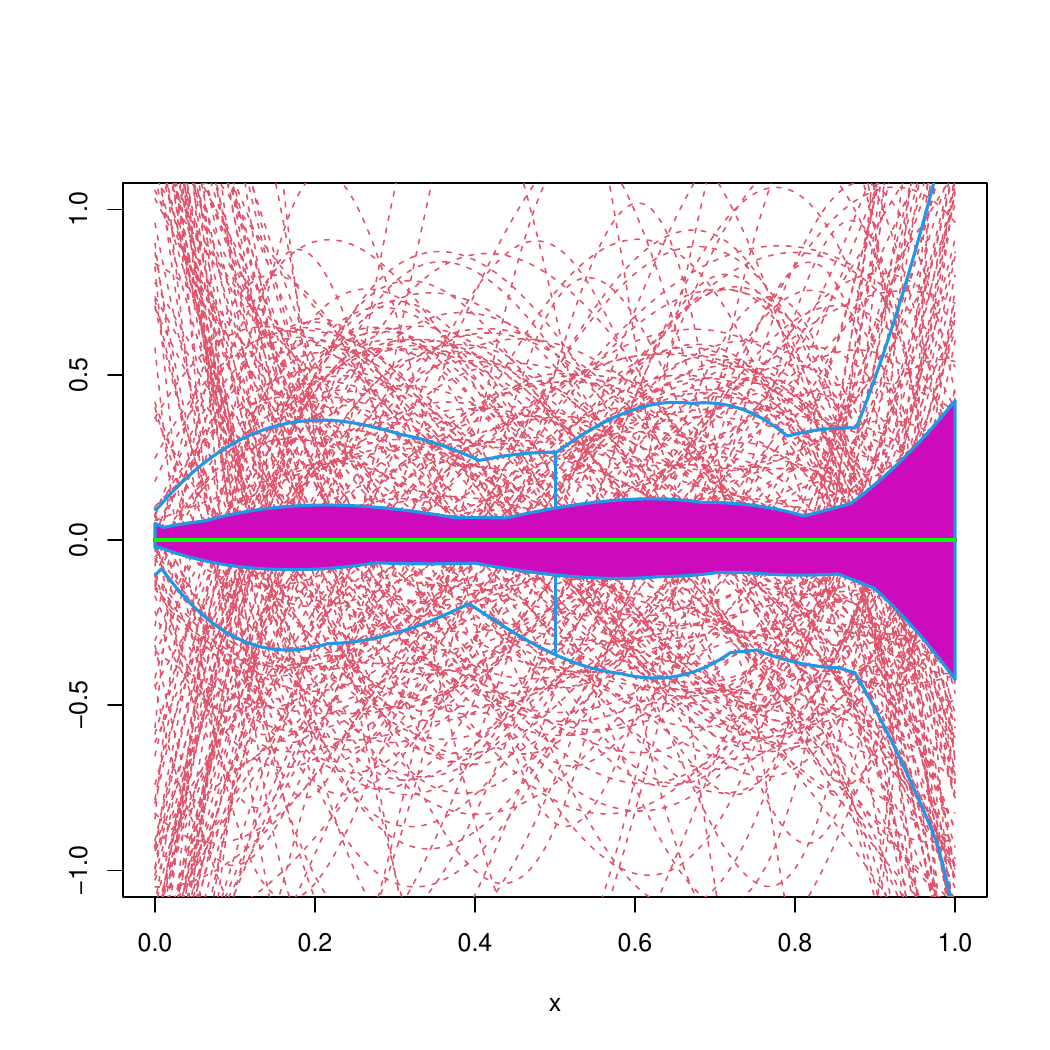} \\
   \includegraphics[scale=0.22]{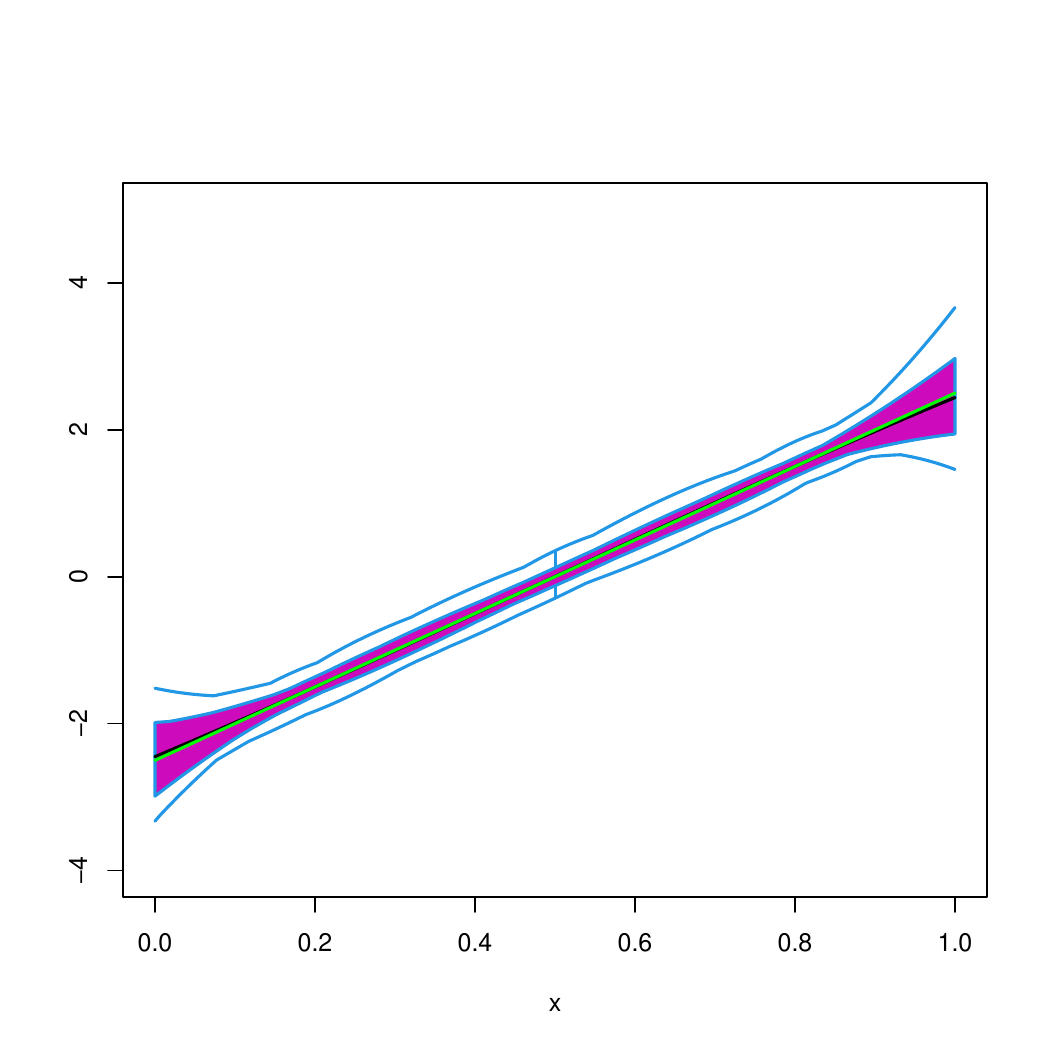} &
 \includegraphics[scale=0.22]{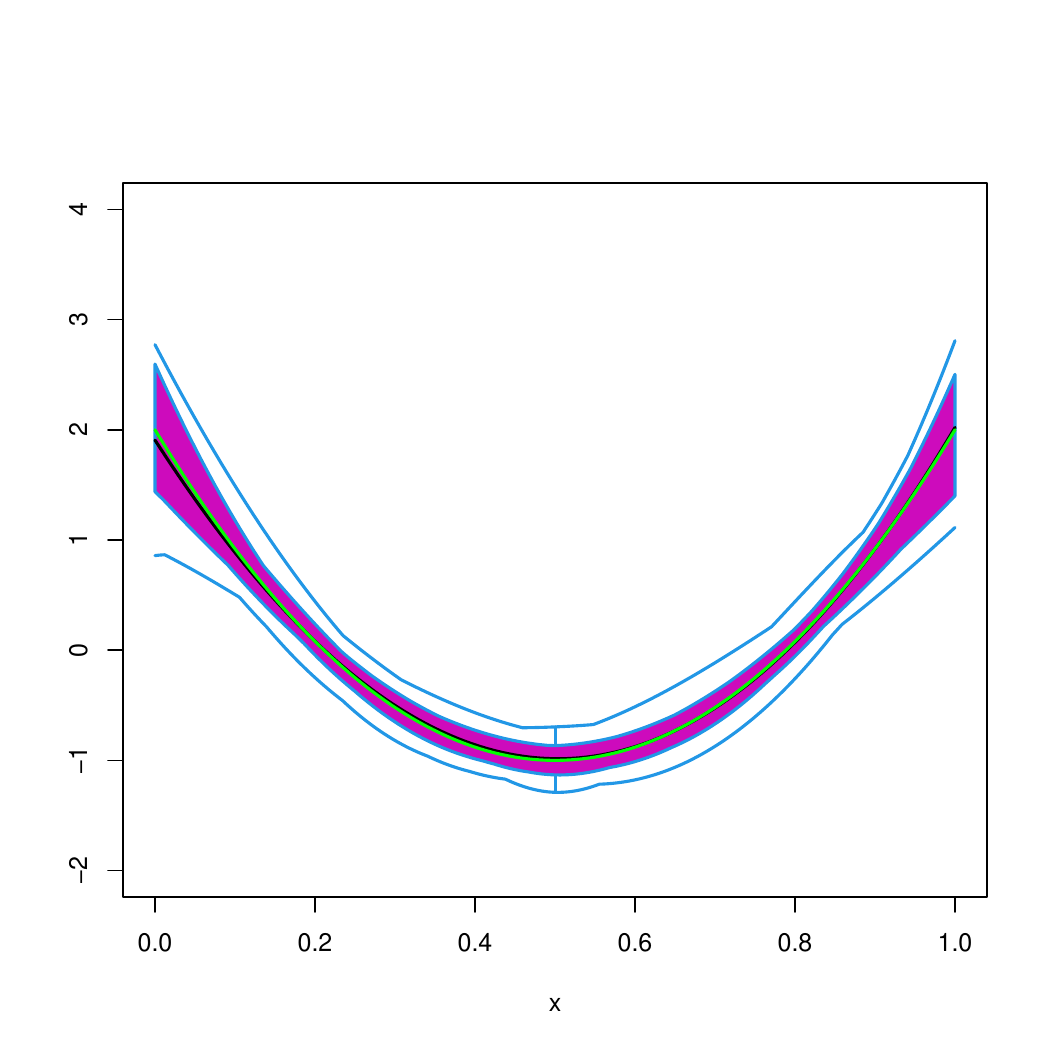} &
\includegraphics[scale=0.22]{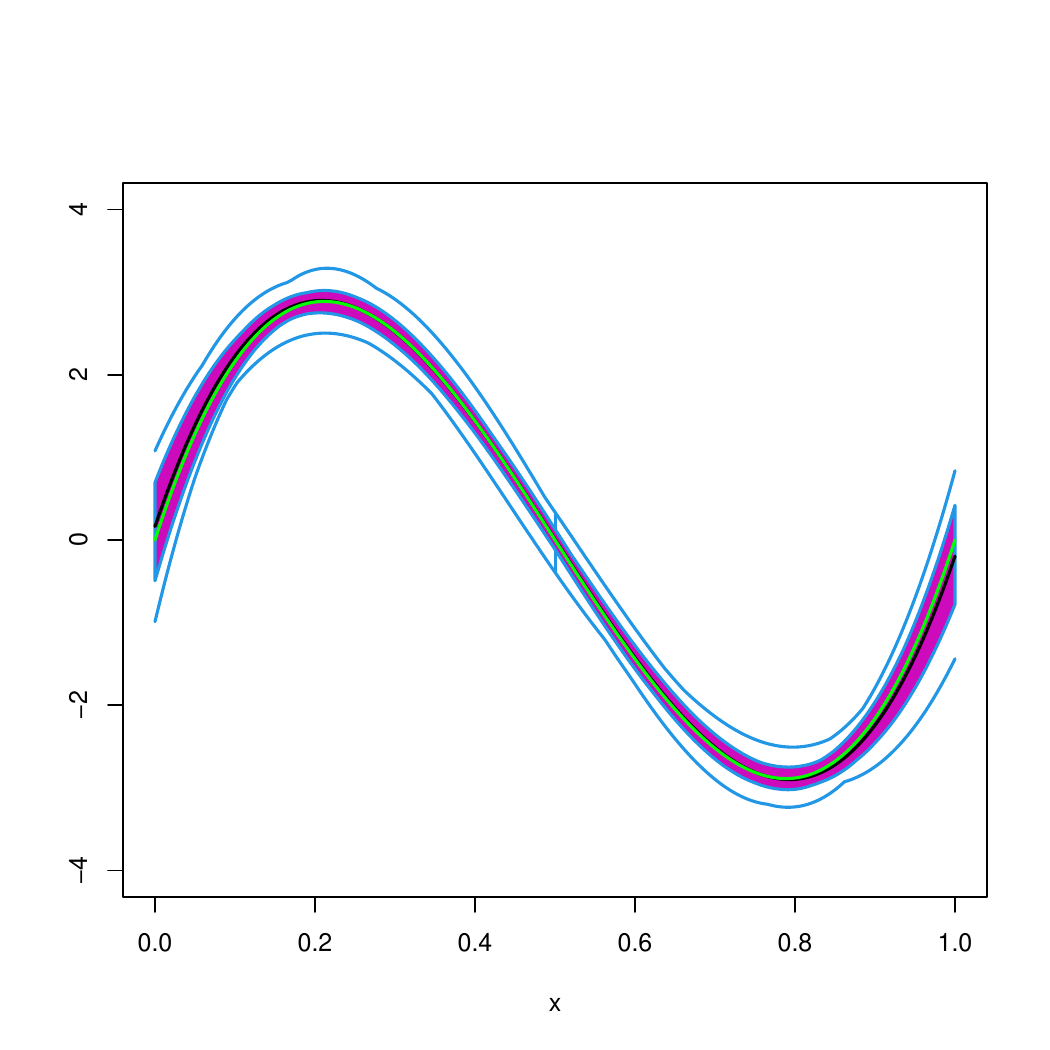}&
\includegraphics[scale=0.22]{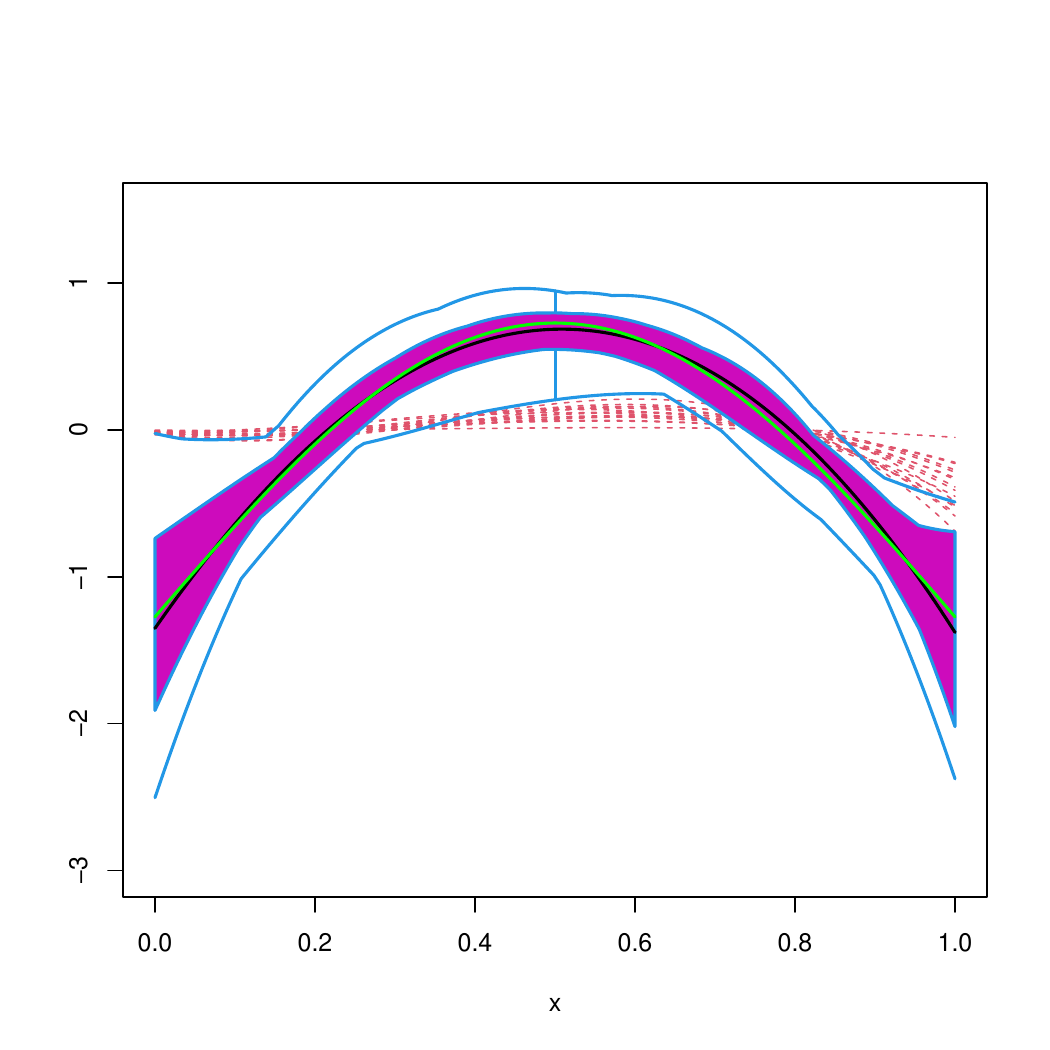}&
\includegraphics[scale=0.22]{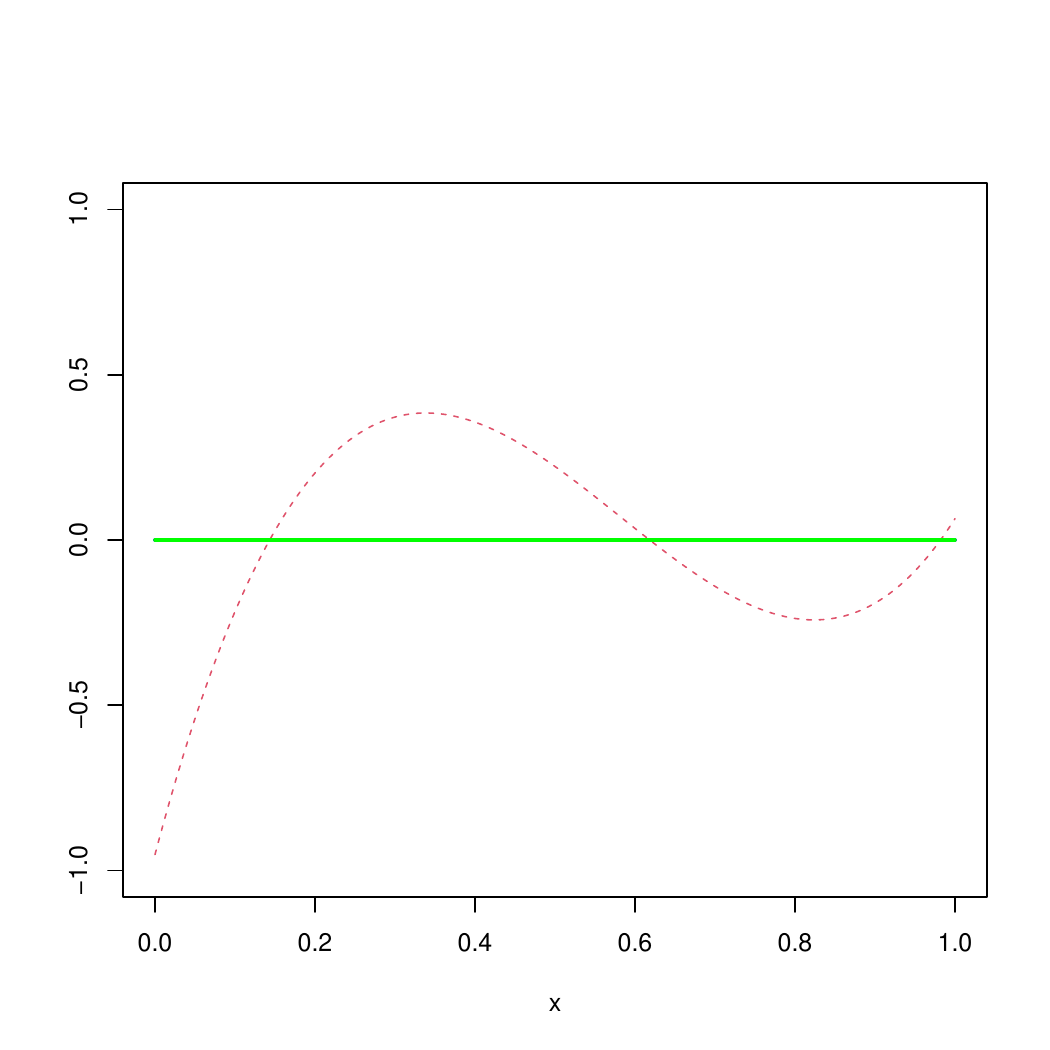}\\
\includegraphics[scale=0.22]{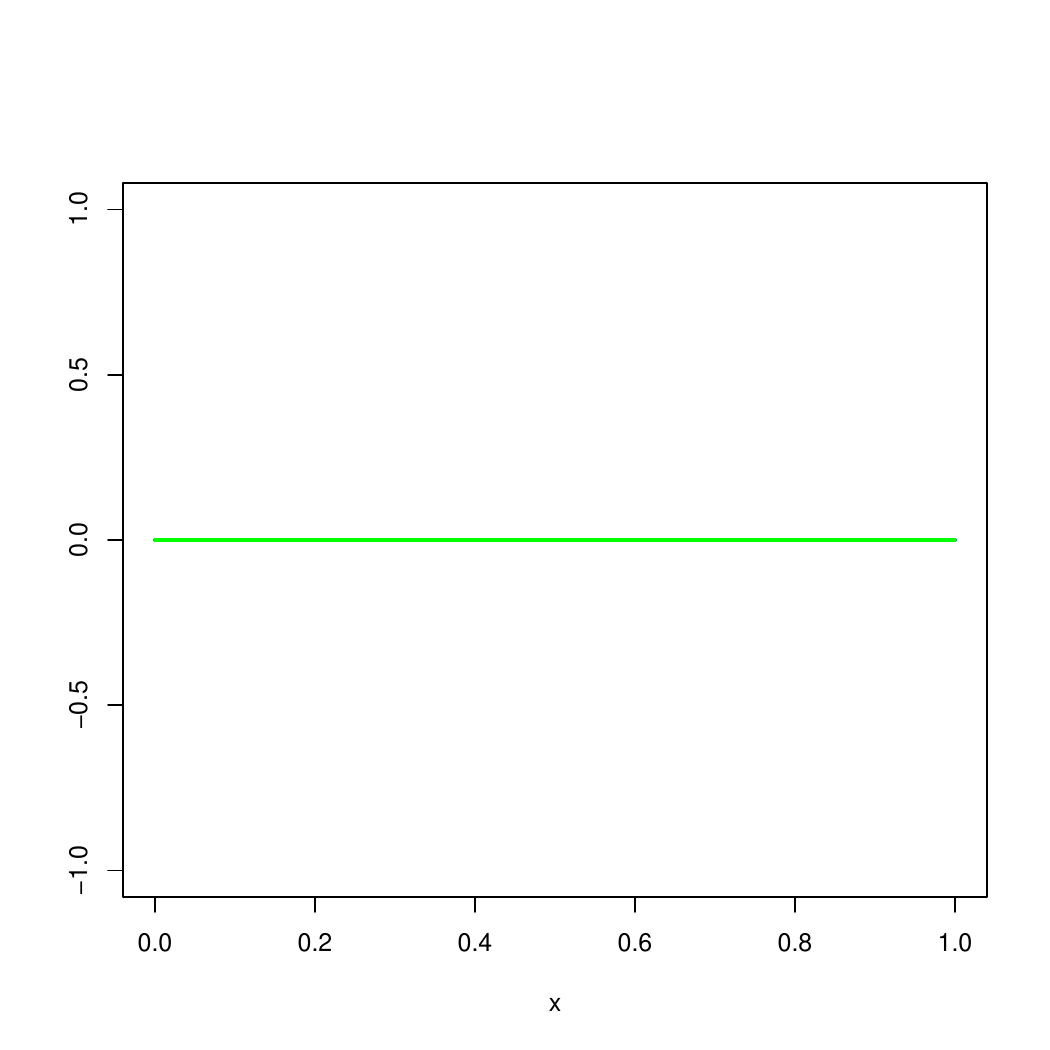} &
 \includegraphics[scale=0.22]{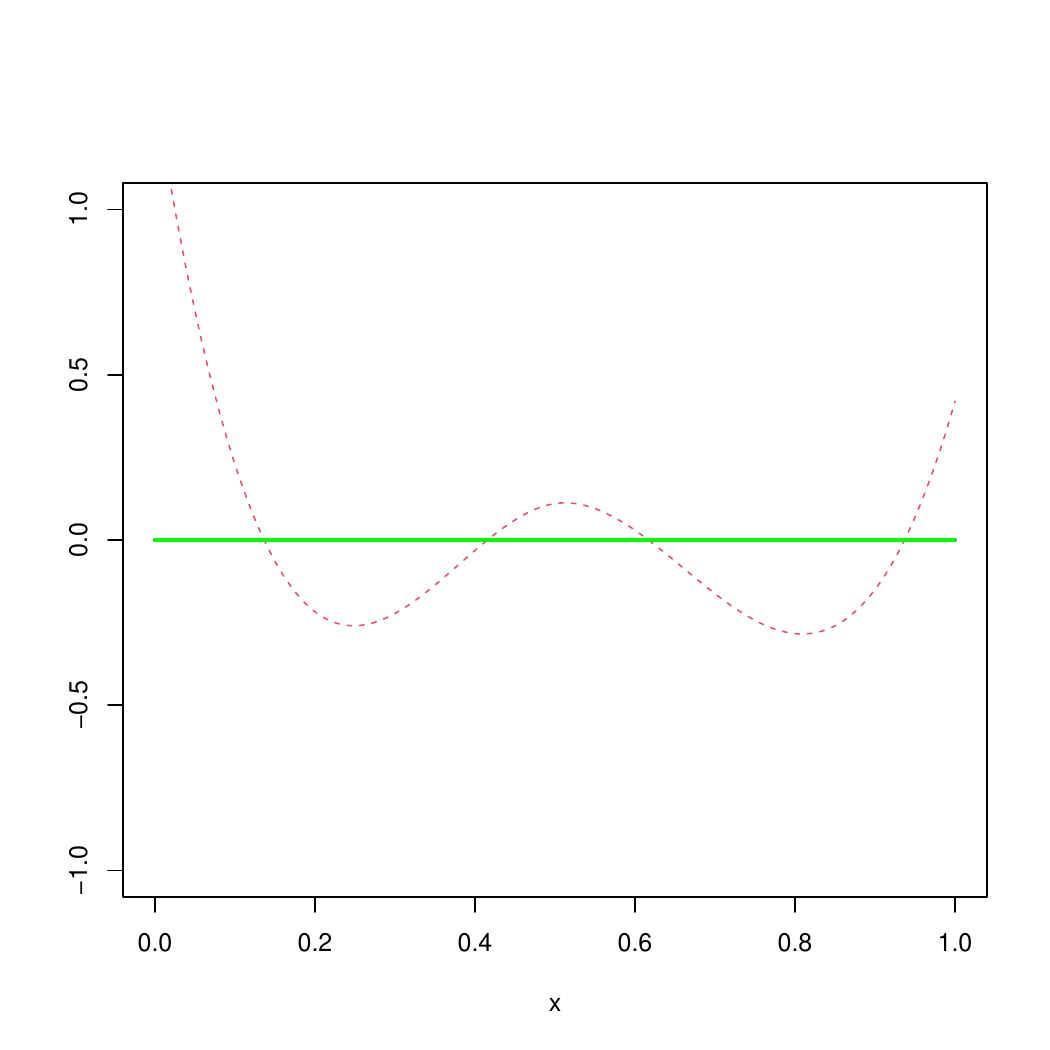}&
 \includegraphics[scale=0.22]{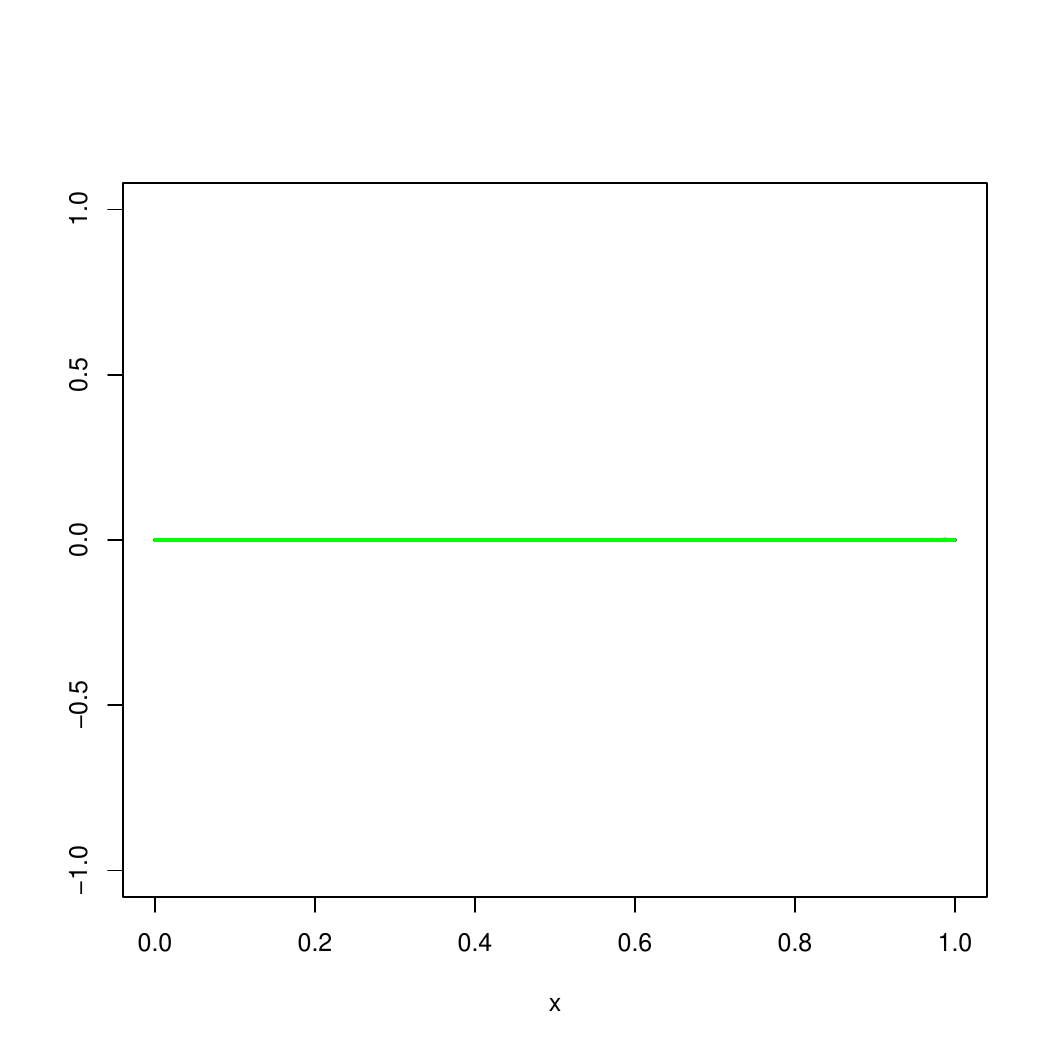}&
  \includegraphics[scale=0.22]{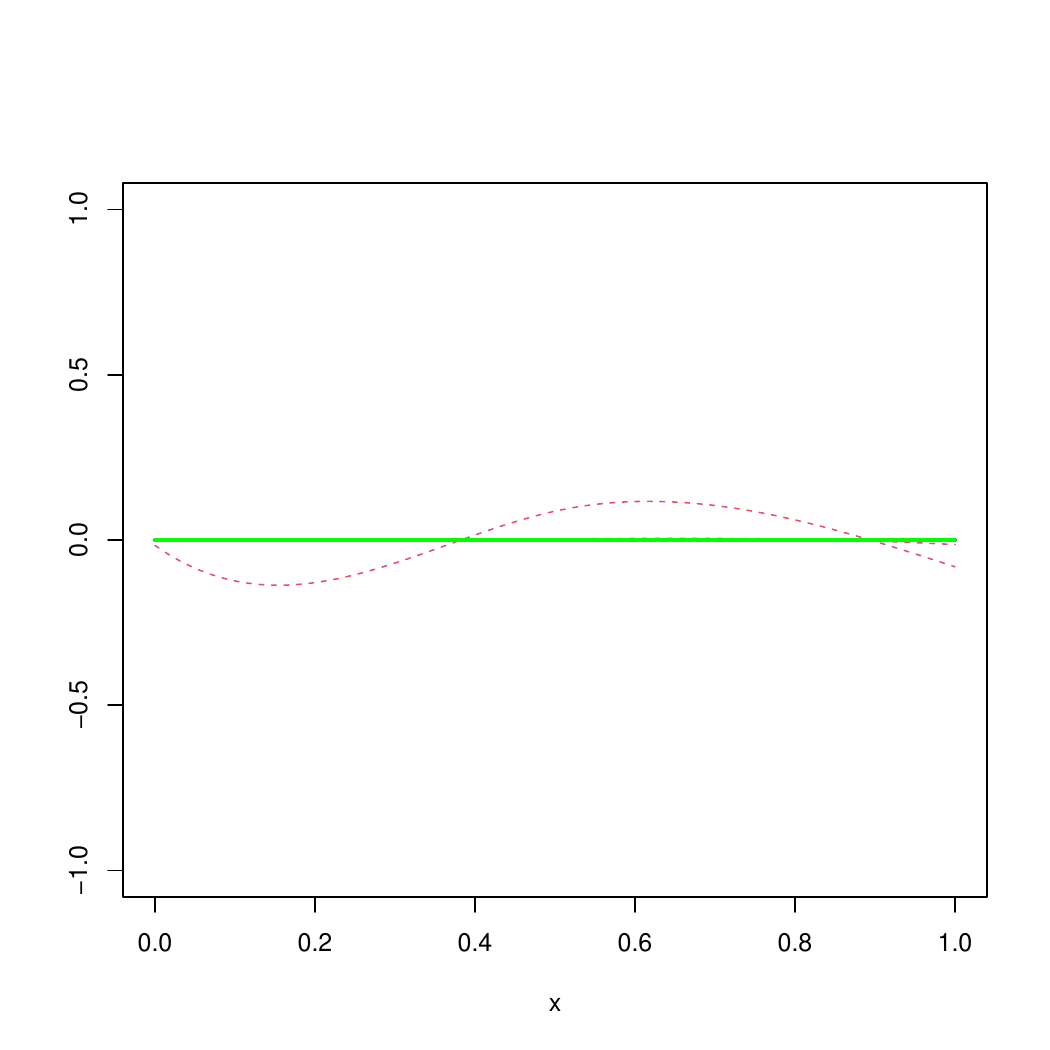}&
   \includegraphics[scale=0.22]{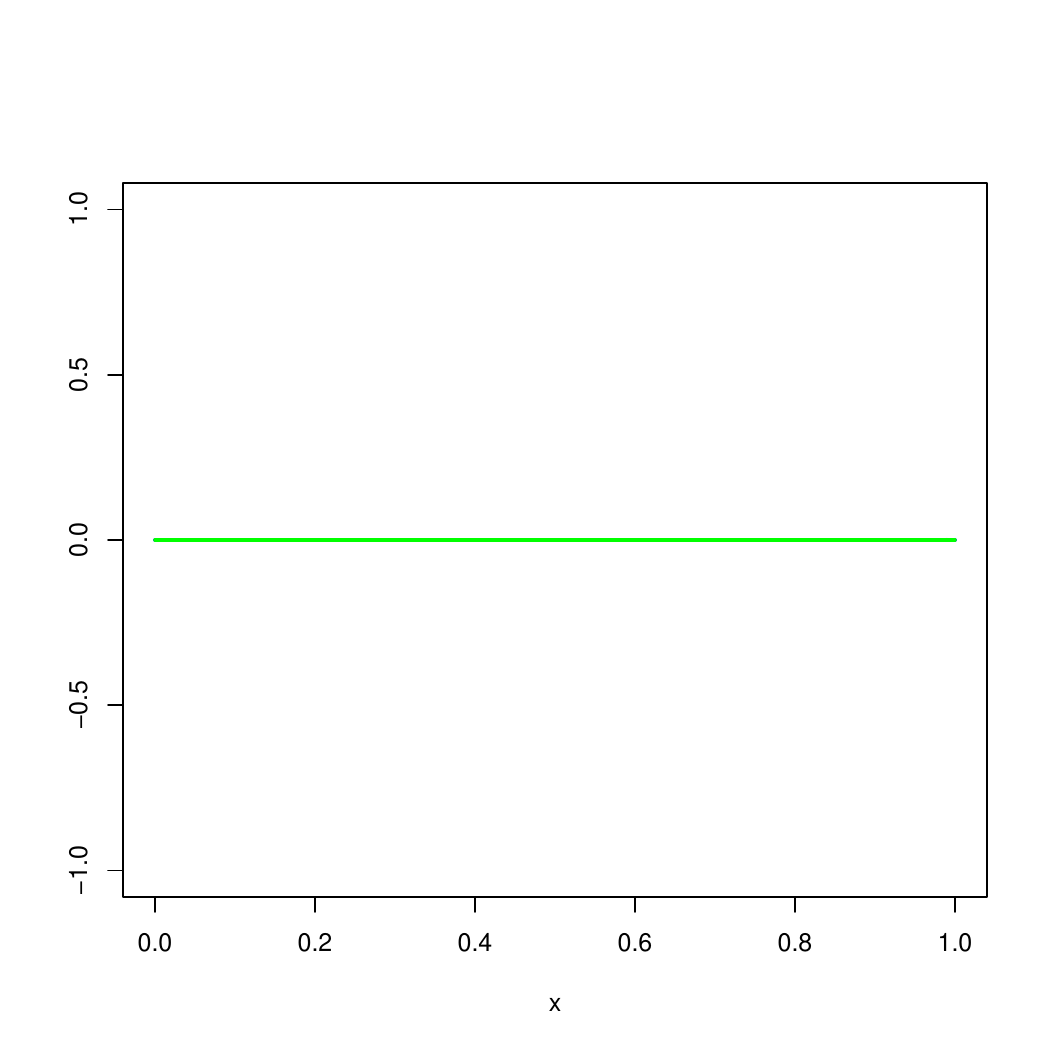}
   \end{tabular}
\caption{\small \label{fig:curvas-cl-rob-C5} The first two rows contain the functional boxplots of the estimated additive functions using the least square-based estimator while the last two rows contain the functional boxplots of the estimated additive functions by the robust approach, for $n=200$ and the contamination setting $C_4$.} 
\end{center}
\end{figure}

\begin{figure}[htbp]
 \begin{center}
\small
\begin{tabular}{ccccc}
\includegraphics[scale=0.22]{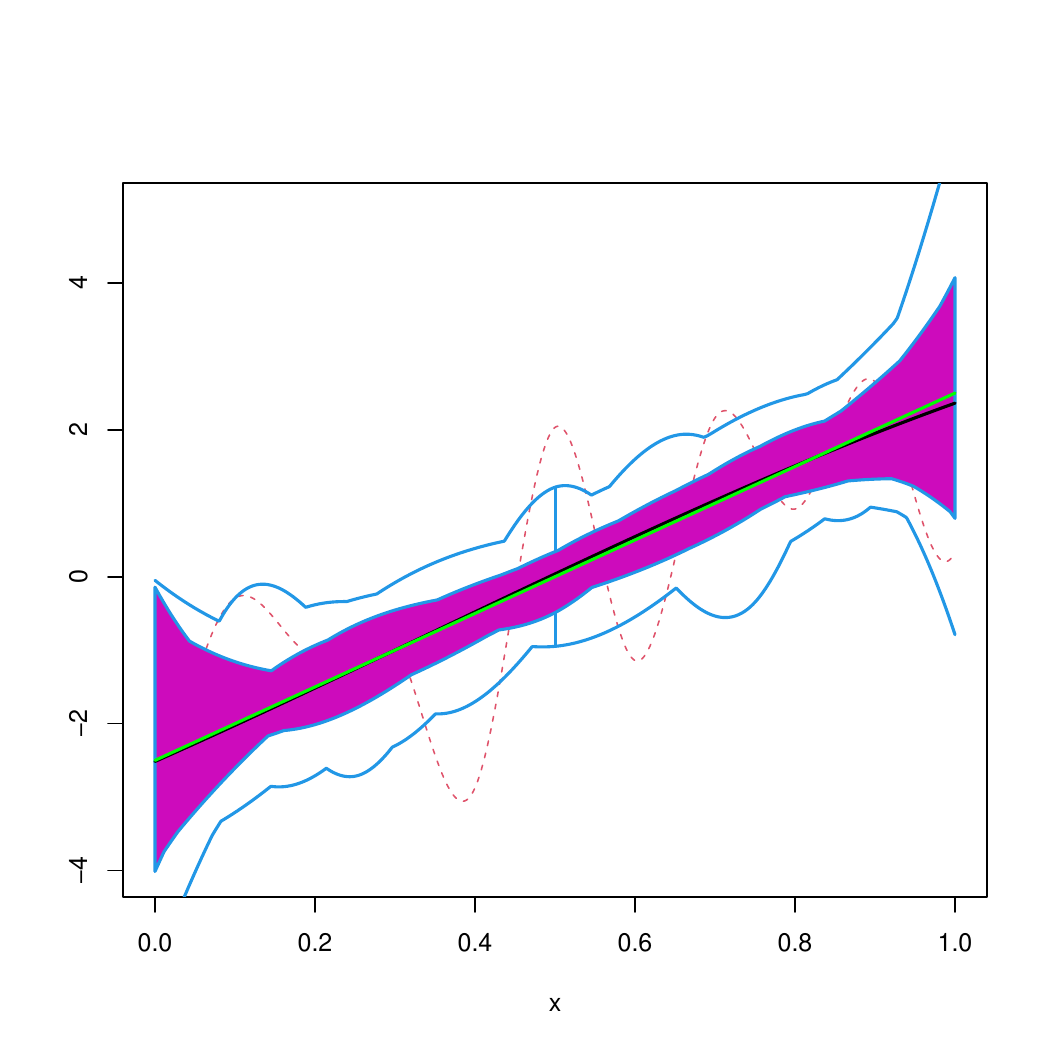} &
 \includegraphics[scale=0.22]{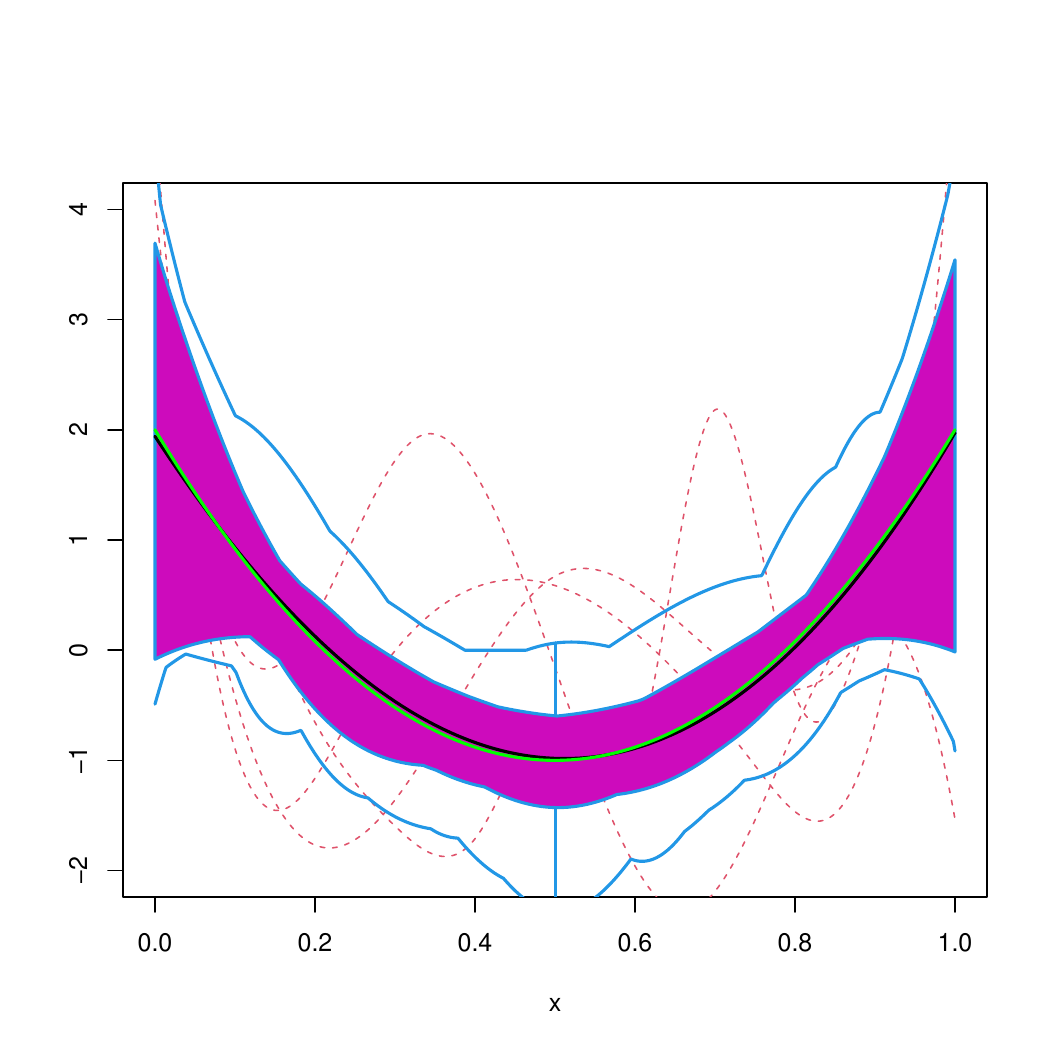} &
\includegraphics[scale=0.22]{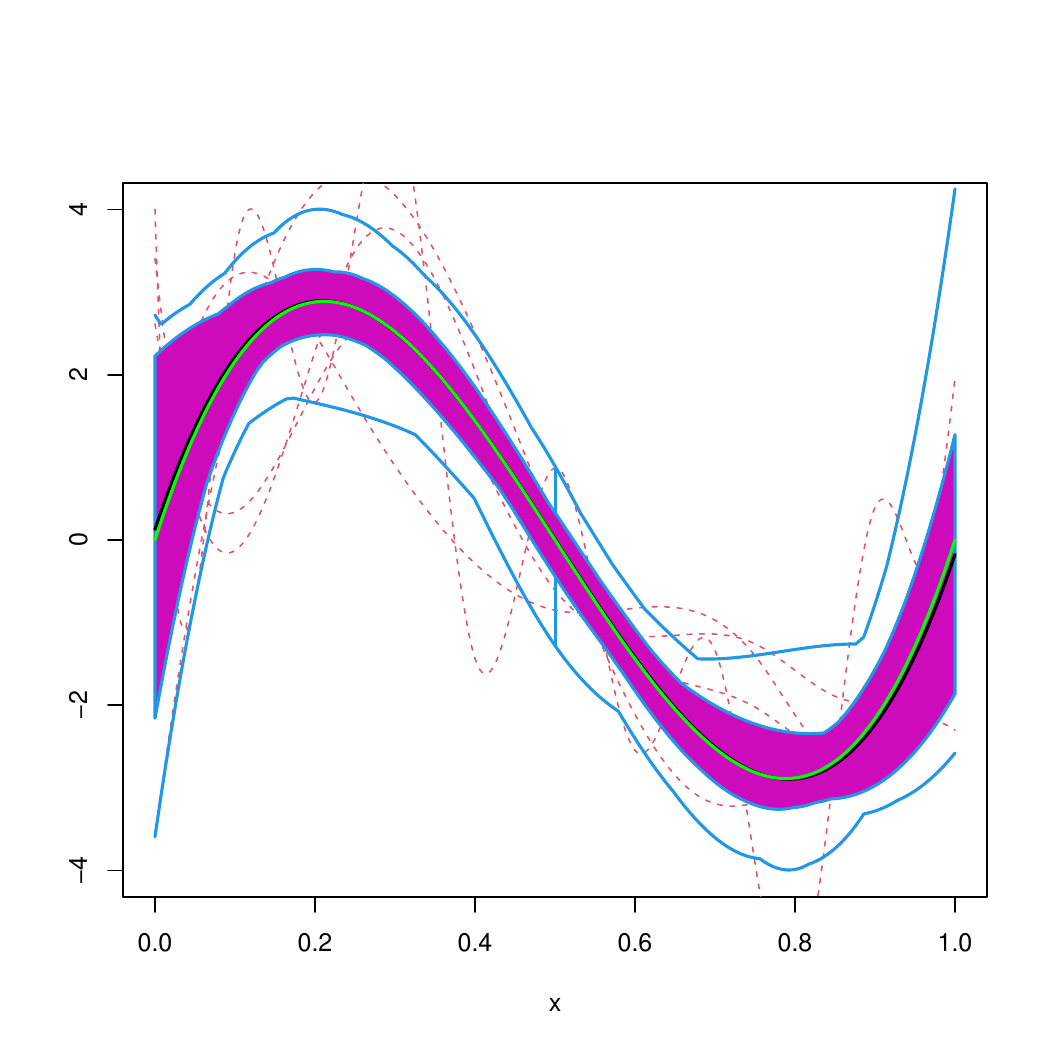}&
\includegraphics[scale=0.22]{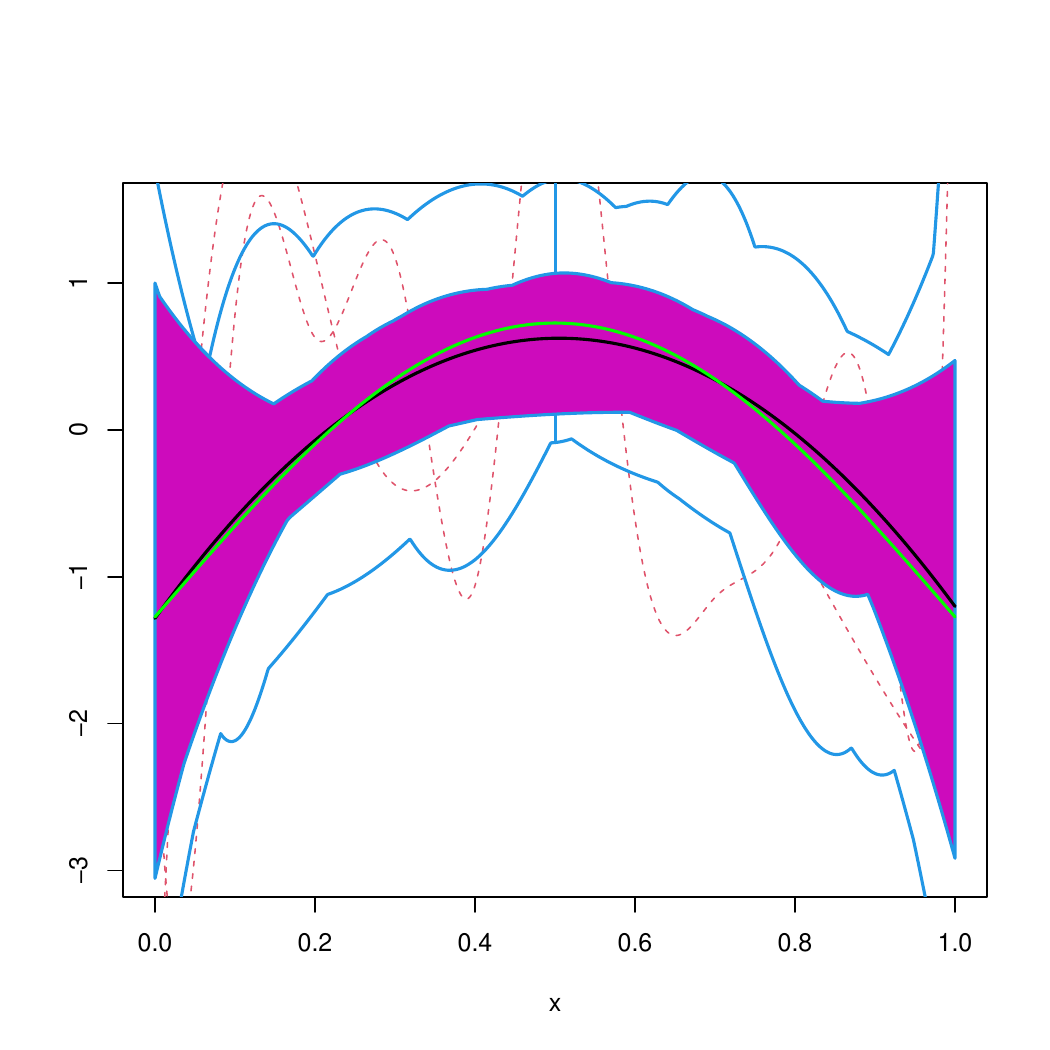}&
\includegraphics[scale=0.22]{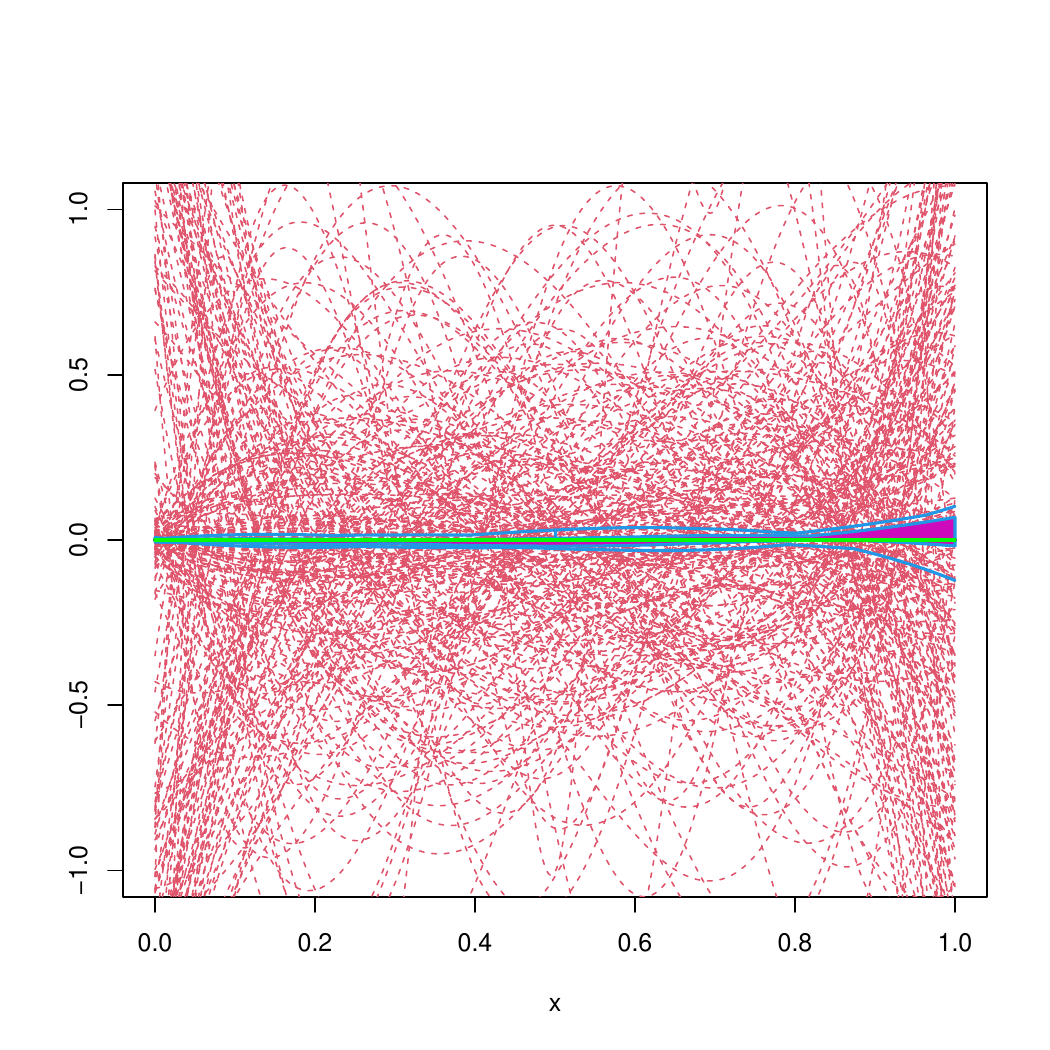}\\
\includegraphics[scale=0.22]{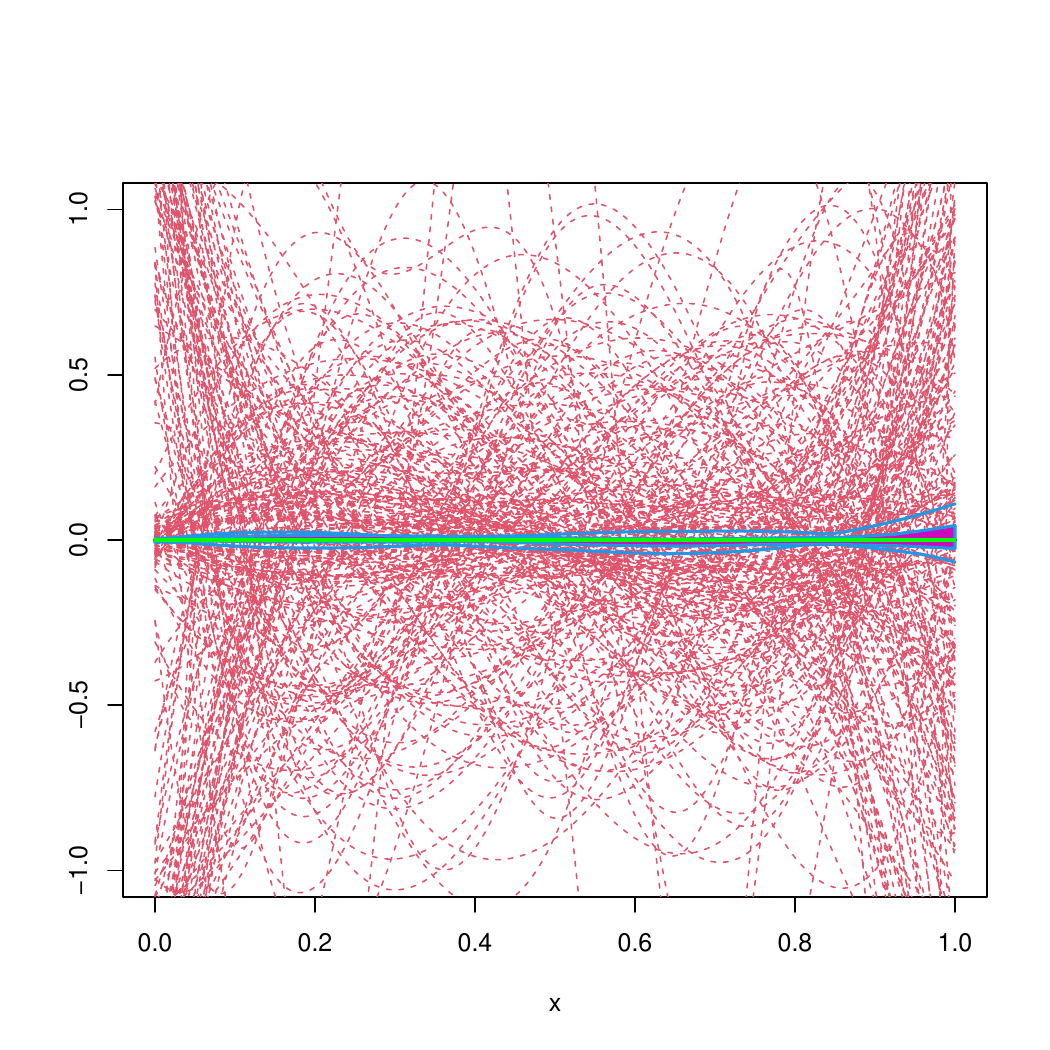} &
 \includegraphics[scale=0.22]{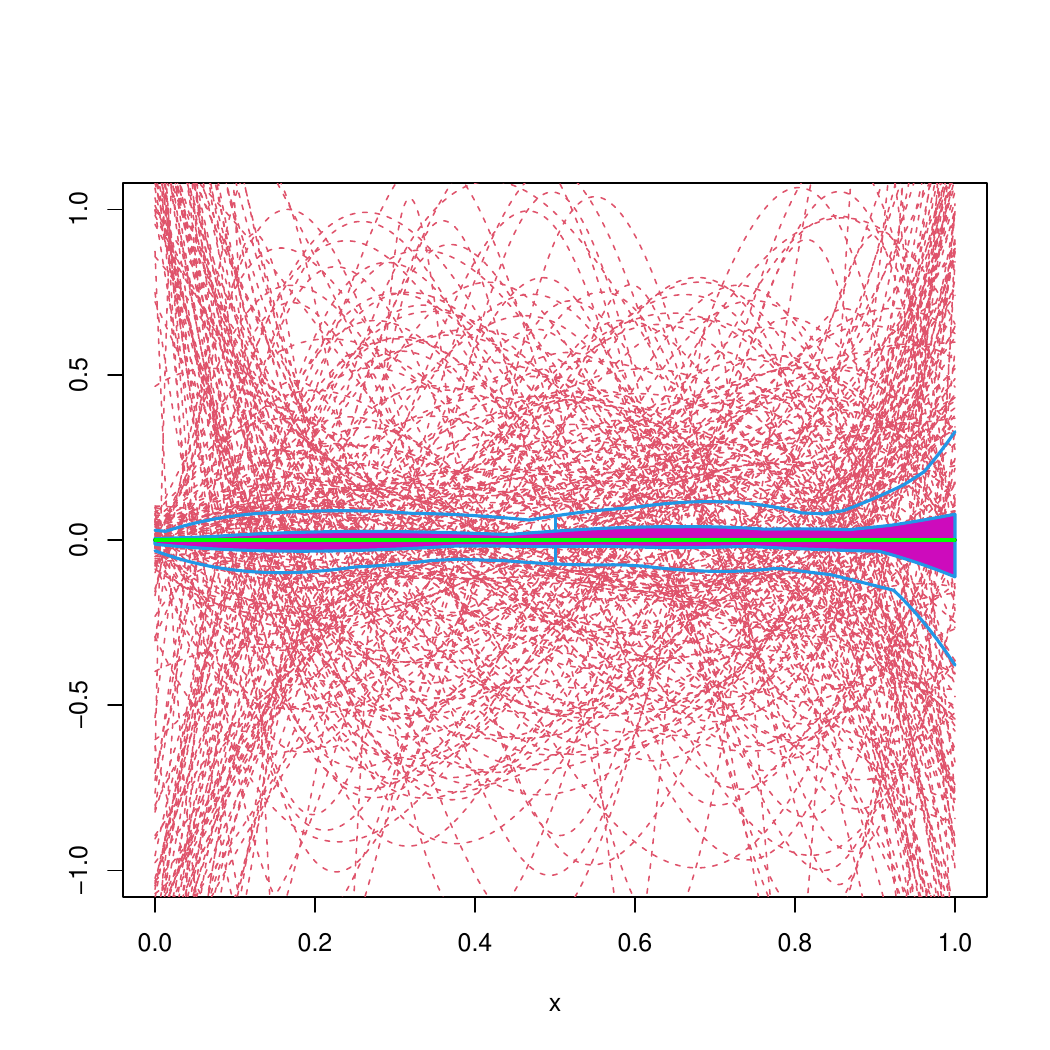}&
 \includegraphics[scale=0.22]{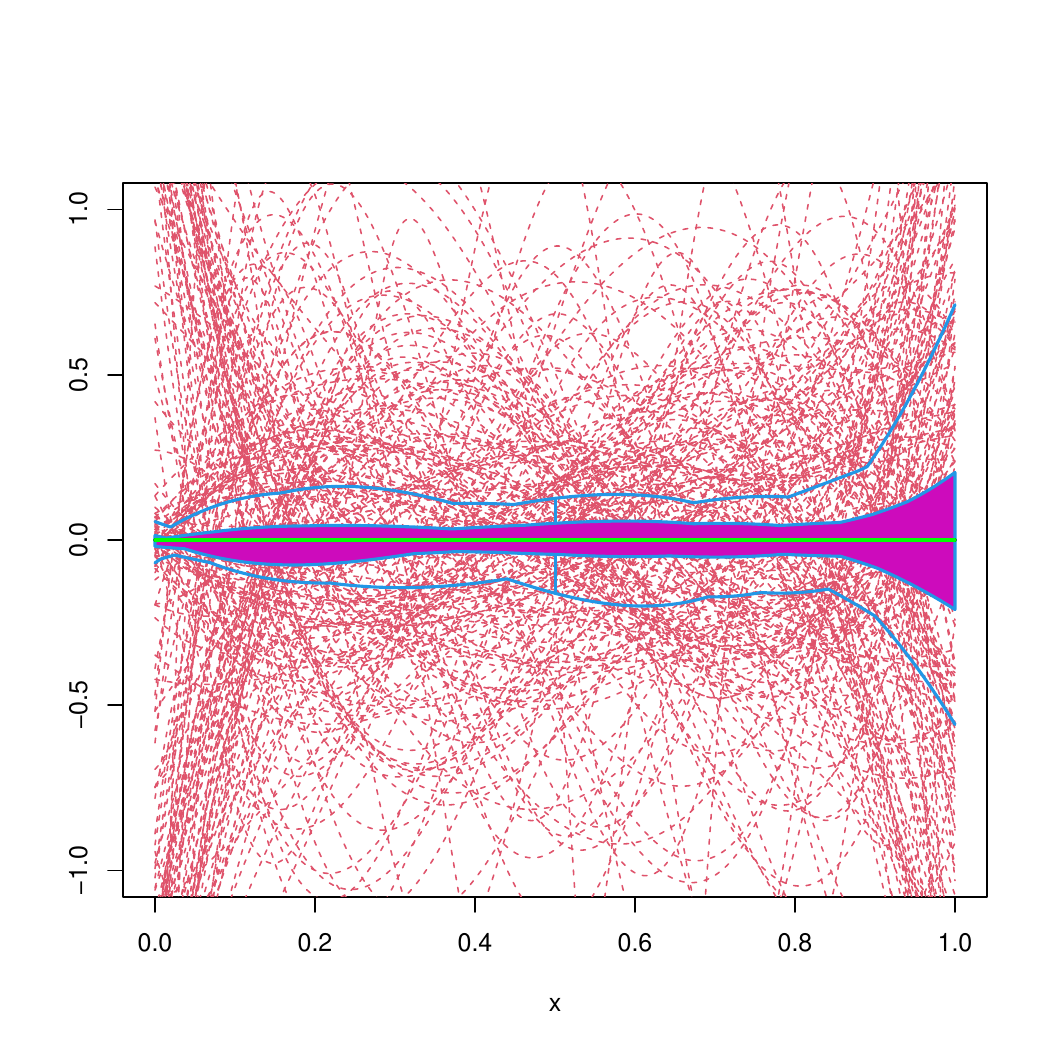}&
  \includegraphics[scale=0.22]{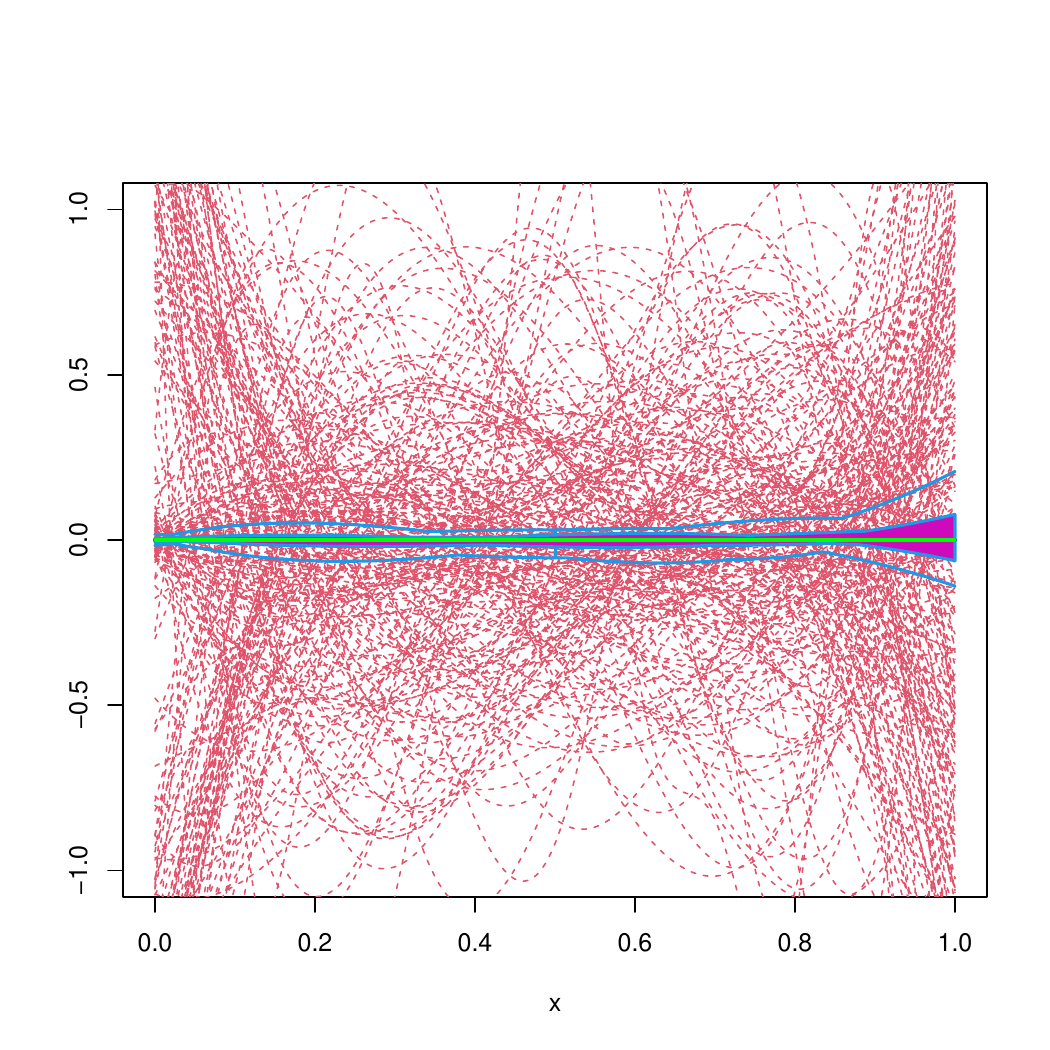}&
   \includegraphics[scale=0.22]{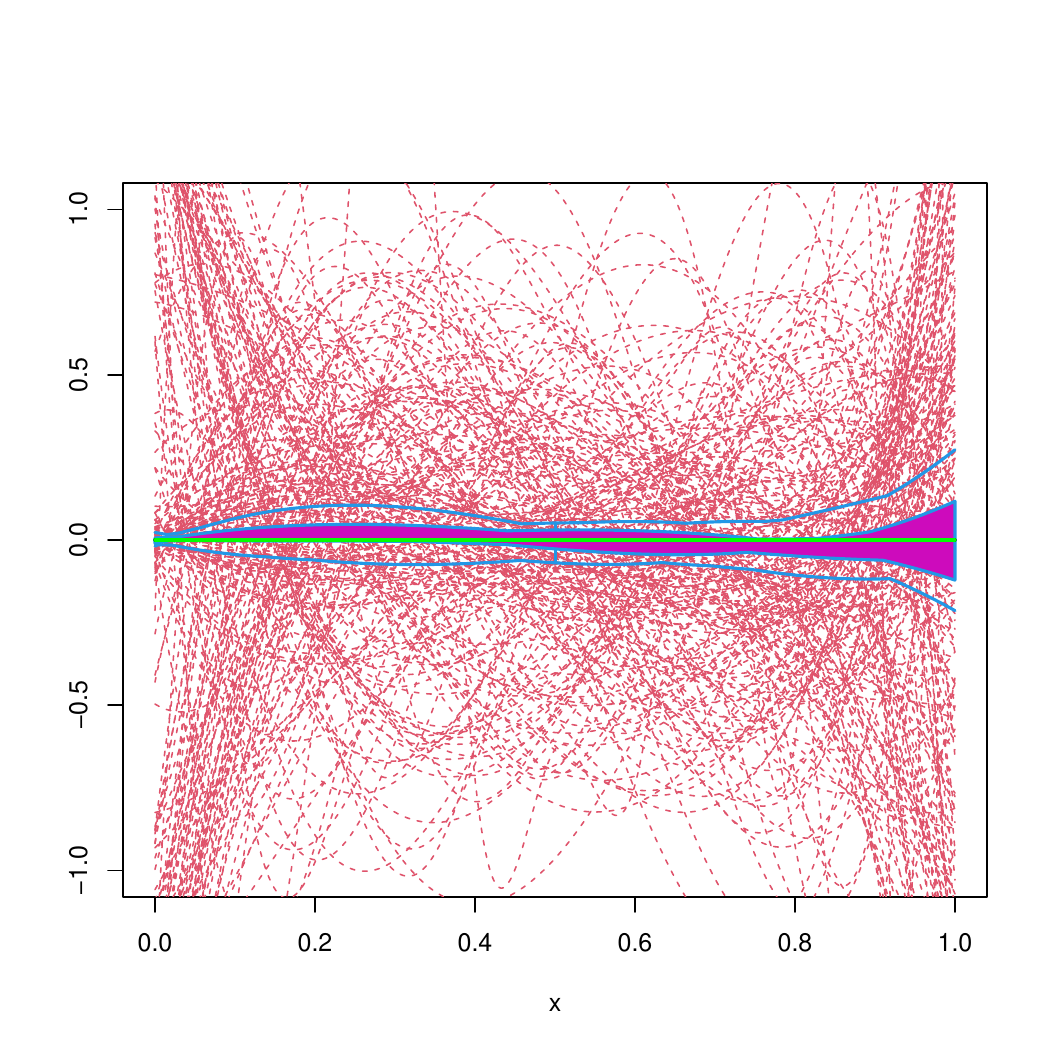} \\
   \includegraphics[scale=0.22]{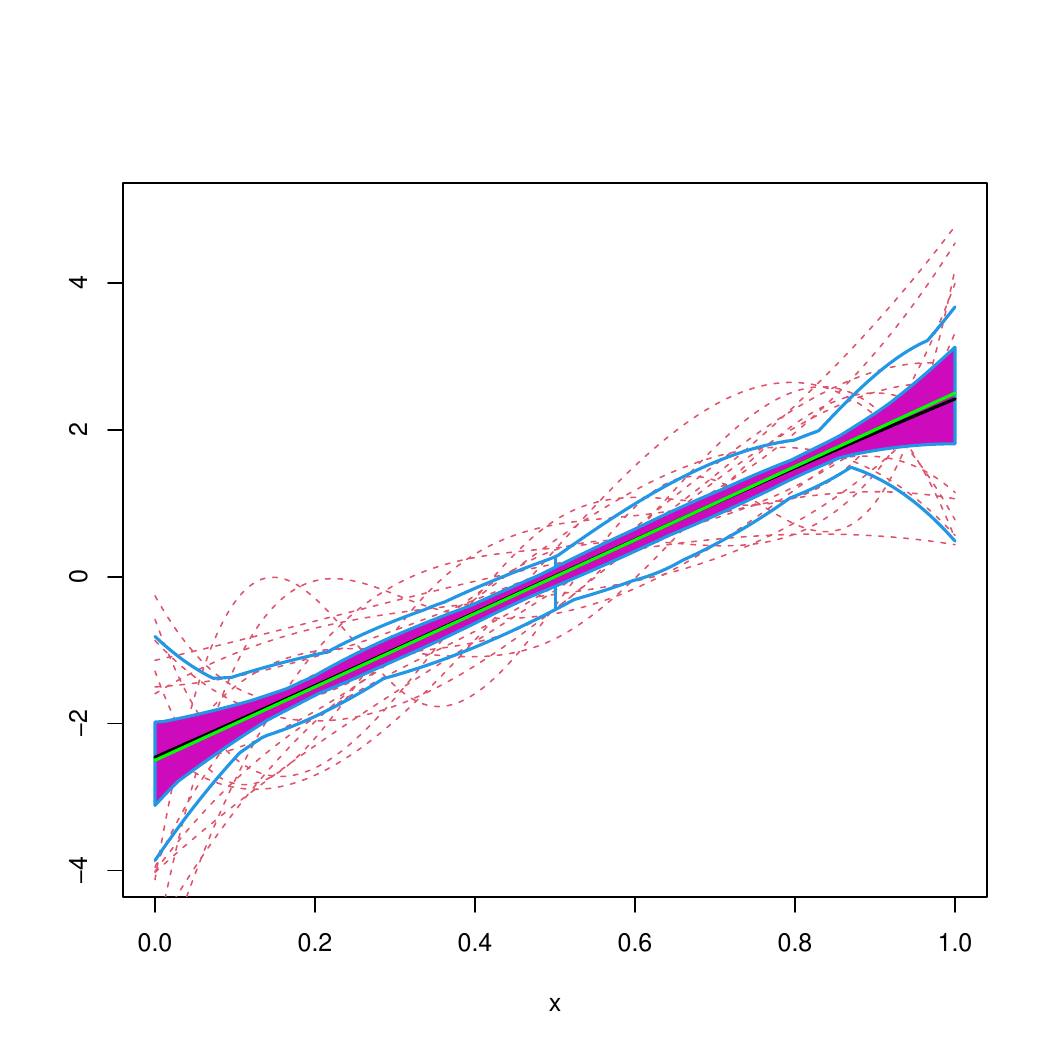} &
 \includegraphics[scale=0.22]{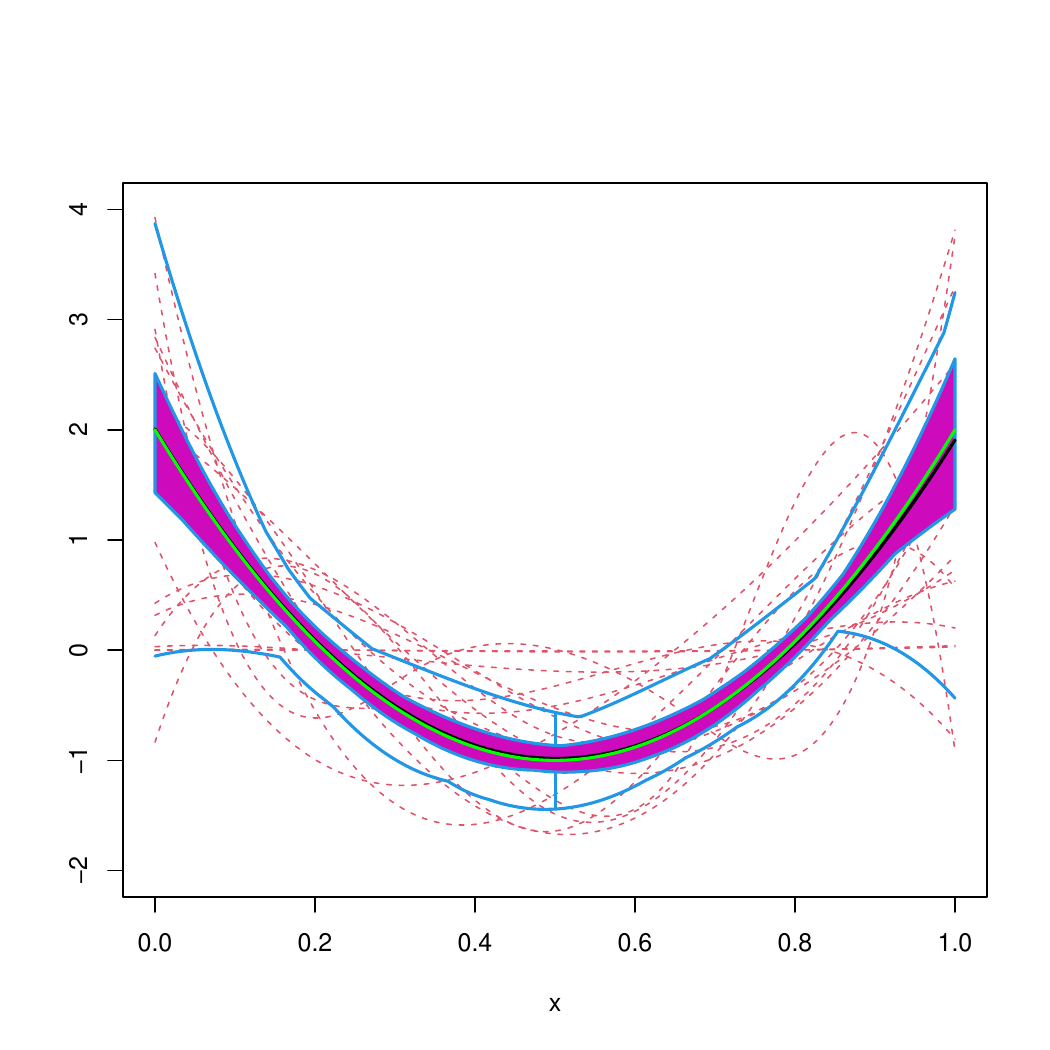} &
\includegraphics[scale=0.22]{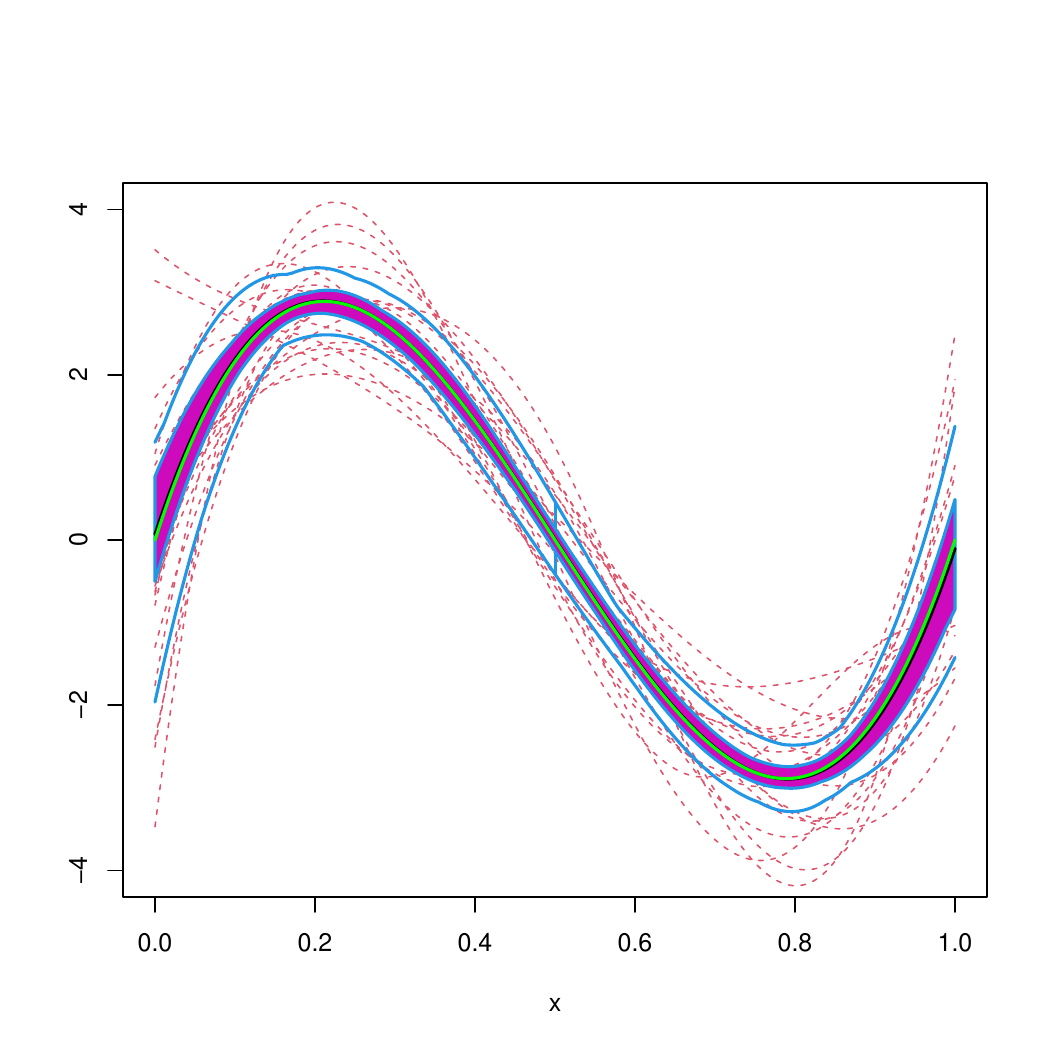}&
\includegraphics[scale=0.22]{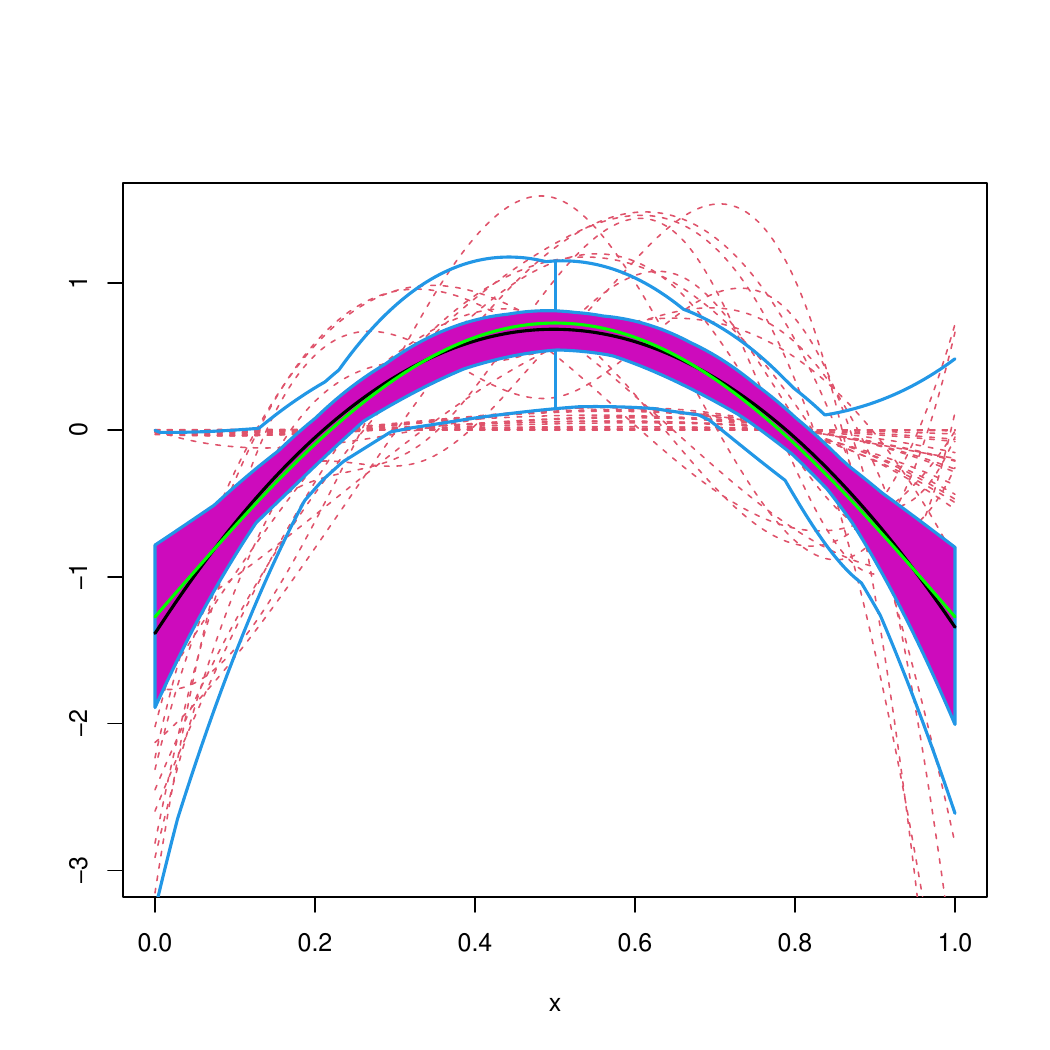}&
\includegraphics[scale=0.22]{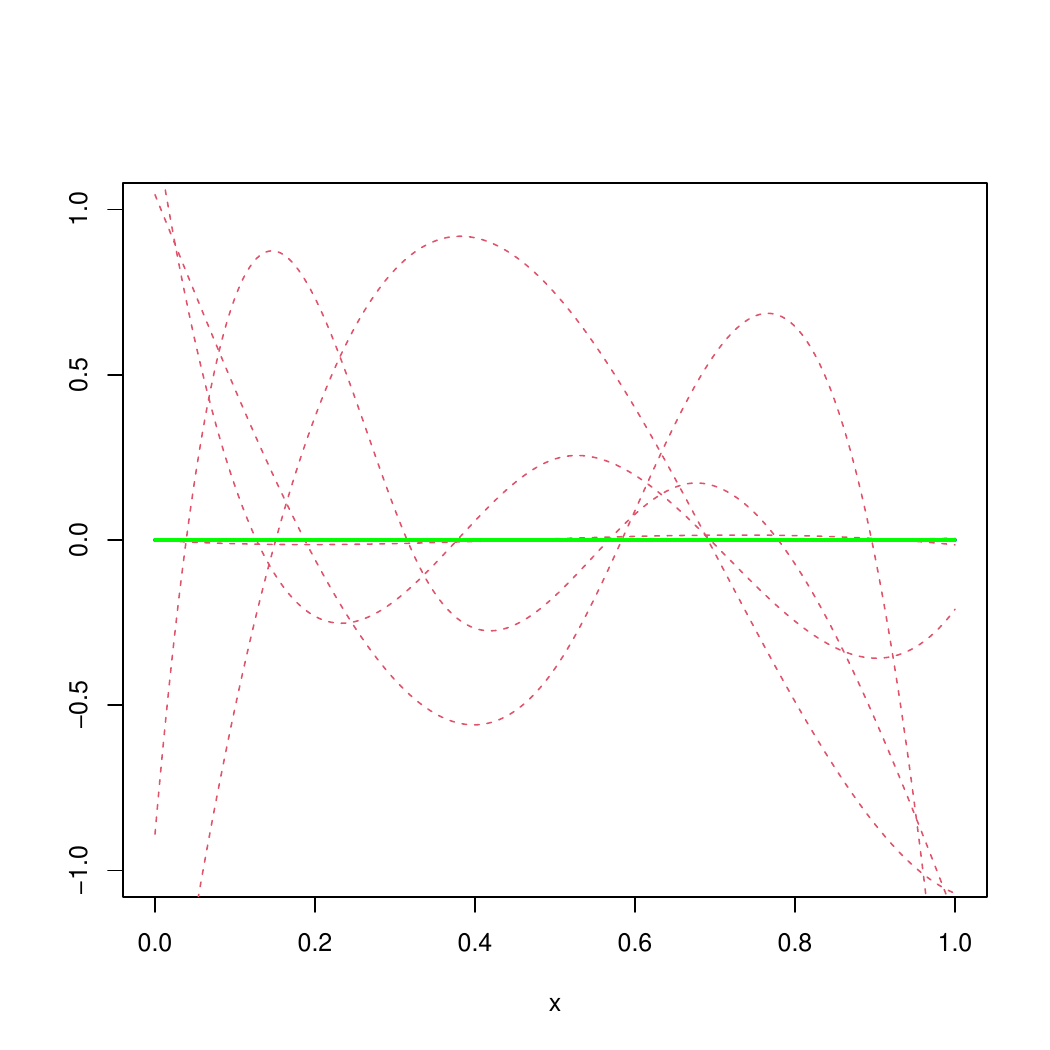}\\
\includegraphics[scale=0.22]{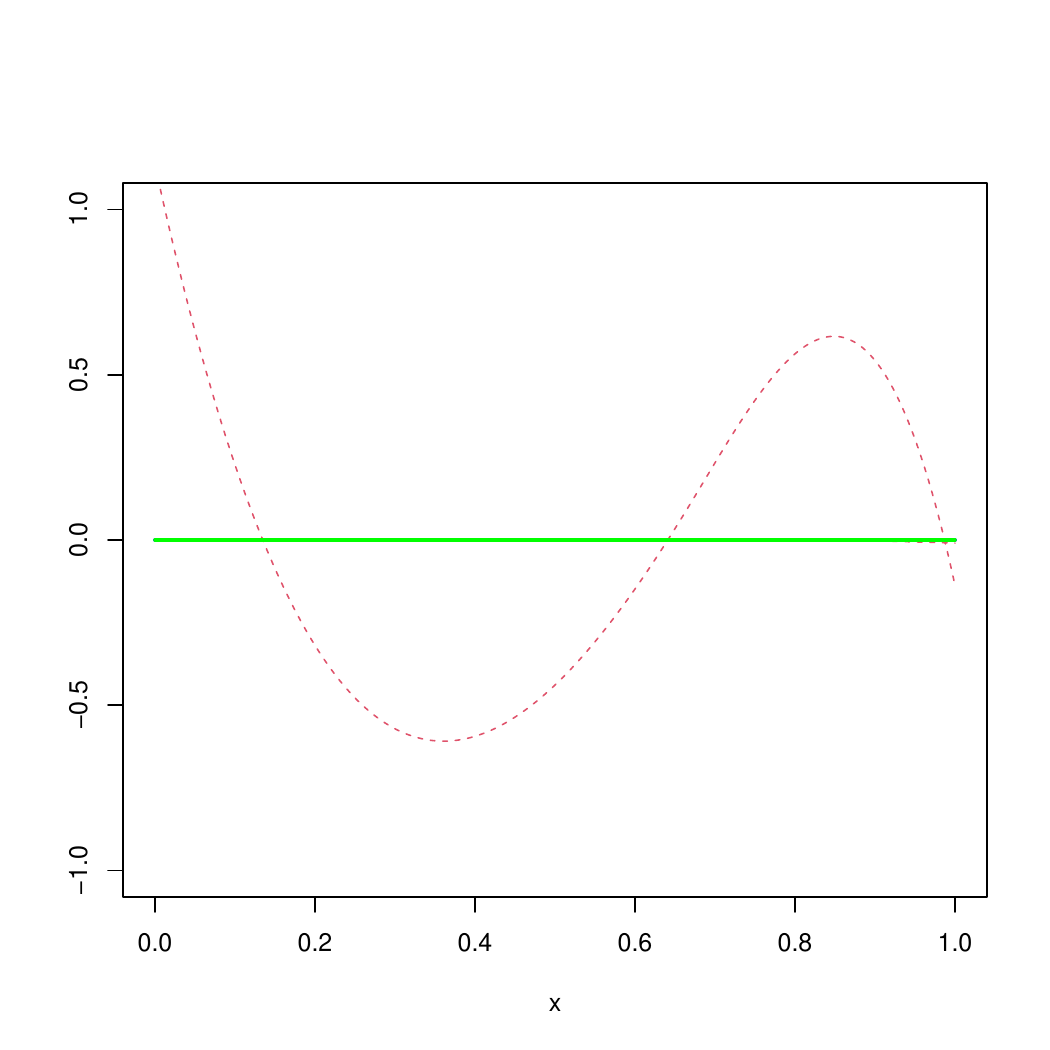} &
 \includegraphics[scale=0.22]{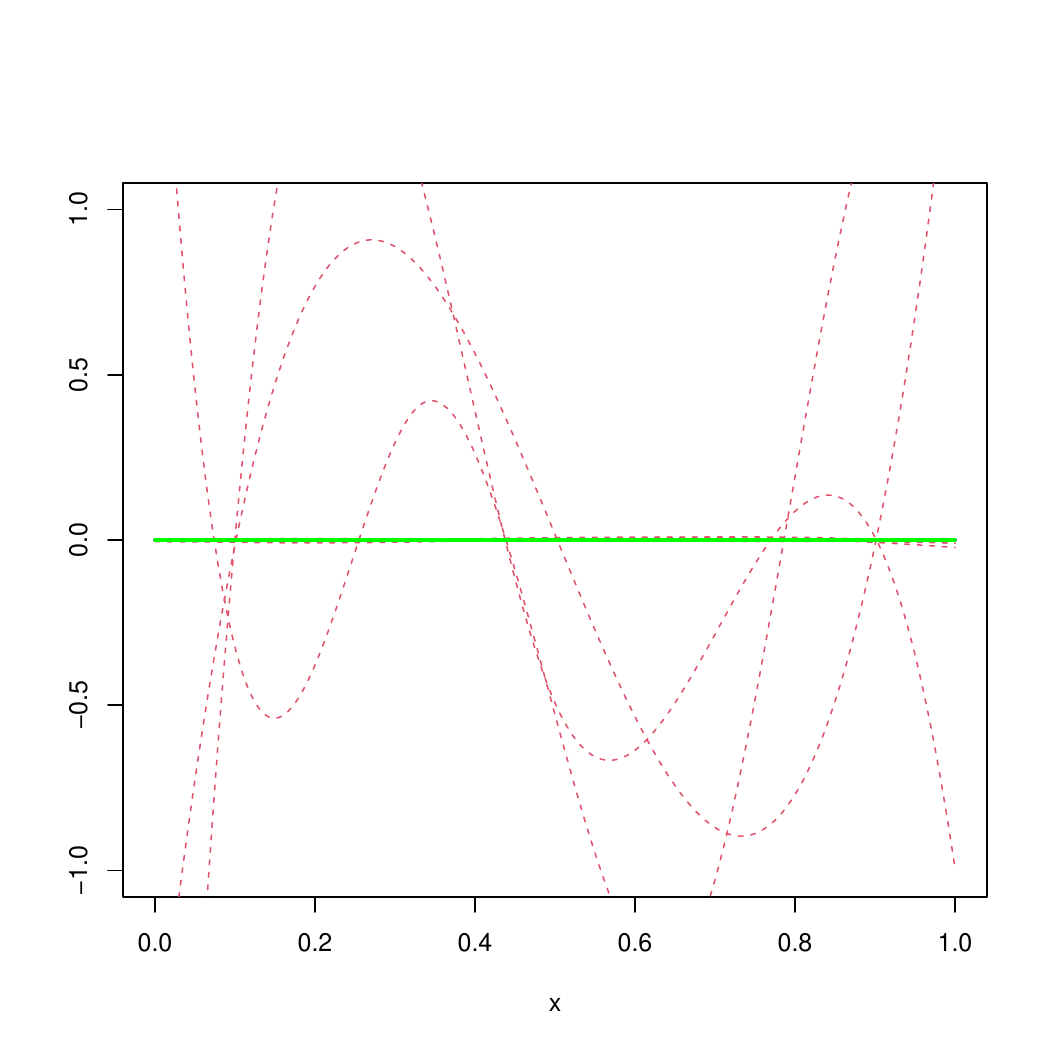}&
 \includegraphics[scale=0.22]{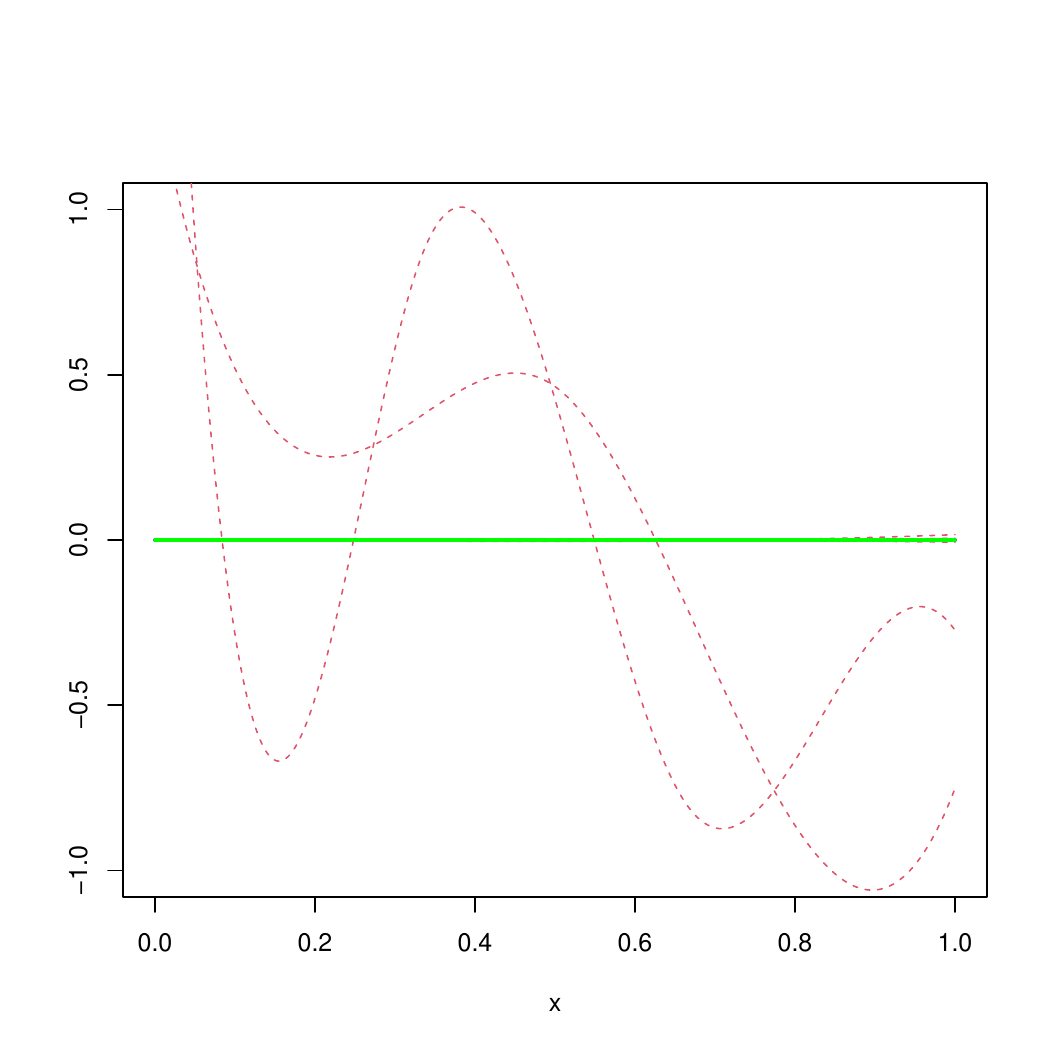}&
  \includegraphics[scale=0.22]{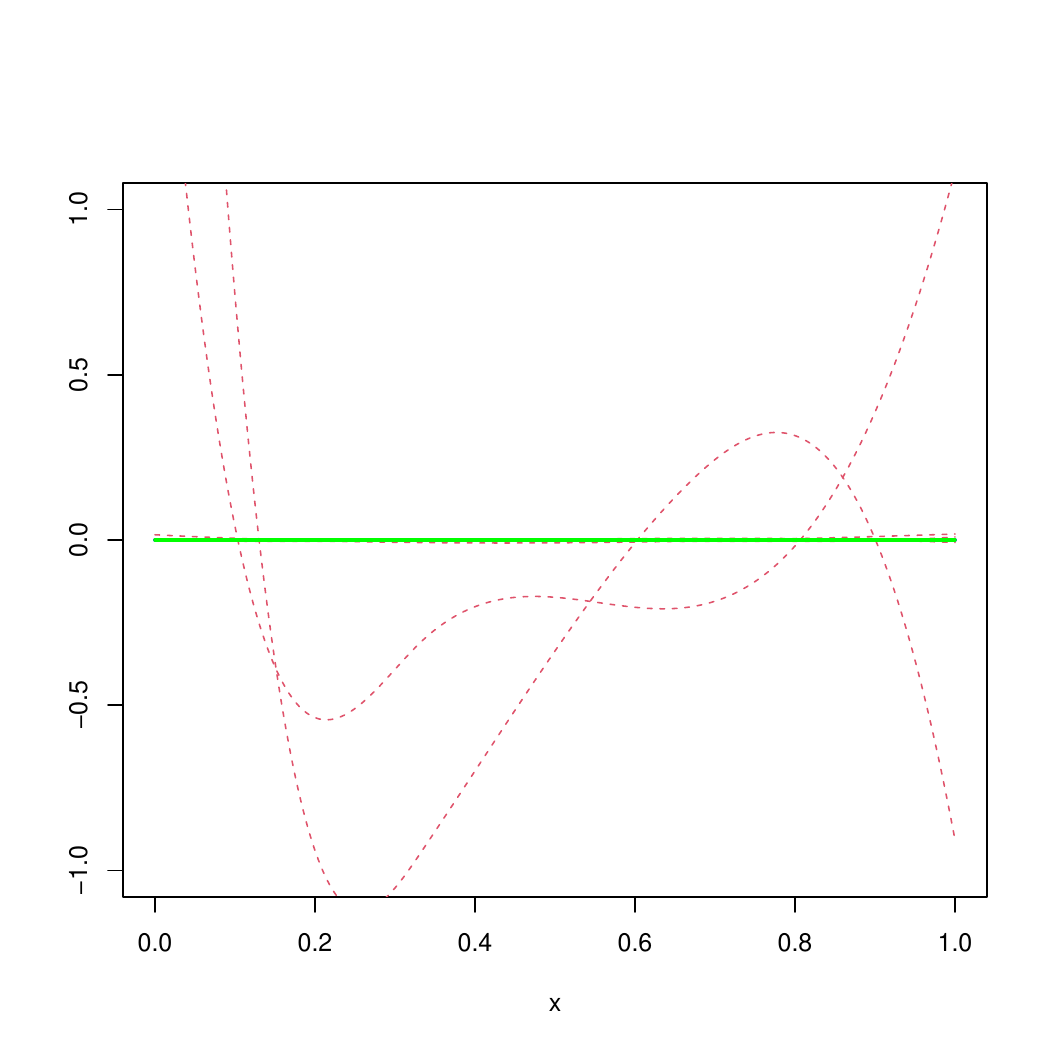}&
   \includegraphics[scale=0.22]{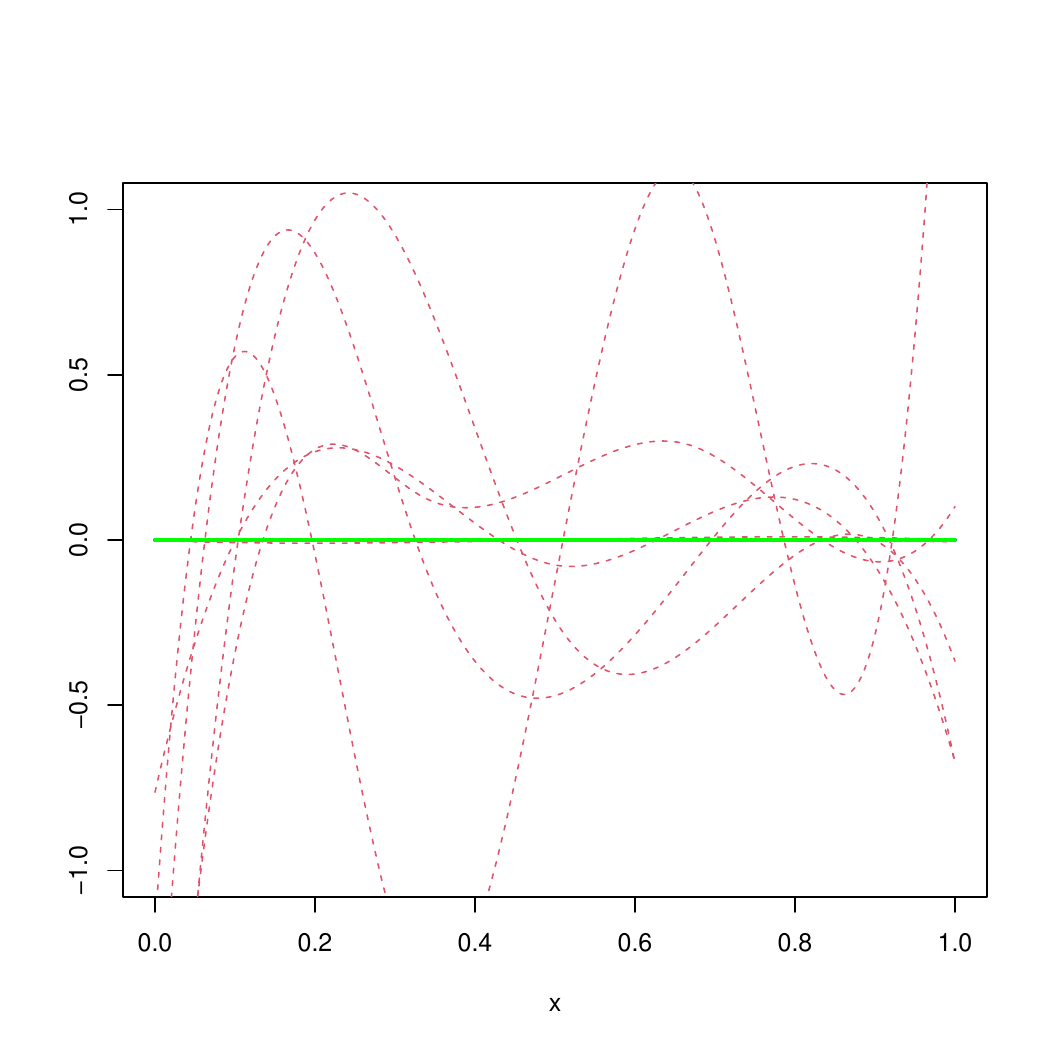}
   \end{tabular}
\caption{\small \label{fig:curvas-cl-rob-C7} The first two rows contain the functional boxplots of the estimated additive functions using the least square-based estimator while the last two rows contain the functional boxplots of the estimated additive functions by the robust approach, for $n=200$ and the contamination setting $C_7$.} 
\end{center}
\end{figure}

\end{landscape}

%\begin{figure}[H]
%\begin{center}
%\includegraphics[scale=0.4]{Cantidad-Outliers-100Muestras.pdf}
%\end{center}
%\caption{\label{fig:outliers} \textcolor{red}{MODIFICAR}.}
%\end{figure}

It is interesting to see which are the $\wtlam_1$ and $\wtlam_2$ selected for each sample and each penalized estimator across the contamination settings. Figures \ref{fig:la1-n200} and \ref{fig:la2-n200} show the boxplots of the $\wtlam_1$ and $\wtlam_2$ show the boxplots of the selected when $n=200$, respectively, while Figures \ref{fig:la1-n400} and \ref{fig:la2-n400} show the selected parameters under $n=400$ and Figures \ref{fig:la1-n600} and \ref{fig:la2-n600} for $n=600$. It can be appreciated that under all contamination scenarios, for a fixed sample size, the robust approach tends to select similar values of the parameters while the least squares estimators vary the selected values yielding to very different boxplots. When $n=200$ and for selecting $\wtlam_1$, except for contamination $C_5$, the sizes of the boxes of the robust approach are quite similar. All the boxplots present whiskers to both sides and the median value is located at the same number. A different behaviour is observed for contamination $C_5$, where the top whisker is much shorterr than in the other settings, the box is less than a half of the otherss, the median is located at a  lower position and there are right-side outliers. On the other hand, the boxes of the boxplots of the selected $\wtlam_1$ by the least squares estimator are located at different values of $y-$axis: at the bottom of the plot ($C_1$, $C_2$, $C_3$ and $C_4$), in the middle of the plot ($C_0$, $C_6$ and $C_7$) and at the top of the plot ($C_5$), leading to different locations of the median values. 

With respect to the selection of $\wtlam_2$, Figure \ref{fig:la2-n200} shows a similar behavior. While the values selected by the robust approach are quite consistent, reflected in similar boxplots with comparable boxes, medians, and whisker lengths, the least squares estimator displays a different pattern. Although the boxes are of similar sizes (except in the $C_5$ case), the medians vary slightly, and the whiskers range from shorter ones (in $C_0$ to $C_3$) to longer ones (in $C_4$, $C_6$, and $C_7$). Under contamination setting $C_5$, the LS$-$estimator seems to select values in a certian what that the box spans nearly the entire grid range for $\wtlam_2$.
%With respect to the selection of $\wtlam_2$, Figure \ref{fig:la2-n200} shows a similar behaviour. While the selected values by the robust approach are quite similar, appreciated in similar boxplots, that is, similar boxes, median and lengths of whiskers, the least squares estimator shows a different behaviour. Even though the boxes have similar sizes, except for the $C_5$ case, the median is located at different (but quite similar) values and the whiskers vary from small ones ($C_0$ to $C_3$) to larger ones ($C_4$, $C_6$ and $C_7$). Under the contamintion setting $C_5$, the LS$-$estimator seems to be selecting all the values randomly since the box almost cover the whole set of values of the grid for $\wtlam_2$. 
For $n=400$, shown in Figures \ref{fig:la1-n400} and \ref{fig:la2-n400}, the more stable behavior of the robust estimator across scenarios is evident for both tuning parameters, particularly for $\widetilde{\lambda}_1$, where all boxplots are positioned at the lower end of the plot. A distinct behavior appears under $C_5$, where the box is reduced to the value $0.05$. For the selection of $\widetilde{\lambda}_2$, the boxplots across contaminated settings are quite similar to that of the uncontaminated scenario $C_0$. In contrast, the values of $\lambda_1$ under $C_0$ for the LS$-$estimator show a small box with whiskers on both sides and some right-side outliers, while other settings show much larger boxes, differing medians, and different lengths of whiskers. A similarly unpredictable pattern is seen in the least-squares approach for selecting $\wtlam_2$; under $C_0$, the boxplot has its median at $0.3$, centered with whiskers on both sides, but the other settings display larger medians and a range of box sizes, from narrower (e.g., $C_1$ or $C_4$) to much larger ones (e.g., $C_5$).
%When $n=400$, as it is shown in Figures \ref{fig:la1-n400} and \ref{fig:la2-n400}, the more stable behaviour of the robust estimator across scenarios can be appreciated for both tuning parameters, but especially for $\widetilde{\lambda}_1$, where all the boxplots are located at the bottom of the plot. A different behaviour appears under $C_5$ where the box is reduced to the number $0.05$. For the selection of $\widetilde{\lambda}_2$, the boxplots obtained under the different contaminated settings are quite similar to that of the noncontaminated scenario $C_0$. On the contrary, the selected values of $\lambda_1$ under $C_0$ for the LS$-$estimator shows a boxplot a small box a whiskers to both sides and right-side outliers, while the boxes of the others are much larger, some of them with different medians, others with no left whiskers and/or long right ones. A similar unpredicted behaviour can be observed by the least-squares approach when selecting $\wtlam_2$. Even though under $C_0$ the boxplot has its median at $0.3$, almost in the middle of the box and with whiskers to both sides, the other boxplots show medians larger than that of the $C_0$ case and the boxes vary from narrower ones as under $C_1$ or $C_4$ and much larger ones as in $C_5$.
Finally, for $n=600$ and selecting $\wtlam_1$ (see Figure \ref{fig:la1-n600}), the robust estimator consistently selects the same value, $0.05$, except for some outliers. However, the least squares approach shows greater variability, with larger boxes. As in the previous sample sizes, box sizes, median locations, and whisker lengths vary across contamination schemes. For selecting $\wtlam_2$ (Figure \ref{fig:la2-n600}), the robust approach continues to show consistent boxplots, with boxes of similar size, whiskers on both sides, and medians at close values. The LS$-$based approach, in contrast, behaves differently. Under $C_0$, the box is centered on the $y$-axis with symmetrical whiskers and a median at $0.25$. For other contamination scenarios, except $C_4$, the bottom whisker is slightly longer, and medians are equal or greater than $0.4$. In $C_3$ to $C_7$, the top whiskers are missing, in $C_4$ the box is narrower and the left whisker shorter.
%Finally, when $n=600$ and for selecting $\wtlam_1$ (see Figure \ref{fig:la1-n600}), the robust proposal seems to be selecting almost always the same value, which is $0.05$. However, the least squares approach varies the selection throwing boxplots with larger boxes. Besides, similarly to the other sample sizes considered, the sizes of the boxes, the locations of the medians and the lengths of the whiskers are different among the contamination schemes. For selecting $\wtlam_2$ (Figure \ref{fig:la2-n600}), the robust approach still throws similar boxplots, with boxes of similar widths, whiskers to both sides and median located at to close values, while the approach based on LS does not. Under $C_0$, the box is in the middle of the $y-$axis with whiskers of equal length and the median at $0.25$, for the rest of the contamination scenarios, except for $C_4$, the bottom whiskers a little larger and the medians are greater or equal to $0.4$, and for $C_4$ the top whisker disappears, as in scenarios $C_3$, $C_5$, $C_6$ and $C_7$, the box is narrower and the bottom whisker and there are some outliers.

\begin{landscape}

\begin{figure}[htbp]
\begin{center}
\small
	 \renewcommand{\arraystretch}{-0.5} %0.4
 \newcolumntype{M}{>{\centering\arraybackslash}m{\dimexpr.1\linewidth-1\tabcolsep}}
   \newcolumntype{G}{>{\centering\arraybackslash}m{\dimexpr.2\linewidth-1\tabcolsep}}
\begin{tabular}{M G G G}
\textsc{LS} & \includegraphics[scale=0.6]{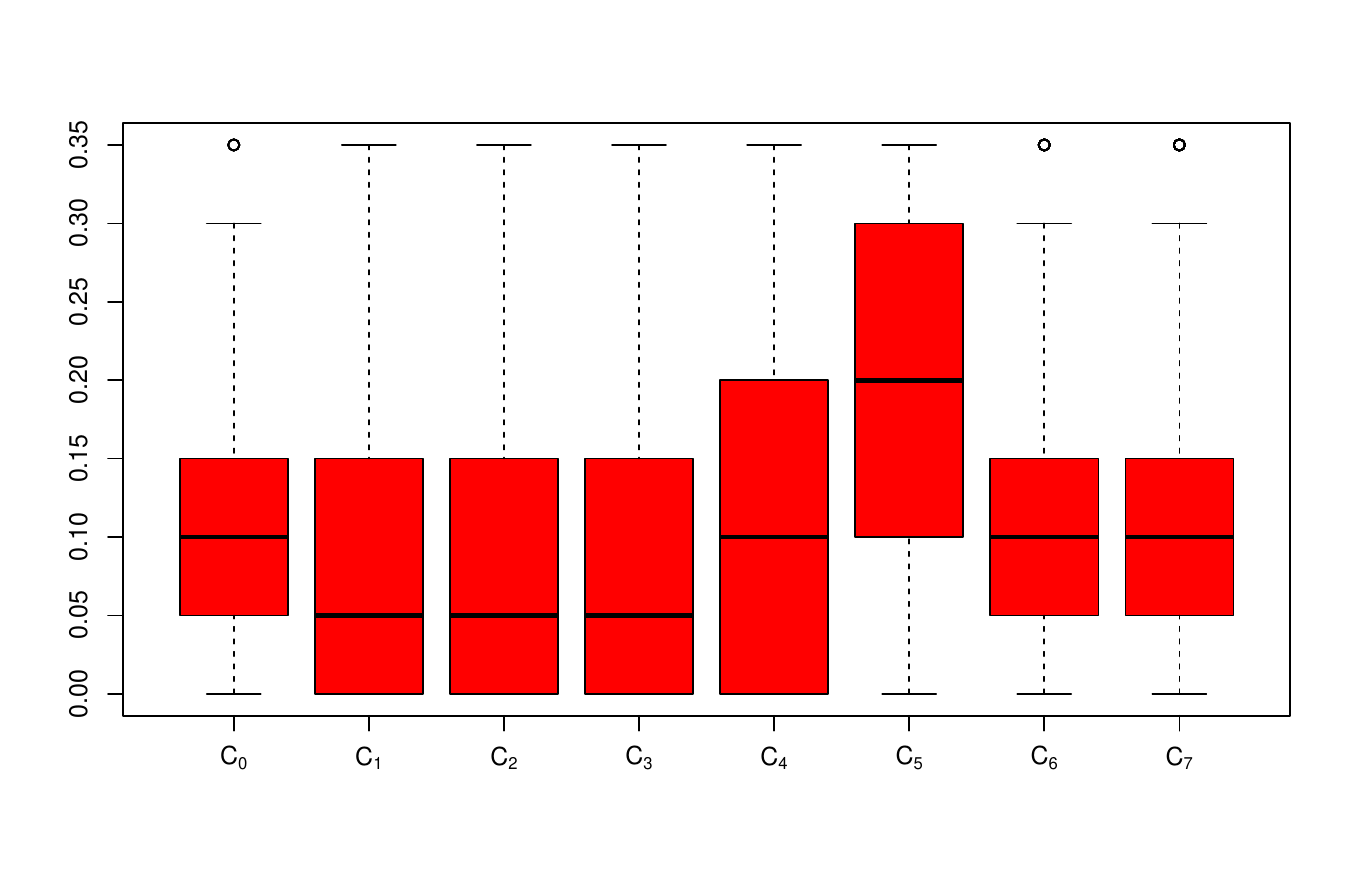} & & \\
\textsc{ROB} & \includegraphics[scale=0.6]{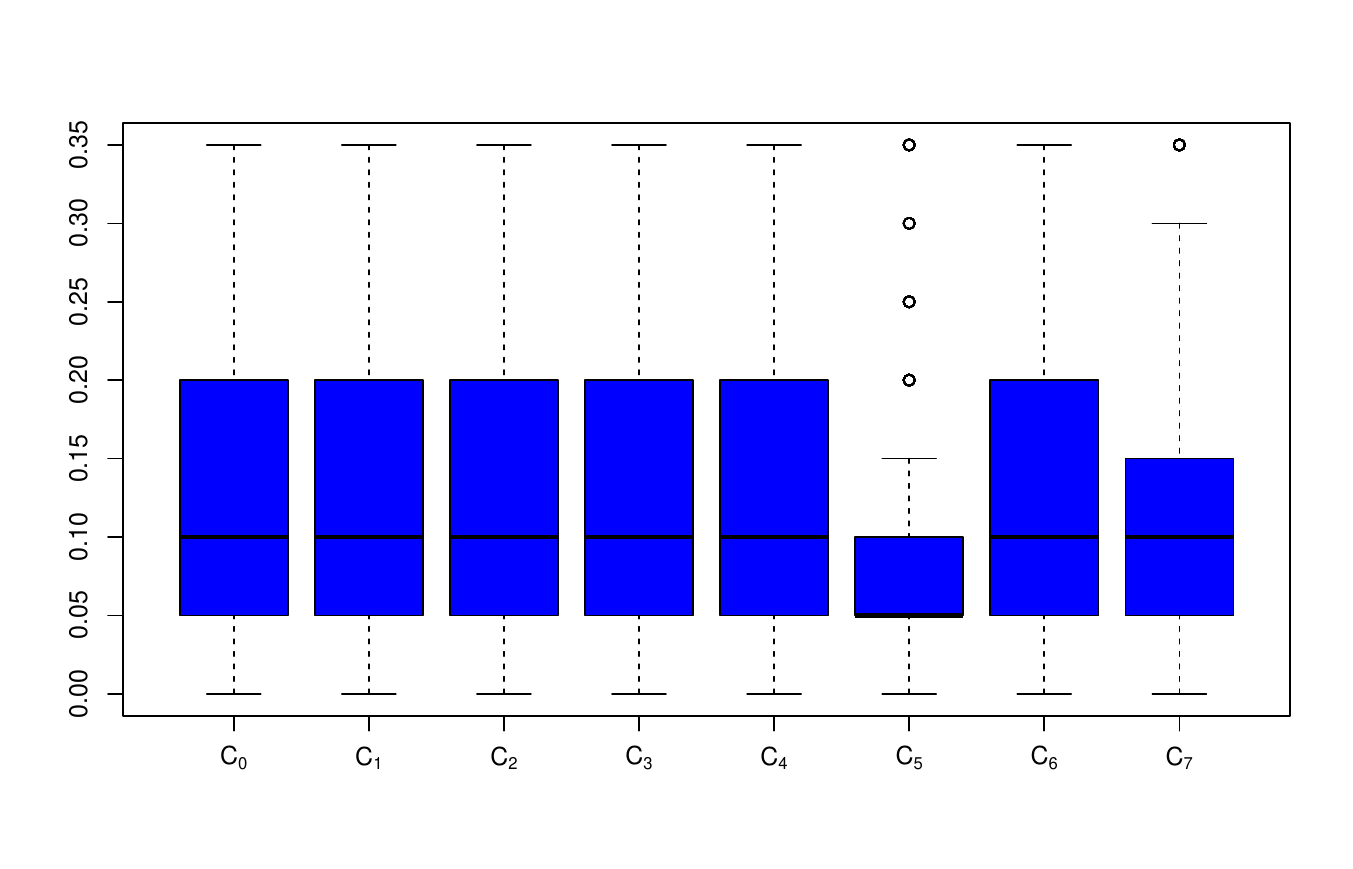} & &\\
\end{tabular}
\end{center}
\vspace{-1.4cm}
\caption{\label{fig:la1-n200} Boxplots of the $\wtlam_1$ selected under different contamination schemes for both least squares and robust estimators, when $n=200$.}
\end{figure}

\begin{figure}[htbp]
\begin{center}
\small
	 \renewcommand{\arraystretch}{-0.5} %0.4
 \newcolumntype{M}{>{\centering\arraybackslash}m{\dimexpr.1\linewidth-1\tabcolsep}}
   \newcolumntype{G}{>{\centering\arraybackslash}m{\dimexpr.2\linewidth-1\tabcolsep}}
\begin{tabular}{M G G G}
\textsc{LS} & \includegraphics[scale=0.6]{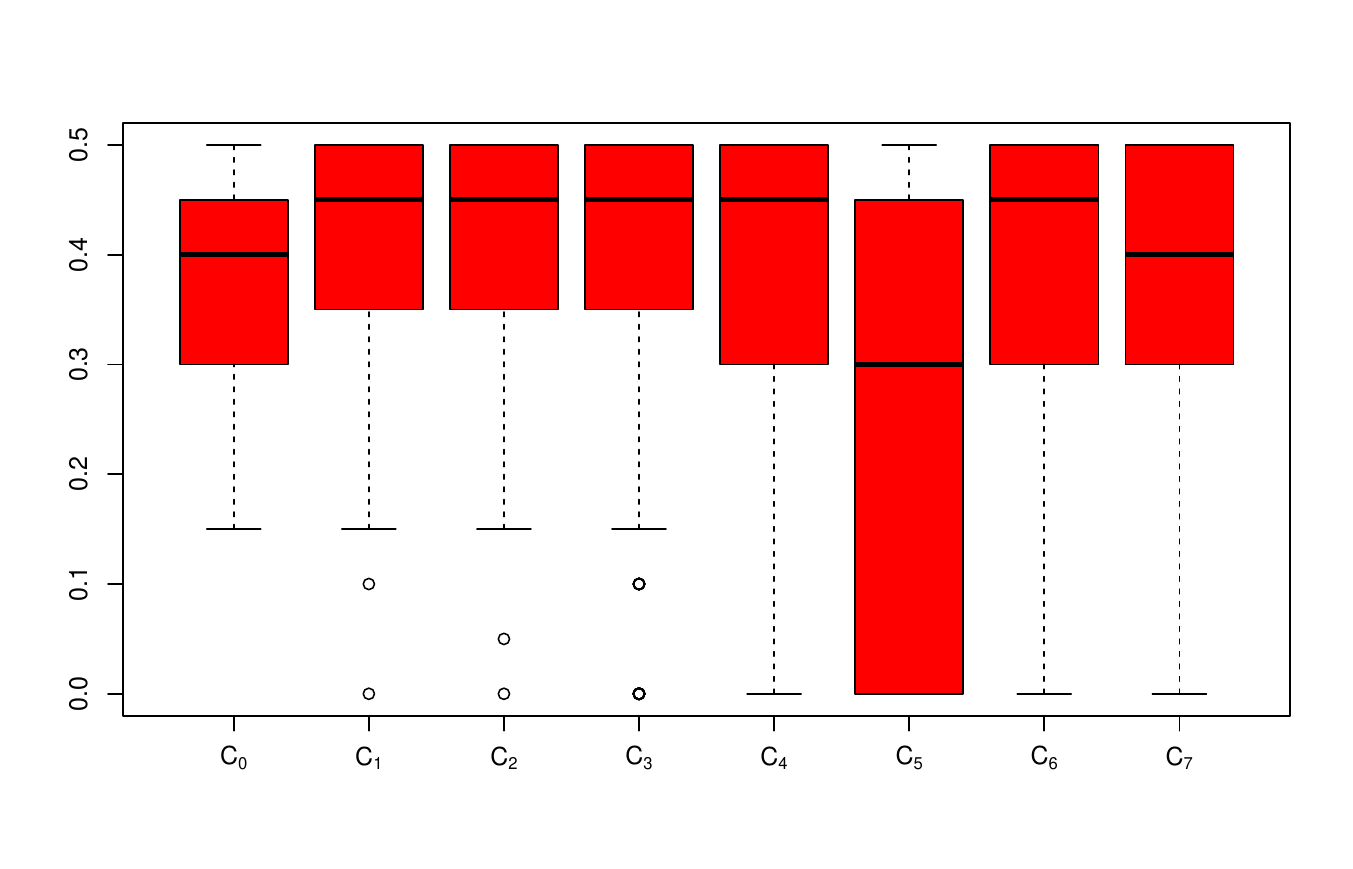} & & \\
\textsc{ROB} & \includegraphics[scale=0.6]{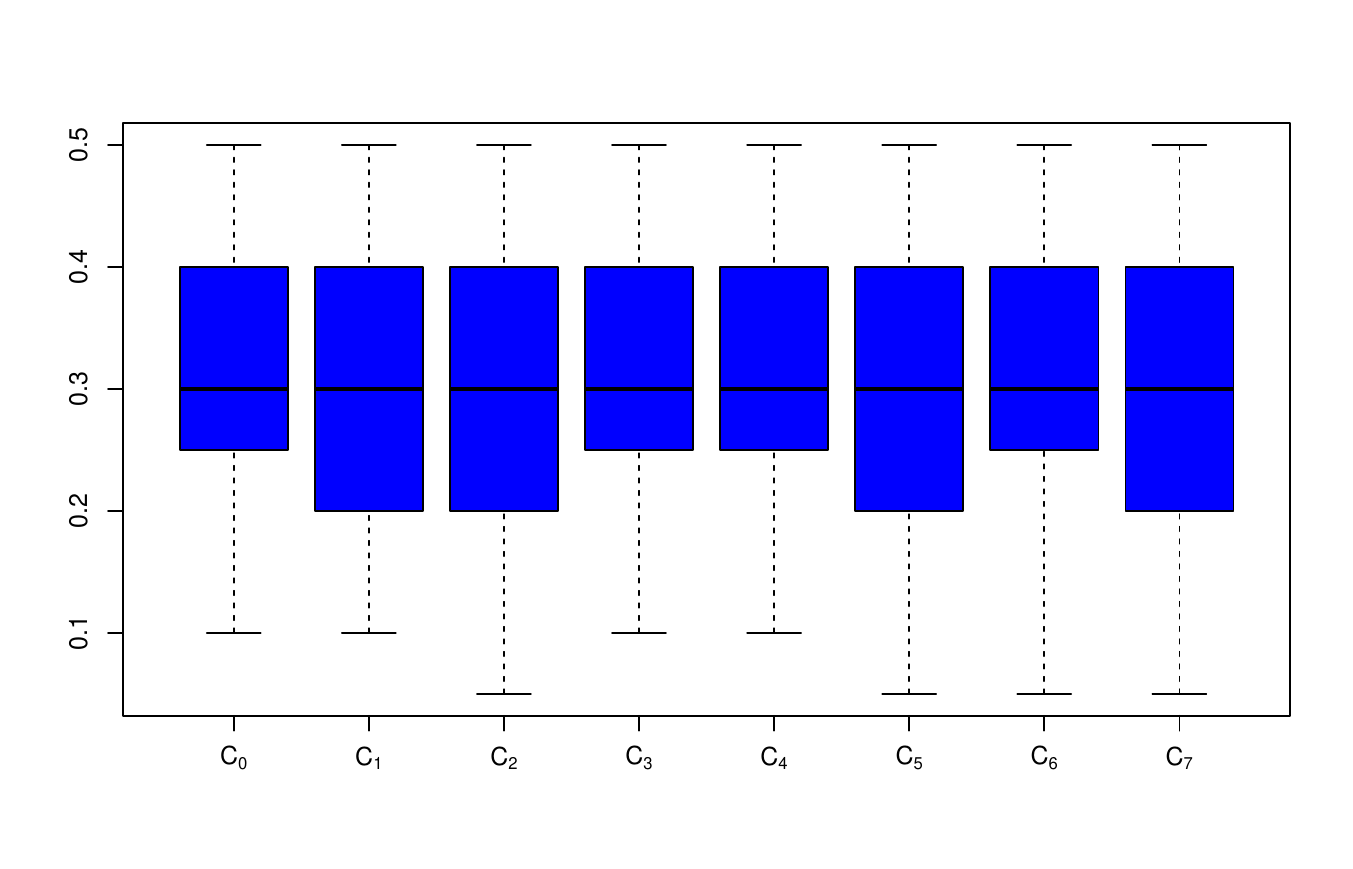} & &\\
\end{tabular}
\end{center}
\vspace{-1.4cm}
\caption{\label{fig:la2-n200} Boxplots of the $\wtlam_2$ selected under different contamination schemes for both least squares and robust estimators, when $n=200$.}
\end{figure}

\begin{figure}[htbp]
\begin{center}
\small
	 \renewcommand{\arraystretch}{-0.5} %0.4
 \newcolumntype{M}{>{\centering\arraybackslash}m{\dimexpr.1\linewidth-1\tabcolsep}}
   \newcolumntype{G}{>{\centering\arraybackslash}m{\dimexpr.2\linewidth-1\tabcolsep}}
\begin{tabular}{M G G G}
\textsc{LS} & \includegraphics[scale=0.6]{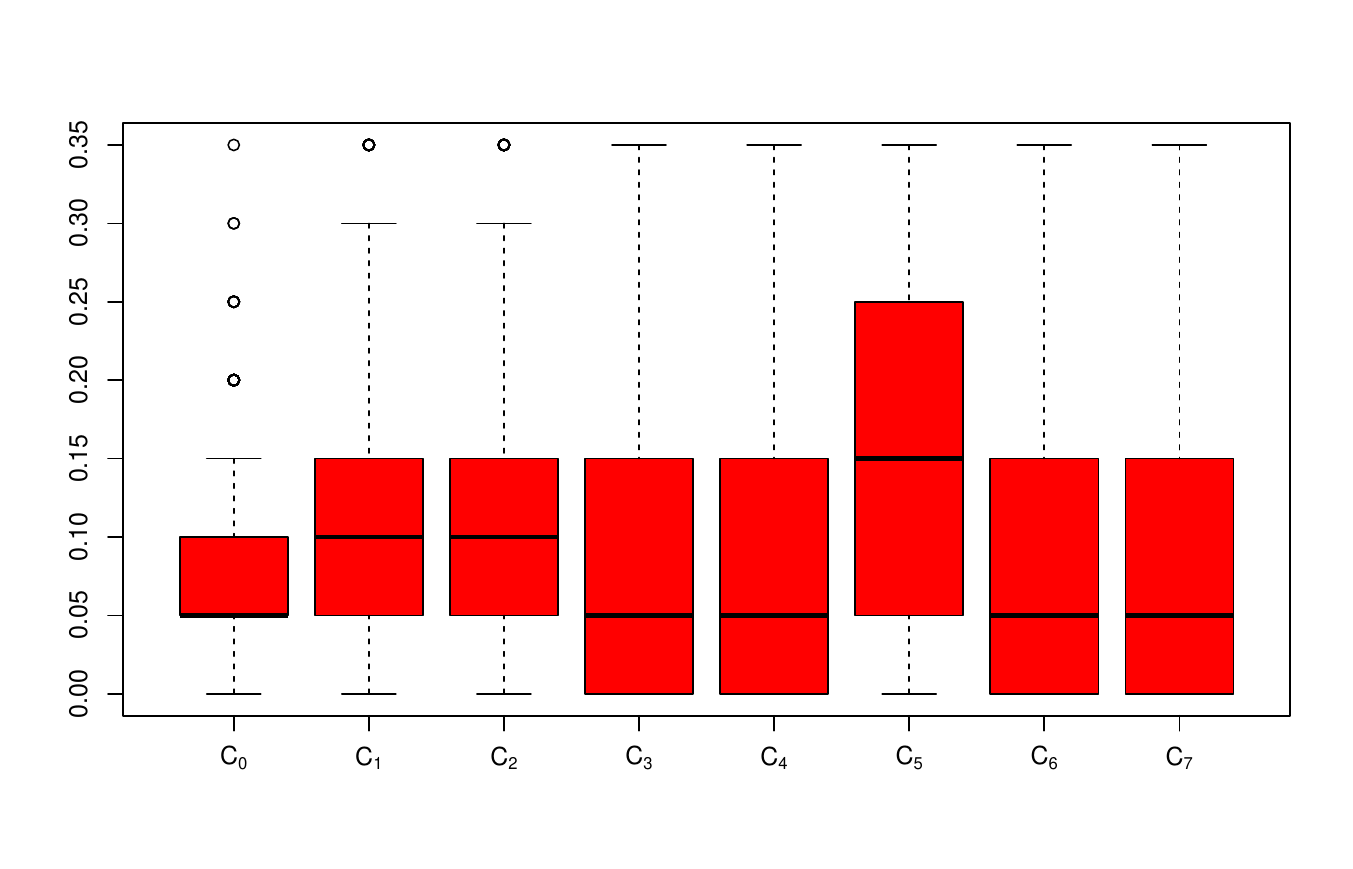} & & \\
\textsc{ROB} & \includegraphics[scale=0.6]{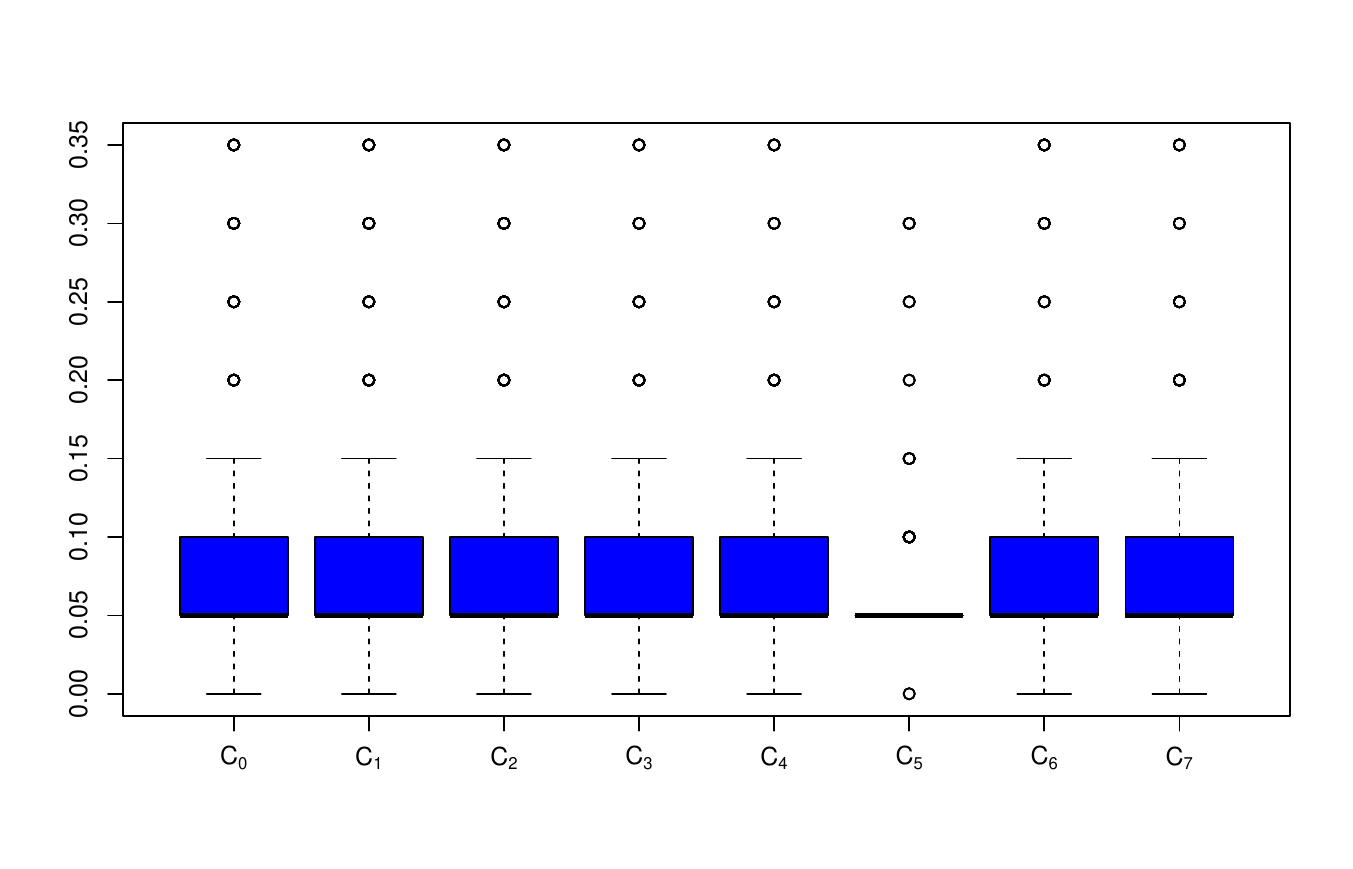} & &\\
\end{tabular}
\end{center}
\vspace{-1.4cm}
\caption{\label{fig:la1-n400} Boxplots of the $\wtlam_1$ selected under different contamination schemes for both least squares and robust estimators, when $n=400$.}
\end{figure}

\begin{figure}[htbp]
\begin{center}
\small
	 \renewcommand{\arraystretch}{-0.5} %0.4
 \newcolumntype{M}{>{\centering\arraybackslash}m{\dimexpr.1\linewidth-1\tabcolsep}}
   \newcolumntype{G}{>{\centering\arraybackslash}m{\dimexpr.2\linewidth-1\tabcolsep}}
\begin{tabular}{M G G G}
\textsc{LS} & \includegraphics[scale=0.6]{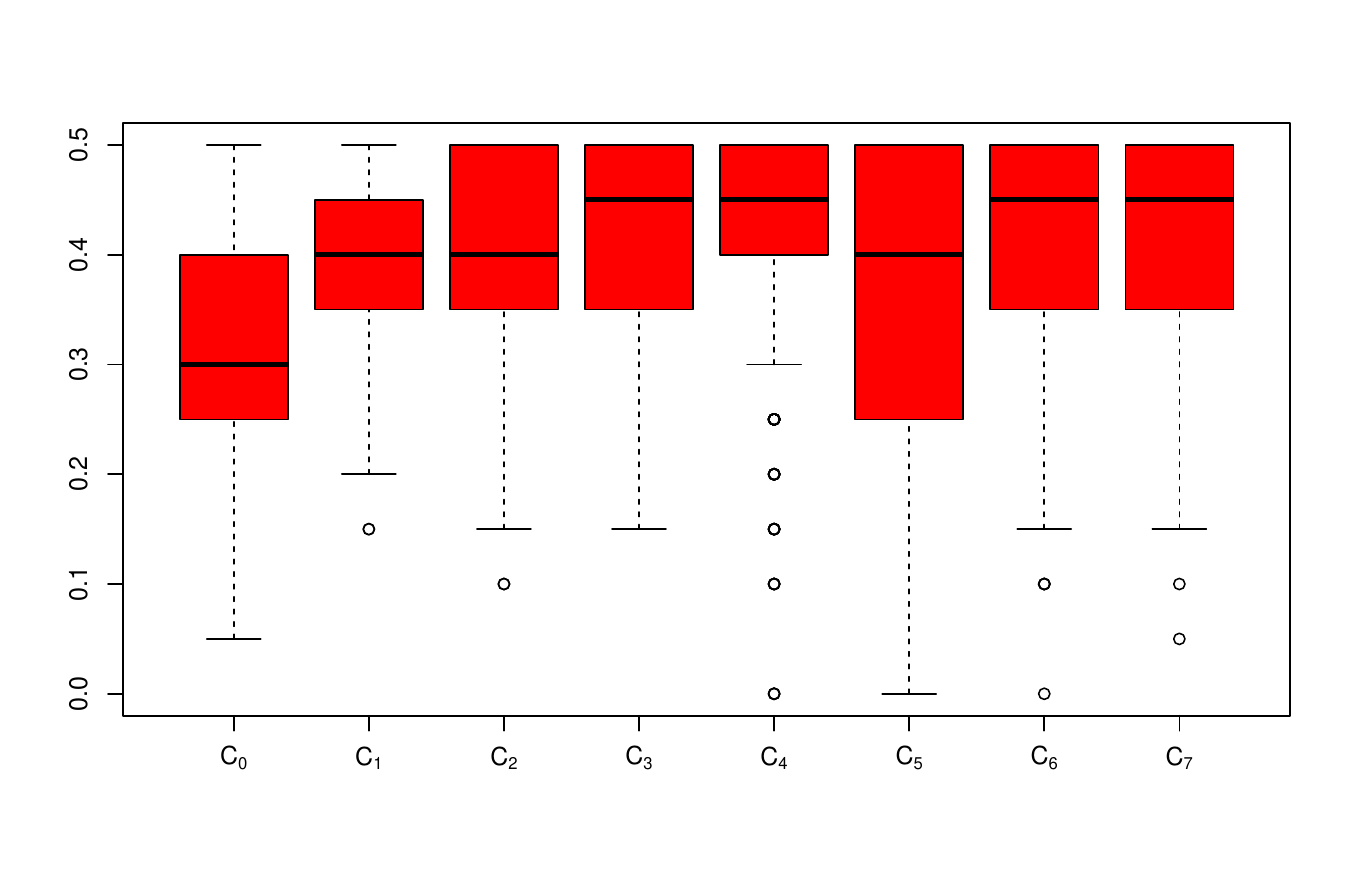} & & \\
\textsc{ROB} & \includegraphics[scale=0.6]{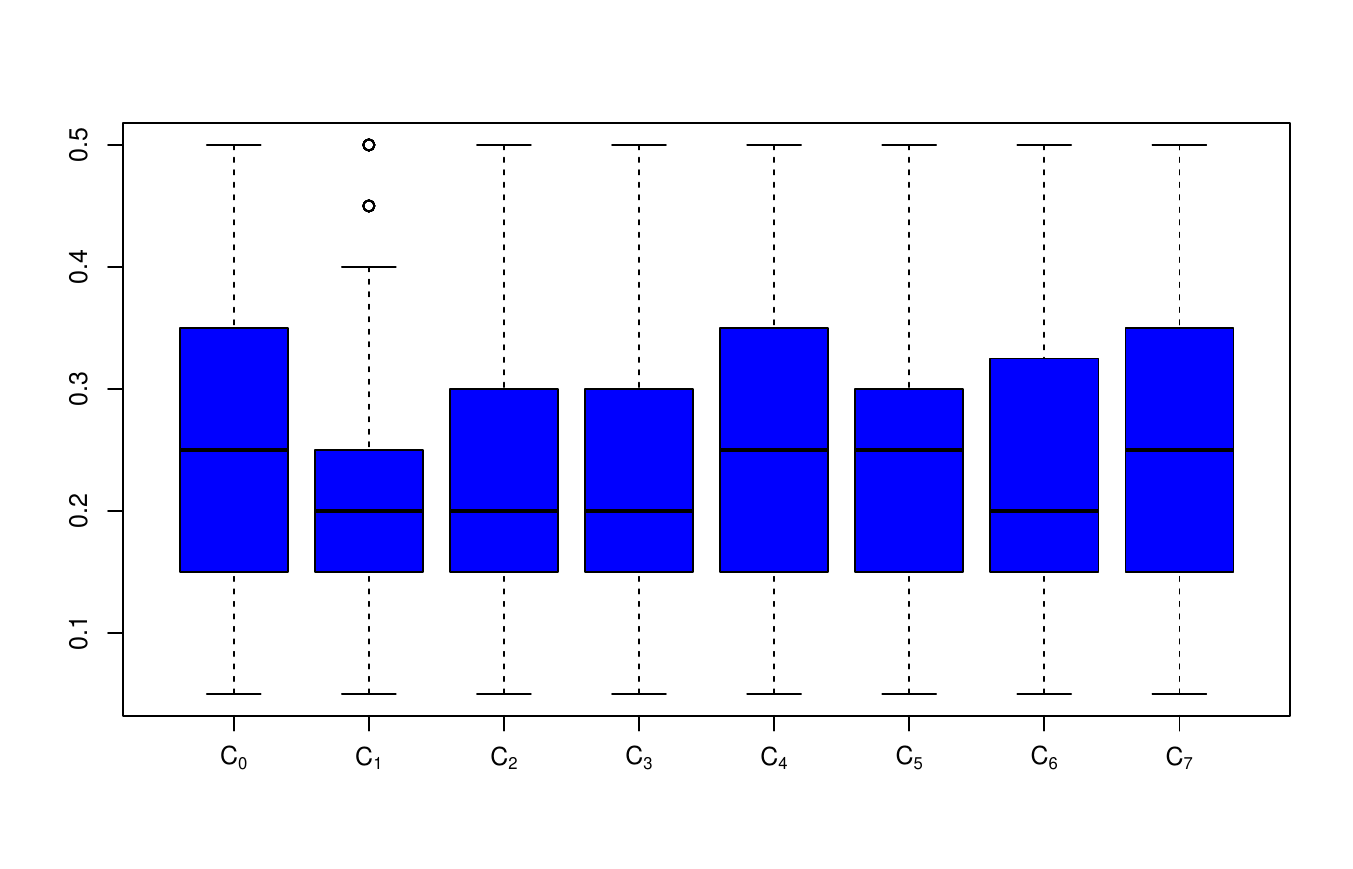} & &\\
\end{tabular}
\end{center}
\vspace{-1.4cm}
\caption{\label{fig:la2-n400} Boxplots of the $\wtlam_2$ selected under different contamination schemes for both least squares and robust estimators, when $n=400$.}
\end{figure}

\begin{figure}[htbp]
\begin{center}
\small
	 \renewcommand{\arraystretch}{-0.5} %0.4
 \newcolumntype{M}{>{\centering\arraybackslash}m{\dimexpr.1\linewidth-1\tabcolsep}}
   \newcolumntype{G}{>{\centering\arraybackslash}m{\dimexpr.2\linewidth-1\tabcolsep}}
\begin{tabular}{M G G G}
\textsc{LS} & \includegraphics[scale=0.6]{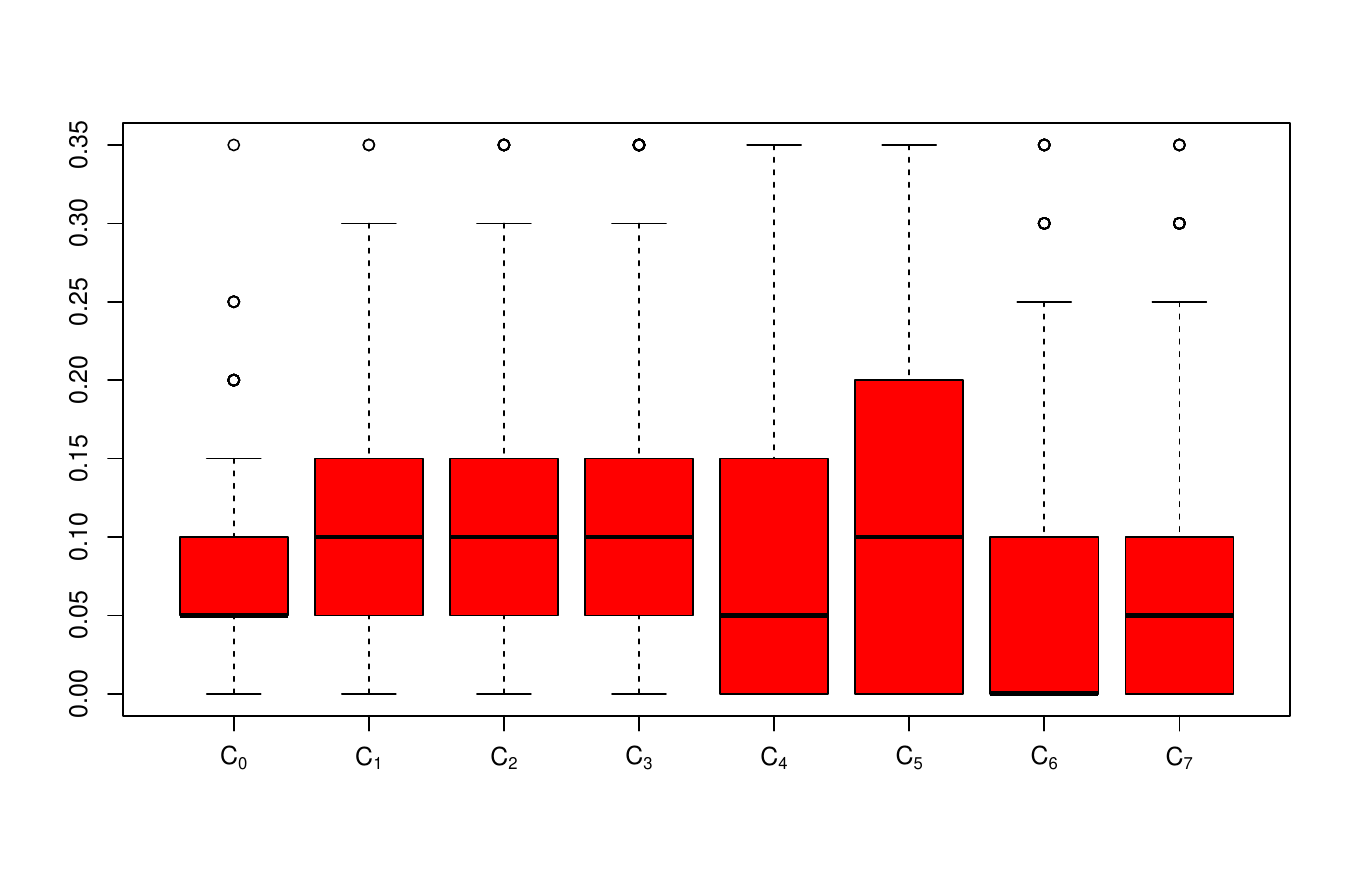} & & \\
\textsc{ROB} & \includegraphics[scale=0.6]{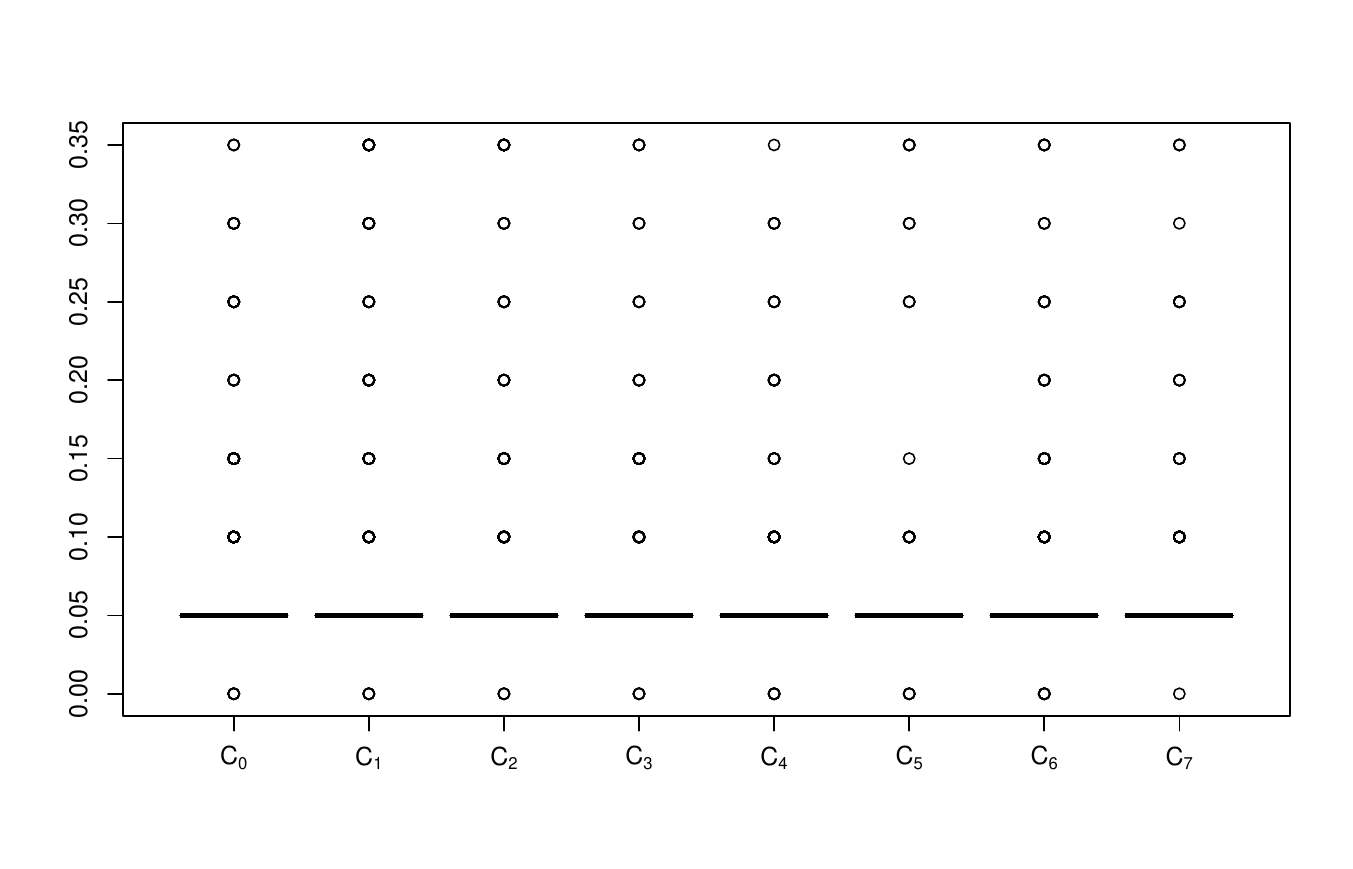} & &\\
\end{tabular}
\end{center}
\vspace{-1.4cm}
\caption{\label{fig:la1-n600} Boxplots of the $\wtlam_1$ selected under different contamination schemes for both least squares and robust estimators, when $n=600$.}
\end{figure}

\begin{figure}[htbp]
\begin{center}
\small
	 \renewcommand{\arraystretch}{-0.5} %0.4
 \newcolumntype{M}{>{\centering\arraybackslash}m{\dimexpr.1\linewidth-1\tabcolsep}}
   \newcolumntype{G}{>{\centering\arraybackslash}m{\dimexpr.2\linewidth-1\tabcolsep}}
\begin{tabular}{M G G G}
\textsc{LS} & \includegraphics[scale=0.6]{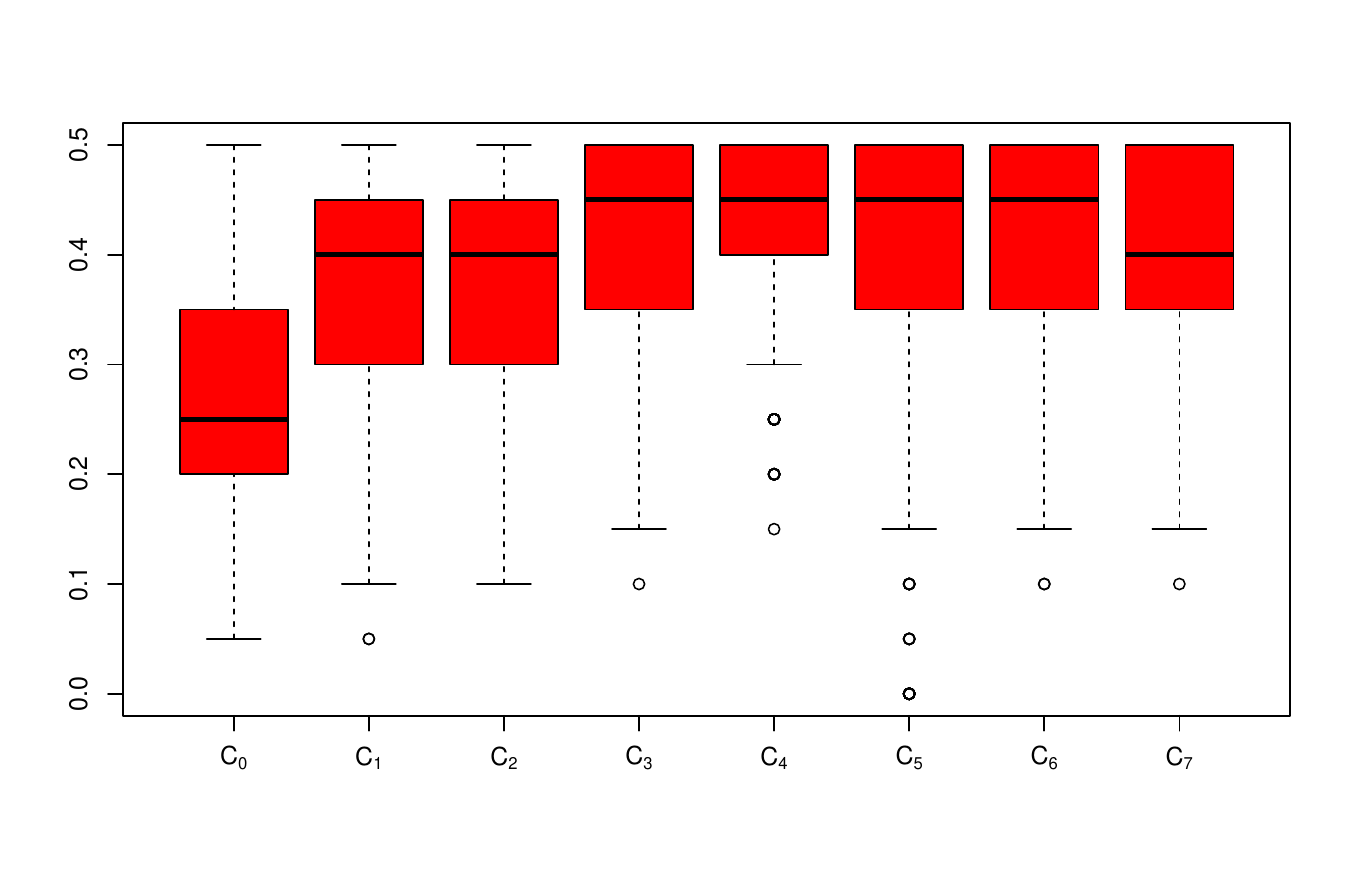} & & \\
\textsc{ROB} & \includegraphics[scale=0.6]{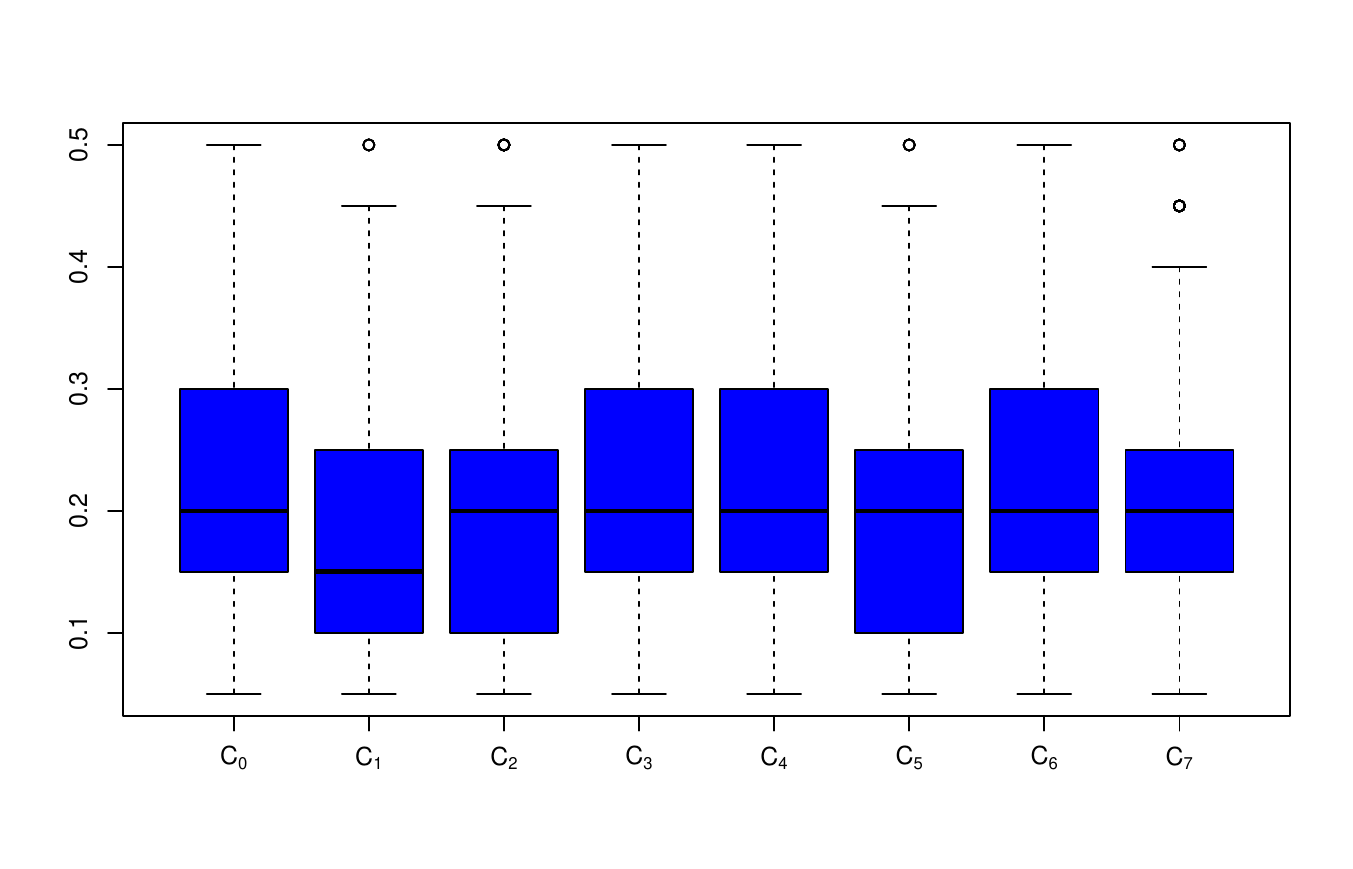} & &\\
\end{tabular}
\end{center}
\vspace{-1.4cm}
\caption{\label{fig:la2-n600} Boxplots of the $\wtlam_2$ selected under different contamination schemes for both least squares and robust estimators, when $n=600$.}
\end{figure}

\end{landscape}

\section{A real data example}\label{sec:realdata}

%An analysis of a real data set is carried on in this section where it is a considered  the well-known plasma beta-carotene level data set, collected by \citet{Nierenberg:etal:1989}. Its study is an important issue since it has been shown a direct relationship between beta-carotene and certain types of cancer \citep[see][]{fairfield:fletcher:2002}. This data set consists of 315 observations and 14 variables and it is available in {\tt{R}} as the set {\tt{plasma}} of the library {\tt{gamlss.data}}. 
%It has been widely studied by several authors specially in the context of semiparametric models. For instance, \citet{zhao:etal:2014} considered a partially linear varying coefficient model while \citet{liu:etal:2011}, \citet{Banerjee:Ghosal:2014}, \citet{jeong:etal:2022}, \citet{Guo:etal:2013}, and \citet{lv:yang:guo:2017} considered a partially linear additive model. However, there are some differences among their approaches that we are going to explain later.

In this section, we analyze a real dataset: the well-known plasma beta-carotene level dataset collected by \citet{Nierenberg:etal:1989}. This dataset is of significant interest as studies have shown a direct relationship between beta-carotene levels and certain types of cancer \citep[see][]{fairfield:fletcher:2002}. It contains 315 observations and 14 variables and is available in {\tt{R}} as the dataset {\tt{plasma}} within the library {\tt{gamlss.data}}.

The aim is to explore the relationship between plasma beta-carotene levels and various regulatory factors, including age, sex, dietary intake, smoking status, and alcohol consumption. The response variable, plasma beta-carotene (measured in ng/ml) and denoted as BETAPLASMA, will be modeled using the following covariates:
%It is of interest the relationship between the plasma beta-carotene level and some regulatory factors such as age, sex, dietary intake, smoking status and alcohol consumption. The response variable is the plasma beta-carotene in ng/ml, denoted BETAPLASMA, and will be modelled using the following:
\begin{itemize}
 \item[] AGE: age (years);
\item[] SEX: sex (1=male, 2=female);
\item[] SMOKSTAT: smoking status (1=never, 2=former, 3=current smoker);
\item[] QUET: quetelet (weight/(height)$^2$);
\item[] VITUSE: vitamin use (1=yes, fairly often, 2= yes, not often, 3=no);
\item[] CALORIES: number of calories consumed per day;
\item[] FAT: grams of fat consumed per day;
\item[] FIBER: grams of fiber consumed per day;
\item[] ALCOHOL: number of alcoholic drinks consumed per week;
\item[] CHOL: cholesterol consumed (mg/day);
\item[] BETADIET: dietary beta-carotene consumed (mcg/day).
%\item[] RETDIET: dietary retinol consumed (mcg/day);
%\item[] RETPLASMA: plasma retinol (ng/ml).
\end{itemize}
As it is pointed out by \citet{liu:etal:2011} and \citet{Guo:etal:2013}, there is one extremely high leverage point in ALCOHOL which, for obvious reasons, we will not remove. The categorical variables were divided into dummy variables before being included in the model. Thus, SMOKSTAT enters the model as SMOK1 (1=former smoker, 0=other) and SMOK2 (1=current smoker, 0=other) and VITUSE is separated into VIT1 (1=fairly often, 0=other) and VIT2 (1=not often, 0=other). Besides, all variables except binary ones are standarized using the \textsc{median} as the center and the \textsc{mad} as the deviation, assuming that some covariates may present high-leverage points.

This dataset has been widely studied by several authors, especially in the context of semiparametric models and, in particular, in partially linear additive models.
For instance, in \citet{liu:etal:2011}, an approach based on least squares is used to model the logarithm of BETAPLASMA with nine covariates: QUET, CALORIES, FAT, FIBER, BETADIET, SEX, ALCOHOL, SMOK1 and SMOK2 entering in the linear part, and AGE and CHOL in the additive part. The same model was considered in \citet{Banerjee:Ghosal:2014} using a Bayesian model selection method, while \citet{jeong:etal:2022} also used a Bayesian model selection approach but to study the specific effect of the CHOL covariate.
On the other hand, \citet{Guo:etal:2013} proposed a composite quantile regression approach with a regularization procedure to select variables from the linear component, modeling BETAPLASMA without transforming it. They included ten variables in the linear part: SEX, SMOK1, SMOK2, QUET, VIT1, VIT2, CALORIES, FAT, ALCOHOL and BETADIET, while AGE, CHOL and FIBER were included in the additive part. \citet{lv:yang:guo:2017} considered a modal regression approach to model BETAPLASMA with ten covariates entering in the linear part: SEX, SMOKSTAT, QUET, VITUSE, CALORIES, FAT, ALCOHOL, BETADIET and two other covariates named RETDIET and RETPLASMA, and the same three covariates entering in the additive component of the model: AGE, FIBER and CHOL. While SMOKSTAT was treated as a discrete variable in their analysis, converting it into dummy variables might align better with standard modeling practices. %\textcolor{red}{Te parece bien, Graciela, que diga esto? Intent\'e suavizarlo un poco pero la idea era decir que era muy extra\~no lo que hab\'{\i}an hecho con esa covariable.} 
%It is interesting to observe that SMOKSTAT is a categorical variable and so treat it like a discrete variable with no conversion into dummy variables it doesn't seem too appropiate. \textcolor{red}{Te parece bien, Graciela, que diga esto?} 
Finally, \citet{boente:martinez:2024} considered the same data set and proposed the same model as in \citet{Guo:etal:2013}, but with variable selection applied only tothe linear component of the model.

For this reason, we adopt the same model as \citet{Guo:etal:2013}, where AGE, CHOL and FIBER enter to the model in the additive part, while the remainging variables are considered to have a linear relationship with plasma beta-carotene levels. More precisely, the partially linear additive model considered is the following:
\begin{eqnarray*}
\mbox{BETAPLASMA} &=&\mu+\beta_1\mbox{SEX}+\beta_2\mbox{SMOK1}+\beta_3\mbox{SMOK2}+\beta_4\mbox{QUET}+\beta_5\mbox{VIT1}+\beta_6\mbox{VIT2}\\
&&+\beta_7\mbox{CALORIES}+\beta_8\mbox{FAT}+\beta_9\mbox{ALCOHOL}+\beta_{10}\mbox{BETADIET}+\eta_1(\mbox{AGE})\\
&&+\eta_2(\mbox{CHOL})+\eta_3(\mbox{FIBER})+ \sigma \,\varepsilon,
\end{eqnarray*}
where the errors $ \varepsilon$ are assumed to be independent of the covariates, with symmetric distribution and scale 1. With the notation given in equation \eqref{plam}, $\bZ=(\mbox{SEX}, \mbox{SMOK1}, \mbox{SMOK2}, \mbox{QUET}, \mbox{VIT1}, \mbox{VIT2}, \linebreak \mbox{CALORIES}, \mbox{FAT}, \mbox{ALCOHOL}, \mbox{BETADIET})\trasp\in\real^{10}$
and $\bX=(X_1,X_2, X_3)\trasp=(\mbox{AGE}, \mbox{CHOL}, \linebreak \mbox{FIBER})\trasp\in\real^3$.

For the analysis, two penalized estimators are computed: the robust estimator and the one based on least squares, in a similar way to what has been done in the simulation study. For the robust approach, it is used the Tukey's bisquare loss function as $\rho-$function with tuning constant $c=4.685$, while for the LS$-$estimator, the quadratic function $\rho(t)=t^2$. Cubic $B-$splines are used to estimate the additive components. When estimating $\eta_j$, the internal knots are taken as the $\ell/(k + 1) 100$\% quantiles, $\ell= 1,\dots,k$, of the observed values of the covariate $X_j$. The number of internal knots is the same for all three covariates included in the nonparametric part of the model, i.e. $k_j=k$ for $j=1, 2, 3$. The regularization parameter $k$ and the auxiliary penalty parameters $\wtlam_1$ and $\wtlam_2$ are automatically selected using the $\mbox{RBIC}_{\bk}$ and $\mbox{RBIC}_{\blach}$ criteria, respectively, as described in this work. For the robust estimators, the Tukey's bisquare loss function is used, while for the LS$-$estimator it is used $\rho(t)=t^2$. The grid considered for $\wtbla=(\wtlam_1,\wtlam_2)$ is $\itG=\left\{(\ell_1,\ell_2)\,:\, \ell_1\in\{0; 0.01; 0.02; \dots; 0.1\}, \ell_2\in\{0; 0.1; 0.2; \dots; 2\}\right\}$, as this grid is used for both robust and least-squares approaches.

In order to compare the behaviour of the estimators, 100 observations are randomly selected for prediction purposes, while the remaining 215 observations are used for estimation and variable selection. This procedure is repeated 50 times.
For each iteration, the prediction performance of an estimator is measured through the median absolute prediction error (\textsc{mape}), calculated as the median of $\{|y_\ell-\widehat{y}_\ell|\,,\, \ell=1,\dots,100\}$ where $(y_\ell, \bz_\ell\trasp, \bx_\ell\trasp)\trasp$, $1\leq \ell\leq 100$ represents the observations reserved for prediction and $\widehat{y}_\ell$ is the predicted value at $(\bz_\ell\trasp, \bx_\ell\trasp)\trasp$. The final number of this error is calculated as the mean of the 50 \textsc{mape} values. To evaluate the variable selection procedure, for each iteration and each estimator (robust and least-squares), the number of covariates identified as important and included in the model is recorded. These counts are averaged across the 50 replications to obtain the final number, named \textsc{av.size}.  
%it is calculated the number of covariates selected as important and that should be included in the model. Then, the 50 replications are averaged to obtain the final number, named \textsc{av.size}.

All these measures are calculated for both the penalized robust approach and its least squares counterpart (named \textsc{penalized}). Additionally, unpenalized estimators that do not select variables are computed. These correspond to the estimators defined in \citet{boente:martinez:2023} for the robust approach and to the least squares approach when using $\rho(t)=t^2$ as $\rho-$function.
Note that, since unpenalized estimators do not perform variable selection, their \textsc{av.size} are $13$.

Table \ref{tab:plasma-fig1} summarizes the mean of the \textsc{mape} values obtained over the 50 replications together with the average number of covariates selected (\textsc{av.size}). Although the unpenalized robust estimator achieves the lowest \textsc{mape} value, the robust approach that selects variables slightly increases the prediction error while reducing the average number of selected variables to 5.32, compared to the 13 variables included in the model by the unpenalized approach. On the other hand, both LS$-$estimators yield higher \textsc{mape} values than their robust counterparts. Moreover, for the penalized LS$-$estimator, the average size of the selected model is 7.54, which is notably larger than that of penalized robust approach.

\begin{figure}[ht!]
\begin{center}
\includegraphics[scale=0.45]{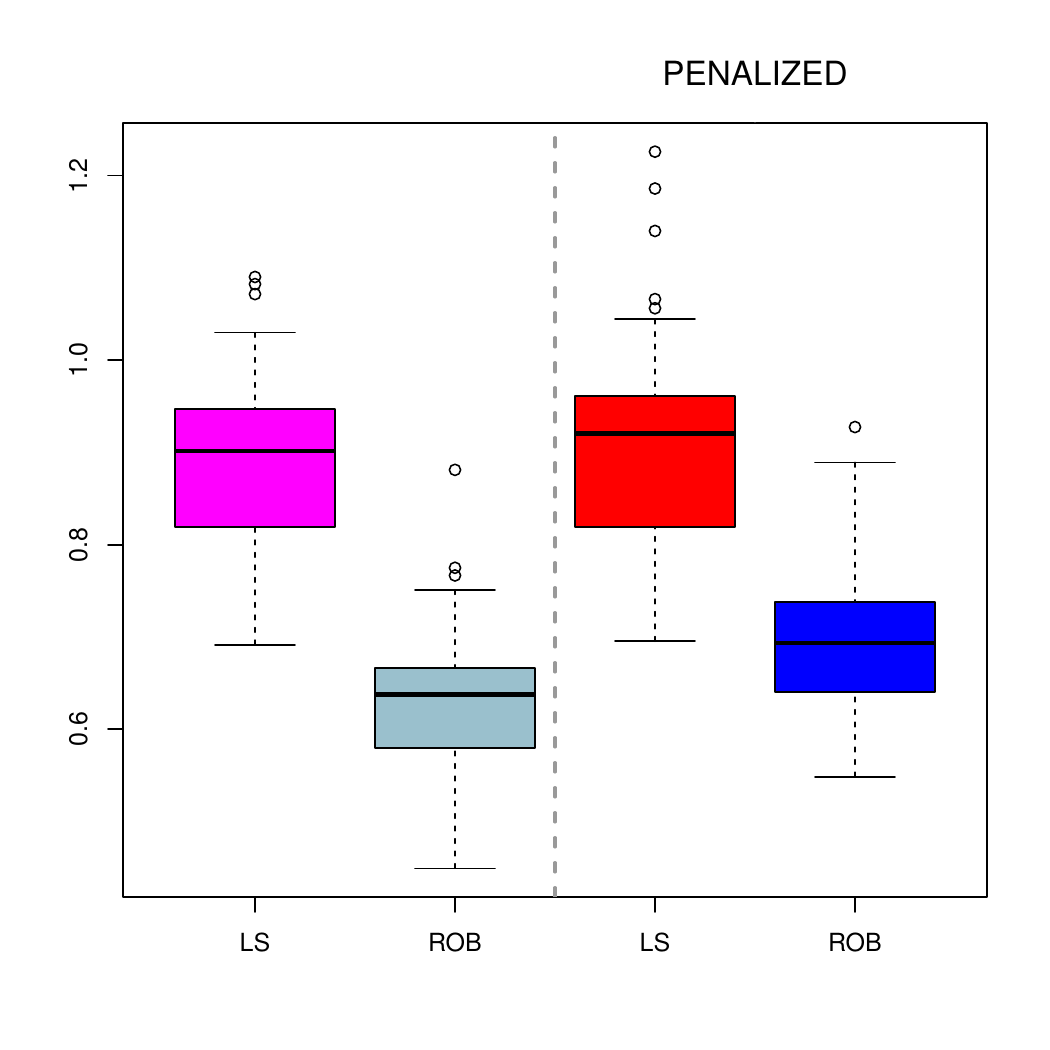}
\vskip-0.2in
\caption{\label{fig:MAPE-measure} Adjusted boxplots for the \textsc{mape} values  obtained for the unpenalized estimators (on the left) and for the penalized ones (on the right).}
\end{center}
\end{figure}

For better visualization, Figure \ref{fig:MAPE-measure} displays adjusted boxplots  for the 50 \textsc{mape} values obtained from each of the four estimators. The two boxplots on the left correspond to the unpenalized estimators (those that do not perform variable selection), while the two on the right correspond to the penalized estimators that incorporate regularization. It can be appreciated that the boxplots for the robust approaches are consistenly lower than those of the least squares estimators. For instance, more than 75\% of the \textsc{mape} values obtained by the robust estimators are smaller that at least 75\% of the values obtained by their LS$-$counterparts. Additionally, as previously noted, the adjusted boxplot for the penalized robust estimator is spligthly higher than that of its unpenalized competitor.

\begin{table}[ht!]
\begin{center}
\scriptsize
\begin{tabular}{|l|c|c|c|}
  \hline
\textsc{method} &  \textsc{mape} & \textsc{av.size} \\ 
  \hline
\textsc{penalized ls}   & 0.9113  &  7.54\\
\textsc{penalized rob}  & 0.6987  &  5.32\\
\textsc{ls} & 0.8920 &  13\\
\textsc{rob}  &  0.6365 & 13\\\hline
\textsc{penalized ls$^{(-\textsc{out})}$} & 0.6598 & 5.96\\
   \hline
\end{tabular}
\caption{\label{tab:plasma-fig1}\footnotesize \textsc{mape} and \textsc{av.size} for the penalized least squares and robust approaches, for the unpenalized least squares and robust estimators in the first two rows and, in the last row, for the LS$-$estimator with the atypical observations removed.} 
\end{center}
\end{table}

To determine which variables are more frequently selected by each method, Tables \ref{tab:plasma-fig2} and \ref{tab:plasma-fig3} present the frequencies with which each variable is identified as non-zero. Table \ref{tab:plasma-fig2} corresponds to the ten variables in the parametric component, while Table \ref{tab:plasma-fig3} focuses on the three covariates in the nonparametric part. By considering a threshold of $0.5$, the least squares estimator selects the variables SEX, SMOK2, QUET, VIT1, VIT2, FAT and BETADIET from the linear component, as well as FIBER from the additive component. In contrast, the robust approach selects SEX, SMOK2, QUET, VIT1 and BETADIET from the linear component, and none from the additive component.

%\textcolor{red}{Vos en esta tabla (Tabla \ref{tab:plasma-fig1}) me pregunt\'as: Me asombra que no d\'e mas chico. Si haces un boxplot de los residuos detectas outliers? Cuantos? Respuesta: Como esto lo hago 100 veces, por cada vez que lo hago detecta una cantidad de outliers: en promedio detecta 5.55 y su mediana es 5. Como tengo 215 datos cada vez, 5 de 215 representa un 2.3\% de los datos. Adem\'as, te pego el histograma de cantidad de outliers detectados, por si ves algo raro.}

\begin{table}[ht!]
\begin{center}
\scriptsize
\begin{tabular}{|l|c|c|c|c|c|c|c|c|c|c|}
  \hline
\textsc{penalized} & \mbox{SEX} & \mbox{SMOK1} & \mbox{SMOK2} & \mbox{QUET} & \mbox{VIT1} & \mbox{VIT2} & \mbox{CALORIES} & \mbox{FAT} & \mbox{ALCOHOL} & \mbox{BETADIET}\\ 
  \hline
  \textsc{ls}      & \textbf{0.78} & 0.22 & \textbf{0.90} & \textbf{1.00} & \textbf{1.00} & \textbf{0.80} & 0.32 & \textbf{0.56} & 0.14 & \textbf{0.94} \\
\textsc{rob}     & \textbf{0.66} & 0.40 & \textbf{0.62} & \textbf{0.88} & \textbf{0.54} & 0.42 & 0.32 & 0.24 & 0.02 & \textbf{0.72} \\\hline
\textsc{ls$^{(-\textsc{out})}$} & \textbf{0.88} & 0.34 & \textbf{0.88} & \textbf{1.00} & \textbf{0.76} & \textbf{0.52} & 0.16 & 0.38 & 0.00 & \textbf{0.98}\\ 
   \hline
\end{tabular}
\caption{\label{tab:plasma-fig2} Selected frequencies as non-zero for the ten covariates of the parametric component, for the least squares approach in the first row, the robust proposal in the second row and, in the last row, the least squares approach computed without the atypical observations.} 
\end{center}
\end{table}

\begin{table}[ht!]
\begin{center}
\scriptsize
\begin{tabular}{|l|c|c|c|}
  \hline
\textsc{penalized} & \mbox{AGE} & \mbox{CHOL} & \mbox{FIBER}\\ 
  \hline
  \textsc{ls}    & 0.24 & 0.04 & \textbf{0.60}  \\
\textsc{rob}     & 0.06 & 0.04 & 0.40\\\hline
\textsc{ls$^{(-\textsc{out})}$} & 0.02 & 0.02 & 0.02\\
   \hline
\end{tabular}
\caption{\label{tab:plasma-fig3} Selected frequencies as non-zero for the three covariates of the additive component, for the least squares approach in the first row, the robust proposal in the second row and, in the last row, the least squares approach computed without the atypical observations.} 
\end{center}
\end{table}

\begin{figure}[ht!]
\begin{center}
\includegraphics[scale=0.35]{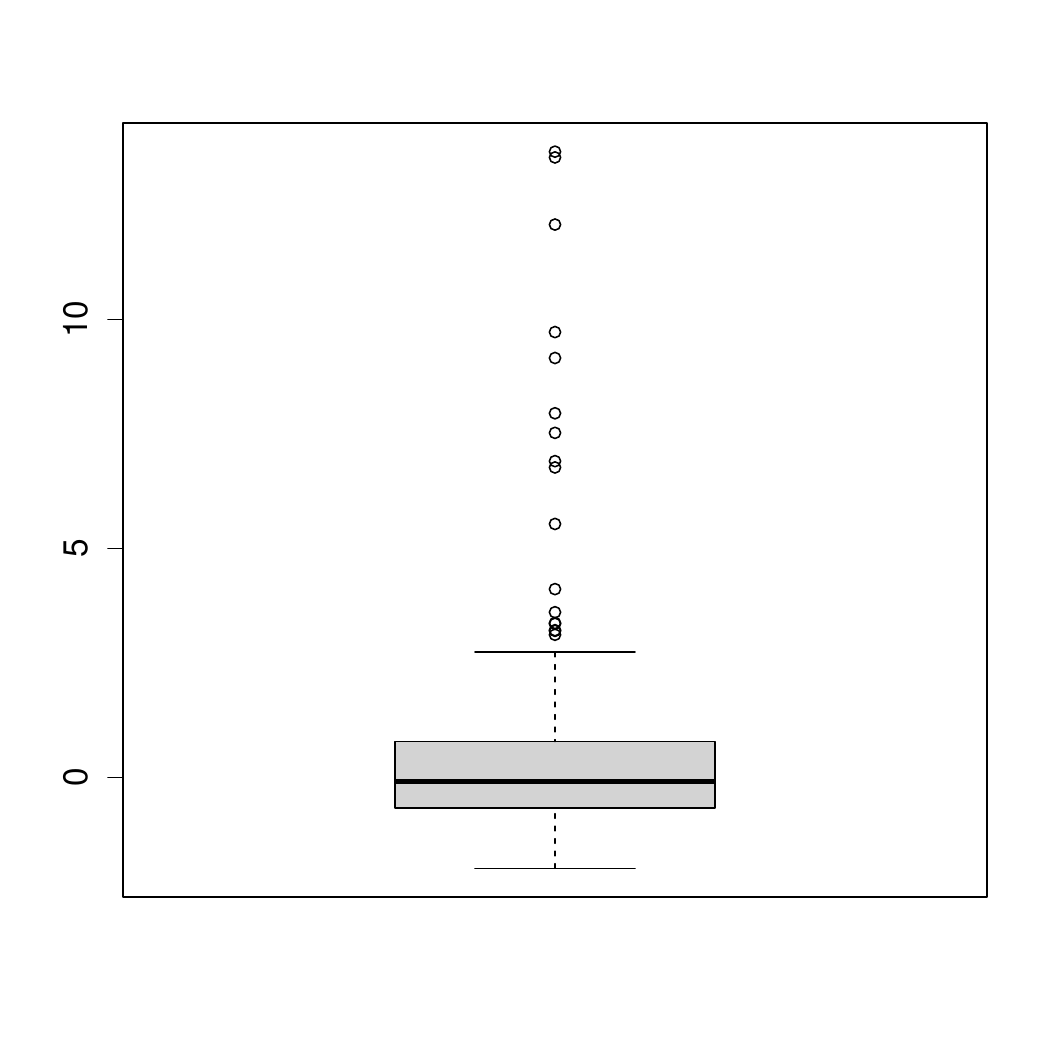}
\vskip-0.2in
\caption{\label{fig:boxplot-resid} Boxplot of the residuals obtained when fitting the complete data set using the penalized robust estimator.}
\end{center}
\end{figure}

Since outlying observations may influence the estimator based on least squares, resulting in higher \textsc{mape} values, the penalized robust approach is computed on the whole dataset to identify large residuals. The boxplot of residuals obtained using the robust method, shown in Figure \ref{fig:boxplot-resid}, reveals the presence of 17 vertical outliers, that is, 17 observations detected with large residuals.

\begin{figure}[ht!]
\begin{center}
\includegraphics[scale=0.45]{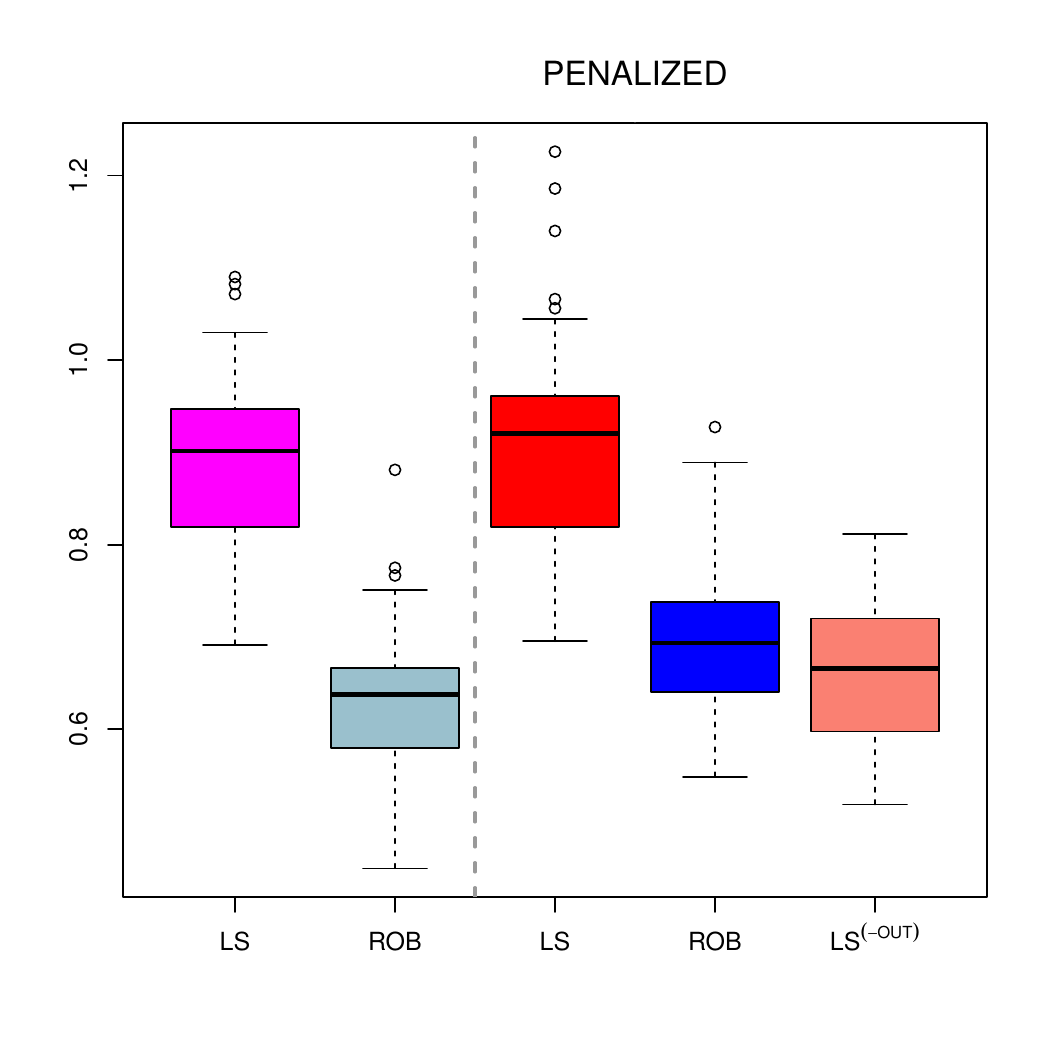}
\vskip-0.2in
\caption{\label{fig:MAPE-measure-lostres} Adjusted boxplots for the  \textsc{mape} measures obtained for the estimators without penalization (on the left) and for the penalized (on the right) estimators. The one corresponding to the penalized least squares estimator computed without the outliers is labelled \textsc{ls$^{(-\textsc{out})}$}.}
\end{center}
\end{figure}

Subsequently, the procedure is repeated for the penalized LS$-$estimator after excluding the 17 vertical outliers and also the high leverage observation in ALCOHOL identified by \citet{liu:etal:2011} and \citet{Guo:etal:2013}. That is, for each of the 50 iterations, out of a total of 297 observations, 100 are randomly excluded for testing, while the 197 remaining are used for training the estimator. The \textsc{mape} and \textsc{av.size} values are then calculated. The results, shown in the last row of Table \ref{tab:plasma-fig1} (labeled with an upperscript $(-\textsc{out})$), show the average \textsc{mape} and model size values. Figure \ref{fig:MAPE-measure-lostres} shows the adjusted boxplots of the \textsc{mape} values obtained by the same four estimators as before but with the recently computed LS$-$based estimator with fewer observations. 

As it can be appreciated in Table \ref{tab:plasma-fig1}, the averaged \textsc{mape} and \textsc{av.size} for this estimator are now quite similar to those obtained by the penalized robust approach.
Additionally, Figure \ref{fig:MAPE-measure-lostres} displays adjusted boxplots for the \textsc{mape} values of the same four estimators as before, now including the LS$-$estimator computed on the reduced sample, which is in fact quite similar to those obtained by the robust estimators. Finally, last rows of Tables \ref{tab:plasma-fig2} and \ref{tab:plasma-fig3} indicate that the proportion of times each covariate is selected by this reduced-sample LS$-$estimator aligns closely with the selection frequencies of the robust method. Except for VIT2, both the robust approach and the LS$-$estimator on the reduced sample select the same covariates.

\section{Concluding remarks}{\label{sec:comments}}

Partially linear additive models (\textsc{plam}) are a suitable choice for modeling a response variable using a set of covariates that may have either a linear or  nonparametric relationship with the response. Even when the relationship of a covariate is unknown, the linear part of the model allows for the inclusion of a certain type of covariates, such as categorical or discrete ones, which cannot be easily accommodated by fully nonparametric models.  

In practice, practitioners typically include all available variables in the model, assiging them to the linear or additive components based on their expertise or the intrinsic nature of the covariates. However, including too many covariates can lead to issues such as lack of parsimonity or reduced predicted performance. Variable selection, therefore, becomes crucial, and much research has focused on it, especially on selecting variables from the linear part of the model. However, the excess of covariates can arise not only in linear component but also in the additive part of a \textsc{plam}, making this an interesting and relatively unexplored are of investigation.

On the other hand, since atypical data may arise, robust procedures are needed not only for accurately estimating the relationship between the response variable and the covariates but also for correctly identifying and selecting significant variables while discarding irrelevant ones.

To address these challenges, this paper introduces a robust method for estimating and selecting variables in sparse partially linear additive models. The proposal includes automatic selection of regularization and penalty parameters, achieved through generalized BIC criteria. A general algorithm is developed to compute the proposed estimator, with the option to incorporate an adaptive procedure for penalty parameters. A Monte Carlo study, conducted with three sample sizes and eight contamination scenarios (including a non-contaminated case), is used to compare the robust proposal with its least squares counterpart. Finally, the approach is evaluated using the well-known plasma beta-carotene level dataset. The results from both the simulation study and the real-data analysis highlight the advantages of the robust proposal, particularly in the presence of atypical data.

\noi\textbf{\small Acknowledgements.} {\small  This research was partially supported by  grants Proyecto Interno CD-CBLUJ 204/19 from the Departamento de Ciencias B\'asicas, Universidad Nacional de Luj\'an, PICTO-2021-UNLU-00016 from \textsc{anpcyt} and Universidad Nacional de Luj\'an, and PICT-2021-I-A-00260 from \textsc{anpcyt}, Argentina.} %\textcolor{red}{Graciela, me parece que debo tener algo desactualizado. Puede ser que no est\'e en ning\'un proyecto tuyo?}  

\vspace{0.5cm}

%%%%%%%%%%%%%%%%%%%%%%%%%%%%%
\small
%\nocite{*}

%\nocite{*}
\bibliographystyle{apalikeesp}
 
\bibliography{referencias2}

\end{document}